\documentclass[12pt,a4paper]{article}
\usepackage[utf8]{inputenc}
\usepackage[english]{babel}
\usepackage{amsmath}
\usepackage{amsfonts}
\usepackage{amssymb}
\usepackage{dsfont}
\usepackage[all,cmtip]{xy}
\usepackage{makeidx}
\usepackage{authblk}
\usepackage{bm}
\usepackage{graphicx}
\usepackage[left=1cm,right=1cm,top=3cm,bottom=3cm]{geometry}
\author[@]{David T. Heider\thanks{david.heider@tum.de}}
\author[@]{J. Leo van Hemmen\thanks{lvh@tum.de}}
\affil[@]{Physik Department T35 and Department of Mathematics, Technische Universit\"{a}t M\"{u}nchen, 85747 Garching bei M\"{u}nchen, Germany}

\title{Geometric Perturbation Theory for a Class of Fiber Bundles}

\begin{document}
\maketitle

\begin{abstract}
A systematic study of small, time-dependent, perturbations to geometric wave-equation domains is hardly existent. Acoustic enclosures are typical examples featuring locally reacting surfaces that respond to a pressure gradient or a pressure difference, alter the enclosure's volume and, hence, the boundary conditions, and do so locally through their vibrations. Accordingly, the Laplace-Beltrami operator in the acoustic wave equation lives in a temporally varying domain depending on the displacement of the locally reacting surface from equilibrium. The resulting partial differential equations feature nonlinearities and are coupled though the time-dependent boundary conditions. The solution to the afore-mentioned problem, as presented here, integrates techniques from differential geometry, functional analysis, and physics. The appropriate space is shown to be a (perturbation) fiber bundle. In the context of a systematic perturbation theory, the solution to the dynamical problem is obtained from a combination of semigroup techniques for operator evolution equations and metric perturbation theory as used in AdS/CFT. Duhamel's principle then yields a time-dependent perturbation theory, called geometric perturbation theory. It is analogous to, though different from, both Dirac's time-dependent perturbation theory and the Magnus expansion. Specifically, the formalism demonstrates that the stationary-domain approximation for vibrational acoustics only introduces a small error. Analytic simplifications methods are derived in the framework of the piston approximation. Globally reacting surfaces (so-called pistons) replace the formerly locally reacting surfaces and reduce the number of independent variables in the underlying partial differential equations. In this way, a straightforwardly applicable formalism is derived for scalar wave equations on time-varying domains. 	
\end{abstract}

\section{Introduction}
\label{intro}

The motivation of the present paper stems from a concrete and rather typical example originating from acoustics. The mathematical algorithm leading to high-precision perturbative solutions has been explained in details elsewhere \cite{david1} and is essential to mathematically understanding azimuthal sound localization in more than half of the terrestrial vertebrates \cite{ice-editorial}. What we do here is providing the foundations in terms of fiber-bundle theory. 

\begin{figure}
\centering
\includegraphics[width=.99\linewidth]{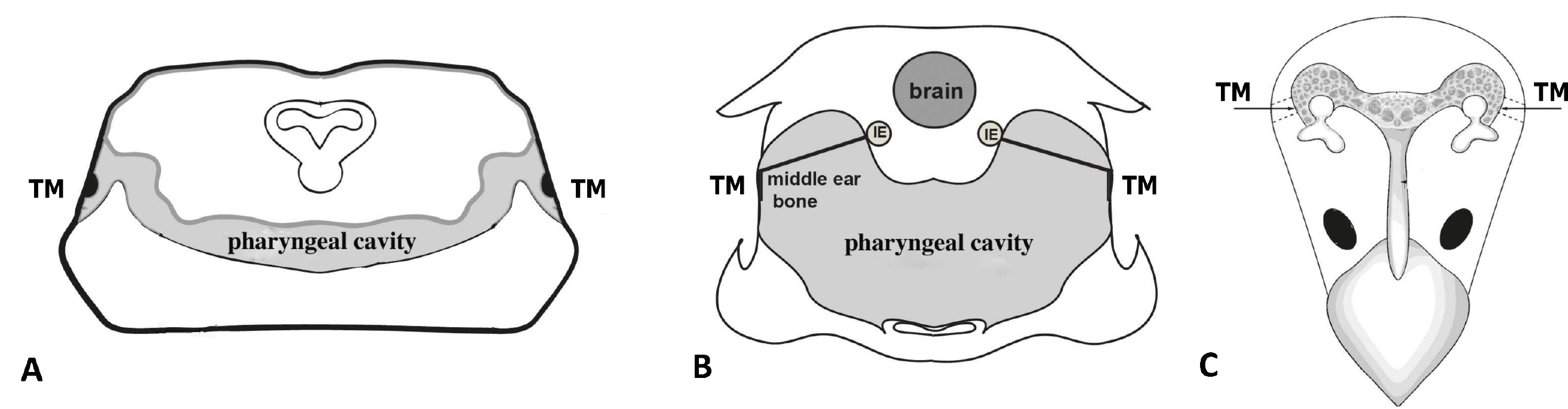}
\caption{Schematic representation of Internally Coupled Ears (ICE) in frogs (a), lizards (b), and birds (c). The bird in (c) is seen from the top, the other two in (a) and (b) show a cross section, and all three exhibit the interaural cavity as a gray tube. 
Figures 1 (a)--(c) have been taken from van Hemmen et al. \cite{ice-editorial}}
\label{iTD}
\end{figure}

As shown by Figure~\ref{iTD}, both frogs and lizards and birds and, hence, more than half of the terrestrial vertebrates have left and right eardrums that are connected by an air-filled, interaural, cavity in between. This setup realizes the notion of internally coupled ears (ICE). Land-living vertebrates perform azimuthal sound by neuronally determining the time difference between left and right eardrum and what they measure at the eardrums is a key to understanding the ensuing auditory processing. 

Let us assume that the $x$-axis is through the center of and orthogonal to the two parallel eardrums and that the latter are positioned at $x=0$ and $x=L$.  An external sound source, which is usually far (enough) away from the two ears and depending on the time $t$, generates a time-varying pressure, which is uniform at the eardrums and represented by $p_{\rm ext}(0,t)$ and $p_{\rm ext}(L,t)$. The latter bring the eardrums into motion. Sound, loud as it may seem, leads to eardrum motion with extremely small amplitudes (nm). On the other hand, $L$ being of the order of cm, the perturbed dynamics seems, and is \cite{david1},  accessible to perturbation theory. 

To localize a sound source, the auditory system uses the time difference between left and right eardrum. Given the interaural distance $L$, this time difference equals $L \sin (\theta)$ where $\theta=0$ is straight ahead. In passing we note that there is a degeneracy $L \sin (\pi/2 \pm \epsilon)$ but that is a universal problem and not the issue here.  

Because of the internal cavity with pressure $p$, which effectively couples the two eardrums,  the actual force on each of them is in fact the difference $\Psi(x=0/L,t) \equiv [p (x=0/L,t) - p_{\rm ext}(0/L,t)]$. The resulting dynamical system consists in the present case of a system of three, coupled, partial differential equations,
\begin{align}
\dfrac{\partial^2 p}{\partial t^2}-c^2\Delta_t p &=0 \ , 
\label{3DW} \\
\dfrac{\partial^2 u_{0/L}}{\partial t^2}+2\alpha\dfrac{\partial u_{0/L}}{\partial t}-c^2_m\Delta_2 u_{0/L}&= \pm \left. ( p_{\rm{ex}}-p) \right\vert_{x=u(t,\Gamma_{0/L})} \ .
\label{2DW}
\end{align}
The former is the wave equation for $p$ in the three-dimensional interaural cavity, which up to the two fluctuating eardrums is fixed. The latter two equations labeled by $x=0/L$ refer to the two-dimensional damped wave equations for left and right eardrums with Laplacian $\Delta_2$, deviations $u_{0/L}$ parallel to the $x$-axis, damping constant $\alpha > 0$, $c_m$ as material constant characterizing the tympanic membranes, and $\pm$ stands for a $-$ at $x=0$ and a $+$ at $x=L$. The other material constants can be found elsewhere \cite{david1}. The eardrums constitute through $u_{0/L}(t)$ the time-dependent part of the two-dimensional manifolds $\Gamma_{0/L}(t)$ that are the boundary and, hence, provide the time-dependent boundary conditions for the cavity's three-dimensional Laplacian $\Delta_t$, where $t$ succinctly indicates this time dependence. The nature of $\Delta_t$ will soon be analyzed in detail. 

As for the boundary conditions themselves, they are no-slip \cite[chapter 2]{Temkin}, which means that the velocity of the membrane equals that of the air attached to it. We start here with the (linearized) Navier-Stokes equation without viscosity ($\mu =0$), which is Euler,  and require 
\begin{align}
\label{boundcon}
\partial_t(\hat{\mathbf{n}}\mathbf{v})=-\dfrac{(\hat{\mathbf{n}}\nabla)p}{\rho_0}\text{ on }\partial\Omega(t).
\end{align}
Here $\rho_0$ is the density of air and $\hat{\mathbf{n}}$ denotes the outward unit normal vector to the boundary $\partial\Omega(t)$ of the cavity $\Omega(t)$ with $\Gamma_{0/L}(t) \subset \Omega(t)$ and $\mathbf{v}$ as the acoustic fluid velocity, viz., air, so that $\hat{\mathbf{n}}\mathbf{v}$ is the normal component of $\mathbf{v}$. Because of the no-slip boundary condition \cite[chapter 2]{Temkin}, we can take the normal velocity $\hat{\mathbf{n}}\mathbf{v}$ at both eardrums equal to $\partial u_{0/L}/\partial t$. That is,   $\hat{\mathbf{n}}\mathbf{v} = \partial u_{0/L}/\partial t$ in our concrete example. Except for the possibly swinging eardrums, we have fixed boundaries for (\ref{3DW}) so that there we are left with Neumann boundary conditions as $\mathbf{v} =0$. Since the borders of the eardrums are taken to be fixed, the membrane deflections $u_{0/L}$ in (\ref{2DW}) satisfy Dirichlet boundary conditions on the borders of what we take here to be circular sectors: $u_{0/L}=0$ on $\partial\Gamma_{0/L}$. 

Sloppily formulated, by putting the deflection $u_{0/L} \equiv 0$ as it was so small but keeping nonzero (\ref{boundcon}) on $\Gamma_{0/L}$, one arrives at the  acoustic boundary conditions introduced (ABC) by Beale and Rosencranz \cite{beale1,beale2}. What we do here is analyzing in full mathematical detail and generalizing the combination of (\ref{boundcon}) with the dynamics (D) incorporated by (\ref{3DW}) \& (\ref{2DW}). That is, we generalize the acoustic boundary-condition dynamics (ABCD) that has been introduced elsewhere \cite{david1} to far more general manifolds and in its mathematically natural context of fiber bundles.

\paragraph{First summary} The above arguments show the need for combining the acoustic boundary conditions (ABC) with their dynamics (D). That is, symbolically 
\begin{align*}
\boxed{\text{ABC}+\text{D}=\text{ABCD}}
\end{align*}
Here we present a formalism based on the combination of geometrical and analytic methods in the context of vibrational acoustics for models similar to the model of internally coupled ears. To this end, we first skim through the background of the approach that is going to appear. 

\paragraph{Our problem} Our problem can be described in non-technical manner as follows. Suppose, you are given a suitably well-behaved volume with flexible boundaries. Suppose further that there is a small demon inside the volume that enacts by emission of a pressure wave on each point of the boundary from the interior a force density, i.e., pressure, if and only if you enact from the exterior of the boundary a force density, i.e., pressure, on that point on the boundary by emission of a (plane) wave. The boundaries start to vibrate due to the net difference of force per unit area enacted on the boundary from the exterior and interior. The overall question is on how the pressure in the interior of the enclosed volume evolves in time assuming that only the vibrations of the boundaries trigger the demon to emit a pressure wave?

 In order to answer the above question while keeping physical applications in mind, we have to find answers to the following three formal questions. First, what is the appropriate geometrical and analytical formalism to account for the local, but non-global perturbations of the volume by deformation of its boundaries? Second, how can we quantify the smallness of those perturbations and relate them to solving the model equations in the unperturbed state of the system? Third, how can we account for time dependency of the perturbations in an operator-theoretic perturbation formalism. 
 
 Again from the viewpoint of applications in the sciences, it is desirable to find a \emph{geometrical} answer to the question of how we may handle weakly curved deformations of the boundaries. Namely, we ask under what conditions is it appropriate to neglect the local curvature properties of the vibrating boundaries with respect to the equilibrium boundaries and simply consider their normal displacement from the unperturbed state?

\paragraph{Strategy} Our strategy consist of three parts. A geometrical one for the geometrical setup, a physical one for finding the appropriate model, and an analytical one for finding a perturbation theory to solve the model equations.
\begin{itemize}
\item\textbf{Geometry} The first part is mostly of geometrical flavor spiced with a bit of topology. It concerns the question of what the appropriate formalism for the geometry of the problem is. This question has not been addressed in the vibrational-acoustics literature so far. We start from considering sub-manifolds of a surrounding Euclidean space as models for the volume under consideration. Specifically, we model the time-dependency of the problem by using smooth $1$-parameter families of such manifolds. 

One family is just the constant family $t\to\Omega_0$ mapping each point in time to he unperturbed volume. The other family is $t\to\Omega_t$ which maps the time $t$ to the deforming volume. The reason why we start from this geometrical construction is that the time-evolution of the manifolds, i.e., $t\to\Omega_t$ can be observed in experiments most easily. We then ask where a wave equation for the acoustic pressure lives geometrically. In the textbook literature on partial differential equations, the acoustic wave equation is usually considered to live on a product manifold $\mathbb{R}\times\Omega_0$ consisting of a base manifold $\mathbb{R}$ as a model for the time coordinate $t$ and a manifold $\Omega_0$ embedded in $\mathbb{R}^n$ which is the living space of the the spatial dependencies of the acoustic pressure, i.e., the solution to the acoustic wave equation. 

In a geometrical language, the living space of the solution is a product manifold which is an example of a fiber bundle. Likewise, we can endow under suitable constraints on the topology of $t\to\Omega_t$,namely that it is unaffected by the perturbations, the object $\mathcal{M}^\circ:=\bigcup_{t>0}\lbrace t\rbrace\times\Omega_t$ with a fiber bundle structure. Intuitively, a fiber bundle is just a manifold which admits locally a product structure of a base manifold, the time axis $\mathbb{R}^+$ in our setup and a fiber manifold $\Omega_t$ glued to each point $\lbrace t\rbrace\subset\mathbb{R}^+$ of the time axis. 

Geometrically, we then only need to relate $\mathcal{M}_0:=\mathbb{R}^+\times\Omega_0$, the stationary fiber bundle to the perturbation bundle $\mathcal{M}$. The requirement to be made is that the topology of $\mathcal{M}_0$ and $\mathcal{M}$ and of the fibers $t\to\Omega_0$ and $t\to\Omega_t$ are left invariant by the perturbations aids at relating by means of bundle and manifold diffeomorphisms $\mathcal{M}\simeq\mathcal{M}_0$ and $\Omega_0\simeq\Omega_t$ globally. 

Using the imbedding space for the bundles, we give using the topological constraints and the therefore global Gauss map for the fibers $t\to\Omega_0$ and $t\to\Omega_t$ the Lorentzian metric originating from restriction of the Lorentzian metric of the imbedding space $\mathbb{M}^{n+1}$, the $(n+1)$-dimensional Minkowskian space. Combing the metric and the diffeomorphism, we are able to reduce the problem to comparing the reference bundle $(\mathcal{M}_0,G_0)$ with time-independent metric to the pull-back bundle $(\mathcal{M}_0,G=G(t))$ with time dependent metric. This will be the starting point for the derivation of the perturbation operator in the model equations.

\item\textbf{Physics} Physically, the theories relevant to our present modeling belong to continuum field theory. On the one hand, the acoustic pressure is quantity of the acoustic limit of fluid dynamics assuming irrotational, isentropic and inviscid fluid flow. We start by a modification of a literature action functional of this theory of fluid dynamics to curved space-time. Using the theory of differential forms, we can derive the Euler equations and by acoustic linearization we recover a scalar wave equation in curved space-time - the acoustic wave equation. The dynamics of the boundaries, i.e., the boundary vibrations, can be well described by the motion of a massive membrane-like structure comparable to bio-membranes which we have in mind as prototypical applications of our theory. 

Our starting point is an action functional including the variation of area of the boundaries and including by an approach inspired by Chern-Weyl theory or, more physically, (Dirac-)Born-Infeld electrodynamics, curvature effects, we arrive at a differential equation which is of second order in time and fourth order in spatial variables. The equation reproduces in suitable limits the bending membrane equation derived before and the conservative flexible membrane equation. 

To include dissipation, we transfer the concept of time-lapsing from general relativity to continuum field theory and modify the derivation of the boundary vibrations equation accordingly. In the end, we discuss how to include boundary and initial conditions as well as source terms to our model equations. A Cauchy-Kovalevskaya argument shows that we can convert boundary conditions in source terms and vice versa. We answer the question of how to model the practically more relevant case of localized boundary vibrations: Localized boundary vibrations are boundary vibrations which are non-zero only on one or more mutually dis-connected sub-manifolds of the boundary of the unperturbed fiber $\Omega_0$.

\item\textbf{Analysis} The analytical part is the bridge between the geometrical and the physical part. Physically, we are interested in solving the model equations and a convenient tools is perturbation theory as developed for quantum mechanics by Dirac \cite{Dirac1,dirac}, Dyson \cite{dyson1,dyson3} (see also Reed and Simon \cite[Section X.12]{ReedSimon2}), and a decade later later by Magnus \cite{magnus1, magnus2}, whose method has been explained nicely by Blanes et al. \cite{magnus10} and has been refined recently by Fern\'andez \cite{magnus11}.  

We will present a detailed derivation of a time-dependent perturbation theory in the spirit of Magnus, which then is applied by using Duhamel's principle and Banach's fixed-point theorem in the spirit of Dirac's perturbation ansatz. The relevant mathematical theory is the theory of semigroups of operators, operator evolution equations, and perturbation theory for linear operators applied to ``classical" partial differential equations. As a check of our formalism, we compare our results with the literature. Particularly, for the conceptual basis of the analytical perturbation theory, with the theories of Brillouin \cite{brillouin}, Fr\"{o}hlich \cite{frohlich}, and Cabrera \cite{nc,cabrera} as well as Feshbach \cite{feshbach}, which partially date back to the thirties of last century. 

From the result-oriented point of view, we compare our approach with the perturbation theory developed by Deng and Li \cite{DengLi} more recently using the seminal work of Beale, which has since then undergone refinement by Casarino et al. \cite{beale4}, Gal et al. \cite{beale5}, once more Casarino et al.  \cite{beale6}, and many more who are to remain unnamed here. 

In our perturbation theory, we use a decoupling argument that implicitly assumes the smallness of a certain parameter. Upon formulating our model, we will see that this is the case in a large class of vibrational acoustics models; particularly, those used in bio-acoustics and (classical) vibrational acoustics. The final consistency check will be performed by comparing the detailed results we are going to obtain in the general formalism with those that have been achieved in our previous papers \cite{christine,anupam1,anupam2}. In doing so, we will verify as to whether the semigroups obtained in this paper agree with the semigroups we found before when specializing to the model of internally coupled ears, ICE for short. 

In order to address the convergence issues of the semigroup, we use the concept of analytic vectors to re-obtain a result on the convergence of the Magnus series comparable to a result proved recently by Batkai et al. \cite{magnus3} from the Nagel group. However, we are not interested in initial value problems but rather in inhomogeneous problems so that we could not use the aforementioned result directly. Furthermore, we modify a result obtained by Fernandez \cite{magnus4,magnus5} concerning the convergence of the Magnus series for bounded, normal operators. We sketch a proof that for weakly perturbed symmetric operators, the same criterion can be applied.
\end{itemize}
A small Lie- and operator-algebraic digression is needed during the treatment to derive some tools that we need for the further solution of the problem.

\paragraph{Background literature} Since we do not assume that the reader is familiar with all of the pre-requisites that we use in this paper, we give a physically inspired list of textbooks and articles that we found useful in understanding the underlying mathematics. Typically, we start from a few general references \cite{hassani1, hassani2, goldbart} and pave our way (see below) to the more rigorous treatments.

\begin{itemize}
\item\textbf{Geometry and topology} A good start to learn classical differential geometry is \cite{struik}. A modern approach including a discussion of differential forms to the extent we need it is found in \cite{morita}. The classic reference in Riemannian geometry with emphasis on an analytic approach is found in \cite{jost2}. A nice presentation of the theory of fiber bundles up to Chern-Weyl theory is given by Baum \cite{baum}. The interplay between geometry and physics is surveyed from the perspective of a mathematician by Jost \cite{jost3} and from the perspective of a physicist by Nakahara \cite{nakahara}. 

The topological preliminaries can be found in condensed version, mostly omitting detailed proofs but giving detailed examples in \cite{sato} - we think that the reference is appropriate given the limited amount of topology we use. In order to see physicists using differential geometry in a hands-on-manner, we refer the mathematical audience to \cite{metric1, metric2, metric3} which combines solution strategies to differential equations on manifolds with high-energy physics. The theory presented in the above references has some formal analogies to our theory and one of the authors has also worked in this fascinating field.

\item\textbf{Partial and ordinary differential equations} The theory of partial differential equations is summarized concisely in \cite{jost1}, an applied approach on solutions in the Green's functions formalism can be found in the encyclopedic works of Polyanin \cite{polyanin1, polyanin2, polyanin3}. Practical and theoretical ordinary differential equations can be found in Polyanin et al. \cite{polyanin4, emmerich} and Emmerich \cite{emmerich} to the extent we need it.

\item\textbf{Functional analysis and Magnus expansion} Functional analysis can be abstract such that we used the applied introductions in \cite{zeidler1, zeidler2}. The Magnus series is still absent in the graduate physics-textbook literature so that we refer to the pedagogic introduction \cite{magnus10} and the more intensive survey \cite{magnus2} as well as the original paper \cite{magnus1}. Convergence issues have been addressed elsewhere \cite{magnus12, magnus13}.

\item\textbf{Operator evolution equations and acoustic boundary conditions} Operator evolution equations have been expounded in an introductory manner by Emmerich \cite{emmerich}, in a more advanced way by Reed and Simon  \cite{ReedSimon2}, and in full detail by Engel and Nagel \cite{engel}. The foundations of acoustic boundary conditions (ABC without dynamics) has been treated in detail by Beale \cite{beale1, beale2, beale3}. 

\item\textbf{Quantum mechanics and perturbation theory} We simply refer to Dirac's classic  \cite[\S\S44-46]{dirac} for a physical introduction to perturbation theory. A more modern, and lucid, account of the Dyson series has been given by Zeidler \cite{zeidler2}. One may consult the older literature \cite{dyson1, dyson3} for the original motivation. The theoretical physics up to the point we use it as guideline for the mathematical formalism is presented in a more mathematical manner by Scheck and others \cite{scheck1, scheck2, scheck3, scheck4, scheck5}.

\item\textbf{Acoustics, hydrodynamics and elasticity} We refer to the textbooks by Howe \cite{howe1, howe2, howe3, howe4} covering the acoustics and hydrodynamics we need. For membrane elasticity, we refer to introductions published elsewhere \cite{membranleo, membranegeometry}. For an investigation of acoustic applications of parts of our formalism we like to mention the articles \cite{DengLi, pan1, pan2}. 


\item\textbf{Internally coupled ears and bio-acoustics} Bio-acoustics has been surveyed by Fletcher \cite{fletcher} on the basis of linear systems theory combined with a purely harmonic input varying like $\exp(i\omega t)$ and impedances as fit parameters. A quick introduction to the ideas underlying the model of internally coupled ears (ICE) is available \cite{siamleo}. The internally coupled ears model has been both presented physically \cite{christine, anupam1, anupam2} and analyzed mathematically from the point view of perturbation theory \cite{david1}.
\end{itemize}
The reader interested in the biology of the ICE model is referred to elsewhere \cite{manley1, manley2, chris1, chris2, chris3, chris4,ice-editorial}.

\paragraph{Technical prolegomena} Technically, we will be as rigorous as possible but without sacrificing the applicability of the theory. The main guideline is that physical insight is more valuable than overly pedantic mathematical rigor. For instance, we do not prove that the Laplace-Beltrami operator is a normal operator, which would be needed for one-hundred-percent completeness but is a rather straightforward exercise in functional analysis applied to partial differential equations. 

In general, we do not prove results that other authors of textbooks or research literature have obtained, unless we think the argument provides insightful to the theory developed in this paper. Rather we provide a reference. Furthermore, we will not adopt the dull definition-lemma-proof (non-)example style pervading (pure) mathematics textbooks but conform to the more applied literature that physicists and engineers use instead. Nevertheless, mathematical rigor will not be sacrificed fully. 

We are confident that the applied mathematicians will feel comfortable with this style. Because of the length of the present paper, we refer the reader for applications to the first  \cite{david1} in this series of articles, which focuses on acoustic boundary-conditions dynamics (ABCD).

\section{Settings and Perturbation-Bundle Theory}

\paragraph{Conventions} \emph{We use big Latin indices $I,J,K,L,...$ to denote local orthornormal coordinates stemming from $\mathcal{M}_{ref}$ on $\mathcal{M}$ and $\mathcal{M}_0$, small Greek indices $\mu,\nu,\kappa,\lambda,...$ to denote the local orthornormal coordinates on the second component of the fiber bundles, $\Omega_t$ and $\Omega_0$ stemming from $B^n_1(\mathbf{0})$ and small Latin indices $i,j,k,l,...$ to denote local orthornormal coordinates on $\partial\Omega_t$ and $\partial\Omega_0$ stemming from $S^n_1(\mathbf{0})$.}

\paragraph{Unperturbed bundle} Let $(\mathcal{M}_0,\pi_0,\mathbb{R}^+_0,\Omega_0)$ be an oriented trivial smooth fiber bundle over $\mathbb{R}^+_0$ such that $\mathcal{M}_0$ is imbedded in the $(n+1)$-dimensional flat Minkowskian space $(\mathbb{M}_{n+1},\eta)$ with $\eta =\text{diag}(-1,1,...,1)$. Let the fiber $\Omega_0$ be an $n$-dimensional Riemannian manifold $(\Omega_0,g_0)$ such that $\Omega_0$ is smooth, compact, retractible and oriented. Let $\Omega_0$ have a smooth, compact, oriented, closed and $(n-2)$-connected topological boundary $\partial\Omega_{0}$ such that $\Omega_0$ is imbedded diffeotopically in $\mathbb{R}^n$ and $\partial\Omega_0$ is a smooth $(n-1)$-dimensional Riemannian submanifold of $\mathbb{R}^{n}$ and $\Omega_0$ with the properties listed above. We call $\mathcal{M}_0$ for short the fiber bundle $(\mathcal{M}_0,\pi_0,\mathbb{R}^+_0,\Omega_0)$ and call it \emph{unperturbed bundle.}

\paragraph{Perturbations} We assume at the moment that $(\Omega_t)_{t> 0}$ are known. Thus, also the fiber bundle $\mathcal{M}$ is known. Since $\Omega_t$ is imbedded in $\mathbb{R}^n$, we can interpret every $\mathbf{x}\in\Omega_t$ by means of the imbedding $\iota_t$ as a point in $\mathbb{R}^n$. We choose arbitrary $t_1$ and and consider the fiber $\Omega_{t_1}$ in $\mathbb{R}^n$. For every $t_2>0$ we take the fiber $\Omega_{t_2}=(\text{proj}_2\circ\pi^{-1})(\lbrace t_2\rbrace)$. We move $\Omega_{t_2}$ in $\mathbb{R}^n$ by an orientation preserving motion, i.e. an affine mapping $M_{t_1,t_2}:\mathbf{x}\to\mathsf{M}\mathbf{x}+\mathbf{b}$ with $\mathsf{M}\in\text{SO}(n),\,\mathbf{b}\in\mathbb{R}^n$ such that $\text{Vol}_n(\Omega_{t_1}\cap M(t_1,t_2)(\Omega_{t_2}))=\text{max.}!$ with respect to the Lebesgue-Borel integration measure $\text{Vol}_n:\mathcal{B}(\mathbb{R}^n)\to\mathbb{R}^+_0$ where $\mathcal{B}(\mathbb{R}^n)$ denotes the Borel-$\sigma$-algebra on $\mathbb{R}^n$. It is generated by the topology on $\mathbb{R}^n$ induced by the Euclidean norm $\Vert .\Vert_2:\mathbb{R}^n\to\mathbb{R}^+_0$. This process ensures that for all $t>0$ the embeddings $\iota_t$ map $\Omega_t$ to $\text{im}(\Omega_t)\subsetneq\mathbb{R}^n$ such that we maximize the $n$-dimensional volume enclosed in the two bounded manifolds. In particular, we can now start comparing $\mathbf{x}_1\in\Omega_{t_1}$ and $\mathbf{x}_1\in\Omega(t_2)$ in $\mathbb{R}^n$ by means of the Euclidean norm $\Vert .\Vert_2:\mathbb{R}^n\to\mathbb{R}^+_0$. We further relate $\Omega_0$ to $(\Omega_t)_{t\geq 0}$. By smoothness of $t$-dependence of $\Omega_t$ for $t>0$, the volume $\text{Vol}_n(\Omega_t\cap\Omega_{t'})$ depends smoothly on $t,t'>0$ where we denote by $\Omega_t,\Omega_{t'}$ the moved copies of the original $(\Omega_t)_{t>0}$. Now, we are able to define $\Omega_0:=\lim_{t\to 0^+}(\Omega_t)$ where the limit is to be understood in the sense that $\Omega_0 = \lim_{t\to 0^+}(\chi_t(B^n_1(\mathbf{0})))$ where we recall $\chi_t = \psi_t^{-1}:B^n_1(\mathbf{0})\to\Omega_t$ denotes the smooth $1$-parameter family of (global) parameterizations of $\Omega_t$ for all $t>0$ w.r.t. the coordinates on $B^n_1(\mathbf{0})$. Now we define the notion of a perturbation bundle.\

\paragraph{Perturbation bundle} Let $(\mathcal{M},\pi,\mathbb{R}^+,(\Omega_t)_{t> 0})$ be the oriented smooth fiber bundle with total space $\mathcal{M}=\bigcup_{t>0}\lbrace t\rbrace\times\Omega_t$ over $\mathbb{R}^+$ such that $\mathcal{M}$ is imbedded in the $(n+1)$-dimensional flat Minkowskian space $(\mathbb{M}_{n+1},\eta)$ with $\eta = \text{diag}(-1,1,...,1)$. Let the fibers $(\Omega_t)_{\mathbb{R}^+_0\ni t> 0}$ be a family  of $n$-dimensional Riemannian manifolds $(\Omega_t,g_0)$ with the metric $g_0$ from the fibers $(\Omega_0,g_0)$ of the unperturbed bundle $\mathcal{M}_0$, parameterized smoothly by $t$ in the base space $\mathbb{R}^+$ and  imbedded diffeotopically in $\mathbb{R}^n$, such that $(\Omega_t)_{t>0}$ are smooth, compact, retractible and uniformly oriented. Let $\Omega_t$ for all $t>0$ have smooth, compact, oriented and $(n-2)$-connected topological boundary $\partial\Omega_t$ for all $t\in\mathbb{R}^+$ which are Riemannian submanifolds of $\mathbb{R}^n$ and $\Omega_t$ of dimension $n-1$. 

We call $\mathcal{M}$ for short fiber bundle and call it \emph{perturbed bundle} if there is a real $\kappa$ where $0<\kappa < 1$ such that for $\Omega_-=\text{im}(\kappa\mathds{1}_n:\Omega_0\to\mathbb{R}^n)$ and $\Omega_+=\text{im}(\kappa^{-1}\mathds{1}_n:\Omega_0\to\mathbb{R}^n)$, $\vert\kappa^{-1}-1\vert< \epsilon$ and $\vert\kappa -1\vert < \epsilon$ for a real $\epsilon \ll 1$, it holds that $\Omega_-\subsetneq\Omega_t\subsetneq\Omega_+$. We call $\epsilon$ the \emph{pertubation strength}.

\paragraph{Visualization} The definition of a perturbation bundle is visualized in Fig.~2.

\begin{figure}
\begin{center}
\includegraphics[width = 0.8\textwidth]{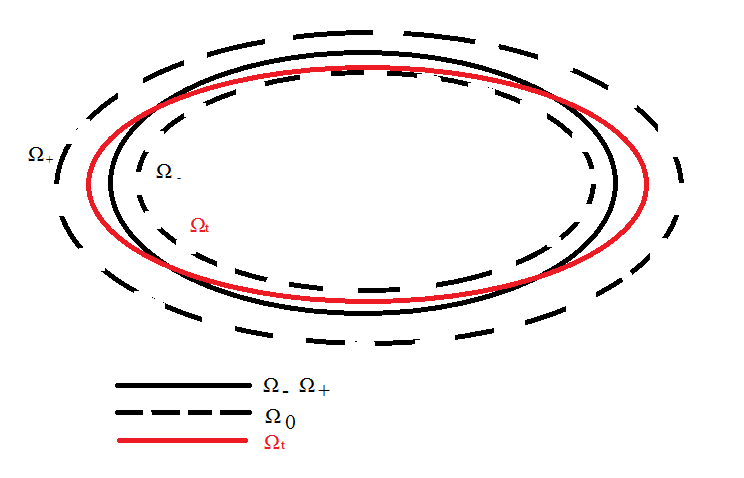}
\caption{Visualization of the definition of a \emph{perturbation bundle}. The perturbed fiber, $\Omega_t$, depicted as the region bounded by the solid red line, is contained for all $t>0$ in the manifolds $\Omega_+$ visualized by the region bounded by outer dashed black line. Further the perturbed fiber $\Omega_t$ contains for all $t>0$ the manifold $\Omega_-$, i.e. the region enclosed by the inner dashed black line. $\Omega_+,\Omega_-$ are by definition smaller resp. bigger sized copies of $\Omega_0$ which is the region enclosed by the solid black line.}
\end{center}
\end{figure}

\paragraph{Bundle structure of $\mathcal{M}$} Observe that $\mathcal{M}$ is indeed a fiber bundle over $\mathbb{R}^+$. We introduce on $\mathbb{R}^+$ the norm topology $\mathcal{T}_{\vert\vert}$ such that $U\subset\mathcal{T}_{\vert\vert}$ is the topology generated by all open intervals $]a,b[\subset\mathbb{R}^+$. The projection map is given by $\pi: \mathcal{M}\to \mathbb{R}^+, (t,\mathbf{x}_t)=t$. By diffeotopy of $\Omega_t$, the imbeddings $\iota_t:\Omega_t\to\mathbb{R}^n\simeq\lbrace t\rbrace\times\mathbb{R}^n$ give rise to a diffeomorphism $\upsilon_t:\Omega_t\to\iota_t(\Omega_t)\subsetneq\mathbb{R}^n$ with non-equality by $\partial\mathbb{R}^n =\emptyset\neq\partial\Omega_t$. 

Since each $\Omega_t$ is retractible, smooth and bounded topologically by $\partial\Omega_t$, we have further a diffeomorphism $\psi_t: B^n_1(\mathbf{0})\to\Omega_t$. Comparing dimensions, we can choose $\psi_t$ to be a proper imbedding. That is, $\psi_t(\partial\Omega_t)\to S^n_1(\mathbf{0})$, where $S^n_1(\mathbf{0})$ is the unit sphere in $\mathbb{R}^n$. Introduce $\chi_t\equiv\psi_t^{-1}: B^n_1(\mathbf{0})\to\Omega_t$ in order to obtain after localization $n$-dimensional polar coordinates on $\Omega_t$. We have $\mathcal{M}\simeq\mathbb{R}^+\times\mathbb{B}^n_{1}(\mathbf{0})$ via a diffeormorphism $\Phi:\bigcup_{t\in ]a,b[} \lbrace t\rbrace\times\Omega_t\to ]a,b[\times B^n_1(\mathbb{0}), p=(t,\mathbf{x}_t)\to (\pi(p))=t,\psi_{\pi(p)}(\text{proj}_2(p))\in ]a,b[\times B^n_1(\mathbf{0})$. 

Differentiability of $\Phi$ follows from the assumptions on smoothness of $(\Omega_t)_{t\geq 0}$ and bijectivity follows from bijectivity of the projection map on the first component, differentiability of $\Phi^{-1}$ is a consequence of the inverse function theorem. Even more,we have globally $\Phi:\mathcal{M}\to\mathbb{R}^+\times\mathbb{B}^n_1(\mathbf{0})\equiv\mathcal{M}_{ref}$ because $\mathcal{M}=(\text{id}_{\mathbb{R}^+}\times\chi_{\text{id}_{\mathbb{R}^+}})\left(\mathbb{R}^+,B^n_{1}(\mathbf{0})\right)$ and we can choose $\Omega_t$ to have the same orientation for all $t>0$ by uniform orientedness. I.e., $\mathcal{M}$ is a trivial bundle over $\mathbb{R}^+$.

\paragraph{Manifold structure of $\mathcal{M}$: }Since $\mathcal{M}$ is a trivial bundle,it inherits from $\mathbb{R}^+$ and $(\Omega_t)_{t>0}$ the product manifold structure. Since further $\iota_t:\Omega_t\to\mathbb{R}^n$ by assumption and $\text{id}:\mathbb{R}^+\to\mathbb{R}^+$, we can define a metric of Lorentzian signature on $\mathcal{M}$ by $G = \Phi^{\ast}\eta\vert_{\mathcal{M}_{ref}}$, i.e., by pulling back on $\mathcal{M}$ the Minkowski metric induced on $\mathcal{M}_{ref}$ by restriction of the standard Minkowskian metric on $\mathbb{M}_{n+1}$. In components, we have
\begin{align}
G_{0,IJ}dx^Idx^J = -dt^2+(\left(\psi_0^{\ast}dr\right)^2 + \left(\psi_0^{-1}(r)\right)^2\psi_0^{\ast}(d\Omega_{n-1}))
\end{align}
where $d\Omega_{n-1}$ denotes the metric element of $S^n_1(\mathbf{0})$. In matrix representation,
\begin{align}
G_0 = \left(\begin{array}{cc}-1 & \mathbf{0}^T_n\\ \mathbf{0}_n & g_0)\end{array}\right),
\end{align}
where $g_0$ is the metric on $\Omega_0$, given by $g=\left(\psi_0^{\ast}dr\right)^2 + \left(\psi_0^{-1}(r)\right)^2\psi_0^{\ast}(d\Omega_{n-1})$. I.e., $g_0$ is the pullback of the $n$-dimensional Euclidean metric on $B^n_1(\mathbf{0})$ to $\Omega_0$. Since the $n$-dimensional spherical coordinates can be equally taken to form a coordinate system of $\mathbb{R}^n$, we can extend the metric $g_0$ from $\Omega_0$ to all of $\mathbb{R}^n$.

\paragraph{Link between $\mathcal{M}_0$ and $\mathcal{M}$} In this paragraph, we explain Fig.~3. The diffeomorphism between the fiber bundles and the fibers of the respective fibers are depicted. Right column: Because of the topological requirements on the $(\Omega_t)_{t >0}$ an the $\Omega_0$, each $\Omega_t$, $t\geq 0$ is properly diffeomorphic to the closed unit ball $B^n_1(\mathbf{0})$ in $\mathbb{R}^n$. By means of the global diffeomorphism $\psi_0:\Omega_0\to B^n_1(\mathbf{0})$ and $\psi_t:\Omega_t\to B^n_1(\mathbf{0})$, the fibers of the two bundles $\mathcal{M}_0$ and $\mathcal{M}_{ref}$ are properly diffeormorphic to each other, $\psi_t^{-1}\circ\psi_0:\Omega_t\to\Omega_0$. Left column: Because $\mathcal{M}_{ref}$ and $\mathcal{M}_0$ are fiber bundles, they are at least locally diffeomorphic to $\mathbb{M}_{n+1,+}=\mathbb{R}^+\times\mathbb{R}^n$. Because of topological obstructions, namely the boundary$\partial\Omega_t\neq\empty$ for all $t\geq 0$, these diffeomorphisms are only local. 

By orientedness of the fiber spaces $\Omega_t,\,t\geq 0$ and orientedness of the total spaces $\mathcal{M},\mathcal{M}_0$ of the 
bundles, a global diffeomorphism, even more bundle biffeomorphisms can be defined: $\Phi:\mathcal{M}\to\mathcal{M}_{ref}$, $\Phi_0:\mathcal{M}_0\to\mathcal{M}_{ref}$. These can be composed using bijectivity of the bundle morphism to obtain a bundle morphism $\Phi^{-1}\circ\Phi_0:\mathcal{M}\to\mathcal{M}_0$. By means of the projection maps $\pi:W_0\subseteq\mathcal{M}\to U_0\subseteq\mathbb{R}^+$ and $\pi_0:W\subseteq\mathcal{M}_0\to U\subseteq\mathbb{R}^+$, open subsets of the total spaces $\mathcal{M}_0,\mathcal{M}$ can be mapped to open subsets of the base space $\mathbb{R}^+$. Since by assumption, $\mathcal{M},\mathcal{M}_0$ and $\Omega_t,\,t\geq 0$ are imbedded in a Euclidean space $\mathbb{R}^{n+1}$ and $\mathbb{R}^n$ of equal dimension as the total spaces resp. fibers, and thus are equipped with the corresponding relative topologies, the single point $\lbrace t\rbrace$ can be obtained by means of the intersection of all its $\delta$-neighborhoods $U_\delta(\lbrace t\rbrace)$ in $\mathbb{R}^+$ starting at small enough $\delta> 0$. 

Inverting the projections $\pi,\pi_0$, the fiber bundles "evaluated" at the point $t$, $\lbrace t\rbrace\times\Omega_t\subset\mathcal{M}$ and $\lbrace t\rbrace\times\Omega_0\subset\mathcal{M}_0$ is obtained. The reference bundle $\mathcal{M}_{ref}=\mathbb{R}^n_+$ is, as a product manifold, a trivial bundle over $\mathbb{R}^+_n$ with fiber $B^n_1(\mathbf{0})$, such that $\lbrace t\rbrace\times B^n_1(\mathbf{0})$ can be included in a canonical way in the total space $\mathcal{M}_{ref}$. By means of the projection on the second component, $\text{proj}_2$, of the product spaces $\lbrace t\rbrace\times\Omega_t\,(t>0),\,\lbrace t\rbrace\times\Omega_0,\,\lbrace t\rbrace\times B^n_1(\mathbf{0})$, the manifolds $\Omega_t\,(t>0),\,\Omega_0,\,B^n_1(\mathbf{0})$ are obtained directly from the fiber bundles $\mathcal{M},\,\mathcal{M}_0,\,\mathcal{M}_{ref}$. Fig. 4 contains Fig. 3 and the Minkowski space $\mathbb{M}_{n+1}$ as the overall imbedding space of the bundles $\mathcal{M}_0,\mathcal{M}_{ref}$ and $\mathcal{M}$.
\begin{figure}
\begin{center}
\includegraphics[width = 1\textwidth]{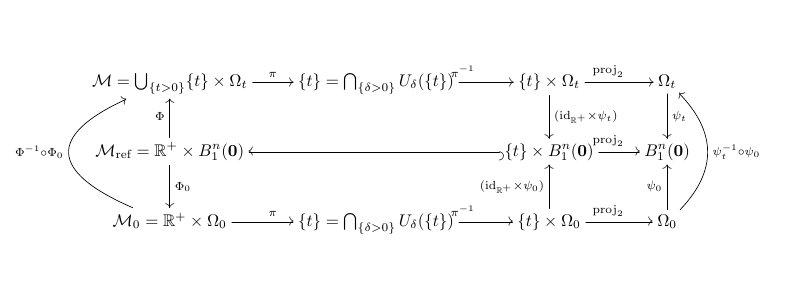}
\end{center}
\caption{Transformations between $\mathcal{M}_0,\mathcal{M}$ and $\mathcal{M}_{ref}.$ See the main text for explanations.}
\end{figure}
\begin{figure}
\begin{center}
\includegraphics[width = 1\textwidth]{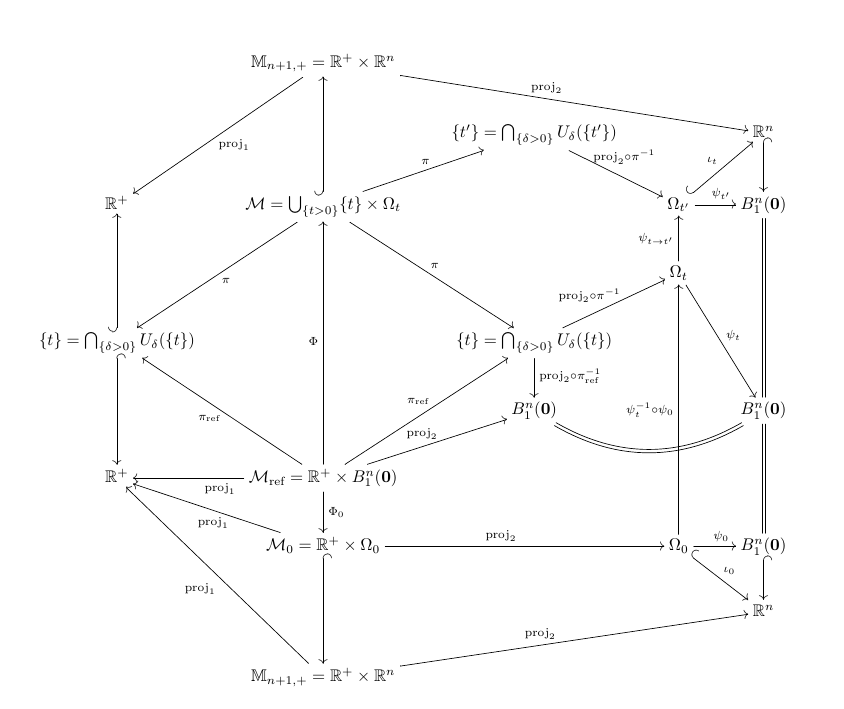}
\end{center}
\caption{Overview of the geometric ingredients used in the perturbation theory and the relations between them. See the main text for explanations.}
\end{figure}

\paragraph{$\infty$-periodicity} We call a perturbation bundle \emph{$\infty$-periodic} if we have $\Omega_0=\lim_{t\to 0^+}$ and $\lim_{t\to\infty^-}=\Omega_\infty = \Omega_0$ with $\mathcal{C}^1$ regularity as subsets of $\mathbb{R}^n$ such that the fiber bundle $$\mathcal{M}'=\left(\bigcup_{t\in\mathbb{R}^+\cup\lbrace\infty\rbrace}\lbrace t\rbrace\times\Omega_t\right)\cup\left(\lbrace 0\rbrace\times\Omega_0\right)\cup\left(\bigcup_{t\in\mathbb{R}^+\cup\lbrace \infty\rbrace}\lbrace -t\rbrace\times\Omega_t\right)$$has a quotient bundle $\bar{\mathcal{M}'}=\lim_{T\to\infty}(\mathcal{M}'/T\mathbb{Z})$ such that $\bar{\mathcal{M}'}=\mathcal{M}$, i.e., is the original perturbation bundle. The quotient operation is understood to act on the base space $\bar{\mathbb{R}}$ of $\mathcal{M}'$.

\paragraph{Explanation} Let use define the bundle $\mathcal{M}'$ including the time-boundaries $\lbrace 0\rbrace\times\mathcal{M}$ and $\lbrace\infty\rbrace\times\Omega_\infty$ as follows
\begin{align*}
\mathcal{M}'=\left(\bigcup_{t\in\mathbb{R}^+\cup\lbrace\infty\rbrace}\lbrace t\rbrace\times\Omega_t\right)\cup\left(\lbrace 0\rbrace\times\Omega_0\right)\cup\left(\bigcup_{t\in\mathbb{R}^+\cup\lbrace \infty\rbrace}\lbrace -t\rbrace\times\Omega_t\right)
\end{align*}
using the \emph{$\infty$-periodicity condition} $\lim_{\delta\to 0^+}\Omega_\delta=\lim_{\delta^{-1}\to\infty^-}\Omega_{\delta^{-1}}=\Omega_0$ with $\mathcal{C}^1$ regularity. The bundle $\mathcal{M}'$ has the advantage that when we consider later on a wave equation with boundary and initial condition, we can regard the initial conditions as boundary conditions on $\lbrace 0\rbrace\times\Omega_0$ and $\lbrace\infty\rbrace\times\Omega_0$ using the isomorphy $\lbrace 0\rbrace\cup\pi(\mathcal{M})=\lbrace 0\rbrace\cup\mathbb{R}^+=\mathbb{R}^+_0=\lim_{T\to\infty}(\bar{\mathbb{R}}/T\mathbb{Z})$ where $\bar{\mathbb{R}}=\mathbb{R}\cup\lbrace-\infty,\infty\rbrace$. By means of this identification, we identify $\lbrace 0\rbrace\times\Omega_0\equiv\lbrace\infty\rbrace\times\Omega_0\Leftrightarrow t=0\equiv\infty$. By smoothness of $\mathcal{M}$, thus of $\mathcal{M}'$ we obtain smoothness of 
\begin{align*}
\bar{\mathcal{M}}'=\lim_{T\to\infty}\mathcal{M}'/(T\mathbb{Z})
\end{align*}
except possibly at $t=0\equiv\infty$. However, the bundle is at least continuous by the $\infty$-periodicity condition. By the setting, we further have $\mathcal{C}^1$ regularity of the limit $\lim_{t\to 0^+}\Omega_t=\Omega_0=\lim_{t\to\infty^-}\Omega_t$. Thus $\bar{\mathcal{M}}'$ is a $\mathcal{C}^1$ differentiable fiber bundle. From the theory of differentiable manifolds and bundles, we know that we can choose for $\mathcal{C}^k,k\geq 1$ manifolds or bundles even a $\mathcal{C}^\infty$ atlas. Thus, $\mathcal{M}_\infty$ can be turned again in a $\mathcal{C}^\infty$ bundle. By definition, we can identify this with $\mathcal{M}$, namely $\bar{\mathcal{M}}',\mathcal{M}$ up to a Lebesgue null-set w.r.t. the Lebesgue-Borel integration measure $\text{Vol}_{n+1}:\mathcal{B}((\mathbb{M}_{n+1},\eta))\to\mathbb{R}^+_0$. The workflow is depicted in Fig. 4.\begin{figure}
\includegraphics[width = 1\textwidth]{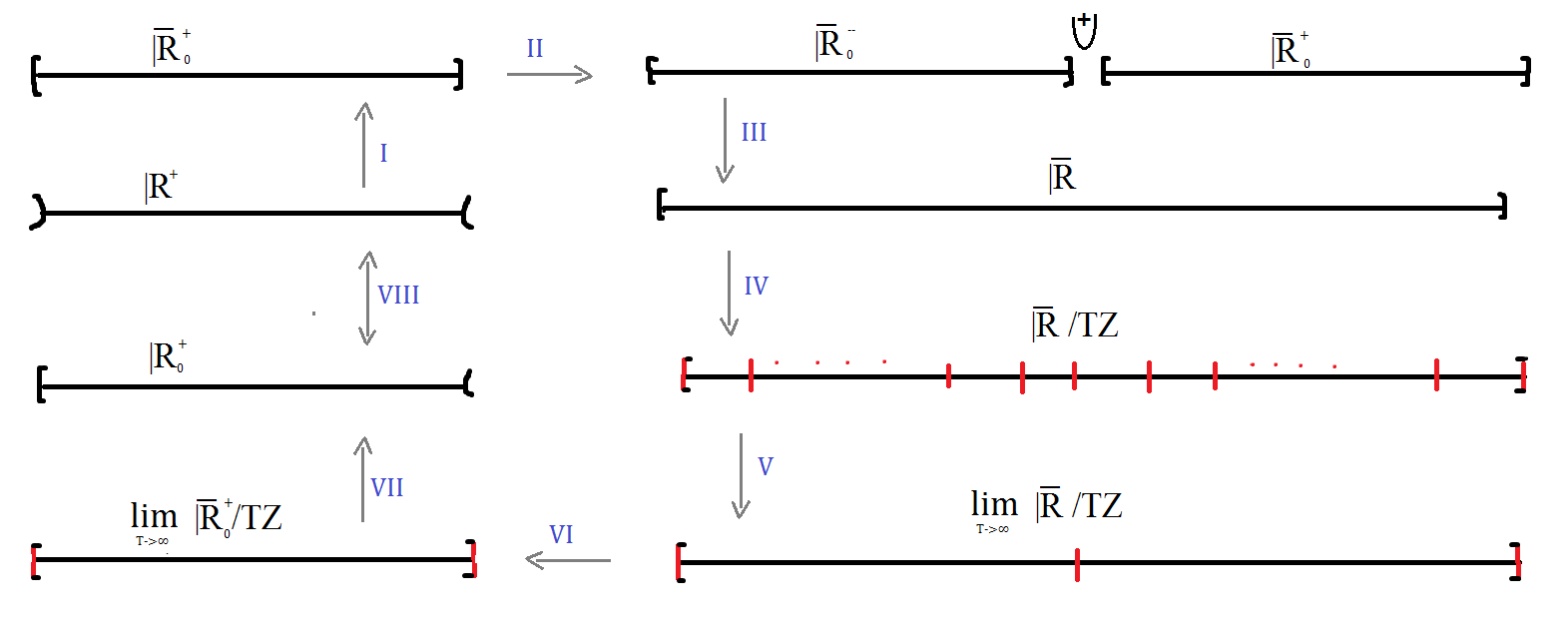}
\caption{The workflow used to include the initial conditions of a wave equation on $\mathcal{M}$. The Roman numbers in blue correspond to the individual steps explained in the main text.}
\end{figure}
\newline
\newline
\textbf{$\infty$-periodicity and physical relaxation: }A short comment on the interpretation of this identification is in order: Physically, the fibers start deforming smoothly at $t=0$ from $\Omega_0$ and approach $\Omega_0$ again at $t=\infty$. The identification $0\equiv\infty$ tells us that the bundle is ready again to go through the entire deformation process. In view of the ICE model \cite{anupam1,anupam2, christine, siamleo, ice-editorial, david1}, this means that after one exposition of the gecko to sound stimulus, the gecko can be exposed to another sound stimulus such that the hearing system of the gecko has forgotten that there was a prior sound stimulus (no echo). The requirement $\Omega_0=\Omega_\infty$ which was needed for the identification means that the deformations relax again, i.e., are damped in some sense. In the ICE model \cite{anupam1,anupam2, christine, siamleo, ice-editorial, david1} this corresponds to the deformations that the cylinder undergoes by membrane vibration. Since the membranes are damped because the gecko should ultimately stop hearing a certain stimulus from $t=0$ at $t\to\infty$, the cylinders return at $t=\infty$ to their equilibrium shapes at $t=0$, i.e., $\Omega_0=\Omega_\infty$.

\paragraph{Proper perturbation bundles} We call a perturbation bundle \emph{proper} if $(\phi_{0\to t}-\text{id}_{\partial\Omega_0})(\mathbf{y})\parallel \nabla_{\mathbf{X}}(\phi_{0\to t}-\text{id}_{\partial\Omega_0})(\mathbf{y})$ for all $\mathbf{y}\in\partial\Omega,\,\mathbf{X}\in T_{\mathbf{y}}\Omega_0$ and $t>0$ in the imbedding space $\mathbb{R}^n$ and $\text{span}_{\mathbb{R}}\left\lbrace d_{\mathbf{y}}(\phi_{0\to t}-\text{id}_{\partial\Omega})\right\rbrace\perp T_{\mathbf{y}}\partial\Omega$ for all $\mathbf{y}\in\Omega_0$. Here $\nabla\in\text{Hom}(\Gamma(T\mathbb{R}^n)\to\Gamma(T\mathbb{R}^n)\ast\otimes\mathbb{R})$ denotes the Levi-Civita connection on $\mathbb{R}^n$ induced by the metric $g_0$. It is now time for a few motivating explanations. 

If $n=3$, $\text{dim}_{\mathbb{R}}(\partial\Omega)=2=\dim_{\mathbb{R}}(\partial\Omega_t)$. The definition formalizes the intuition that given local coordinates $(u,v,w)$ in $\mathbb{R}^3$ and $\partial\Omega_0\supseteq U = \lbrace (u,v,w)\vert w = 0\rbrace$ whereas $\partial\Omega_t \supseteq V = \lbrace (u,v,w)\vert w = f(u,v;t)\rbrace$ with a pointwisely non-zero function $f$ depending on the local coordinates $(u,v)$ on $U\subseteq \partial\Omega$ and, in general, the time-parameter $t$. Indeed, $T_{(u_0,v_0)}\partial\Omega = \mathbb{R}^2\times\lbrace 0\rbrace$, whereas $d_{(u_0,v_0)}(\psi_{0\to t}\vert_{\partial\Omega_0}-\text{id}_{\partial\Omega_0}) = \hat{e}_w(\partial_u f + \partial_v f)$ spans $\lbrace\mathbf{0}\in\mathbb{R}^2\rbrace\times\mathbb{R}$. Thus, we have the decomposition of $T_{(u_0,v_0,f(u_0,v_0))}\mathbb{R}^3\simeq\mathbb{R}^2 = \mathbb{R}^2\times\lbrace 0\rbrace\oplus\lbrace\mathbf{0}\in\mathbb{R}^2\rbrace\times\mathbb{R}=T_{(u_0,v_0)}\partial\Omega_0\oplus\text{span}\lbrace d_{\mathbf{y}}(\psi_{0\to t}-\text{id}_{\partial\Omega_0})\rbrace$. We visualize the situation as modeled in the following piece of graphics, Fig. 5. Notice that $\phi_t: \partial\Omega_t\to S^n_1(\mathbf{0})$ and $\phi_0:\partial\Omega_0\to S^n_1(\mathbf{0})$ are nothing but the Gaussian mapping from $(n-1)$-dimensional (imbedded), oriented and closed submanifolds $\partial\Omega_0,\partial\Omega_t\hookrightarrow\mathbb{R}^n$ to the oriented unit sphere $S^n_1(\mathbf{0})\hookrightarrow\mathbb{R}^n$.

\begin{figure}
\begin{center}
\includegraphics[width = 0.8\textwidth]{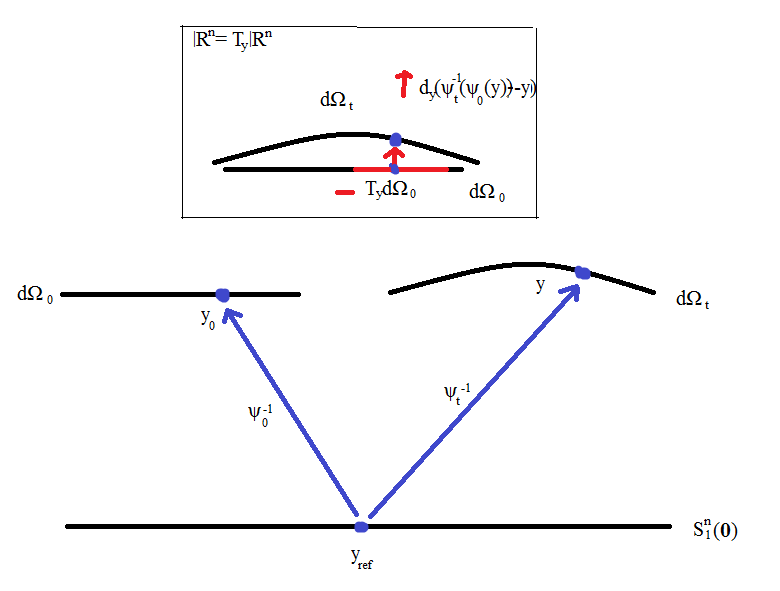}
\end{center}
\caption{Visualization of the definition of a \emph{proper} perturbation bundle. The vector $\mathbf{y}':=\psi_t^{-1}\circ\psi_0(\mathbf{y}), \mathbf{y}\in\partial\Omega_0$ is perpendicular to $\partial\Omega$ for all $t>0$ and $\mathbf{y}\in\Omega_0$ in the target space $\mathbb{R}^n$ of the imbeedings $\iota_t,\iota_0$ for $\Omega_t,\Omega_0\to\mathbb{R}^n$.}
\end{figure}

\paragraph{Fibratable perturbation bundles} We call a proper perturbation bundle \emph{fibratable}, if the diffeomorphism $\sigma_{ref}:S^n_1(\mathbf{0})\times [0,1]\to B^n_1(\mathbf{0}),\,(\mathbf{y},s)\mapsto \mathbf{y}- s\mathbf{n}_{S^n_1(\mathbf{0})}\Vert\mathbf{y}\Vert_{B^n_1(\mathbf{0})}$ induce diffeomorphisms $\sigma_0:\partial\Omega_0\times [0,1]\to\Omega_0,\,(\mathbf{y},s)\mapsto \mathbf{y}- s\mathbf{n}_{\partial\Omega_0}\Vert\mathbf{y}\Vert$ and $\sigma_t:\partial\Omega_t\times[0,1]\to\Omega_t\to (\mathbf{y},s)\mapsto \mathbf{y}- s\mathbf{n}_{\partial\Omega_t}\Vert\mathbf{y}\Vert_{g_0}$ such that $\sigma_t$ is smooth in $t$. The symbols $\mathbf{n}_{\heartsuit}$ denote the unit normal vectors to $S^n_1(\mathbf{0}),\,\partial\Omega_0$ and $\Omega_t$.

By means of the Gauss maps $\phi_0:\partial\Omega_0\to S^n_1(\mathbf{0})$ and $\phi_t:\partial\Omega_t\to S^n_1(\mathbf{0})$ every fiber $\Omega_t$, $t\geq 0$, allows the diffeomorphisms in the above equation by pulling back $\sigma_{ref}$ by the Gauss maps. Thus, every proper perturbation bundle is fibratable. That is, can be fibrated. Let is quickly make a digression to what fibratable means in the present context. 
For the unit ball $B^n_1(\mathbf{0})=\lbrace\mathbf{x}\in\mathbb{R}^n\vert\Vert\mathbf{x}\Vert_2\leq 1\rbrace$ the $n$-sphere $S^n_1(\mathbf{0})$ gives rise to a fibration of $B^n_1(\mathbf{0})\setminus\lbrace\mathbf{0}\rbrace$ the following way. Let us denote an element of $S^n_1(\mathbf{0})$ by $\mathbf{y}$ and define $\sigma_{ref}$ by
\begin{align}
\sigma_{ref}:[0,1)\times S^n_1(\mathbf{0})\to B^n_1(\mathbf{0});\, (s,\mathbf{y})\mapsto \sigma_{ref}(s,\mathbf{y})=\mathbf{y}-s\hat{e}_r\Vert\mathbf{y}\Vert_2, 
\end{align}
where $\hat{e}_r=\hat{\mathbf{n}}$ is the outward unit normal vector to $S^n_1(\mathbf{0})$ as a $\mathbb{R}^n$-submanifold of co-dimension $1$. In order to ensure bijectivity, we had to exclude the center of $B^n_1(\mathbf{0})$, i.e., $\mathbf{0}\in\mathbb{R}^n$. However, $\lbrace\mathbf{0}\rbrace$ is a Lebesgue-Borel null-set w.r.t. the Lebesgue-Borel measure $\text{Vol}_n:\mathcal{B}(\mathbb{R}^n)\to\mathbb{R}^+$ restricted to $B^n_1(\mathbf{0})\subset\mathbb{R}^n$ such that this does not hinder our further progress. Now, $\phi_0=\psi_0\vert_{\partial\Omega_0}:\partial\Omega_0\to S^n_1(\mathbf{0})$ induces an analogous fibration on $\Omega_0$ if
\begin{align}
\sigma_0:&[0,1)\times\partial\Omega_0\to\Omega_0\setminus\lbrace\mathbf{0}\rbrace;\\
&(s,\mathbf{y}_0)\mapsto\sigma_0(s,\mathbf{y}_0)=\mathbf{y}_0-s\hat{\mathbf{n}}_{\partial\Omega_0}(\mathbf{y}_0)\Vert\mathbf{y}_0\Vert_2 = (\text{id}_{[0,1)}\times\phi_0)^\ast\sigma_{ref}(s,\mathbf{y}).
\end{align}
In the previous equation, a point on the boundary is denoted $\mathbf{y}_0$ and a point in $\Omega_0\setminus\lbrace\mathbf{0}\rbrace$ by $\mathbf{x}_0$. Analogously, $\phi_t:\partial\Omega_t\to S^n_1(\mathbf{0})$ induces a fibration on $\Omega_t$ is required by the definition if
\begin{align}
\sigma_t:&[0,1)\times\partial\Omega_t\to\Omega_t\setminus\lbrace\mathbf{0}\rbrace;\\
&(s,\mathbf{y}_0)\mapsto\sigma_t(s,\mathbf{y}_t)=\mathbf{y}_t-s\hat{\mathbf{n}}_{\partial\Omega_t}(\mathbf{y}_t)\Vert\mathbf{y}_t\Vert_2 = (\text{id}_{[0,1)}\times\phi_t)^\ast\sigma_{ref}(s,\mathbf{y}),
\end{align}
where $\mathbf{y}_t\in\partial\Omega_t$ and $\mathbf{x}_t$ denotes a generic point in $\Omega_t\setminus\lbrace\mathbf{0}\rbrace$. Denoting symbolically $\sigma_0(s,\partial\Omega_0):=(1-s)\cdot\partial\Omega_0$ and similarly $\sigma_t(s,\partial\Omega_t):=(1-s)\cdot\partial\Omega_t$, we can write
\begin{align}
\begin{split}
\Omega_0\setminus\lbrace\mathbf{0}\rbrace &\equiv \biguplus_{s\in [0,1)}(1-s)\cdot\partial\Omega_0\equiv\biguplus_{s\in[0,1)}\sigma_0(s,\partial\Omega_0)\, ,\\
\Omega_t\setminus\lbrace\mathbf{0}\rbrace &\equiv \biguplus_{s\in [0,1)}(1-s)\cdot\partial\Omega_t \equiv \biguplus_{s\in[0,1)}\sigma_0(s,\partial\Omega_t)\, .
\end{split}
\end{align}
The geometric intuition stored in the definitions of $\sigma_0,\,\sigma_t\text{ (and }\sigma_{ref})$ is illustrated in the subsequent figure (Fig. 6).
\begin{figure}
\begin{center}
\includegraphics[width = 0.67\textwidth]{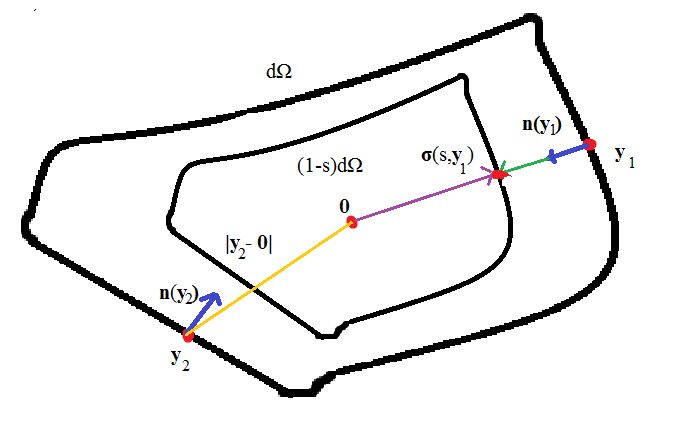}
\end{center}
\caption{The unperturbed domain $\Omega$ and its boundary $\partial\Omega$ together with the smaller sized $(1-s)\partial\Omega=\sigma(s,\partial\Omega)$ for a fixed $s\in[0,1)$. The points $\mathbf{y}_1,\mathbf{y}_2$ are elements of $\partial\Omega$. $\mathbf{n}(\mathbf{y}_i)$ denotes for $i\in\lbrace 1,2\rbrace$ the \emph{inward} unit normal vector at $\mathbf{y}_i$ on $\partial\Omega$ (blue vector). Staring from $\mathbf{y}_1\in\partial\Omega$, we move by an amount of $s\vert\mathbf{y}_1,\mathbf{n}(\mathbf{y}_1)\vert$ along $\mathbf{n}(\mathbf{y}_1)$ \emph{inwards} $\Omega$ (green vector). The result is $\sigma(s,\mathbf{y}_1)$ depicted by the violet vector. The necessity to take $s\vert\langle\mathbf{y},\mathbf{n}\rangle\vert$ instead of simply $s\vert\mathbf{y}\vert$ for the amount we move \emph{inwards} $\Omega$ is depicted by the golden line. $\mathbf{y}_2\in\partial\Omega$ has in general a non-zero tangential component, where tangential means tangential to $\partial\Omega$. For a general $\Omega\neq B^n_R(\mathbf{0}),R>0$, the position vector $\mathbf{y}_2$ points to $\partial\Omega$ but, regarded as vector field, is neither element of $T_{\mathbf{y_2}}\partial\Omega$ nor $(T_{\mathbf{y_2}}\partial\Omega)^{\perp}$ but only $\mathbf{y}_2\in T_{\mathbf{y}_2}\Omega = T_{\mathbf{y_2}}\partial\Omega\oplus(T_{\mathbf{y_2}}\partial\Omega)^{\perp}$ where the orthogonal complement of the sub vector space $T_{\mathbf{y}_2}\partial\Omega$ is to be taken in $T_{\mathbf{y}_2}\Omega$.}
\end{figure}

\paragraph{Physical perturbation bundles} We call a proper perturbation bundle \emph{physical} if there is a (dimensional) constant $C$ satisfying $C/[C]>0$ and $C/c^2=\mathcal{o}(\epsilon^{2})$ such that in the metric $G_0$ continued to $\mathcal{M}_{n+1}\vert$ $\Vert\nabla_{\partial_t}(\phi_{0\to t}-\text{id}_{\Omega_0})\Vert_{g_0}^2\leq C\Vert\nabla_{\partial_i}(\phi_{0\to t}-\text{id}_{\Omega_0})\Vert_{g_0}^2$, \emph{conservative} if the above inequality is an equality and \emph{dissipative} if the above inequality is strict.


\textbf{Explanation: } The property of a perturbation bundle to be physical 
states that the internal energy density $U$ of the perturbations is non-increasing. We consider the boundary $\partial\Omega_t$ to be a membrane. In the physical picture, membranes can be thought of as a continuum of harmonic oscillators coupled to each other. This gives rise to the potential energy density $\mathcal{V}=T_0 g_{(0)}^{ij}\nabla_i u\nabla_j u$ where $T_0$ is a physical constant, the membrane tension. The vibration of the membranes on the other hand gives rise to a kinetic energy density $\mathcal{T}=-\sigma_m G_0^{tt}\partial_t u\partial_t u$ noting that $G_0^{tt}<0$ if $G_0$ is brought into diagonal form. $\sigma_m$ denotes a physical constant, the surface mass density of the membrane. The property ``conservative" states that each oscillator can convert the entire potential energy into kinetic energy. The property ``dissipative" expresses the physical reality that there are heat losses due to internal friction effects between the individual oscillators and external friction due to the membrane interacting with its environment which hinder the conversion of all potential energy $\mathcal{V}$ into kinetic energy $\mathcal{T}$ on a local level, i.e., for each oscillator. I.e., $\mathcal{T}<\mathcal{V}$. Equating $C\equiv c^2_m = T_0/\sigma_m$ we recover the definition. The condition $C\ll c^2$ now turns into $c_m/c=\mathcal{o}(\epsilon)$. For the values given in \cite{anupam1} it can be fulfilled by choosing a suitable $\epsilon\ll 1$. 

The condition $\epsilon\ll 1$ expresses as the admission of the boundary waves to be 
transmitted without information loss through the cavity. This requires that the speed of propagation of the membranes vibration, $c_m$, is much smaller than the speed of propagation $c$ of the wave inside the gecko's interaural cavity, $\Omega_t$. Otherwise, the membranes would vibrate so fast that not all cavity eigenmodes within the audible frequency range of the gecko $<5\,\text{kHz}$ have the chance to transmit the information of membrane vibration. Physiologically, it is only the lowest eigenmodes that the gecko can perceive by its hearing system.\newline
\newline
\textbf{Convention: }\emph{Henceforth, we let $(\mathcal{M},\mathbb{R}^+,\pi)$ denote a proper and physical perturbation bundle with fibratable fibers $(\Omega_t)_{t>0}$ around $(\mathcal{M}_0,\mathbb{R}^+,\pi_0)$.}\newline
\newline
\textbf{Connection to fluid mechanics: }The original ICE model which serves as a physical role model for our mathematical structures has been formulated in terms of physical quantities, namely the acoustic pressure $p$ and the membrane displacements $u$. It is natural to ask how the geometric approach relates to the acoustic quantities. At first, we take the diffeomorphism $\psi_{0\to t}=\psi_t^{-1}\circ\psi_0$ and perform a rewriting using bijectivity and smoothness of diffeomorphisms as well as the convenience of the definition of a perturbation bundle that the relevant fibers $\Omega_0$ and $\Omega_t$ are imbeded submanifolds of $\mathbb{R}^n$. We let $\mathbf{x}\in\Omega_0$ and consider the subsequent equation as a vectorial equation in $\mathbb{R}^n$:
\begin{align}\label{fluiddiffeos}
\begin{split}
\psi_{0\to t}(\mathbf{x}) &= \psi_t^{-1}\circ\psi_0(\mathbf{x})\\
& = (\psi_t^{-1}-\psi_0^{-1}+\psi_0^{-1})\circ\psi_0(\mathbf{x})\\
& = \text{id}_{\Omega_0}(\mathbf{x})+(\psi_t^{-1}-\psi_0^{-1})\circ\psi_{0}(\mathbf{x}).
\end{split}
\end{align}
We use this equation to make contact to fluid dynamics in a twofold way. Firstly, we isolate the second term in the equation and restrict it to the boundary $\partial\Omega_0$. We recall the previously introduced notation $\psi_t\vert_{\partial\Omega_0}=\phi_t$ and $\psi_0\vert_{\partial\Omega}=\phi_0$. Further,we use that $\psi_t,\psi_0$ are proper in order to justify for the notation $\psi_t^{-1}\vert_{S^n_1(\mathbf{0})}=\phi_t^{-1}$ and $\psi_0^{-1}\vert_{S^n_1(\mathbf{0})}=\phi_0^{-1}$. Using regularity of the perturbation bundle, we have for all $\mathbf{y}\in\partial\Omega_0$,
\begin{align}
(\phi_t^{-1}-\phi_0^{-1})\circ\phi_0 = \mathbf{n}_{\partial\Omega_0}(\mathbf{y})\Vert(\phi_t^{-1}-\phi_0^{-1})\circ\phi_0(\mathbf{y})\Vert_2,
\end{align}
because $(\phi_t^{-1}-\phi_0^{-1})\circ\phi_0(\mathbf{y})$ is normal to $\partial\Omega_0$ for all $\mathbf{y}\in\partial\Omega_0$ by regularity.
The Euclidean norm $\Vert .\Vert_2:\mathbb{R}^n\to\mathbb{R}^+_0$ of the boundary manifolds diffeomorphisms $\phi_0: S^n_1(\mathbf{0}),\phi_t: S^n_1(\mathbf{0})$ however is nothing else than the deviation of the perturbed boundary $\partial\Omega_t$ from the unperturbed boundary $\partial\Omega_t$ orthogonal to the boundary $\partial\Omega_0$. It is parameterized w.r.t. local coordinates $\mathbf{y}$ on $\partial\Omega_0$. This allows the identification
\begin{align}
u = \Vert,(\phi_t^{-1}-\phi_0^{-1})\circ\phi_0(\mathbf{y})\Vert_2,
\end{align}
i.e., the non-identity contribution in restriction of (\ref{fluiddiffeos}) can be interpreted as the membrane displacement $u$. It has been argued by authors \cite{anupam1, anupam2, christine} that at the boundary the fluid normal velocity $\langle\mathbf{n}(\partial\Omega_t),\mathbf{v}\rangle_{\mathbb{R}^n}$ is equal to the membrane velocity $\partial_t u$ using the no-slip boundary condition to the Navier-Stokes equations resp. their high Reynolds ($\text{Re}\gg 1$) limit - the Euler equations. This allows the further identification,
\begin{align}
\langle\mathbf{n}_{\partial\Omega_t},\mathbf{v}\rangle_{\mathbb{R}^n}(\mathbf{y}_t) = (\partial_t\Vert,(\phi_t^{-1}-\phi_0^{-1})\circ\phi_0\Vert_2)\circ(\phi_{0\to t}^{-1})(\mathbf{y}_t),
\end{align}
where $\mathbf{y}_t\in\partial\Omega_t$ and we used $\phi_{0\to t}^{-1}:\partial\Omega_t\to\partial\Omega_0$ to transform the domain of the object inside Euclidean norm from $\partial\Omega_0$ to $\partial\Omega_t$. Since we have assumed fibratable fibers $\Omega_t$, we can extrapolate the equation from $\partial\Omega_t$ to $\Omega_t\setminus\lbrace\mathbf{0}\rbrace$ to find,
\begin{align}
\langle\mathbf{n}_{(1-s)\cdot\partial\Omega_t}(\mathbf{x}_t),\mathbf{v}\rangle_{\mathbb{R}^n}(\mathbf{x}_t) = \left(\partial_t\langle\mathbf{n}_{(1-s)\partial\Omega},(\psi_t^{-1}-\psi_0^{-1})\circ\psi_0\rangle_{\mathbb{R}^n}\right)\circ(\psi_{0\to t}^{-1})(\mathbf{x}_t).
\end{align}
We use  the notation once again that $\mathbf{x}_t\in\Omega_t\setminus\lbrace\mathbf{0}\rbrace$. This allows the identification of the normal part of the acoustic\footnote{The attribute acoustic refers to the acoustic linearization Ansatz to Euler's equations. If we denote by $\mathbf{V}$ the fluid velocity field in Euler's equations, acoustic linearization $\mathbf{V}=\mathbf{v}_0+\mathbf{v}$, where the acoustic fluid velocity $\mathbf{v}$ is a small quantity, i.e., must be considered only in linear order. $\mathbf{v}_0$ is the background velocity velocity which is set equal to $\mathbf{0}$ in the fluid rest frame. This has been employed in the ICE model, \cite{anupam1, anupam2, christine}. Indeed, taking the time derivative of the component normal to $(1-s)\cdot\partial\Omega_0$ (or evaluating the $0$-component of the Levi-Civita connection induced by $G_0$ on $\mathcal{M}_0$ which is precisely $\partial_t$), we see that $\partial_t\langle\mathbf{n}_{(1-s)\partial\Omega_0}(\psi_{0\to t})\rangle_{\mathbb{R}^n}=\mathbf{v}$ since $\text{id}_{\partial\Omega}$ is time-independent.} (!) fluid velocity $\mathbf{v}_{\mathbf{n}}$ in the notation of \cite{christine} with the non-identical term in (\ref{fluiddiffeos}) pulled back to $\Omega_t$ by means of $\psi_{0\to t}^{-1}:\partial\Omega_t\to\partial\Omega_0$.\newline
\newline
\textbf{Properties of $\psi_{0\to t}$: }We investigate the diffeomorphisms $\psi_{0\to t}$ in order to obtain alternative representations which are more useful for the practical calculations. At first, we would like to bound $\psi_{0\to t}-\text{id}_{\Omega_0}$ from above. Let $\mathbf{x}\in\Omega_0$ be arbitrary. Then we have w.r.t. the metric $g_0$ continued from $\Omega_0$ to the imbedding space $\mathbb{R}^n$,
\begin{align*}
&\Vert(\psi_{0\to t}-\text{id}_{\Omega_0})(\mathbf{x})\Vert_{g_0}\\
&\leq \Vert(\kappa^{-1}\text{id}_{\Omega_0}-\kappa\text{id}_{\Omega_0})(\mathbf{x})+(\psi_{0\to t}-\text{id}_{\Omega_0})(\mathbf{x})\Vert_{g_0}\\
&\leq\Vert(\kappa^{-1}\text{id}_{\Omega_0}-\psi_{0\to t})(\mathbf{x})\Vert_{g_0}+\Vert(\kappa\text{id}_{\Omega_0}-\text{id}_{\Omega_0})(\mathbf{x})\Vert_{g_0}\\
&\leq \Vert(\kappa^{-1}\text{id}_{\Omega_0}-\kappa\text{id}_{\Omega_0})(\mathbf{x})\Vert_{g_0}+\Vert(\kappa\text{id}_{\Omega_0}-\text{id}_{\Omega_0})(\mathbf{x})\Vert_{g_0}\\
&= \Vert((\kappa^{-1}\text{id}_{\Omega_0}-\text{id}_{\Omega_0})+(text{id}_{\Omega_0}-\kappa\text{id}_{\Omega_0}))(\mathbf{x})\Vert_{g_0}+\Vert(\kappa\text{id}_{\Omega_0}-\text{id}_{\Omega_0})(\mathbf{x})\Vert_{g_0}\\
&\leq \Vert(\kappa^{-1}\text{id}_{\Omega_0}-\text{id}_{\Omega_0})(\mathbf{x})\Vert_{g_0}+\Vert(\text{id}_{\Omega_0}-\kappa\text{id}_{\Omega_0}))(\mathbf{x})\Vert_{g_0}+\Vert(\kappa\text{id}_{\Omega_0}-\text{id}_{\Omega_0})(\mathbf{x})\Vert_{g_0}\\
&=\vert\kappa^{-1} - 1\vert\Vert\mathbf{x}\Vert_{g_0}+\vert\kappa -1\vert\Vert\mathbf{x}\Vert_{g_0}+\vert\kappa - 1\vert\Vert\mathbf{x}\Vert_{g_0}\\
&< 3\epsilon\Vert\mathbf{x}\Vert_{g_0}\\
\Leftrightarrow \Vert\psi_{0\to t}-\text{id}_{\Omega_0}\Vert &= \dfrac{\Vert(\psi_{0\to t}-\text{id}_{\Omega_0})(\mathbf{x})\Vert_{g_0}}{\Vert\mathbf{x}\Vert_{g_0}} < 3\epsilon
\end{align*}
Furthermore, we have for the linear approximation w.r.t. the Levi-Civita connection $\nabla: T\Omega_0\to (T\Omega_0)^\ast\times T\Omega_0$ the norm estimate $\Vert\nabla_{\mathbf{X}}(\psi_{0\to t}-\text{id}_{\Omega_0})(\mathbf{x}) \Vert_{g_0}<3\epsilon$ for all $\mathbf{X}\in T_{\mathbf{x}}\Omega_0$ with $\Vert\mathbf{X}\Vert_{g_0}=1$. In other words, we can approximate for $\mathbf{x}\in\Omega_0$,
\begin{align}
(\psi_{0\to t}-\text{id}_{\Omega_0})(\mathbf{x}) = U(t,\mathbf{x})\mathbf{x},
\end{align}
with a function $U:\mathcal{M}_0\to\mathbb{R}$, such that in the maximum norm $\Vert U\Vert_{\infty}<3\epsilon$ and
\begin{align}
U(t,\mathbf{x})=\langle\mathbf{n}_{\partial (1-s)\Omega_0},\psi_{0\to t}-\text{id}_{\Omega_0}\rangle_{\Omega_0}.
\end{align}
We allow $U$ to be in $H^{1,2}_0(\mathcal{M}_0)$. By properness of the bundle, we have that $\psi_{0\to t}-\text{id}_{\Omega_0}\parallel\mathbf{n}_{(1-s)\cdot\partial\Omega_0}\parallel\mathbf{n}_{\partial\Omega_0}$ in the imbedding space $\mathbb{R}^n$. Denoting the normal component of $\mathbf{x}$ as $\mathbf{x^s}$ and setting $s=1$ on $\partial\Omega_0$ by fibratability of the fibers $\Omega_0$ of the bundle $\mathcal{M}_0$, we find the easier expression,
\begin{align}
(\psi_{0\to t}-\text{id}_{\Omega_0})(\mathbf{x})=U(t,\mathbf{x})x^s\mathbf{n}_{\partial(1-s)\Omega_0}.
\end{align}
We want to express the function $U$ by a different object that is defined for arguments in $\mathbb{R}^+_0\times\partial\Omega_0$. We recall that using diffeomorphisms $\sigma_0:[0,1]\times\partial\Omega_0\to\Omega_0,\,\sigma_t:[0,1]\times\partial\Omega_t\to\Omega_t$ we can express by choosing proper imbeddings $\psi_0:\Omega_0\to S^n_1(\mathbf{0}),\psi_t:\partial\Omega_t\to S^n_1(\mathbf{0})$ the diffeomorphisms $\psi_{0\to t}:\Omega_0\to\Omega_t$ by their restrictions to the boundary, $\phi_{0\to t}=\psi_{0\to t}\vert_{\partial\Omega_0}:\partial\Omega_0\to\partial\Omega_t$. We calculate the explicit expression
\begin{align*}
\psi_{0}&=\sigma_{ref}\circ(\text{id}_{[0,1]}\times\phi_{0})\circ\sigma_{0}^{-1}\\
\psi_{t}&=\sigma_{ref}\circ(\text{id}_{[0,1]}\times\phi_{t})\circ\sigma_t^{-1}\\
\Rightarrow\psi_{0\to t}&=\psi_{t}^{-1}\circ\psi_0\\
&=(\sigma_{ref}\circ(\text{id}_{[0,1]}\times\phi_{t})\circ\sigma_t^{-1})^{-1}\circ(\sigma_{ref}\circ(\text{id}_{[0,1]}\times\phi_{0})\circ\sigma_{0}^{-1})\\
&= \sigma_t\circ(\text{id}_{[0,1]}\times\phi_t^{-1})^{-1}\circ\sigma^{-1}_{ref}\circ\sigma_{ref}\circ(\text{id}_{[0,1]}\times\phi_0^{-1})\circ\sigma_0^{-1}\\
&= \sigma_t\circ(\text{id}_{[0,1]}\times\phi_t^{-1})\circ(\text{id}_{[0,1]}\times\phi_0)\circ\sigma_0^{-1}\\
&= \sigma_t\circ((\text{id}_{[0,1]}\circ\text{id}_{[0,1]})\times(\phi_t^{-1}\circ\phi_0))\circ\sigma_0^{-1}\\
&=\sigma_t\circ(\text{id}_{[0,1]}\times\phi_{0\to t})\circ\sigma_0^{-1}.
\end{align*}
Likewise, we can express $\text{id}_{\Omega_0}$,
\begin{align}
\text{id}_{\Omega_0}=\sigma_0\circ(\text{id}_{[0,1]}\times\text{id}_{\partial\Omega_0})\circ\sigma_0^{-1}.
\end{align}
Taking the difference for $\mathbf{x}=(\mathbf{y},x^s)$, we find using properness of the perturbation bundle $\mathcal{M}$,
\begin{align}
(\psi_{0\to t}-\text{id}_{\Omega_0}) &= \phi_{0\to t}(\mathbf{y})-\mathbf{y} -s\mathbf{n}_{\partial\Omega_t}\Vert\phi_{0\to t}(\mathbf{y})\Vert_{g_0}+s\Vert\mathbf{y}\Vert_{g_0}\mathbf{n}_{\partial\Omega_0}\\
&=\phi_{0\to t}(\mathbf{y})-\mathbf{y}-s(\phi_{0\to t,\ast}(\mathbf{n}_{\partial\Omega_0})\Vert\phi_{0\to t}(\mathbf{y})\Vert_{g_0}-\mathbf{n}_{\partial\Omega_0}\Vert\mathbf{y}\Vert_{g_0})\\
&=(1-s)\Vert\phi_{0\to t}(\mathbf{y})-\text{id}_{\Omega_0}\Vert_{g_0}\mathbf{n}_{\partial\Omega_0}+\mathcal{O}(\epsilon^2),
\end{align}
since $\mathbf{n}(\partial\Omega_t)=\mathbf{n}_{\partial\Omega_0}+\mathcal{O}(\epsilon)$.  We define the \emph{boundary vibrations}\footnote{They are precisely the membrane displacements in the ICE model.}
\begin{align}
\langle\mathbf{n}_{\partial\Omega_0},\phi_{0\to t}-\text{id}_{\partial\Omega_0}\rangle_{g_0}=u(t,\mathbf{y})\mathbf{n}_{\partial\Omega_0}
\end{align}
Thus, we have
\begin{align}
\psi_{0\to t}-\text{id}_{\Omega_0}=(1-s)u(t,\mathbf{y})\mathbf{n}_{\partial\Omega_0}.
\end{align}
The claim $\mathbf{n}(\partial\Omega_t)=\mathbf{n}_{\partial\Omega_0}+\mathcal{O}(\epsilon)$ is verified by noticing that given a local orthonormal coordinate system on $\partial\Omega_0$ and setting the orthogonal coordinate $x^s$, we can express $\partial\Omega_t$ as the graph of $u=u(t,\mathbf{y}):\mathbb{R}^+_0\times\partial\Omega_0\to\mathbb{R}$. We have
\begin{align}
\partial\Omega_t = \text{Graph}(u(t,\mathbf{y})) = \left\lbrace (\mathbf{y},u(t,\mathbf{y})):\mathbf{y}\in\partial\Omega_0\right\rbrace .
\end{align}
The outward unit normal to $\partial\Omega_0$ in $\mathbb{R}^n$ is given by $\mathbf{n}_{\partial\Omega_0}=\partial_{x^s}=\partial_s$. For $\partial\Omega_t$, we calculate
\begin{align}
\mathbf{n}_{\partial\Omega_t} &=\dfrac{\nabla^{\mu}(x^s-u)\partial_\mu}{\Vert\nabla^\mu(z-u)\Vert_{g_0}}\\
&= \dfrac{\partial_s}{\sqrt{1 + \partial_i u\partial^i u}}+\dfrac{\partial^i u\partial_i}{\sqrt{1+\partial_i u\partial^i u}}\\
&= \partial_s + \mathcal{O}(\epsilon).
\end{align}
Let us express $\psi_{0\to t}$ w.r.t. the basis $\lbrace\partial_{\mu}\rbrace_{1\leq\mu\leq n}$.
\begin{align}
\psi_{0\to t}^\mu\partial_\mu = x^j\partial_j + (x^s+(1-s)u(t,\mathbf{y}))\partial_s
\end{align}
Ultimately, we arrive at a practical expression for the deviation in $\mathbb{R}^n$ of the diffeomorphism $\psi_{0\to t}$ to be the identity $\text{id}_{\Omega_0}$
\begin{align}
(\psi_{0\to t}^\mu-x^\mu)\partial_\mu = (1-s)u(t,\mathbf{y})\partial_s .
\end{align}
We want to use this equation in order to obtain the pull-back from $\Omega_t$ of the metric $\psi_{0\to t}^\ast g_0=g_t$ on $\Omega_0$. In terms of local coordinates $\rbrace x^\mu\lbrace_{1\leq\mu\leq n}$ on $\Omega_0$ and $\lbrace X^\mu\rbrace_{1\leq\mu\leq n}$ where we choose the \"radial" coordinate $x^s = s\in[0,1]$, we have
\begin{align*}
g &= ((g_0)_{\mu\nu}\circ\psi_{0\to t})\psi_{0\to t}^\ast(dX^\mu)\psi_{0\to t}^\ast(dX^\nu)\\
&= (g_0)_{\mu\nu}(x^\mu+(\psi_{0\to t}-x^\mu))dx^\mu dx^\nu\\
&= (g_0)_{\mu\nu}dx^\mu dx^\nu + (g_0)_{\lambda\kappa}\dfrac{\partial(\psi_{0\to t}^\kappa-x^\kappa)}{(\partial x^\nu)}\dfrac{\partial(\psi_{0\to t}^\lambda -x^\lambda)}{\partial(x^\mu)}dx^\mu dx^\nu\\
&= (g_0)_{\mu\nu}dx^\mu dx^\nu+u^2g^{(0)}_{ss}dx^s dx^s -2(1-s)u\partial_{i}u g^{(0)}_{s s}dx^{i}dx^s+(1-s)^2\partial_{i}u\partial_{j}u g^{(0)}_{ss}dx^{i}dx^j\\
&\equiv g_0 + \delta g.
\end{align*}
The object $\delta g$ is a metric perturbation and has a symmetric associated matrix,
\begin{align}
(\delta g_{\mu\nu})_{1\leq\mu ,\nu\leq n} = g^{(0)}_{ss}\left(\begin{array}{cc}u^2 & -(1-s)u\partial_i u\\ -(1-s)u\partial_i u & (1-s)^2\partial_i u\partial_j u\end{array}\right).
\end{align}
For the metric $G_0$ on the total space of the fiber bundle $\mathcal{M}$, we can use the bundle morphism $\Phi^{-1}\circ\Phi_0$ to obtain the pull-back metric. We obtain for the metric perturbation $\delta G \equiv (\Phi^{-1}\circ\Phi_0)^\ast G_0 -G_0$ in terms of coordinates $x^J, x^K$ on $\mathcal{M}_0=\mathbb{R}^+_0\times\Omega_0$ the associated matrix
\begin{align}
(\delta G)_{JK}=g^{(0)}_{ss}\left(\begin{array}{ccc}+(1-s)^2\partial_t u\partial_t u & -(1-s)u\partial_t u & -(1-s)^2\partial_tu\partial_j u\\-(1-s)u\partial_t u & u^2 & -(1-s)u\partial_i u\\-(1-s)^2\partial_t u\partial_i u & -(1-s)u\partial_i u & (1-s)^2\partial_i u\partial_j u\end{array}\right)
\end{align}
Since $G_0$ is a metric, we can choose the local coordinate system $\lbrace x^J\rbrace_{0\leq J\leq n}$ on $\mathcal{M}_0$ such that $G_0$ is diagonal. We note down the Taylor expansion of the volume element around the diagonal unperturbed metric $G_0$ because $\Vert\delta G\Vert_{Frob,\infty}$ is of order $\epsilon$ in the Frobenius norm for quadratic matrices with the norm for the coefficients not being the modulus, but the maximum norm $\Vert .\Vert_{\infty}$. Since $\Vert u\Vert_{\infty} \leq \Vert U\Vert_{\infty} < 3\epsilon$, the assumption that the perturbation bundle is physical yields indeed $\delta G = \mathcal{O}(\epsilon)$.
\begin{align*}
\sqrt{-\vert G\vert} &= \sqrt{-\vert G_0+\delta G\vert}\\
&= \sqrt{-\vert G_0\vert\vert\mathds{1}_{n+1}+G_0^{-1}G\vert}\\
&= \sqrt{-\vert G_0\vert}\left(1+\dfrac{1}{2}\text{Tr}(G_0^{-1}\delta G)\right)\\
&= \sqrt{-\vert G_0\vert}(1+\dfrac{1}{2}G^{IJ}\delta G_{IJ}).
\end{align*}
Integrated over $\mathcal{M}_0$, we obtain a constant contribution $\text{Vol}_{n+1}(\mathcal{M}_0)$ which diverges due to $t\in\mathbb{R}^+_0$ but can be neglected. The physically relevant information are stored in the change of volume in the $G_0$ metric of $\mathcal{M}$ and $\mathcal{M}_0$ relative to the volume of $\mathcal{M}_0$ which cancels the $\infty$'s. In equations, we have
\begin{align*}
\dfrac{\Delta V}{V}&=\dfrac{\text{Vol}_{G_0}(\mathcal{M})-\text{Vol}_{G_0}(\mathcal{M}_0)}{\text{Vol}_{G_0}(\mathcal{M}_0)}\\
&=\int_{\mathcal{M}}d\text{Vol}_{G_0}(\mathcal{M})-\int_{\mathcal{M}_0}d\text{Vol}_{G_0}(\mathcal{M}_0)\\
&=\int_{\mathcal{M}_0}d(\text{Vol}_G(\mathcal{M}_0)-\text{Vol}_{G_0}(\mathcal{M}_0))\\
&= \dfrac{1}{2}\int_{\mathcal{M}_0}d^{n+1}x\,\sqrt{-\vert G_0\vert}G^{IJ}\delta G_{IJ}.
\end{align*}
Apart from some minor fixes concerning units, this can be used as a part of an action functional $S_{geom}[\delta G]$ to look for a partial differential equation satisfied by $u$!\newline
\newline
\textbf{Digression - a matrix identity: }Let $\mathsf{M}=\mathds{1}+\delta\mathsf{M}$ be a non-degenerate symmetric and quadratic matrix. We have 
\begin{align*}
\det\mathsf{M}&=\exp\log\det\mathsf{M}\\
&= \exp\text{Tr}\log\mathsf{M}\\
&= \exp\text{Tr}\log\left(\mathds{1}+\delta\mathsf{M}\right)\\
&= \exp\text{Tr}\left(-\sum_{j=1}^\infty\dfrac{(-1)^j(\delta\mathsf{M})^j}{j}\right)\\
&=\exp\left(\sum_{j=1}^\infty\dfrac{(-1)^{j+1}\text{Tr}((\delta\mathsf{M})^j)}{j}\right)\\
&= \sum_{k=0}^\infty\dfrac{1}{k!}\left(\sum_{j=1}^\infty\dfrac{(-1)^{j+1}\text{Tr}((\delta\mathsf{M})^j)}{j}\right)^k\\
&= 1+\text{Tr}(\delta\mathsf{M})+\mathcal{O}((\delta\mathsf{M})^2).
\end{align*}
For $g$ and $G$ this results together with $\sqrt{1+x}=1+x/2+\mathcal{O}(x^2)$ in the equations
\begin{align}
\sqrt{-\vert G\vert}-\sqrt{-\vert G_0\vert} &= \dfrac{\text{Tr}(G_0\delta G)}{2}\\
\sqrt{\vert g\vert}-\sqrt{\vert g_0\vert} &=\dfrac{\text{Tr}(g_0\delta g)}{2}.
\end{align}
The first equation has been used during the above calculation, the second equation will be useful in the next section.
\newline
\newline
\textbf{Reconstruction: }We recall that the acoustic pressure $p$ is actually defined on $\mathcal{M}$. We can pull it back to $\mathcal{M}_0$ by composing with $\psi_{0\to t}$ from the right. Using the coordinate expressions for $\psi_{0\to t}$ from above, we have
\begin{align}
p(t,X^\mu) &= p(t,x^\mu + (\psi_{0\to t}^\mu(\lbrace x^\nu\rbrace)-x^\nu))\\
&= p(t,x^\mu)+\dfrac{\partial(\psi_{0\to t}^\mu-x^\mu)}{\partial x^\nu}\partial_\nu p + \mathcal{O}(\epsilon^2)
\end{align}
We can solve the acoustic wave equation on $\mathcal{M}_0$
\begin{align}
\partial_t^2P - c^2\Delta_{g_0} P = c^2\mathsf{V}[P](\text{Source term})
\end{align}
instead of the acoustic wave equation on $\mathcal{M}$
\begin{align}
\partial_t^2p - c^2\Delta_{g_0} p = (\text{Source term})
\end{align}
transforming boundary and initial conditions accordingly and afterwards either set $p=P$ at the cost of an error linear in the perturbation strength, i.e., of order $\epsilon$ relative to $p=\mathcal{O}(\varepsilon)$ or transform back by means of the diffeomorphisms $\psi_{0\to t}$. This fact is commonly exploited in engineering and physical vibrational acoustics. The object $\mathsf{V}$ is the \emph{perturbation operator} which will be defined properly in the next section.  In the next section, we will use this observation to show that we can take the eigenfunctions of the Laplacian $\Delta_{g_0}$ on $\Omega_0$ instead of the eigenfunctions of $\Delta_g$ in order to obtain the eigenfunctions of $\Delta_{g_0}$ on $\Omega_t$.

\section{Acoustics Wave Equation and Properties of the Laplace-Beltrami-Operators}
\textbf{Introduction: }The overall goal of the paper is to examine small perturbations to the (acoustic) wave equation on $\mathcal{M}$ stemming from "metric perturbations". Later on, we will define the notion of a perturbation bundle such that we can specify the terminology metric perturbations. We discuss the method for pedagogic reasons in a familiar setting, namely a wave equation on $\mathbb{M}_{n+1,+}$.
\begin{itemize}
\item\textbf{Generalities on the wave equation: }Recall that in $\mathbb{M}_{n+1,+}=\mathbb{R}^+\times\mathbb{R}^n$ the acoustic wave equation reads
\begin{align}
(\partial_t^2 - \Delta_n)p=\left(\partial_t^2 -c^2\sum_{i=1}^n\partial_i^2\right)p=0,
\end{align}
with the appropriate constraints on the asymptotics of $p$ similar to boundary conditions at infinity. The factor $c^2$ is there for physical reasons to assign $\Delta_n$ a dimension of $[\Delta_n]=1\text{ s}^{-2}$. The solution $p$ lives in a Hilbert space, more precisely a Sobolev space over the product space $\mathbb{R}^+\times\mathbb{R}^n\ni(t,\mathbf{x})$,
\begin{align}
p \in &H^{1,2}_0(\mathbb{R}^+\times\mathbb{R}^n)
\equiv \left\lbrace f\in L^2(\mathbb{R}^+\times\mathbb{R}^n):\left(\int_{0}^\infty dt\int_{\mathbb{R}^n}d^nx\,\left(\vert p\vert^2+\vert\partial_t p\vert^2+\vert c\nabla p\vert^2\right)\right)^{\frac{1}{2}} < \infty, \lim_{\vert\mathbf{x}\vert\to 0}\mathbf{x}\nabla p = 0 = \lim_{\vert\mathbf{x}\vert\to 0}p\right\rbrace .
\end{align}
The multiple integral in the definition of the Sobolev space $H^{1,2}_0(\mathbb{R}^+\times\mathbb{R}^n)$ is the so-called Sobolev norm $\Vert\Vert_{1,2}:H^{1,2}_0(\mathbb{R}^+\times\mathbb{R}^n)\to\mathbb{R}^+_0$. \item\textbf{Euclideanization of time $t$ - the imaginary time $\tau$: }Sobolev spaces can be defined over (pseudo-)Riemannian manifolds as well. We can rewrite our equation using the standard Minkowski metric, $\eta_{IJ}=\text{diag}(-1,1,...,1)$, where the $1$ enters $n$ times,
\begin{align}
\eta^{IJ}\partial_I\partial_J p = 0,
\end{align}
absorbing a factor of $c$ in the definition of $\partial_J$ for $1\leq J\leq n$. Now transform the coordinates by a Wick rotation, $t = i\tau$ $\mathsf{U}\in\text{SU}(n+1)$, $\mathsf{U}=\text{diag}(i,1,1,...,1)$. The time $t$ has become a Euclidean time $\tau$ now.
\begin{align}
\Vert p\Vert_{1,2}^2&= i\int_{-i\cdot0}^{-i\infty} d\tau\int_{\mathbb{R}^n}d^nx\,\left(\vert p\vert^2+\vert\partial_\tau p\vert^2+\vert \partial_i p\vert^2\right).
\end{align}
The norm square now features an imaginary $i$, but its modulus is still a non-negative real number. 
Setting $p'=p'(\tau,\mathbf{x})=p(i\tau,\mathbf{x})$, the wave equation is reformulated as
\begin{align}
\delta_{IJ}\partial_I\partial_Jp' = 0,
\end{align}
with $\partial_0=\partial_\tau$ now. We can reformulate the equation once more in terms of the Laplacian $\Delta_{n+1}=\sum_{I=0}^{n}\partial_I^2$,
\begin{align}
\Delta_{n+1}p'=0.
\end{align}
The general theory of Laplace's equation \cite{jost1}
tells us that a solution with
\begin{align}
\Vert p' \Vert_{1,2}^2=\int_{0}^{\infty} d\tau\int_{\mathbb{R}^n}d^n x\,\left(\vert p\vert^2+\vert\partial_\tau p\vert^2+ \sum_{i}\vert\partial_i p\vert^2\right) < \infty
\end{align}
exists. However, it is unclear whether the norm survives letting $\tau\in (-i)\mathbb{R}\subset\mathbb{C}$ instead $\tau\in\mathbb{R}$. Let us go therefore one step further.
\item\textbf{Mapping to Euclidean $\mathbb{R}^{n+2}$: }We introduce $z=q_1+iq_2$, where $q_1\in\mathbb{R}^+, q_2\in\mathbb{R}^-$ and solve Laplace's equation in $(n+2)$ dimensions,
\begin{align}
\Delta_{n+2}p''=0 
\end{align}
where $\Delta_{n+2}=\partial_z\partial_{\bar{z}}+\sum_{i=0}^n\partial_i^2$. Appropriate asymptotic boundary conditions are understood. The solution $p''$ now lives in $H^{1,2}_0(\mathbb{R}^+\times\mathbb{R}^{-}\times\mathbb{R}^n)$ and depends on $z$. We introduce polar coordinates $(\rho,\phi)\in (0,\infty)\times [3\pi/2, 2\pi)$ and write 
\begin{align}
\partial_z\partial_{\bar{z}}=\rho^{-1}\partial_\rho(\rho\partial_\rho)+\rho^{-2}\partial_\phi^2 \, .
\end{align}

\item\textbf{Back transforming to the wave equation on $\mathbb{M}_{n+1,+}$: }After having solved the above equations we use $2q_1 =z +\bar{z},\, 2iq_2 = z-\bar{z}$ and $\rho = \sqrt{q_1^2+q_2^2},\phi = -\arccos(q_2/\sqrt{q_1^2+q_2^2})$ so as to transform back to $z$ and $\bar{z}$. Given suitable boundary conditions, i.e., prescribed asymptotic behavior of $p''$ at spatial infinity, $\Vert\mathbf{x}\Vert_2\to\infty$ and initial-boundary conditions of $p''(z=0),\partial_zp''(z=0)=0+i\cdot 0$, the general theory of Laplace's equation \cite{jost1} tells us that such a solution exists. We recall $q_2 <0$, define $z=z(q_2)=iq_2=\tau$ and set $p'(\tau,\mathbf{x})=p''(z(q_2),\mathbf{x})$. By definition of $p'$, we then find $p(t,\mathbf{x})=p''(t,\mathbf{x})$, i.e., $p\in H^{1,2}_0(\mathbb{R}^+\times\mathbb{R}^n)$ exists. At first sight, the derivation looks unnecessarily complicated, however it has an advantage over the classical existence proofs relying on the validity of the expansion theorem in $\mathbb{M}_{n+1,+}$ \cite{jost1}:, We just need an appropriate domain in $\mathbb{R}^{n+2}$ to derive existence of $p$. On contrast, the expansion theorem relies heavily on the assumption that the spatial domains, i.e., $\mathbb{R}^n$, stays also locally the same for all $t\in\mathbb{R}^+$. Our spatial domains, i.e., the fibers $\Omega_t$ vary locally smoothly in $t$. We use this methodology to derive existence of a solution to a general wave equation on $\mathcal{M}$.
\end{itemize}
\textbf{Wave equation on $\mathcal{M}$: }As stated in the introduction, we want to examine a vibro-acoustic system containing an acoustic wave equation on the total space of perturbation bundle $\mathcal{M}$.
\begin{itemize}
\item\textbf{Inclusion of a curvy Laplacian: }The equation which we will derive below is given by
\begin{align}
\partial_t^2p-c^2\Delta_{g_0}p = 0.
\end{align}
$p$ denotes the acoustic pressure, $c\approx 343\text{ ms}^{-1}$ and $\Delta_{g_0} = \sqrt{\vert g_0\vert}^{-1}\partial_\mu\left(\sqrt{\vert g_0\vert}g_0^{\mu\nu}\partial_\nu\right)$ is the Laplace-Beltrami operator on $\Omega_t$, $\Delta_{g_0}: H^{1,2}_0(\Omega_t)\to L^2(\Omega_t)$.
\item\textbf{Boundary conditions: }Further, we specify the following boundary and initial data to be satisfied by $p$: For the boundary conditions, we amend to $\Delta_g$ Robin boundary conditions  with the inhomogeneity given by $f\in L^2(\mathcal{M})$ such that $\lim_{t\to 0^+}f = 0 = \lim_{t\to\infty^-}f$
\begin{align}
n^\mu_{\partial\mathcal{M}}(t,\mathbf{x})\partial_\mu p(t,\mathbf{x}) + n_\mu m^\mu p(t,\mathbf{x}) = f(t,\mathbf{x})\text{ on }\partial\mathcal{M}=\bigcup_{t>0}\lbrace t\rbrace\times\partial\Omega_t.
\end{align}
$\mathbf{n}_{\partial\Omega}=n^\mu_{\partial\mathcal{M}}\partial_\mu$ is the spatial part of the outward by orientedness of $\mathcal{M}$ unit normal vector to $\mathcal{M}$ in $\mathbb{M}_{n+1,+}$. The index $\mu\in\lbrace 1,...,n\rbrace$. The vector $m=m^0\partial_0+m^\mu\partial_\mu = m^0\partial_0$ with $m^0=1$ . Notice that $n_\mu m^\mu=0$ such that the Robin boundary conditions reduce to Neumann boundary conditions.
\item\textbf{Initial conditions: } Last, we specify homogeneous initial conditions $\partial_0 p(t=0,\mathbf{x}) = \partial_t p(t=0,\mathbf{x})=0$ on $\mathcal{M}$.
\item\textbf{Conversion into a Laplace equation on $\mathcal{M}$: }The advantage of the bundle-theoretic interpretation of the acoustic wave equation becomes clear now: The boundary data on $\partial\mathcal{M}$ vary smoothly in $t$ because of the smooth $t$-dependence of the $1$-parameter family $(\Omega_t)_{t>0}$.
\item\textbf{Preparatory step - Quasi-periodicity of $\mathcal{M}$: }In order to establish the initial conditions as boundary conditions on $\partial\mathcal{M}$ in $n+1$ dimensions, we proceed as follows. By definition, we have $\Omega_0=\lim_{t\to 0}\Omega_t$. We append $\lbrace 0\rbrace\times\Omega_0$  and $\lbrace\infty\rbrace\times\Omega_\infty $ to the fiber bundle $\mathcal{M}$ to form a new bundle $\mathcal{M}'$. This does not alter regularity properties of the solution because we only added a Lebesgue null set w.r.t. to the integration measure $\text{Vol}_{n+1}:\mathcal{B}(\mathbb{M}_{n+1,+})\to\mathbb{R}^+.$. Let us now set $n_{\partial\mathcal{M}'\setminus\partial\mathcal{M}}=n^0\partial_0+n^\mu\partial_\mu$ with $n^0_{\partial\mathcal{M'}\setminus\partial\mathcal{M}}=1$. Further identify $\lbrace 0\rbrace$ and $\lbrace\infty\rbrace$ such that $t=0\Leftrightarrow t\to\infty$. Then $n^\mu=0$ for $1\leq\mu\leq n$ and $n^0=-\partial_t$ if $t\to 0^+$ and $n^0=\partial_t$ if $t\to\infty^-$.
\item\textbf{Lifting of the boundary and initial data to $\mathcal{M}$: } Since $f$ is only defined for $t\in\mathbb{R}^+$, the boundary data at $\lbrace 0\rbrace\times\Omega_0$ give us $-\partial_0p(t=0,\mathbf{x})+p(t=0,\mathbf{x})=0$. On the other hand, at $t\infty$, i.e., on $\lbrace\infty\rbrace\times\Omega_{\infty}=\lbrace\infty\rbrace\times\Omega_0$, we have $\lim_{t\to\infty}(\partial_tp(t,\mathbf{x})+p(t,\mathbf{x}))=0$. By the above identification $0=t\Leftrightarrow t=\infty$, we can equally put the boundary conditions at $t=0$ and $t=\infty$ on $\lbrace t=0\rbrace\times\Omega_0$ and forget about the $\lbrace\infty\rbrace\times\Omega_0$ contribution to $\partial\mathcal{M}'$ at $t=\infty$. Then we have the linear equation system on $\lbrace 0\rbrace\times\Omega_0$,
\begin{align}
0 &= -\partial_t p(0,\mathbf{x})+p(0,\mathbf{x})\\
0 &= \partial_t p(0,\mathbf{x})+p(0,\mathbf{x}).
\end{align}
This returns the initial conditions $p(t=0,\mathbf{x})=0$ and $\partial_t p(t=0,\mathbf{x})=0$ on $\lbrace 0\rbrace\times\Omega_0$.
\item\textbf{Laplace's equation on $\mathcal{M}$: }Absorbing the $c^2$ in the local coordinates, i.e., $x^\mu\to cx^\mu$, and using the definition of the metric $G$ on $\mathcal{M}$, we have
\begin{align}
\Delta_{G_0} p = 0\text{ on }\mathcal{M}
\end{align}
together with the Robin boundary data $n^I_{\partial\mathcal{M}}\partial_I p + n_I m^Ip = f$. This is a Laplace equation on the $(n+1)$-dimensional manifold $\mathcal{M}$ with Robin boundary conditions on $\partial\mathcal{M}$.
\item\textbf{Euclideanization of $t$: }So far, we have arrived at Laplace's equation as before. Now, we can make $\delta_G$ elliptical by switching to Euclidean time, $t=i\tau$. The fiber bundle then reads $\mathcal{M}=\biguplus_{\tau\in i\bar{\mathbb{R}}^-_0}\lbrace i\tau\rbrace\times\Omega_{i\tau}$.
\item\textbf{Complexification of Euclidean time $\tau$: } Then we complexify the time component by introducing $\mathcal{M}'_{\mathbb{C}}=\bigcup_{z\in (0,1)\times i\cdot\bar{\mathbb{R}}^-_0}\lbrace z=q_1+iq_2\rbrace\times\Omega_{iq_2}$. This has the same properties as $\mathcal{M}$ except that the base space now is $(0,1)\times\bar{\mathbb{R}}^-_0\hookrightarrow\mathbb{R}^{n+2}$. The differential equation for $p$ turns into the following equation for $p''$
\begin{align}
(\partial_z\partial_{\bar{z}}+\Delta_{g_0})p''=0,
\end{align}
and we specify the asymptotic condition $\lim_{q_1\to 0}p''=0\lim_{q_1\to1}p''$. In order to ensure dependence on real quantities again, we notice that we could introduce polars $\rho = \sqrt{q_1^2+q_2^2}$ and $\phi = -\arccos(q_2/\sqrt{q_1^2+q_2^2})$ as we did already before.
\item\textbf{$\mathcal{M}\hookrightarrow\mathcal{M}'\hookrightarrow\mathcal{M}_{\mathbb{C}}\subset\mathbb{R}^{n+2}$ - Deriving existence of $p\in H^{1,2}_0(\mathcal{M})$: }By regularity of $\mathcal{M}_{\mathbb{C}}$ in $\mathbb{R}^{n+2}$ a solution to the $n+2$-dimensional Laplace equation on $\mathcal{M}_{\mathbb{C}}$ exists and is in $H^{1,2}_0(\mathcal{M}_{\mathbb{C}})$. Choosing the parameterization $z=z(\tau)=i\tau= t$, we see that $p(t,\mathbf{x})=p''(z=t,\mathbf{x})\in H^{1,2}_0(\mathcal{M})$ because of independence of $p''(z=t,\mathbf{x})$ from $\Re [z]$.
\end{itemize}
In Fig. 3, the complexification process is depicted.
\begin{figure}
\begin{center}
\includegraphics[width = 0.67\textwidth]{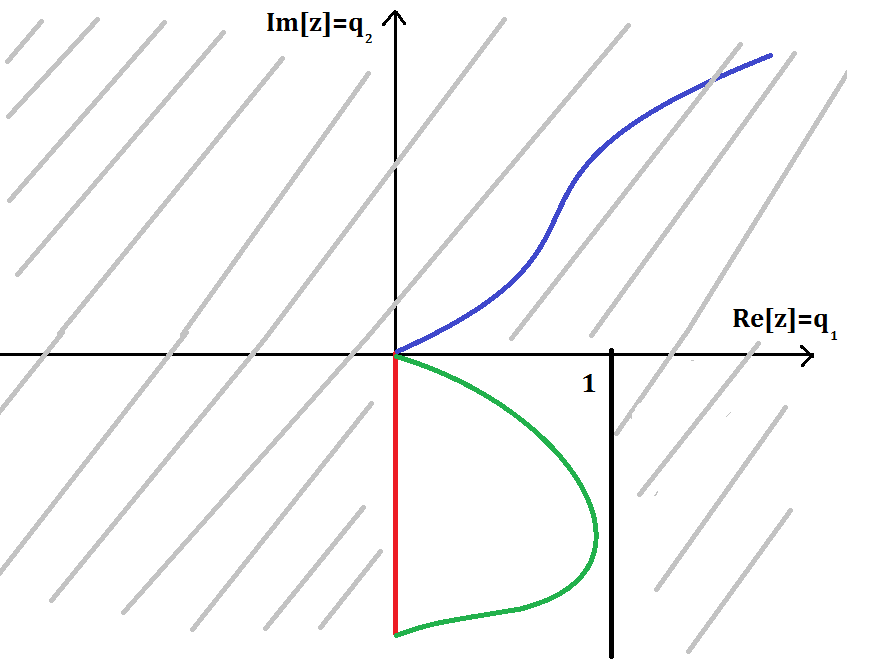}
\end{center}
\caption{The complexification process of $t$ is depicted. The Euclidean time $\tau$ lives on the lower imaginary semi-axis painted in red. The grey lines denotes what parts of $\mathbb{C}$ we want to exclude. The blue path is a time-path in the forbidden region, the path in green depicts a path in the the allowed region, starting at $t=0$ and ending asymptotically at $t=\infty$.}
\end{figure}
In the next paragraph, we notice two important properties of the Neumann-Laplacians $\Delta_g$. The $\Delta_g$'s are indeed Neumann because $n_\mu m^\mu=0$ s.t. the original Robin boundary conditions reduce to Neumann boundary conditions.\newline
\newline
\textbf{Properties of $\Delta_g$: }We comment on some of the properties of $\Delta_g$ on $\Omega_0$ for fixed $t\geq 0$, i.e., fixed $g=g(t)$. The case $g_0=g(t=0)$ is included.
\begin{itemize}
\item\textbf{Self-adjointness: }Firstly, $\Delta_g$ is formally self-adjoint on $H^{1,2}_0(\Omega_0)$. Let $\mathcal{S}_g=\mathcal{S}_g(\Omega_0\to\mathbb{C})$ be the Schwarz space of functions on $\Omega_0$,
\begin{align}
\mathcal{S}_g&=\left\lbrace f\in C^\infty(\Omega_0\to\mathbb{C})\vert\forall\vec{\alpha}\in\lbrace 1,2,...,n\rbrace^n\forall k\in\mathbb{N}_0\exists C_{\vec{\alpha},k}\in\mathbb{R}^+_0\right.\\
&\left.:\vert\partial_{\vec{\alpha}}f\vert \leq C_{\vec{\alpha},k}\left(1+\sqrt{x^\mu x_\mu}\right)^{-k}\right\rbrace .
\end{align}
We have the inclusions $C^\infty_0(\Omega_0\to\mathbb{C})\subset\mathcal{S}_g\subset L^2(\Omega_0\to\mathbb{C})$ where $C^\infty_0(\Omega_0\to\mathbb{C})$ denotes the space of smooth functions with compact support $\text{supp}(f)\equiv\overline{\lbrace\mathbf{x}\in\Omega_0=\text{D}(f): f(\mathbf{x})\neq 0\rbrace}\subset\Omega_0$. By definition of $L^2(\Omega_0\to\mathbb{C}),$ the space of smooth complex-valued functions on $\Omega_0$ is dense in $C^\infty_0(\Omega_0\to\mathbb{C})$. By the inclusion, we have that $\mathcal{S}_g$ is dense in $L^2(\Omega_0\to\mathbb{C})$. Assuming the reader's familiarity with $L^2$-spaces, we just note down the inner product on the $L^2(\Omega_0\to\mathbb{C})$ and sketch in a minimalist way the construction of the $L^2$-spaces. Let $f_1,f_2\in C^\infty_0(\Omega_0\to\mathbb{C})$,
\begin{align}
\langle f_1\vert f_2\rangle_{L^2_g}\equiv \int_{\Omega_0}d^nx\sqrt{\vert g\vert}\bar{f}_1(\mathbf{x})f_2(\mathbf{x}).
\end{align}
By setting $f_1=f_2$ and taking the square root of the $\langle f_1,f_1\rangle_{L^2_g}$, we obtain in a standard way a norm $\Vert f_1\Vert_{L^2_g}$ for $f_1\in C^\infty_0(\Omega_0\to\mathbb{C})$. The Lebesgue space $L^2(\Omega_0\to\mathbb{C})$ now originates upon completion of the normed (thus metric) space $(C^\infty_0(\Omega_0\to\mathbb{C}),d(f_1,f_2)=\Vert f_1-f_2\Vert_{L^2_g})$ w.r.t. $d$ and then forms a Banach space and by definition of $\langle ., .\rangle_{L^2_g}$ even a Hilbert space choosing only functions modulo functions the support of which is a Lebesgue-Borel null-set w.r.t. the Lebesgue-Borel measure $\text{Vol}_n(\text{supp}(f))\neq 0$ on $\Omega_0$. Now, we have to relate $\mathcal{S}_g$ to the Sobolev spaces $H^{1,2}_0(\Omega_0\to\mathbb{C})$. We notice that we can define a Fourier transformation on $\mathcal{S}$ using compactness of $\Omega_0$ and local trivializations. We define on $\mathcal{S}_g$ the norm $\Vert .\Vert_{1,2}$ by
\begin{align}
\Vert f\Vert_{1,2}^2 &= \int_{\Omega_0}d^nx\sqrt{\vert g\vert}\sum_{k=0}^1\sum_{\alpha\in\lbrace 1,...,n\rbrace^k}\vert(\partial_{\alpha} f)(\partial^\alpha f)\vert\\
&= \int_{\Omega_0}d^nx\sqrt{\vert g\vert}\left(\vert f\vert^2+\sum_{\alpha\in\lbrace 1,...,n\rbrace}\vert(\partial_{\alpha} f)(\partial^\alpha f)\vert\right) .
\end{align}
Since $H^{1,2}_0(\Omega_t\to\mathbb{C})$ originates from $C^\infty_0(\Omega_0\to\mathbb{C})$ by completion w.r.t. the usual $H^{1,2}_0$-norm and $\mathcal{S}_g\supset C^\infty_0(\Omega_0\to\mathbb{C}),$ we obtain $H^{1,2}_0(\Omega_0)$ as the completion of $\mathcal{S}_g$ w.r.t. to the Fourier transform of the usual $H^{1,2}_0$-norm. We notice that $\langle .,\rangle_{L^2_g}$ defines an antilinear isometric isomorphism between $\mathcal{S}_g$ and $\mathcal{S}^\ast_g=C^\infty_0(\mathcal{S}_g\to\mathbb{C})$ and thus between $H^{1,2}_0(\Omega_0\to\mathbb{C})$ and $H^{-1,2}(\Omega_t\to\mathbb{C})$ where $H^{-1,2}(\Omega_0\to\mathbb{C})$ is the completion of $\mathcal{S}^\ast_g$ w.r.t. $\Vert .\Vert_{-1,2}$ since by non-degeneracy of $L^2_g$ inner product, $H^{-1,2}_0(\Omega_0) = (H^{1,2}_0(\Omega_0))^\ast$. Now, let $\Delta_g:\mathcal{S}_g\to\mathcal{S}_g$ and
\begin{align*}
\langle f_1\vert\Delta_g f_2\rangle_{L^2_g}&=\int_{\Omega_0}d^nx\sqrt{\vert g\vert}\, \bar{f}_1\Delta_g f_2\\
&= \int_{\Omega_0}d^nx\sqrt{\vert g\vert}\bar{f}_1\sqrt{\vert g\vert}^{-1}\partial_\mu(\sqrt{\vert g\vert}g^{\mu\nu}\partial_\nu f_2)\\
&\stackrel{\text{i.b.p}^2}{=}\int_{\Omega_0}d^nx\sqrt{\vert g\vert}\sqrt{\vert g\vert}^{-1}\partial_\mu(\sqrt{\vert g\vert}g^{\mu\nu}\partial_\nu \bar{f}_1)f_2\\
&= \int_{\Omega_0}d^nx\sqrt{\vert g\vert}\overline{\sqrt{\vert g\vert}^{-1}\partial_\mu(\sqrt{\vert g\vert}g^{\mu\nu}\partial_\nu f_1)}f_2\\
&=\langle\Delta_g f_1\vert f_2\rangle_{L^2_g}\\
&=\langle\Delta_g^\ast f_1\vert f_2\rangle_{L^2_g},
\end{align*}
i.e. $\Delta_g=\Delta_g^\ast$ on $\mathcal{S}_g$. By completion w.r.t. $\Vert .\Vert_{1,2}$ we have even $\Delta_g=\Delta_g^{\ast}$ with $\Delta_g: H^{1,2}_0(\Omega_0\to\mathbb{C})\to H^{-1,2}(\Omega_t\to\mathbb{C})\simeq H^{1,2}_0(\Omega_0\to\mathbb{C})$ by the previous derivation and $\Delta_g^\ast: H^{-1,2}_0(\Omega_0\to\mathbb{C})\to H^{1,2}_0(\Omega_t\to\mathbb{C})$.
\item\textbf{Perturbations of the spectrum $\sigma(\Delta_g)$: }The second property concerns an estimate how much the spectrum $\sigma(\Delta_g)$ deviates from $\sigma(\Delta_{g^{(0)}})$ in the sense that if $-\lambda_n(t),\,\lambda_n$ denotes the $n$-th eigenvalue of $\Delta_g,\,\Delta_{g^{(0)}}$ on $\Omega_0$ or equivalently in terms of diffeometries of the $n$-th eigenvalue of $\Delta_{g^{(0)}}$ on $\Omega_0$ and $\Omega_t$ we ask of what order in the perturbation strength $\epsilon$ $\lambda_n(t)-\lambda_n$ scales. For this purpose, define the Dirichlet-Laplacians $\Delta_g^D$ and $\Delta^D_{g^{(0)}}$ on $\Omega_t$ and $\Omega_0$. By assumption, $\Omega_0$ and $\Omega_t$ are in particular simply-connected and compact in $\mathbb{R}^n$, i.e.,domains. Now suppose we can find $\Omega_-,\Omega_+\subset\mathbb{R}^n$ such that $\Omega_-\subsetneq\Omega_t\subsetneq\Omega_+$ for all $t\geq 0$ and that $\Omega_-$ and $\Omega_+$ are at least $C^2$-bounded domains in $\mathbb{R}^n$. By domain monotonicity of the eigenvalues for the Dirichlet-Laplacians, we have for the Dirichlet eigenvalues $-\lambda_{n,t}^D$ of $\Delta_{g}$,
\begin{align}
\lambda_{n,-}^D\geq \lambda_{n,t}^D \geq \lambda_{n,+}^D
\end{align}
for all $n\in\mathbb{N}$. Unfortunately, this property does not carry over to the Neumann-Laplacian. Consider e.g. the Neumann-Laplacian $\partial_x^2+\partial_y^2$ on $\Omega_1:=[0,a]\times[0,b]\subset\mathbb{R}^2$ with $a>b$ and the Neumann Laplacian on $\Omega_2:=[0,1]^2\subset\mathbb{R}^2$. If we let $1 < a < \sqrt{2}$ and $b$ sufficently small, we can use an affine transformation to achieve $\Omega_1\subset\Omega_2$. However, for the eigenvalues $\lambda_{n,m;1}$ and $\lambda_{n,m;2}$ of the Neumann Laplacians on $\Omega_1$ and $\Omega_2$,
\begin{align}
\lambda_{n,m;1}=\dfrac{n^2\pi^2}{a^2}+\dfrac{n^2\pi^2}{b^2}\text{ and }\lambda_{n,m;2}=n^2\pi^2+m^2\pi^2 .
\end{align}
Then by the assumption $\Omega_1\subset\Omega_2$ the first $3$ eigenvalues of the two Neumann-Laplacians read,
\begin{align}
\lambda_{0;1}=0,\,\lambda_{1;1}=\pi^2 a^{-2},\,\lambda_{2;1}=\pi^2 b^{-2}\text{ and }\lambda_{0;2}=0,\, \lambda_{1;2}=\pi^2,\, \lambda_{2;2}=\pi^2.
\end{align}
Then we have $\lambda_{1;1}<\lambda_{1;2}$ but $\lambda_{2;1}>\lambda_{2;2}$. Thus, domain monotonicity does - in general - not hold for the Neumann-Laplacians. We note, there is a weaker version of domain monotonicity. Namely, we have for all $k\in\mathbb{N}$ that there is a $k$-independent constant $c\geq 1$ such that $\lambda_k^D < \lambda_k < c\cdot\lambda_k^D$. I.e., the $k$-th Neumann eigenvalue is quenched between the $k$-th Dirichlet eigenvalue and its multiple by $c$. The currently existent estimates on Neumann eigenvalues and Dirichlet eigenvalues are stronger than we need them. We approach the spectral properties of $\Delta_g$ by means of the perturbation formalism: Firstly, Lichnerowicz' theorem guarantees the existence of eigenvalues $\lbrace\lambda_{n}(t)\rbrace_{n\in\mathbb{N}}$ for the Neumann Laplacian $\Delta_g$ on $\Omega_0$ for a fixed time $t$ and corresponding complete set of eigenfunctions $\lbrace\Psi_{n}(t,\mathbf{x})\rbrace_{n}$ again for fixed $t$. Since $\Omega_0$ is by means of the imbedding property $C^2$ bounded domain in $\mathbb{R}^n$ for all $t\geq 0$, already the expansion theorem in Euclidean space tells us that we can the $\lbrace\Psi_{n}(t)\rbrace_{n\in\mathbb{N}}$ to be orthonormal Suppressing the spatial argument, at all times $t$, we can write in Dirac notation
\begin{align}
\Delta_g = -\sum_{n\in\mathbb{N}}\vert\Psi_{n}(t)\rangle\lambda_n(t)\langle\Psi_n(t)\vert
\end{align}
It is further known that $\lbrace(\lambda_{n}(t),\Psi_n(t))\rbrace_{n\in\mathbb{N}}$ depend at least continuously on the geometry of $\Omega_t$. But $\Omega_t=\psi_{0\to t}(\Omega_0)$, i.e., all $t$-dependencies of the eigenfunctions stem form $\psi_{0\to t}$. Using mollification of the eigenvalues and eigenfunctions, we can assume this dependence to be $C^\infty$ because $C^\infty_0(\Omega_t)$ is dense in $H^{1,2}_0(\Omega_t)$ for all $t$ and $\Omega_t=\psi_{0\to t}(\Omega_0)$ is smooth such that $C^\infty_0(\Omega_t)=psi_{0\to t}^\ast C^\infty_0(\Omega_0)$ by properness of $\psi_{0\to t}$. Now, $\Psi_n(t,\mathbf{x})$ still lives in a Sobolev space over $\Omega_t$. By means of pulling $\Psi_{n}(t)$ back, we obtain functions on $\Omega_0$,
\begin{align}
\Psi_{n}(\psi_{0\to t},\psi_{0\to t}(\mathbf{x}))
\end{align}
The additional $\psi_{0\to t}$-argument reflects the fact that not only arguments $\mathbf{x}\in\Omega_t$ need to be transformed by $\psi_{0\to t}$ in general also normalization constants and eigenfrequencies which are constants in $\Omega_t$ but depend on $t$ nonetheless. Below, we will clarify this point in an example. Using the regularity considerations from mollification, we can apply Taylor's theorem and expand around $\delta\psi_{0\to t}=\psi_{0\to t}-\text{id}_{\Omega_0}$. Likewise, we can expand into eigenfunctions $\lbrace\Psi_n(\mathbf{x})\rbrace_{n\in\mathbb{N}}$ on $\Omega_0$ because this set of eigenfunctions is complete on $\Omega_0$. In total we have two representation for $\Psi_{n}(t)$ pulled back to $\Omega_0$,
\begin{align}
\Psi_{n}(\psi_{0\to t},\psi_{0\to t}(\mathbf{x})) &= \sum_{m\in\mathbb{N}}c_m\Psi_{m,0}\\
\Psi_{n}(\psi_{0\to t},\psi_{0\to t}(\mathbf{x})) &= \sum_{k\in\mathbb{N}_0}(k!)^{-1}\partial_{\mathbf{x}}\Psi_{n}(\psi_{0\to t},\psi_{0\to t})\vert_{\delta\psi_{0\to t}=\mathbf{0}}(\mathbf{x})\delta\psi_{0\to t}.
\end{align}
Equating and comparing coefficients, we see that $\Psi_{n}(\psi_{0\to t},\psi_{0\to t}(\mathbf{x}))=\Psi_{n}(\mathbf{x})+\mathcal{O}(\delta\psi_{0\to t})$. We have $\Vert\delta\psi\Vert_{\infty}<\epsilon$. Thus, as a starting point for perturbation theory, we can simply choose 
\begin{align}
\Psi_{n}(t,\mathbf{x})=\Psi_{n}(\psi_{0\to t}^{-1}(\mathbf{x}))+\mathcal{O}(\epsilon)
\end{align}
This equation means that for sufficiently small deviations of $\Omega_0$ from $\Omega_t$ in $\mathbb{R}^n$, the time-dependencies of $\Psi_{n}(t)$ as elements of $H^{1,2}_0(\Omega_0)$ are exclusively due to pulling back by means of the diffeomorophism $\psi_{0\to t}:\Omega_0\to\Omega_t$. What does this imply on the eigenvalues of $\Delta_g$? By Lichernowicz' theorem, we further have
\begin{align*}
\lambda_n(t)&=\langle\Psi_n(t)\vert\Delta_g\vert\Psi_n(t)\rangle_g\\
&=\int_{\Omega_0}d^nx\sqrt{\vert g\vert}\bar{\Psi}_n(t)\Delta_g\Psi_{n}(t)\\
&=\int_{\Omega_0}d^nx\sqrt{\vert g_0\vert}\bar{\Psi}_n\Delta_g\Psi_n+\mathcal{O}(\epsilon)\\
&=\int_{\Omega_0}d^nx\sqrt{\vert g_0\vert}\bar{\Psi}_n\left(\Delta_{g^{0}}+\left(\Delta_{g}-\Delta_{g^{(0)}}\right)\right)\Psi_n\\
&= \int_{\Omega_0}d^nx\sqrt{\vert g_0\vert}\bar{\Psi_n}\Delta_{g^{(0)}}\Psi_n + \int_{\Omega_0}d^nx\sqrt{\vert g\vert}\bar{\Psi}_n\mathsf{V}\Psi_n\\
&= \lambda_{n}+ \int_{\Omega_0}d^nx\sqrt{\vert g_0\vert}\bar{\Psi}_n\mathsf{V}\Psi_n
\end{align*}
Rearranging the equation, we find
\begin{align}
\delta\lambda_{n}\equiv \lambda_{n}(t)-\lambda_n = \int_{\Omega_0}d^nx\sqrt{\vert g_0\vert}\bar{\Psi}_n\mathsf{V}\Psi_n. = \langle\Psi_n\vert\mathsf{V}\vert\Psi_n\rangle_{g^{(0)}}.
\end{align}
If we manage to show that $\langle\Psi_n\vert\mathsf{V}\vert\Psi_n\rangle_{g^{(0)}}\sim\mathcal{O}(\epsilon)$ (or better), we have not only found a perturbation theory, but also shown that \emph{Small changes in the fibers $\Omega_t$ from the reference fiber $\Omega_0$ lead to small changes in the spectra of the Neumann-Laplacians $\Delta_g$ on $\Omega_0$ and $\Delta_{g^{(0)}}$ on $\Omega_0$}. A similar result has already been sketched in a less general setting than ours in [DengLi].
\end{itemize}
We turn to the investigation of $\Delta_{G_0}$, the Robin-Laplacian on $\mathcal{M}$.
\newline
\newline
\textbf{Self-adjointness of $\Delta_{G_0}$: }The Robin-Laplacian $\Delta_{G_0}$ satisfies self-adjointness of $H^{1,2}(\mathcal{M})$. Since we have already proven that $\Delta_g^0$ is self-adjoint on $\Omega_t$ by diffeometric equivalence $(\Delta_g,\Omega_0)\simeq(\Delta_{g^0},\Omega_t)$, we can simply take the expression $\partial_t^2-\Delta_{g_0}$ absorbing again the $c^2$ factor into $\Delta_{g_0}$ and construct the Sobolev-spaces $H^{1,2}_0(\mathcal{M})$ and $H^{-1,2}(\mathcal{M})$ just as we did before. As we have seen before, the boundary conditions to $\Delta_{G_0}$ are equivalent to the boundary and initial conditions chosen for the wave equation on $\mathcal{M}$. Since the self-adjointness is verified by considering only the homogeneous initial and boundary conditions, we can use integration by parts freely. We let $F_1\in H^{1,2}(\mathcal{M}),F_2\in H^{-1,2}(\mathcal{M})\simeq H^{1,2}(\mathcal{M})$ with the antilinear isometric isomorphy $\simeq$ valid again because of the non-degeneracy of the $L^2_G$ inner product on $L^2(\mathcal{M}\to\mathbb{C})$ and calculate
\begin{align*}
\langle F_1\vert\Delta_{G_0} F_2\rangle_{L^2_{G_0}}&=\int_{\mathcal{M}}d^nx\sqrt{\vert G_0\vert}\, \bar{F}_1\Delta_{G_0} F_2\\
&= \int_{\mathcal{M}}d^nx\sqrt{\vert G_0\vert}\bar{F}_1\sqrt{\vert G_0\vert}^{-1}\partial_I(\sqrt{\vert G_0\vert}G_0^{IJ}\partial_J f_2)\\
&\stackrel{\text{i.b.p}^2}{=}\int_{\Omega_t}d^nx\sqrt{\vert G_0\vert}\sqrt{\vert G_0\vert}^{-1}\partial_I(\sqrt{\vert G_0\vert}G_0^{IJ}\partial_J \bar{F}_1)F_2\\
&= \int_{\mathcal{M}}d^nx\sqrt{\vert G_0\vert}\overline{\sqrt{\vert G_0\vert}^{-1}\partial_I(\sqrt{\vert G_0\vert}G_0^{IJ}\partial_J F_1)}F_2\\
&=\langle\Delta_{G_0} F_1\vert F_2\rangle_{L^2_{G_0}}\\
&=\langle\Delta_{G_0}^\ast F_1\vert F_2\rangle_{L^2_{G_0}}.
\end{align*}
In total, we have established self-adjointness, $\Delta_{G_0} = \Delta_{G_0}^\ast$ on $H^{1,2}(\mathcal{M})$.\newline
\newline
\textbf{Perturbation theory: }The previously demonstrated existence of solutions is nice, but not adequate for practical physical calculations.
\begin{itemize}
\item\textbf{The geometry of $\Omega_0$: }Let $\Omega_0=\lim_{t\to 0^+}\Omega_t=\lim_{t\to\infty^-}\Omega_t$ be an imbedded $n$-dimensional orientable, diffeotopic, retractible,compact and smooth submanifold of $\mathbb{R}^n$ with smooth $(n-2)/2$-connected, and orientable boundary $\partial\Omega_0$ such that $\partial(\partial\Omega_0)=\emptyset$ and define the following trivial bundle $\mathcal{M}_0=\mathbb{R}^+\times\Omega_0$ over $\mathbb{R}^+$. There is, as explained above, a diffeomorphic proper imbedding $\psi_0:\Omega_0\to B^n_1(\mathbf{0})\subsetneq\mathbb{R}^n$. Further, the manifold structure on $\mathbb{R}^+$ and $\Omega_0$ give rise to a product manifold structure on $\mathcal{M}$. Thus, the Minkowskian metric $\eta$ on $\mathbb{M}_{n+1,+}$ can be pulled back by means of the canonical inclusion $B^n_1(\mathbf{0})\hookrightarrow \mathbb{R}^n$ to a metric $\eta\vert_{\mathcal{M}_{ref}}$. This metric in turn can be pulled back by means of $\Phi_0=(\text{id}_{\mathbb{R}^+}\times\psi_0):\mathcal{M}_{ref}\to\mathcal{M}_0$ to a metric $G_0\in (T\mathcal{M}\times T\mathcal{M})^{\ast}$ which has the structure
\begin{align}
G_0 = \left(\begin{array}{cc}-dt^2 & \mathbf{0}^T_n\\
\mathbf{0}_n & g_0\end{array}\right),
\end{align}
with $g_0$ the pull-back (via $\psi_0$) of the induced metric on $B^n_1(\mathbf{0})$, i.e.,
\begin{align}
g_0 = (\psi_0^\ast(dr))^2 + (\psi_0^{-1}(r))^2\psi_0^\ast(d\Omega_{n-1}).
\end{align}
The symbol $d\Omega_{n-1}$ denotes again the metric tensor on $S^n_1(\mathbf{0})$.
\item\textbf{Relating $\Omega_0$ to $\Omega_t$: }Since $\psi_0:\Omega_0\to B^n_1(\mathbf{0})$ and $\psi_t:\Omega_t\to B^n_1(\mathbf{0})$ are diffeomorphism and thus invertible, $\psi_{0\to t}=\psi_t^{-1}\circ\psi_0:\Omega_0\to\Omega_t$ is again a diffeomorphism such that $\psi_{0\to t}(\partial\Omega_0)=\partial\Omega_t$ because $\psi_0,\psi_t$ are proper. We assume that the $\Omega_t$'s differ \"only slightly" from $\Omega_0$ for all $t>0$ where the attribute \" only slightly" will be made precise below. \item\textbf{Wave equations on $\mathcal{M}$ and $\mathcal{M}_0$: }We seek to compare the wave equation of $\mathcal{M}$
\begin{align}
\partial_t^2 p - c^2\Delta_{g_{0}} p = 0,
\end{align}
with Neumann boundary conditions on $\partial\mathcal{M}$, $n^\mu\partial_\mu p = f\circ(\text{id}_{\mathbb{R}^+}\times\psi^{-1}_{0\to t}$ for $f\in L^2(\mathcal{M}_0))$ to the equation
\begin{align}
\partial_t^2p' - c^2\Delta_{g^{(0)}}p' =0,
\end{align}
with Neumann boundary conditions $n^\mu\partial_\mu p = f$ on $\partial\mathcal{M}_0$ for $f\in L^2(\mathcal{M}_0))$. The initial conditions for the two equations shall stay the same as well: Firstly, $p(t=0,\mathbf{x})=0=p'(t=0,\mathbf{x}')$  and secondly $\partial_t p(t=0,\mathbf{x})=0=\partial_t p'(t=0,\mathbf{x'})$ for $\mathbf{x}\in\Omega_t$ and $\mathbf{x'}\in\Omega_0$. Since we know $\mathcal{M}$, we can solve the last equation.
\item\textbf{Feasibility of a perturbation theory: }The question is how to obtain a perturbative solution to the first equation from the known solution to the second one? In order to answer the question, we have to problems to fix: Firstly, $\mathcal{M}\neq\mathcal{M}_0$ such that for a comparison we need to ensure that both wave equations live on the same bundle, preferably $\mathcal{M}_0$ by its product structure, $\mathcal{M}_0=\mathbb{R}^+\times\Omega_0$. Secondly, we need to specify comparison. Effectively, the only algebraic-symbolic difference in the two equations is the time dependent $\Delta_{g}$ in the first equation and the time-independent $\Delta_{g^{(0)}}$ in the second equation. However $\mathsf{V}_{try}=\Delta_g-\Delta_{g^{0}}$ is not sensible because the operator can at most be defined on the intersection of domains of $\Delta_g,\Delta_{g^{(0)}}$, i.e., $\text{Dom}(\mathsf{V}_{try})=\text{Dom}(\Delta_g)\cap\text{Dom}(\Delta_{g^{(0)}})$. This definition finally can be excluded because it imposes high restrictions on the functions that $\mathsf{V}_{try}$ could operate on.
\item\textbf{Definition of the perturbation operator $\mathsf{V}$: }We want to emphasize some properties of the metric tensors $g,g_0\in(T^2\Omega_0)^\ast$. Notice that by construction the metric $g$ is invariant w.r.t. pull-backs by means of $\psi_{0\to t}$. We have
\begin{align}
g_0(X,Y)=\psi_{0\to t}^\ast(g_0)(\psi_{0\to t,\ast}X,\psi_{0\to t,\ast}Y) =g(d\psi_{0\to t}\circ X,d\psi_{0\to t}Y)= g(X(t),Y(t)),
\end{align}
where $X,Y\in T\Omega_0=\mathcal{V}(\omega_0)$ and $X(t)=d\psi_{0\to t}\circ X = \psi_{0\to t,\ast}(X)\in T\Omega_t=\mathcal{V}(\Omega_t),\, Y(t)=d\psi_{0\to t}\circ Y = \psi_{0\to t,\ast}(Y)\in T\Omega_t=\mathcal{V}(\Omega_t)$. By definition, we have
\begin{align*}
g_0=g_{0,\mu\nu}dx^\mu dx^\nu &= (\psi_{0}^\ast(dr))^2+ (\psi_{0}^{-1}(r))^2\psi_{0}^\ast d\Omega_{n-1}\\
g=g_{\mu\nu}dx^\mu dx^\nu &= (\psi_{t}^\ast(dr))^2+ (\psi_{t}^{-1}(r))^2\psi_{t}^\ast d\Omega_{n-1}\\
&= ((\psi_{0}\circ\psi_{0\to t})^\ast(dr))^2+(\psi_{0\to t}^{-1}(\psi_0^{-1}(r)))^2(\psi_{0}\circ\psi_{0\to t})^\ast d\Omega_{n-1}\\
&= \psi_{0\to t}^\ast((\psi_0^\ast(dr))^2)+(\psi_{0\to t}^{-1}(\psi_0^{-1}(r)))^2(\psi_{0\to t}^\ast(\psi_0^\ast)d\Omega_{n-1})\\
&= \psi_{0\to t}^\ast (g_{0,\mu\nu}dx^\mu dx^\nu)\\
&= \psi_{0\to t}^\ast g_0
\end{align*}
Notice that the diffeomorphisms $\psi_0$ and $\psi_t$ originated from pulling back the foliation of $B^n_1(\mathbf{0})$ by spheres $S^n_{1-s}(\mathbf{0}),\, 0\leq s < 1$ to $\Omega_0$ by means of $\phi^{-1}_0: S_1^n(\mathbf{0})\to \partial\Omega_0$ and to $\Omega_t$ by means of $\psi_t^{-1}: S^n_1(\mathbf{0})\to \partial\Omega_0$, leaving the "radial variable" $s$ invariant. However, we can further foliate $\mathbb{R}^n\setminus\lbrace\mathbf{0}\rbrace$ by letting $-\infty<s<1$. We can express the Euclidean metric $\delta$ in spherical coordinates and exclude a Lebesgue null-set w.r.t. to the Lebesgue-Borel integration measure $\text{Vol}_n:\mathcal{B}(\mathbb{R}^n)\to\mathbb{R}^+_0$ and call the resultant metric $g_{\mathbb{R}^n}$. Letting $-\infty < s < 1$ in $\psi_0$, we can foliate not only $\Omega_0\setminus\lbrace\mathbf{0}\rbrace$ by means of $\partial\Omega_0$ but also $\mathbb{R}^n$ and accordingly letting $-\infty < s <1$ in $\psi_t$, we can foliate $\mathbb{R}^n\setminus\lbrace\mathbf{0}\rbrace$ by means of $\partial\Omega_t$. These extension of the domain of definition of $\psi_0,\psi_t$ to $-\infty < s <1$ allows us to define the metric $g_0$ on $\mathbb{R}^n\setminus\lbrace\mathbf{0}\rbrace$. This gives $g_0 = \psi_0^\ast(g_{\mathbb{R}^n\setminus\lbrace 0\rbrace})\in (T^2\mathbb{R}^n\setminus\lbrace\mathbf{0}\rbrace)^\ast = \mathbb{R}^n\setminus\lbrace\mathbf{0}\rbrace\times\mathbb{R}^n\setminus\lbrace\mathbf{0}\rbrace\to\mathbb{R}$ as an inner product on the imbedding space of $\Omega_0\setminus\lbrace\mathbf{0}\rbrace$ and $\Omega_t\setminus\lbrace\mathbf{0}\rbrace$, $\mathbb{R}^n\setminus\lbrace\mathbf{0}\rbrace$. We are now ready to leave the realm of \emph{intrinsic geometry}. Since the notion of perturbation bundles abstracts the setting of the ICE model, we take the perspective of the experimentator observing the evolution of the gecko's interaural cavity in time. The experimentator \"measures" the function $t\mapsto\Omega_t$. The perturbations itself, i.e., the fact that $\Omega_t\neq\Omega_{t'}$ for general $t\neq t'$, is observed w.r.t. the length measurement device of the experimentator, i.e., the reference metric tensor $g_0$ which is isometrically equivalent to the flat Euclidean metric $\delta$ on $\mathbb{R}^n$ modulo exclusion of Lebesgue null-sets w.r.t. the Lebesgue-Borel-measure $\text{Vol}_n:\mathcal{B}(\mathbb{R}^n)\to\mathbb{R}^+_0$. Thus, we equip in the experimentator frame the perturbation bundle with the metric tensor $G_0=-dt^2+g_0$ with the advantage of having a constant metric but a variable domain. On the other hand, we can use the $\psi_{0\to t}'s$ on the fiber level or the $\Phi^{-1}\circ\Phi_0:\mathcal{M}_0\to\mathcal{M}$ on the bundle level to pull $G_0$ back. This gives us $G=-dt^2+g$ where $g=\psi_{0\to t}^\ast g_0$ by means of the diffeomorphism $\psi_0,\psi_t$ continued smoothly to $\psi_0,\psi_t:\to\mathbb{R}^n\setminus\lbrace\mathbf{0}\rbrace\to\mathbb{R}^n\setminus\lbrace\mathbf{0}\rbrace$ such as to allow pulling back an object defined on $\Omega_t\hookrightarrow\mathbb{R}^n$. Our metric $g$ is now time-dependent, however it is defined on the same $\Omega_0$ and expressible in terms of local coordinates $\lbrace x^\mu\rbrace_{1\leq \mu\leq n}$ on $\Omega_0$. We emphasize that in the context of perturbation bundles, the perturbation is not a simple change of coordinates which leaves the metric tensor invariant. For the perturbation bundle theory we have the following identifications by means of isometries $\psi_{0\to t}:\Omega_0\to\Omega_t$ of pairs of $n$-dimensional Riemannian manifolds,
\begin{align}
(\Omega_0,g_0)\simeq(\Omega_t,g)\text{ and }(\Omega_0,g)\simeq (\Omega_t,g_0).
\end{align} 
We notice that the pairs $(\Omega_0,g_0)$ and $(\Omega_0,g)$ are not equivalent, the latter being diffeometric by means of the global diffeometry\footnote{A diffeometry is an isometric diffeomorphism between Riemannian manifolds. The attribute diffeometric means correspondingly isometrically diffeomorphic.} $\psi_{0\to t}:\Omega_0\to\Omega_t$. Thus, we can define a metric perturbation
\begin{align}
\delta g = g-g_0 = (\psi_{0\to t}-\text{id}_{\Omega_0})^\ast g_0.
\end{align}
We now have a Laplace-Beltrami operators solely living on $\Omega_0$, namely $\Delta_g\in H^{1,2}_0(\Omega_0)\to L^2(\Omega_0)$ and $\Delta_{g_0}: H^{1,2}_0(\Omega_0)\to L^2(\Omega_0)$ which are self-adjoint w.r.t. the $L^2$ inner products $L^2_{g^0}$ and $L^2_g$ defined by means of the metric tensors $g_0$ and $g$ on $\mathcal{M}$. We are now ready to expand $\Delta_g$. Let $\lbrace x^\mu\rbrace_{1\leq \mu\leq n}$ be a fixed local coordinate system on $\Omega_0$ and assume that $g_0, g$ are expressed in these coordinates. Let further $f\in H^{1,2}_0(\Omega_0,h)$ for the metric tensor $h\in\lbrace g,g_0\rbrace$. By the smallness constraint $\Omega_+\supsetneq\Omega_t,\Omega_0\supsetneq\Omega_-$ and the above two pairs of diffeometrically equivalent Riemannian manifolds, the Sobolev spaces are indifferent to whether $h=g_0$ or $h=g$, they are identical.
\begin{align*}
\Delta_gf &= \sqrt{\vert g\vert}^{-1}\partial_\mu\left(\sqrt{\vert g\vert}g^{\mu\nu}\partial_\nu f\right)\\
&= \sqrt{\vert g_0+\delta g\vert}^{-1}\partial_\mu\left(\sqrt{\vert g+\delta g\vert}(g^{\mu\nu}+\delta g^{\mu\nu}\partial_\nu f\right)\\
&= \sqrt{\vert g_0\vert}^{-1}\partial_\mu(\sqrt{\vert g_0\vert}g_0^{\mu\nu}\partial_\nu f)+ \sqrt{\vert g_0\vert}^{-1}\partial_\mu(\sqrt{\vert g_0\vert}\delta g^{\mu\nu}\partial_\nu f)\\
&+ \dfrac{-1}{2}\sqrt{\vert g_0\vert}^{-1}\text{Tr}(g_0^{-1}\delta g)\partial_\mu(\sqrt{\vert g_0\vert} g^{\mu\nu}_0\partial_\nu f)+  \dfrac{1}{2}\sqrt{\vert g_0\vert}^{-1}\partial_\mu(\sqrt{\vert g_0\vert}\text{Tr}(g_0^{-1}\delta g)g_0^{\mu\nu}\partial_\nu f)+ \mathcal{O}((\delta  g)^2)\\
&= \Delta_{g_0}f + \dfrac{1}{2}\partial_\mu\text{Tr} (g_0^{-1}\delta g)g_0^{\mu\nu}\partial_\nu f + \sqrt{\vert g_0\vert}^{-1}\partial_\mu(\sqrt{\vert g_0\vert}\delta g^{\mu\nu}\partial_\nu f) + \mathcal{O}((\delta  g)^2\\
&= \Delta_{g_0} + \mathsf{V}_{1}[f] \mathcal{O}((\delta g)^2).
\end{align*}
We have defined
\begin{align}
\mathsf{V}_{1}[f] = \dfrac{1}{2}\partial_\mu\text{Tr} (g_0^{-1}\delta g)g_0^{\mu\nu}\partial_\nu f + \sqrt{\vert g_0\vert}^{-1}\partial_\mu(\sqrt{\vert g_0\vert}\delta g^{\mu\nu}\partial_\nu f) .
\end{align}
We recall the definition of the pertubation operator $\mathsf{V}: H^{1,2}_0(\Omega_0)\to H^{1,2}_0(\Omega_0).$ The above calculation shows that we can expand $\mathsf{V}$ in powers of the more \"{}haptic\"{} metric perturbation $\delta g$,
\begin{align}
\mathsf{V}=\sum_{k=1}^{\infty}\mathsf{V}_k
\end{align}
where $\mathsf{V}_k$ contains the metric perturbation $\delta g$ to the power $k$. If we include higher contributions, i.e., considered the tensorial Taylor expansion of the tensor $g$ around $\delta g$ to higher than first order, we can derive higher order correction terms. Since our goal is to stay in lowest non-trivial order, we are allowed to truncate the expansion in first order in $\delta g$
\begin{align}
\mathsf{V}&=\mathsf{V}_1 + \mathcal{O}((\delta g)^2)\\
&= \dfrac{1}{2}\partial_\mu\text{Tr} (g_0^{-1}\delta g)g_0^{\mu\nu}\partial_\nu f + \sqrt{\vert g_0\vert}^{-1}\partial_\mu(\sqrt{\vert g_0\vert}\delta g^{\mu\nu}\partial_\nu f) + \mathcal{O}((\delta g)^2).
\end{align}
We notice an interesting phenomenon. The time-dependencies of $\Delta_g$ are now exclusively stored in $\mathsf{V}_1$. Abbreviating $\mathsf{V}\equiv\mathsf{V}_1$ and indicating the time dependence of the operator $\mathsf{V}=\mathsf{V}(t)$, the acoustic wave equation that we seek to solve turns into
\begin{align}
\partial_t^2p -c^2\Delta_{g_0}[p] = c^2\mathsf{V}(t)[p] + \left(\text{source term}\right).
\end{align}
Let us further choose the local coordinate system $\lbrace x^\mu\rbrace_{1\leq \mu\leq n}$ such that $g_0$ is diagonal. By symmetry of $g_0$ this possible. We denote the coordinate with associated coordinate field $\partial_s$ normal to $T_\mathbf{y}\partial\Omega_0$ by $s$. By invariance of $U=\langle\mathbf{n}_{(1-s)\partial\Omega_0}{\mathbf{x}},(\psi_{0\to t}-\text{id}_{\Omega_0})(\mathbf{x}))\rangle_{g_0}$ as an intrinsic quantity under re-parameterizations, we have
\begin{align}
\text{Tr}(g_0^{-1}\delta g) &= g^0_{ss}\partial_\lambda U\partial^\lambda U
\end{align}
The perturbation operator becomes even more explicit in terms of $U$,
\begin{align*}
\mathsf{V}\left[p\right]&=\dfrac{1}{2}\partial_\mu\left(g^0_{ss}\partial_{\lambda}U\partial^\lambda U\right)g_0^{\mu\nu}\partial_\nu p + \delta g_{\mu\nu}\sqrt{\vert g_0\vert}\partial^{\mu}(\sqrt{\vert g_0\vert}\partial^\nu p) + \partial^\mu\delta g_{\mu\nu}\partial^\nu p\\
&= \dfrac{1}{2}\partial_\mu\left(g^0_{ss}\partial_{\lambda}U\partial^\lambda U\right)g_0^{\mu\nu}\partial_\nu p + g^0_{ss}\partial_{\mu}U\partial_\nu U\sqrt{\vert g_0\vert}^{-1}\partial^\mu(\sqrt{\vert g_0\vert}\partial^\nu p)\\
&+ \partial^\mu(g^0_{ss}\partial_\mu U\partial_\nu U)\partial^\nu p .
\end{align*}
Since $\partial_s U = - u$ with $u=\langle\mathbf{n}_{\partial\Omega_0},\phi_{0\to t}-\text{id}_{\partial\Omega_0}\rangle_{g_0}$ and in general $U=u(1-s)$ we can obtain an even more explicit and longer version of $\mathsf{V}$ in terms of the boundary vibrations $u=u(t,\mathbf{y})$. We continue assuming $g_0$ is in diagonal form,
\begin{align*}
\mathsf{V}\left[p\right] &= \dfrac{1}{2}\partial_i(g^0_{ss}u^2)g_0^{ij}\partial_j p + \dfrac{1}{2}\partial_\mu(g^0_{ss}(1-s)^2\partial_k u\partial^k u)g^0_{\mu\nu}\partial_\nu\\
&+g^0_{ss}u^2\sqrt{\vert g_0\vert}^{-1}\partial^s(\sqrt{\vert g_0\vert}\partial^s p) - g^0_{ss}(1-s)u\partial_i u\sqrt{\vert g_0\vert}^{-1}\partial^{(i}(\sqrt{\vert g_0\vert}\partial^{s)}p)\\
&+ g^0_{ss}(1-s)^2\partial_i u\partial_j u\sqrt{\vert g_0\vert}^{-1}\partial^i(\sqrt{\vert g_0\vert}\partial^j p)\\
&+\partial^s(g^0_{ss}u^2)\partial^s p - \partial^{(s}(g^0_{ss}(1-s)u\partial_i u)\partial^{i)}p+\partial^i(g^0_{ss}(1-s)^2\partial_i u\partial_j u)\partial^j p
\end{align*}
again valid in first order in $\delta g$. The first line is an expanded version of the first contribution in the previous expression of $\mathsf{V}$ in terms of $U$. The second and third line is the expanded version of the second contribution in the expression of $\mathsf{V}$ in terms of $U$. The fourth line is the expansion of the third contribution to $\mathsf{V}$ with $U$ expressed in terms of $u$. We recall the definition of the symmetrization operation $A_{(ij)}=(1-0.5\delta_{ij})A_{ij}+A_{ji}$. Further, we have recalled that $u$ is independent of $s$ since $s$ is the coordinate transversal to $\partial\Omega_0$ in $\mathbb{R}^n$ endowed with the metric $g_0$! This form of the perturbation operator shows the coupling of the vibrations of the boundaries $t\to \partial\Omega_t$ to the acoustic wave equation. We note for the sake of completeness that in the $G_0$ reference frame on $\mathcal{M}$ they are parameterized as follows
\begin{align}
\partial\Omega_0\to\partial\Omega_t:\mathbf{y}\mapsto\text{id}_{\partial\Omega_0}(\mathbf{y})+\mathbf{n}_{\partial\Omega}(\mathbf{y}) u(t,\mathbf{y})
\end{align}
The perturbation operator is quadratic in $u$ and its derivatives. Further, we can obtain the rude bound on the norm of $\mathsf{V}$ relative to the unperturbed Neumann-Laplacian $\Delta_{g_0}$,
\begin{align}
\Vert\mathsf{V}\Vert\equiv\dfrac{\Vert\vert\mathsf{V}[p]\Vert_{2,G_0}\vert}{\vert\langle \Vert\Delta_{g_0} p\Vert_{2,G_0}\vert}.
\end{align}
Checking on the powers of the quantities involved and using that the background metric just contributes a constant factor, partial integration gives us
\begin{align}
\Vert\mathsf{V}\vert p\Vert_{2,g_0}= \text{const.}\Vert u\Vert_{1,2}^2,
\end{align}
i.e., $\Vert\mathsf{V}\Vert\sim\Vert u\Vert_{1,2}^2$. Noting that $u,\partial_i u = \mathcal{O}(\epsilon)$, we have in total
\begin{align}
\Vert\mathsf{V}\Vert = \mathcal{O}(\epsilon^2).
\end{align}
This means that $\mathsf{V}$ is of order $\epsilon^2$ when compared to the contribution that $\Delta_{g_0}$ gives - a result which is convenient for a perturbation theory.
\item\textbf{Temporal derivatives: }The problem with the operator $\mathsf{V}$ is that its does not account for all perturbative contributions but only the spatial part of the perturbations. Recall that the acoustic pressure $P\in H^{1,2}_0(\mathcal{M})$ has arguments $(t,\mathbf{x}_t)\in\mathcal{M}$. by means of the difeomorphism $\Phi^{-1}\circ\Phi_0:\mathcal{M}_0\to\mathcal{M}, (t,\mathbf{x})\to (t,\mathcal{x}_t=\psi_{0\to t}(\mathbf{x}))$, the dependencies on the parameter $t$ parameterizing the fibers $(\Omega_t)_{t\geq 0}$ of $\mathcal{M}$ enters the arguments of $P$. Because we require $P$ to satisfy a wave equation on $\mathcal{M}$, namely
\begin{align}
\partial_t^2 P - c^2\Delta_{g_0}P = c^2\rho_0\partial_t^2 u\circ(\phi_{0\to t}^{-1}),
\end{align}
with homogeneous Neumann boundary conditions and homogeneous initial conditions, the $\partial_t^2$-part of the wave equation also includes derivatives of $\psi_{0\to t}$. This makes the requirement of a smooth parameterization of the fibers, i.e., $t\to\Omega_t$ is smooth, clear:Otherwise, we wouldn't be able to perform differentiation of $\psi_{0\to t}$ w.r.t. $t$. Let us first absorb $c^2$ in the coordinates, i.e., we re-scale the fiber coordinates $x^\mu\to cx^\mu$, leaving the metric $g_0$ unaffected. Then, we can rewrite the right-hand side of the wave equation in terms of the metric $G_0$ on the total space of the fiber bundle $\mathcal{M}$ as
\begin{align}
\sqrt{-\vert G_0\vert}\partial_I(\sqrt{-\vert G_0\vert}G_0^{IJ}\partial_J P)\equiv\square_{G_0}[P].
\end{align}
The D'Alembertian operator w.r.t. the bundle metric $G_0$, $\square_{G_0}$, is self-adjoint on $H^{1,2}_0(\mathcal{M})$ and by means of the global bundle morphism $\Phi_{0\to t}=\Phi^{-1}\circ\Phi_0$, we can pull it back to an operator $\square_{G}$ on $H^{1,2}_0(\mathcal{M}_0)$ which is nothing but the D'Alembertian with respect to the bundle metric $G$. Then setting $p=P(t,\psi_{0\to t}^{-1}(\mathbf{x}_t))$ the pull-back of the wave-equation from $\mathcal{M}$ is given by
\begin{align}
\square_{G}p = \rho_0c^2\partial_t^2 u.
\end{align}
On the left hand side, the pull-back only killed the composition of $(\partial_t^2 u)$ with $\phi_{0\to t}$ since the entire object $\partial_t^2 u$ lives by its derivation from Euler's equation already on $\mathcal{M}_0$ (c.f. the next section)! The most naive and obvious choice for the perturbation operator, $\mathsf{W},$ now is
\begin{align}
\mathsf{W}\equiv\square_G - \square_{G_0}.
\end{align}
In order to assure that $\mathsf{W}$ has homogegenous boundary conditions, we also have to assign $\square_{G_0}$ homogeneous boundary conditions. Likewise, we assign $\square_{G_0}$ homogeneous initial conditions. In other words, $\square_{G_0}$ is just the original D'Alembertian with respect to the unperturbed bundle metric $G_0$ but this time defined on $\mathcal{M}_0$, i.e., $\square_{G_0}: H^{1,2}_0(\mathcal{M}_0)\to H^{1,2}_0(\mathcal{M}_0)$. Notice again that $\mathsf{W}$ is in general not self-adjoint: The $L^2$-product on $\mathcal{M}_0$ requires us to specify one and only one bundle metric. We could either take $G_0$ or $G$. Then, one of the D'Alembertian is self-adjoint w.r.t. such an $L^2$-product, but the other one is in general not self-adjoint w.r.t. this inner product as well. Using the coordinate expressions in terms of a local orthonormal coordinate system w.r.t. $G_0$, we have for an $F\in H^{1,2}_0(\mathcal{M}_0)$
\begin{align}
\mathsf{W}[F]=\sqrt{-\vert G\vert}\partial_I(\sqrt{-\vert G\vert}G^{IJ}\partial_J F)-\sqrt{-\vert G_0\vert}\partial_I(\sqrt{-\vert G_0\vert}G_0^{IJ}\partial_J F).
\end{align}
We recall that at the end of the previous section, we obtained after some work that the pull-back of the metric $G_0$ on $\mathcal{M}$ is the bundle metric $G$ on $\mathcal{M}$ given by $G=G_0+\delta G$. Since we only want to stay in lowest order in $\delta G$, we can use the determinant expansion formulas given at the end of the previous section as well. Last, we notice $G^{IJ}=G_{0}^{IJ}-\delta G^{IJ}$ in first order in $\delta G$, since $\delta_{J}^{I}=G^{IK}G_{KJ}=(G_0^{IK}+\delta G^{IK})(G_{0,KJ}+\delta G_{KJ}) = G_0^{IK}G_{0,KJ}+G_{0}^{IK}\delta G_{KJ}-\delta G^{IK}G_{0,KJ}+\mathcal{O}((\delta G)^2) = G_0^{IK}G^0_{KJ}=\delta_{J}^{I}$ because of symmetry of $\delta G$, $\delta G_{IJ}=\delta G_{JI}$. The derivation is completely analogous to the derivation of $\mathsf{V}$. Hitting in the same vein, we define
\begin{align}
\mathsf{W}=\sum_{k=1}^{\infty}\mathsf{W}_k,
\end{align}
where $\mathsf{W}_k$ contains only the contributions of $\delta G$ in the determinant expansion which scale as $\mathsf{W}_k\sim(\delta G)^k$. Since we are only interested in the lowest order, we have $\mathsf{W}=\mathsf{W}_1+\mathcal{O}((\delta G)^2)$. Identifying $\mathsf{W}\equiv\mathsf{W}_1$ as we did before for $\mathsf{V}$ we can write at first
\begin{align*}
\square_{G}F&=\sqrt{-\vert G\vert}\partial_I(\sqrt{-\vert G\vert}G^{IJ}\partial_J F)\\
&= \sqrt{-\vert G_0+\delta G\vert}\partial_I(\sqrt{-\vert G_0+\delta G\vert}(G_0^{IJ}-\delta G^{IJ})\partial_J F)\\
&= \sqrt{-\vert G_0\vert}\partial_I(\sqrt{-\vert G_0\vert}G_0^{IJ}\partial_J F)-\sqrt{-\vert G_0\vert}\dfrac{1}{2}\text{Tr}(G_0^{-1}\delta G)\partial_I(\sqrt{-\vert G_0\vert}G_0^{IJ}\partial_J F)\\
&+\dfrac{1}{2}\sqrt{-\vert G_0\vert}^{-1}\partial_I(\sqrt{-\vert G_0\vert}\text{Tr}(G_0^{-1}\delta G)G_0^{IJ}\partial_J F)-\sqrt{-\vert G_0\vert}^{-1}\partial_I(\sqrt{-\vert G_0\vert}\delta G^{IJ}\partial_J F)\\
&= \sqrt{-\vert G_0\vert}\partial_I(\sqrt{-\vert G_0\vert}G_0^{IJ}\partial_J F)-\sqrt{\vert g_0\vert}\dfrac{1}{2}\text{Tr}(G_0^{-1}\delta G)\partial_I(\sqrt{-\vert g_0\vert}G_0^{IJ}\partial_J F)\\
&+\dfrac{1}{2}\sqrt{\vert g_0\vert}^{-1}\partial_I(\sqrt{\vert g_0\vert}\text{Tr}(G_0^{-1}\delta G)G_0^{IJ}\partial_J F)-\sqrt{\vert g_0\vert}^{-1}\partial_I(\sqrt{\vert g_0\vert}\delta G^{IJ}\partial_J F)\\
&=\sqrt{-\vert G_0\vert}\partial_I(\sqrt{-\vert G_0\vert}G_0^{IJ}\partial_J F)+\dfrac{1}{2}\partial_I\text{Tr}(G_0^{-1}\delta G)G_0^{IJ}\partial_J F-\sqrt{\vert g_0\vert}^{-1}\partial_I(\sqrt{\vert g_0\vert}\delta G^{IJ}\partial_J F)\\
&=\square_{G_0}[F]+\mathsf{W}[F].
\end{align*}
During the fourth step we have used $\vert g_0\vert = -\vert G_0\vert$ by definition of $G_0$.  Notice that if $I,J\in\lbrace 1,2,...,n\rbrace$, the last three contributions are precisely the prior perturbation operator $\mathsf{V}$. We define a third operator $\mathsf{T}$ which contains all temporal derivatives. $\mathsf{T}$ is defined in all orders in $\delta G$
\begin{align}
\mathsf{T}\equiv\mathsf{W}-\mathsf{V},
\end{align}
where we note that the gradation of $\mathsf{V}$ and $\mathsf{W}$ in terms of $\delta g$ and $\delta G$ naturally lifts to a gradation of $\mathsf{T}$ in terms of powers $(\delta G)^k,\,k\geq 1$. Thus denoting by $\mathsf{T}_1$ the contribution to $\mathsf{T}$ which is of order $(\delta G)^1$ we can identify $\mathsf{T}=\mathsf{T}_1$ in the case of interest again. We can write the acoustic wave equation on the unperturbed bundle in the form of a perturbation equation in the familiar form where we restored the $c$-factors as far as this possible,
\begin{align}
\partial_t^2p- c^2\Delta_g p = \mathsf{T}[p]+c^2\mathsf{V}[p]+\rho_0c^2\partial_t^2 u.
\end{align}
Notice that only the fiber coordinates $\lbrace x^\mu\rbrace_{1\leq \mu\leq n}$ have been rescaled by $c$. Since $\mathsf{T}$ contains the temporal derivatives of  $p$ and $\delta G$, we cannot pull an overall factor out. This notational inconvenience could be worsened by splitting $\mathsf{T}$ up because we introduce even more operators. We only give an expression of $\mathsf{T}$ in terms of the boundary vibrations $u$ and afterwards show that $\mathsf{T}$ has the same scaling behavior in terms of $\epsilon$ as $\mathsf{V}$ has, i.e., $\Vert\mathsf{T}\Vert\sim\mathcal{O}(\epsilon^2)$. We notice as before $\text{Tr}[G_0^{-1}\delta G] = g^{(0)}_{ss}\partial^K U\partial_K U$ where raising and lowering indices takes place w.r.t. $G_0$ analogous to the calculation for $\mathsf{W}$. We derive the explicit expression for $\mathsf{T}$ in terms of $U$,
\begin{align*}
\mathsf{T}[p]&=\mathsf{W}[p]-\mathsf{V}[p]\\
&=\dfrac{1}{2}\partial_{I}(g^0_{ss}\partial_t U\partial^tU)G_0^{IJ}\partial_J p +\dfrac{1}{2}\partial_{(I}(g^0_{ss}\partial_K U\partial^K U)G^{(It)}_0\partial_{t)}p\\
&+g^0_{ss}\partial_{(t}U\partial_{I)}\sqrt{\vert g_0\vert}\partial^{(I}(\sqrt{\vert g_0\vert}\partial^{t)}p)+\partial^{(I}(g^0_{ss}\partial_{(I}U\partial_{t)}U)\partial^{t)}p
\end{align*}
The round brackets around the indices denote symmetrization again, i.e., $\partial_{(I}U\partial_{J)}U=(1-0.5\delta_{IJ})(\partial_I U\partial_J U +\partial_J U\partial_I U)$ and to bracketed indices means that the indices are symmetrized after (!) application of the Einstein summation convention.. We see that $\mathsf{T}$ contains $U$ and its derivatives quadratically in lowest non-trivial order. Using $U=(1-s)u$ as before, we obtain in terms of $u$,
\begin{align*}
\mathsf{T}[p]&=\dfrac{1}{2}\partial_I(g^0_{ss}(1-s)^2(\partial_t u)^2)G_0^{IJ}\partial_J p-\dfrac{1}{2}\partial_{(I}(g^0_{ss}(\partial_tu)^2)G_0^{(It)}\partial_{t)} p\\
&+ \dfrac{1}{2}\partial_{(I}g^{0}_{ss}g^{0}_{ss}u^2)G^{(It)}_0\partial_{t)}p + \dfrac{1}{2}\partial_{(I}g^{0}_{ss}(1-s)^2(\partial_i u)^2g_0^{ii})G^{(It)}_0\partial_{t)}p\\
&-g^0_{ss}(1-s)^2(\partial_t u)^2\sqrt{\vert g_0\vert}\partial^t(\sqrt{\vert g_0\vert}\partial^t p)-g^0_{ss}u\partial_{(t}u\sqrt{\vert g_0\vert}\partial^{(s}(\sqrt{\vert g_0\vert}\partial^{t)}p)\\
&+g^{0}_{ss}(1-s)^2\partial_{(t} u\partial_{i)} u\sqrt{\vert g_0\vert}\partial^{(t}(\sqrt{\vert g_0\vert}\partial^{i)}p)-\partial^t(g^0_{ss}(1-s)^2\partial_t u\partial_t u)\partial^t p\\
&-\partial^{(s}(g^0_{ss}u\partial_t u)\partial^{t)}p+\partial^{(i}(g^0_{ss}(1-s)^2\partial_{(i}u\partial_{t)u}\partial^{t)}p.
\end{align*}
In the expression, we have substituted back to the unperturbed fiber metric $g_0$ as much as possible. The expression is lengthy, but for the bounding the only important thing is that $\mathsf{T}$ is firstly linear in $p$ and secondly does contain the boundary vibrations $u$ and its partial derivatives $\partial_tu$ w.r.t. the time parameter $t$ and the partial derivatives $\partial_iu$ w.r.t. the coordinates $\lbrace x^i\rbrace_{1\leq i\leq n, i\neq s}$ on the boundary $\partial\Omega_0$ of the unperturbed reference fiber $\Omega_0$. As such, we can bound $\mathsf{T}$ in the norm relative to $c^2\Delta_{g_0}$, i.e., $c^2$-times the unperturbed Neumann-Laplacian on $\Omega_0$. We recall the definition of the relative norm $\Vert .\Vert:\text{LinOp}\to\mathbb{R}^+_0$.
\begin{align}
\Vert\mathsf{T}\Vert \equiv \dfrac{\vert\langle p\vert\mathsf{T}\vert p\rangle_{G_0}\vert}{\vert\langle p\vert\Delta_{g_0}\vert p\rangle_{G_0}\vert}
\end{align}
Since $0\leq s\leq 1$, we can set $s=0$ in our estimates. Integrating by parts and using homogeneity of the boundary and initial condition that $p$ satisfies on $\partial\mathcal{M}_0$ and using Hölder's inequality to split the integral featuring product integrand $\sim\partial u\partial u\partial p\partial p$, we obtain two $L^2_{G_0}$ integrals. One over expressions $\sim\partial p\partial p$ which can be canceled against the Laplacian and one expression featuring contributions $\sim u^2,\,u\partial u,\,\partial u\partial u$. We can bound these contributions by $\Vert u\Vert_{1,2,g_0}$ for all combinations. The $x^s=s$ integration in the $u$-dependent integrals does not affect the overall result in terms of giving a divergent result because $0\leq s\leq 1$, so the $x^s=s$ integration just yields a constant finite pre-factor. In total, we have found the bound
\begin{align}
\Vert\mathsf{T}\Vert = C\Vert u\Vert_{1,2,G_0}^2,
\end{align}
with a constant $C$. Notice that our assumptions on the derivatives $\partial_i u$  of $u$ allow us to bound $\vert\partial_i u\vert<3\epsilon$ as we can do for $u$, namely $\vert u\vert < 3\epsilon$. We have assumed that the perturbation bundle is physical. Since $u$ is just a component of the restriction of $\psi_{0\to t}-\text{id}_{\Omega_0}$, the physicality condition also applies to $u$. In particular, we have for the derivative $\vert\partial_t u\vert^2 \leq c^2_m\vert\partial_i u\vert\vert\partial^i u\vert$. Notice that $c^2_m/c^2=\mathcal{O}(\epsilon)$ by physicality. Thus, the derivatives of $u$ w.r.t. the coordinate $t$on the base space of the bundle $\mathcal{M}_0$ contribute in even higher order than the derivatives $\partial_iu$ with respect to coordinates $\lbrace x^i\rbrace_{1\leq i\leq n, i\neq s}$ on the boundary $\partial\Omega_0$ of the unperturbed fiber $\Omega_0$! All in all, $\Vert u\Vert_{1,2;G_0}^2\sim\epsilon^2$ such that
\begin{align}
\Vert\mathsf{T}\vert \sim\mathcal{O}(\epsilon^2).
\end{align}
We notice that the definition of $\mathsf{T}$ can be rearranged to give an expression for $\mathsf{W}$, namely $\mathsf{W}=\mathsf{V}+\mathsf{T}$. The results $\Vert\mathsf{V}\Vert\sim\mathcal{O}(\epsilon^2)$ and $\Vert\mathsf{T}\Vert\sim\mathcal{O}(\epsilon^2)$ from the previous and this sub-paragraph can be condensed in the single scaling-behavior equation for $\mathsf{W}$,
\begin{align}
\Vert\mathsf{W}\Vert\sim\mathcal{O}(\epsilon^2).
\end{align}
When risking an error of order $\epsilon^2$, we can approximate the acoustic wave equation on the perturbation bundle $\mathcal{M}$ by the acoustic wave equation on the unperturbed bundle $\mathcal{M}_0$.
\end{itemize}
\textbf{Lessons learned: }Although the calculations are lengthy, they express two important points. The acoustic wave equation would become non-linear with the non-linearities originating from the vibrations of the boundary, i.e., $u$. Secondly, the temporal dependencies of $\Omega_t$ introduce additional contributions to the perturbation operator $\mathsf{V}$. We had stored them in $\mathsf{T}$. For the Laplacian $\Delta_g$ (!) on $\mathcal{M}_0$ this means that time-dependent deformations of the reference domain $\Omega_0$ introduce additional correction terms for the spectrum $\sigma(\Delta_g)$ by means of $\mathsf{T}$. This is not surprising because the natural setting for the wave equation is the bundle space $\mathcal{M}$ endowed with the unperturbed metric $G_0$ or - diffeometrically equivalent - the bundle space $\mathcal{M}_0$ endowed with the time-dependent metric $G$!

\section{Derivation of the Perturbation Equations}
\textbf{Convention: }\emph{We use big Latin indices $I,J,K,L,...$ to denote local orthornormal coordinates stemming from $\mathcal{M}_{ref}$ on $\mathcal{M}$ and $\mathcal{M}_0$, small Greek indices $\mu,\nu,\kappa,\lambda,...$ to denote the local orthornormal coordinates on the second component of the fiber bundles, $\Omega_t$ and $\Omega_0$ stemming from $B^n_1(\mathbf{0})$ and small Latin indices $i,j,k,l,...$ to denote local orthornormal coordinates on $\partial\Omega_t$ and $\partial\Omega_0$ stemming from $S^n_1(\mathbf{0})$. Further, we will distinguish the perturbation bundle $\mathcal{M}$ from the unperturbed bundle $\mathcal{M}_0$ by \" priming\" the indices for $\mathcal{M}_0,\,\Omega_0,\,\partial\Omega_0$ if ambiguities arise.}\newline
\newline
\textbf{Modelling: }We will derive an abstract ICE model from an action functional $S_{ICE}$ consisting of two parts. A fluid dynamical action $S_{fluid}$ and a geometric action $S_{geom}$.\newline
\newline
\textbf{Fluid action: }We will use a simplified version of the Bateman Lagrangian \cite{batmena, golbart} for irrotational, isentropic non-viscous flow, but put the action functional in the metric background on the perturbation bundle $\mathcal{M}$ given by $G$. We need several steps.
\begin{itemize}
\item\textbf{Fluid velocity normalization: }Recall that $V\in\Gamma(T\mathcal{M})$. Since $V^0 = V^t$ as the "time"-component of the fluid velocity field bears no physical information, we use the normalization $V^0V_0 = -4$, i.e., $V^0=2$, which is conceptually familiar from Einsteinian gravitation theory. That we normalize $V^0=2$ instead of $V^0=1$ has aesthetic reasons that will become clear below. On the other hand, $V\in\mathcal{V}(\mathcal{M})$. By definition of the perturbation bundle $\mathcal{M}$ is compact and the mapping $t\to\Omega_t$ is smooth in $t$. We now levy this definition up to define the map $V\mapsto \mathbf{V}_t=V^\mu_t\partial_\mu$, where the index $t$ indicates that we now have a smoothly parameterized family of vector fields such that $\mathbf{V}_t\in\mathcal{V}(\Omega_t)$ for all positive $t$.
\item\textbf{Hodge decomposition theorem: }Since the dimensions of our setups are not necessarily $n=3$ although we have that as the main application in mind, we can't use the Helmholtz decomposition theorem straight away but need a digression in cohomology theory on Riemannian manifolds. More precisely, we give a minimal account of the Hodge decomposition theorem. Since $\Omega_t$ is compact, the deRham cohomology exact sequence obtained by acting with the cohomology functor $\Omega^\ast$ on $\Omega_t$. For that regard the $\mathbb{Z}_{n+1}$-graded exterior algebra $\Omega(\Omega_t)$ with the exterior derivative $d_k:\Omega^k(\Omega_t)\to\Omega^{k+1}(\Omega_t)$ for all $0\leq k\leq n-1$,
\begin{align}
\lbrace 0\rbrace\hookrightarrow\Omega^0(\Omega_t)\stackrel{d_0}{\rightarrow}\Omega^1(\Omega_t)\stackrel{d_1}{\rightarrow}\Omega^2(\Omega_t)\cdots\Omega^{n-1}(\Omega_t)\stackrel{d_{n-1}}{\rightarrow}\Omega^{n}(\Omega_t)\hookrightarrow\lbrace 0\rbrace,
\end{align}
Treating the spaces $\Omega^k(\Omega_t)$ as vector spaces and using nilpotency of $d_{k+1}\circ d_k$ for all $0\leq k\leq n-2$, we can define the deRham complex
\begin{align}
\lbrace 0\rbrace\hookrightarrow H^0(\Omega_t)\stackrel{d}{\rightarrow}H^1(\Omega_t)\stackrel{d}{\rightarrow}H^2(\Omega_t)\cdots H^{n-1}(\Omega_t)\stackrel{d}{\rightarrow}H^{n}(\Omega_t)\hookrightarrow\lbrace 0\rbrace ,
\end{align}
where $H^k(\Omega_t)\simeq \ker d_k / \text{im} d_{k-1}$. By definition of the deRham groups $H^k(\Omega_t)$, $d_{k-1}$ can be inverted in the sense that $\delta_{k+1}: H^{k+1}(\Omega_t)\to H^k(\Omega_t)$ is the adjoint operator w.r.t. the $L^2$ inner product defined on $\Omega^k(\Omega_t)$ by means of the metric tensor $g$ on $\Omega_t$. The resultant complex is
\begin{align}
\lbrace 0\rbrace\hookleftarrow H^0(\Omega_t)\stackrel{\delta}{\leftarrow}H^1(\Omega_t)\stackrel{\delta_1}{\leftarrow} H^2(\Omega_t)\cdots H^{n-1}(\Omega_t)\stackrel{\delta}{\leftarrow}H^{n}(\Omega_t)\hookleftarrow\lbrace 0\rbrace
\end{align}
Now, given a $k$-form $\omega\in H^k(\Omega_t)$, the Hodge decomposition theorem states that there is a decomposition of the exterior algebra $\Omega(\Omega_t)$  in mutually orthogonal $\mathbb{R}$-vector spaces if $\Omega_t$ is a compact Riemannian manifold. More precisely, it states that
\begin{align}
\Omega^k(\Omega_t)=\text{im} d_{k-1}\oplus \text{im} \delta_{k+1}\oplus \mathcal{H}^k_\Delta(\Omega_t),
\end{align}
where $\mathcal{H}_\Delta^k(\Omega_t)=\lbrace \omega'\in\Omega^k(\Omega_t):\Delta \omega = 0\rbrace$ is the kernel of the Laplacian acting on forms $\Delta:\Omega(\Omega_t)\to\Omega(\Omega_t),\omega'\mapsto \Delta_t\omega':=(d\delta+\delta d)\omega'$. More explicitly, let $\omega\in\Omega^k(\Omega_t)$ be a given $k$-form. Then there is a unique triple $(\alpha,\beta,\gamma)\in H^{k-1}(\Omega_t)\times H^{k+1}(\Omega_t)\times \mathcal{H}_\Delta^k(\Omega_t)$ such that
\begin{align*}
\omega = d\alpha +\delta\beta + \gamma .
\end{align*}
This is remarkably similar to the Helmholtz decomposition theorem in $\mathbb{R}^3$ (which does however not require compactness of $\mathbb{R}^3$) that a vector field $\mathbf{v}$ can be decomposed uniquely as
\begin{align}
\mathbf{v} = \nabla a + \nabla\times\mathbf{b} + \mathbf{c},
\end{align}
where - in physicists' language - $a$ is the scalar potential, $\mathbf{b}$ is the vectorial potential and $\mathbf{c}$ is a harmonic vector field, i.e., $\Delta_{vec}\mathbf{c}=0$.
\item\textbf{Musical isomorphisms and the fluid velocity $1$-form: }Let us turn back to the fluid velocity field, $\mathbf{V}_t\in\mathcal{V}(\Omega_t)$. By means of the musical isomorphism $\mathcal{V}(\Omega_t)\simeq\Omega^1(\Omega_t)$, we can identify $\mathbf{V}_t$ with a unique $1$-form $\omega_t\in\Omega^1(\Omega_t)$ for all $t>0$. Moreover, because of smoothness of the $t$-dependence of $\Omega_t$, the mapping $t\mapsto\omega_t$ is smooth as well. We call $\omega_t$ the \emph{fluid velocity} $1$\emph{-form}. We are ready to translate the irrotationality requirement of the fluid flow in the language of geometry and global analysis. We call the flow \emph{irrotational} if the fluid velocity $1$-form allows a Hodge decomposition $\omega_t = d\upsilon_t$ with the scalar potential $\upsilon_t$ modulo $\mathcal{H}^1_\Delta(\Omega_t)$ such that $t\mapsto\upsilon_t$ is smooth again. I.e., only the $\text{im} d_{0}$-part in the decomposition of $\Omega^1(\Omega_t)$ is zero. Using the musical isomorphisms $\mathcal{V}(\Omega_t)\simeq\Omega^1(\Omega_t)$ again, this translates into the familiar expression $V_t = \partial_\mu \upsilon_t\partial_\mu$. By smoothess and the observation that the covariant resp. because $\upsilon_t\in \Gamma(\Omega_t\to\mathbb{R})$ also partial derivative $\partial_\mu$ does not affect the parameter $t$, we obtain by inversion of the mapping $V\mapsto (V_t)_t$ a decomposition of the form $V = V^0\partial_0 + \partial_\mu\upsilon\partial_\mu$, where $\nu\in\Gamma(\mathcal{M}\to\mathbb{R})$ now. We will call such a vector field $V$ \emph{irrotational flow} and $\nu$ the \emph{hydrodynamic potential}. Since $V^0=1$ by the first sub-paragraph, we have found a Helmholtz-like decomposition of $V$ on $\mathcal{M}$ to model irrotationality in our setting. We will further impose the Dirichlet boundary condition $\upsilon = 0$ on $\partial\mathcal{M}$ as a gauging of the hydrodynamic potential. Notice that this does not imply $\mathbf{V}=\mathbf{0}$ on $\partial\mathcal{M}$!
\item\textbf{Isentropy and inviscidity: }For the non-physical audience, the notions isobaricity and inviscidity deserve an explanation. \emph{Inviscidity} is a typical modelling assumption in vibrational acoustics stating that viscous contribution $\mu\text{Tr}\left[\left(\nabla_{\partial_i}\nabla_{\partial_j}-\nabla_{\nabla_{\partial_i}\partial_j}\right)V^\lambda\partial_\lambda\right]=\mu\Delta_{g_0,tensor}\mathbf{X}$ can be neglected if the \emph{Reynolds number} $\text{Re}\equiv \rho_0 U_{fl} L/\mu\gg 1$ where $\rho_0$ is the fluid mass density, $L$ a characteristic geometric length scale, e.g. $\text{L}=\sqrt[n]{\text{Vol}_n(\Omega_0)}$ and $U_{f}$ is the fluid mean velocity and the parameter $\mu$ is the dynamic viscosity of the fluid, \cite{howe2}. Then in dimensionless units, then $\mu\Delta_{g_0,tensor}\mathbf{X}\sim(\text{Re})^{-1}\Delta_{g_0,tensor}\mathbf{X}$ can be neglected against the convection term $V^\mu(\partial_\mu nabla_{\partial_\kappa}) (V^\nu\partial_\nu)$ in the Navier-Stokes' equations. \emph{Correspondingly, the Navier-Stokes' equations simplify to Euler's equations},\cite{howe2} The notion of \emph{isentropy} refers to a physical system for the description of which the \emph{enthalpy} $H$ is the correct thermodynamic potential if we are interested in modeling (almost) isobaric, $dp/p_0 \ll 1$, volume changes. In differential notation the associated Pfaffian $dH$ can be expressed as $dH = TdS+Vdp$ with the two basis forms in thermodynamic state space $(dS,dp)$ \cite{scheck5}. In continuum physics, the enthalpy $H$ is replaced by the \emph{specific enthalpy} $h=H/(M_{fl})$ where the fluid mass $M_{fl}=\int_{\Omega_t} d^n x\sqrt{\vert g_0\vert}\rho_{fl}$ is expressed in terms of the fluid mass density. Then, we have by the means value theorem, $dh = T\rho_{fl}^{-1}ds + \rho_{fl}^{-1}dp$ where $s$ denotes the entropy density. \emph{In the following, we refer to $h$ as the enthalpy because no confusion with the macroscopic quantity $H$ can arise.}
\item\textbf{Action functional: }We have finished the preparations to define the action functional for isentropic, irrotational inviscid flow.
\begin{align}
S_{fluid}[\rho_f,\upsilon,V]\equiv\int_{\mathcal{M}}dtd^nx\,\sqrt{-\vert G_0\vert}\left(\dfrac{\rho_f}{2}V^I\partial_I\upsilon + u(\rho_f)\right).
\end{align}
Using irrotationality of $V$ and $V^0=2$, we obtain a more practical form depending only on the hydrodynamic potential $\upsilon$ and the internal energy density $u(\rho_f)$\footnote{In \cite{goldbart}, the Euclidean $n=3$ action functional has a contribution $+u(\rho_f)$ instead of $-u(\rho_f)$. We don not follow this sign convention since the first two contributions scale as a kinetic energy, whereas the second contribution scales as a potential energy. Interpreting the integrand of the action functional as a Lagrangian density, $\mathcal{L}=\mathcal{T}-\mathcal{V}$, where $\mathcal{T}$ denotes the kinetic energy density and $\mathcal{V}$ denotes the potential energy density, we achieve formally more closeness to the conventional Lagrangian formalism.}, a function of the mass density $\rho_f$ of the fluid:
\begin{align}
S_{fluid}[\rho_f,\upsilon] &= \int_{\mathcal{M}}dtd^nx\,\left(\rho_f\partial_t\upsilon + \dfrac{\rho_f\partial_\mu\upsilon\partial^\mu\upsilon}{2} + u(\rho)\right)\\
&= \int_{\mathcal{M}}dtd^nx\,\left(\rho_f\partial_t\upsilon + \dfrac{\rho_f\mathbf{V}^2}{2} - u(\rho)\right)
\end{align}
Functional differentiation of $S_{fluid}$ with respect to $\rho_f$ yields,
\begin{align}
\partial_t\upsilon + \dfrac{\mathbf{V}^2}{2}-h(\rho_f)=0,
\end{align}
where we recalled the definition of the enthalpy $h(\rho)=d_{\rho}u(\rho)$. The above equation is a time dependent version of Bernoulli's equation. Variation with respect to $\upsilon$ in turn gives us using the Hodge decomposition of $V$ once again,
\begin{align}
\partial_t\rho_f +\sqrt{-\vert G_0\vert}^{-1}\partial_\mu\left(\sqrt{-\vert G_0\vert }G_0^{\mu\nu}\rho_f V_\nu\right)=0.
\end{align}
Now, we reformulate Bernoulli's equation for an intermediate step in terms of the fluid velocity $1$-form $\omega_t$. This is possible only because Bernoulli's equation is covariant equation,i.e., it has no indices that need to be raised and lowered by means of the metric $G$. Defining $\omega_t^2 = G_0^{\mu\nu}(\omega_t)_\mu(\omega_t)_\nu$, we see that $\mathbf{V}^2=\omega_t^2$ and thus
\begin{align}
2\partial_t\upsilon + \omega_t^2 - 2h(\rho_f) = 0
\end{align}
Acting with the exterior derivative $d$ on the equation, we have
\begin{align}
2\partial_t\omega_t + d\omega_t^2-2dh(\rho_f)= 0
\end{align}
Since $\omega_t^2$ is in $\Omega^0(\Omega_t)$ we have in components $d\omega_t^2 = 2\omega_t^\mu d(\omega_t)_\mu = 2(\omega_t^\mu\partial_\mu)\omega^\nu_t$. Because $d$ acts on an element of $\Omega^0(\Omega_t)$, it involves no Christoffel symbol contributions that would stem from the full Levi-Civita connection on $\Omega_t$. Using that $(\omega_t)_{\mu}=V_\mu$ by definition, we find after dividing by the factor $2$,
\begin{align}
\partial_t V_\nu + (V^\mu\partial_\mu)V_\nu - \partial_\nu h = 0.
\end{align}
Notice that $G_0$ is time-independent. Raising the $\nu$ index by means of the musical isomorphisms is allowed. We use the hydrodynamics definition of the enthalpy in terms of the hydrodynamic pressure $P$,
\begin{align}
h(P) = \int_0^P\dfrac{dP'}{\rho_f(P')}.
\end{align}
Since all former $x^\mu$-dependencies of $\rho_f$are now stored in the hydrodynamic pressure $P$, the chain rule gives us $\partial^\mu h = \rho_f^{-1}\partial_\mu P$ which can be rearranged to $\rho_f \partial_\mu h = \partial_\mu p$. Inserting the result in the equation for $V_\nu$, we find
\begin{align}
\rho_f(\partial_t V^\nu + (V^\mu\partial_\mu)V^\nu) = \partial^\nu P.
\end{align}
This is Euler's equation.
\item\textbf{Derivation of the acoustic wave equation: }Since our interest is in acoustics and not in fluid dynamics, we start acoustic linearization. Introducing the linearization parameter $\varepsilon\ll 1$ which is not to be confused with the standard $\epsilon$ in the definition of the perturbation bundle, we make the Ansatz
\begin{align}
P &= p_0 + \varepsilon p\\
V^\mu &= 0 + \varepsilon v^\mu_{ac}\\
\rho_f &= \rho_0 + \varepsilon \rho_{ac}.
\end{align}
This needs to be inserted in continuity equation and Euler's equation which are repeated for the reader's convenience,
\begin{align}
&\text{Continuity: }\partial_t\rho_f +\sqrt{-\vert G_0\vert}^{-1}\partial_\mu\left(\sqrt{-\vert G_0\vert }G_0^{\mu\nu}\rho_f V_\nu\right)=0\\
&\text{Euler's Eq.: }\rho_f(\partial_t V_\nu + (V^\mu\partial_\mu)V_\nu) = -\partial_\nu P
\end{align}
Setting the coefficient in the $\varepsilon$-expansion of these equation equal to zero, we find the equations for the acoustic quantities $v^\mu_{ac},p,\rho_{ac}$. Namely, we obtain
\begin{align}
\partial_t\rho_{ac} + \rho_0\sqrt{-\vert G_0\vert}^{-1}\partial_\mu\left(\sqrt{-\vert G_0\vert }G_0^{\mu\nu} (v_{ac})_\nu\right)&=0\\
\partial_t ((v_{ac})_\nu)&= \partial_\nu P
\end{align}
We will use the thermodynamic relation $\partial_t\rho_{ac}=\rho_0c^{-2}\partial_t p$ in order to replace $\partial_t\rho_{ac}$ by $\rho_0c^{-2}\partial_t p$. In this relation, $c$ is the speed of sound, $c\approx 343\,\text{ms}^{-1}$. We still have to take the partial derivatives $\partial_t$ of the first equation and the divergence on Riemannian manifolds $\sqrt{-\vert G_0\vert}^{-1}\partial_\mu(\sqrt{-\vert G_0\vert}G_0^{\mu\nu} \bigodot)$ w.r.t. the second equation. Using time-independence of $G_0$, this gives
\begin{align}
\rho_0c^{-2}\partial_t^2p + \rho_0\sqrt{-\vert G_0\vert}^{-1}\partial_\mu\left(\sqrt{-\vert G_0\vert }G_0^{\mu\nu}\partial_t (v_{ac})_{\nu}\right) &= 0\\
\sqrt{-\vert G\vert}^{-1}\partial_\mu\left(\sqrt{-\vert G_0\vert }G_0^{\mu\nu}\partial_t (v_{ac})_{\nu}\right)&= \sqrt{-\vert G_0\vert}^{-1}\partial_\mu\left(\sqrt{-\vert G_0\vert }G_0^{\mu\nu} \partial_\nu p\right).
\end{align}
Now, we insert the second equation solved for components of $\mathbf{v}_{ac}$ in the first equation and obtain dividing by $\rho c^{-2}$,
\begin{align}
\partial_t^2 p + c^{2}\sqrt{-\vert G_0\vert}^{-1}\partial_\mu\left(\sqrt{-\vert G_0\vert }G_0^{\mu\nu}\partial_\nu p\right) &= 0.
\end{align}
In terms of the metric $g_{0}$ on the fibers, we find
\begin{align}
\partial_t^2 p - c^{2}\sqrt{\vert g_0\vert}^{-1}\partial_\mu\left(\sqrt{\vert g_0\vert }g_0^{\mu\nu}\partial_\nu p\right) &= 0\\
\Leftrightarrow\partial_t^2 p - c^2\Delta_{g_0,\Omega_t}p &=0\\
\Leftrightarrow \Delta_{G_0} p &= 0
\end{align}
since $-\vert G_0\vert=\vert g_0\vert$ by the definition of $G_0$ given in the beginning. For the equivalence, we have used the definition of the Laplace-Beltrami operator $\Delta_{g_0,\Omega_t}$ given in the beginning of the paper. The last equation describes in physicists' language a scalar field $p$, the acoustic pressure, propagating (as a wave) in curved space-time $\mathcal{M}$ with metric $G$. Mathematically, it is a second-order hyperbolic partial differential operator acting on $p_{ac}\in\Gamma(\mathcal{M}\to\mathbb{R})$.
\item\textbf{Derivation of boundary and initial conditions: }However, we still do not know what the boundary conditions for the wave equation for $p_{ac}$ are. These are obtained by considering an action functional for $p$ which gives the wave equation for $p$. The boundary conditions hydrodynamic potential $\upsilon$ cannot be used here because $\upsilon$ is a hydrodynamic quantity, i.e., the information stored in it contain information about a solution to Euler's equation in all orders in $\varepsilon$, not only the  order $\varepsilon$ piece of information. The relevant action functional is given upon rescaling by a dimensionful constant $C_{dim}$ to ensure the correct physical dimension of an energy
\begin{align}
S_{acous.} = \dfrac{C_{dim}}{2}\int_{\mathcal{M}}dtd^nx\,\sqrt{-\vert G_0\vert}(\nabla_I p\nabla^I p).
\end{align}
We can integrate on $\mathcal{M}$ by parts and obtain
\begin{align}
S_{acous} &= -\dfrac{C_{dim}}{2}\int_{\mathcal{M}}dtd^nx\,\sqrt{\vert - G_0\vert}p\Delta_{G_0} p\\
&+ \dfrac{C_{dim}}{2}\int_{\partial\mathcal{M}}d\text{Vol}_n(\partial\mathcal{M})\,n^I_{\partial\mathcal{M}}p\nabla_I p.
\end{align}
Let us focus on the second contribution, the boundary contribution $S_{acous,\partial}$,
\begin{align}
S_{acous,\partial} &= \dfrac{C_{dim}}{2}\int_{\partial\mathcal{M}}d\text{Vol}_n(\partial\mathcal{M})\,n^I_{\partial\mathcal{M}}p\nabla_I p\\
&= -\dfrac{C_{dim}}{2}\int_{\lbrace 0\rbrace\times\Omega_0}d^nx\,\sqrt{\vert g_0\vert}n^0_{\lbrace 0\rbrace\times\Omega_0}p\partial_0 p\\
&+\dfrac{C_{dim}}{2}\int_{\bigcup_{t>0}\lbrace t\rbrace\times\partial\Omega_t}dtd^{n-1}y\sqrt{\vert g_0\vert_{\partial\Omega_t}\vert}n^\mu_{\partial\Omega_t}p\partial_\mu p\\
&+\dfrac{C_{dim}}{2}\int_{\lbrace \infty\rbrace\times\Omega_0}d^n x\,\sqrt{\vert g_0\vert}n^0_{\lbrace \infty\rbrace\times\Omega_0}p\partial_0 p.
\end{align}
In the second step, the three contributions contain an initial condition at $t=0$ (first line), the boundary conditions (second line) and an initial condition at $t=\infty\equiv 0$ (third line) by means of the $\infty$-periodicity. Since $p\in\Gamma(\mathcal{M}\to\mathcal{M}\times\mathbb{C})$ is a scalar function, the Levi-Civita connections could be replaced by partial derivatives, $\nabla_I\to\partial_I$. By $\infty$-periodicity again, the first and third term cancel with the difference in signs stemming from the observation that the outward \"normal" to $\bar{\mathbb{R}}^+_0$ is $-\partial_t$ at $t=0$ and $\partial_t$ at $t=\infty$. Since contribution one and three cancel, we can set $p(t=0,\mathbf{x})=0=\partial_tp(t=0,\mathbf{x})$. This choice reflects the physical idea of the ICE model that only an external pressure signal should cause an interaural pressure. In other words, the gecko can hear what's going on outside his head, but not what is going on inside his head. The boundary conditions follow from Euler's equation, we have on $\partial\Omega_t$ for fixed $t$
\begin{align}
n^\mu\partial_\mu p = \rho_0\langle\mathbf{n},\partial_t\mathbf{v}_{ac}\rangle_{g_0} = \rho_0\partial_tv_{\mathbf{n}(\partial\Omega_t)},
\end{align}
if $\Vert\partial_t\mathbf{n}\Vert\ll \Vert_2\partial_t\mathbf{v}_{ac}\Vert_2$. We will see below during the variation of the geometrical action that this is the case. In other words, we have by restriction of all quantities to $\partial\Omega_t$
\begin{align}
\partial_{\mathbf{n}(\partial\Omega_t)}p = \rho_0\partial_t\langle\mathbf{n},\mathbf{v}_{ac}\rangle_{g_0}
\end{align}
Thus, we have inhomogeneous Neumann boundary conditions on $\partial\Omega_t$. We may now insert our previous result instead of $n^\mu\partial_\mu p$ in $S_{acous,\partial}$. The total boundary contributions on $\partial\mathcal{M}$ are then given by
\begin{align}
S_{acous,\partial}=C_{dim}\int_{\bigcup_{t >0}\lbrace t\rbrace\times\Omega_t}dtd^{n-1}y\,p(\rho_0\partial_t\langle\mathbf{n},\mathbf{v}_{ac}\rangle_{g_0},
\end{align}
where we had to cancel a factor of $2$ which would have originated anyway by performing the functional derivative w.r.t. $p$ first. We have to fetch the units. Recall that we have absorbed the $c$ in the the coordinates $\mathbf{x}$. In order to reconvert to SI units, we have to consider an additional $c$ entering by transforming $\partial_\mu\to c^{-1}\partial_\mu$ again. Furthermore, the boundary term has dimension $n$ instead of the full $n+1$ dimensions because one spatial dimension dropped out. Reconversion gives an additional factor $c$ which we have to include. This is achieved by regarding the boundary integral as an integral over a \emph{surface-delta-function}. Define $\partial'\mathcal{M}=\bigcup_{t>0}\lbrace t\rbrace\times\partial\Omega_t$. Then the surface-delta-function $\delta((t,\mathbf{x})\in\partial'\mathcal{M})$ defined by
\begin{align}
\int_{\mathcal{M}}d\text{Vol}_{n+1}(\mathcal{M})\delta(\mathbf{x}\in\partial'\mathcal{M})f(t,\mathbf{x})&=\int_{\partial'\mathcal{M}}dtd\text{Vol}_{n-1}(\partial\Omega_t)\,f\vert_{\partial'\mathcal{M}}(t,\mathbf{x})\\
\int_{\mathcal{M}}d\text{Vol}_{n+1}(\mathcal{M})\delta(\mathbf{x}\in\partial'\mathcal{M}) &= \int_{\partial'\mathcal{M}}dtd\text{Vol}_{n-1}(\partial\Omega_t).
\end{align}
Physically, we have after re-conversion to SI units, $\delta(\mathbf{x}\in\partial'\mathcal{M})$ has dimensions $[\delta]=1\text{ m}^{-1}$. In total the acoustic action functional has the structure,
\begin{align}
S_{acous} &= -\dfrac{\text{C}_{dim}}{2}\int_{\mathcal{M}}dtd^nx\sqrt{-\vert G_0\vert}p\Delta_{G_0} p\\
&+ C_{dim}c^2\rho_0\int_{\mathcal{M}}dtd^nx\sqrt{-\vert {G_0}\vert}p\partial_t\mathbf{v}_{\mathbf{n}}\delta(\mathbf{x}\in\partial'\mathcal{M}).
\end{align}
We check the units once again. In $n$ spatial dimensions we have
\begin{align*}
&[\rho_0]=1\text{ kgm}^{-n},\,[p] = 1\text{ Nm}^{-(n-1)}=1\text{ kgs}^{-2}\text{m}^{-(n-2)}\Rightarrow [c]=\sqrt{[p]/[\rho_0]}=1\,\text{ms}^{-1}\\
&[\partial_t^2 p]=[c^2\Delta_G p] = 1\text{ s}^{-2}\cdot 1\text{ kgs}^{-2}\text{m}^{-(n-2)}= 1\text{ kgs}^{-4}\text{m}^{-(n-2)}\\
&[\rho_0c^2\partial_t v_{\mathbf{n}}\delta((t,\mathbf{x}\in\partial'\mathcal{M}))]=1\text{ kgm}^{-n}\cdot 1\text{ m}^2\text{s}^{-2}\cdot 1\text{ms}^{-2}\cdot 1\text{m}^{-1}=1\text{kgs}^{-4}\text{m}^{-(n-2)}.
\end{align*}
Thus the units on both sides agree. Alternatively, one could have worked in SI units, and prove the procedure by considering an inhomogeneous wave equation with homogeneous boundary conditions and a homogeneous wave equation with inhomogeneous boundary conditions that differ by the factor of $c^2$. Using Cauchy-Kowalewskaja and the Green's operators for both equations which agree by their construction, one verifies that the missing factor is precisely $c^2$ in our sign convention. Let us take the functional derivative w.r.t. $p$. We equate this to $0$ in order to find the equations of motion of $p$ from the variational principle. The procedure yields
\begin{align}
0&\stackrel{!}{=}\dfrac{\delta S}{\delta p}\\
\Leftrightarrow 0 &= -\Delta_{G_0} p + \rho_0c^2\partial_t v_{\mathbf{n}}\delta(\mathbf{x}\in\partial'\mathcal{M})\\
\Leftrightarrow  \partial_t^2p - c^2\Delta_{g_0} p &= \rho_0c^2\partial_t \mathbf{v}_{\mathbf{n}}\delta(\mathbf{x}\in\partial'\mathcal{M}).
\end{align}
This is the acoustic wave equation to be used later on.
\end{itemize}
\textbf{A cohomology dessert: }So far, we have postponed the question how $\Omega^k_{sp}(\mathcal{M})$ relates to $\Omega^k(\Omega_t)$. The index $sp$ indicates that only forms in the span of $\lbrace dx^\mu\rbrace_{1\leq\mu\leq n}$ over a $C^\infty(\mathcal{M})$ ring are considered. Moreover, we need to ensure that $\Omega^k(\Omega_t)\simeq\Omega^k(\Omega_t')$ for $0\leq t,t'$ with $t\neq t'$. By the mutual diffeomorphy of the $\Omega_t$'s, we can use the following diagram to establish isomorphy between the exterior algebras of $\Omega_t,\,\Omega_{t'}$ for $0 < t,t'<\infty$ with $t\neq t'$. By smoothness of $\psi_{0\to t}:\Omega_0\to\Omega_t$, we can even  transform forms in $\Omega^k(\Omega_t)$ to forms in $\Omega^k(\Omega_{t'})$ smoothly for all $0\leq t,t' < \infty$.
\begin{figure}
\begin{center}
\includegraphics[width = 0.5\textwidth]{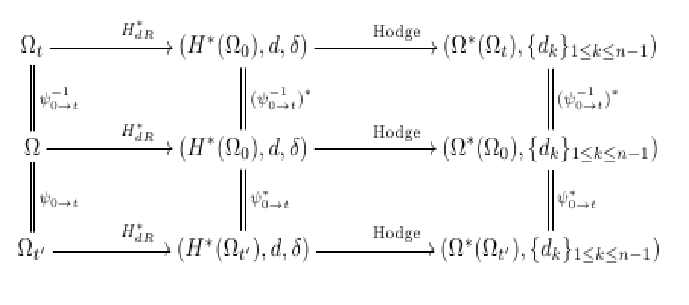}
\end{center}
\caption{Diffeomorphy of $\Omega_t\, ,\Omega_0$ and $\Omega_{t'}$ induces an isomorphism between the de Rham cohomology complex of $\Omega_t,\,\Omega_0$ and $\Omega_{t'}$}
\end{figure}
The second question is answered positively by the method presented in Fig. 7. The first question is somewhat more subtle. Technically, $\mathcal{M}$ is a fiber bundle so we would have to use the entire Serre spectral sequence to reduce the investigation of the deRhamn cohomology complex to the investigation of the deRham cohomology complexes of $\mathbb{R}^+$ and $(\Omega_t)_{t>0}$, \cite{sato}. However, by the topological constraints on $\Omega_t$ due to the setting, we know that a global bundle diffeomorphism between the bundles $\mathcal{M}$ and $\mathcal{M}_{ref}$ exists, $\Phi:\mathcal{M}\to\mathcal{M}_{ref}$. Further, there is a global bundle diffeomorphism between the bundles $\mathcal{M}_{ref}$ and $\mathcal{M}_0$, $\Phi_0:\mathcal{M}_0\to\mathcal{M}_{ref}$. We have composed the bundle diffeomorphisms to obtain a global bundle diffeomorphism between $\mathcal{M}_0$ and $\mathcal{M}$, $\Phi^{-1}\circ\Phi_0: \mathcal{M}_0\to\mathcal{M}$. This can be used to obtain the correspondences depicted in Fig. 10 following the previous logic.
\begin{figure}
\begin{center}
\includegraphics[width = 0.5\textwidth]{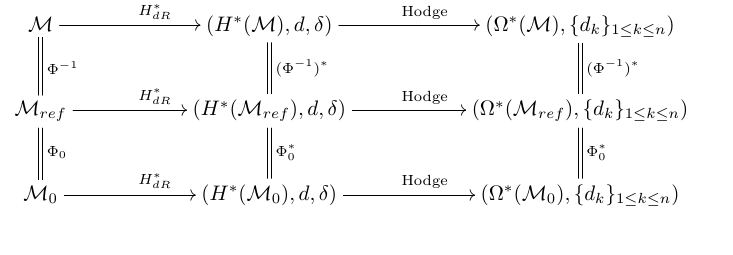}
\end{center}
\caption{Diffeomorphy of $\mathcal{M}\, ,\mathcal{M}_0$ and $\mathcal{M}_{ref}$ induces an isomorphism between the de Rham cohomology complex of $\mathcal{M},\,\mathcal{M}_0$ and $\mathcal{M}_{ref}$}
\end{figure}
The only thing left is to relate the deRham cohomology complex of $\mathcal{M}$ to the cohomology complex of the base space $\mathbb{R}^+$ and $\Omega_t$. By triviality of the bundles $\mathcal{M},\,\mathcal{M}_{ref},\,\mathcal{M}$, the Serre spectral sequence reduces to the Künneth formula for the cohomology of product manifolds. Since the deRham cohomology is a cohomology sequence over the group $\mathbb{R}$, the torsion groups in the general Künneth formula become trivial and can thus be ignored up to isomorphy / diffeomorphy. Since the cohomology complexes of the $(\Omega_t)_{t>0}$ are isomorphic, we have
\begin{align}
H^k(\mathcal{M})&\simeq \bigoplus_{l=0}^kH^l(\Omega_t)\otimes H^{k-1}(\mathbb{R}^+)\\
&\simeq \bigoplus_{l=0}^kH^l(\Omega_0)\otimes H^{k-1}(\mathbb{R}^+)\\
&\simeq (H^{k-1}(\Omega_0)\otimes H^{1}(\mathbb{R}^+))\oplus (H^{k}(\Omega_0)\otimes H^{0}(\mathbb{R}^+))
\end{align}
The group $H^0(\mathbb{R}^+)$ can be thought of as the space of functions $f:\mathbb{R}^+\to\mathbb{R}$ and $H^1(\mathbb{R}^+)$ is just the space of forms $\omega_t dt$ where $\omega_t$ is again a function $\omega_t:\mathbb{R}^+\to\mathbb{R}$. Since we are only interested in $\Omega^k_{sp}(\mathcal{M})$, the first contribution in the above decomposition is the relevant one, i.e., $H^k_{sp}(\mathcal{M})\simeq H^{k}(\Omega_0)\otimes H^{0}(\mathbb{R}^+)$. By the Hodge decomposition theorem, we can obtain from $H^k_{sp}(\mathcal{M})$ the decomposition of a $k$-form $\omega\in\Omega^k_{sp}(\mathcal{M})$ where we have to ignore any contributions including a $1$-form basis vector $dt$. But this reduces by the above derivation to a decomposition of $\omega\in H^k_{sp}(\mathcal{\Omega_t})$ over $H^0(\mathbb{R}^+)\otimes H^k(\Omega_t)$ where only the differential $d_k,\delta_{k-1}$ on $H^\ast(\Omega_t)$ give contributions. Thus, we can reduce the Hodge decomposition of spatial $k$-forms $\omega\in \Omega^k(\mathcal{M})$ to the Hodge decomposition from $ H^0(\mathbb{R}^+)\otimes H^k(\Omega_t)$ where the first factor does not contribute in another than the multiplicative way. In short, also the first question is answered in a positive way. We refer to \cite[chapter 9]{sato}, in particular Example 9.7. for an intuitive approach to the formalism.\newline
\newline
\textbf{Note on the perturbations of geometric quantities: }In this paragraph, we collect some equations from Riemannian geometry that are needed in order to calculate the equations of motion satisfied by $u$.\newline
\newline
\textbf{Geometrical action: }The goal of this paragraph to derive a membrane-plate-equation to model the boundary equations. This is an equation which features unlike the conventional wave equation a polynomial of degree $2$ in $\Delta_{g_0}^\partial$, the Laplace-Beltrami-operator on $\partial\Omega_0$, instead of just the Laplace-Beltrami operator. For this purpose, we partition the action functional $S_{geom}$ into two contributions. A dynamical term, $S_{geom,dyn.}$ containing the derivatives of the boundary vibrations with respect to the base-space coordinate $t$ and a stationary contribution which we think of as a contribution from a potential. I.e., our approach is
\begin{align}
S_{geom}&=S_{geom,dyn}+S_{geom.,stat.}\\
&=\rho_0\int_{\partial\mathcal{M}}dtd\text{Vol}_{n-1}(\partial\Omega_t)\,\mathcal{T}-T_0\int_{\partial\mathcal{M}}dtd\text{Vol}_{n-1}(\partial\Omega_t)\mathcal{V},
\end{align}
with the expressions $\mathsf{T}$ and $\mathsf{V}$ to obtained. Assume that $g_0^\partial$, the metric on $\partial\Omega_0$ is in diagonal form. We will later on use the considerations on how to obtain the differential equations for the boundary vibrations to obtain a working-action as we did for the acoustic pressure which was at first derived from   a fluid dynamical action. 
\begin{itemize}
\item\textbf{Stationary contribution: }The stationary contribution models curvature and volume effects that drive the boundary vibrations. Since the geometrical situation, we are investigating, gives us an imbedding $\partial\Omega_t\hookrightarrow\mathbb{R}^n$ and our external observer measures in the $\mathbb{R}^n$-reference metric $g_0$, we only consider extrinsic curvature effects. Recall that for a Riemannian submanifolds of co-dimension $1$, the relevant extrinsic curvature is the mean curvature, given by $\text{Tr}_{g_0}(\mathsf{II})$. The symbol $\mathsf{II}$ is the second fundamental form, $\mathsf{II}(\mathbf{v},\mathbf{w})=g_0(\nabla_{\mathbf{v}}\mathbf{w},\mathbf{n}(\partial\Omega_t))$, where $\mathbf{v},\mathbf{w}\in T(\partial\Omega_t)$ and $\mathbf{n}\in T(\partial\Omega_t)^\perp$ is the unit normal vector to $\partial\Omega_t$ obtained from the Gaussian map or by treating the oriented co-basis $\lbrace\omega_k\rbrace_{1\leq i\leq n-1}$ to the oriented and locally orthornormal basis $\lbrace\partial_i\rbrace_{1\leq i\leq n-1}$ of $T(\partial\Omega_t)$ as basis vectors in $\Lambda\mathbb{R}^n$ and completing with $\omega_n$ such that $\omega_1\wedge ... \wedge \omega_n = d\text{Vol}_n(\mathbb{R}^n)$. The (outward) unit normal then is given by $1 = \omega_n(\mathbf{n})=g_0(\mathbf{n},\mathbf{n})$. Because $\nabla$ is the Levi-Civita connection on $\Omega_t$, it induces by restriction a Levi-Civita connection on $\partial\Omega_t$ denoted by $\nabla^\partial$. Since $\nabla^\partial$ has vanishing torsion, $\nabla^\partial_\mathbf{v}\mathbf{w}-\nabla^\partial_{\mathbf{w}}\mathbf{v}=[\mathbf{v},\mathbf{w}]\in T(\partial\Omega_t)\perp T(\partial\Omega_t)^\perp$ for vector fields $\mathbf{v},\mathbf{w}\in T(\partial\Omega_t)$, we see that $\mathsf{II}(\mathbf{v},\mathbf{w})=\mathsf{II}(\mathbf{w},\mathbf{v})$, i.e., the second fundamental form is symmetric. By our definition (which uses implicitly that we have a co-dimension one submanifold for all $t\geq 0$), it can be characterized in terms of a matrix with components $\mathsf{II}_{ij}$ using $\mathbb{C}^\infty$-linearity in both arguments by symmetry of $\mathsf{II}$. Using a metricity argument and the Weingarten mapping, extrinsic Riemannian geometry textbooks hand us a formula to calculate the mean curvature, namely
\begin{align}
(n-1)H = \text{Tr}_{g_0\vert_{\partial\Omega_t}}[\mathsf{II}] = g_0\vert_{\partial\Omega_t}^{ij}\mathsf{II}_{ij}
\end{align}
Using that $\phi_{0\to t}:\partial\Omega_0\to\partial\Omega_t$ paramterizes $\partial\Omega_t$ in terms of coordinates on $\partial\Omega_0$ and that $\partial\Omega_t = \text{graph}_{\mathbf{y}}(u(t,\mathbf{y}))$, we obtain the easier expression,
\begin{align}
(n-1)H(u) = \dfrac{\text{Tr}_{g_0}[\nabla^{\partial, i}\nabla^{\partial, j} u]}{\sqrt{1+\nabla^\partial_i u\nabla^{\partial, i}u}} = \text{Tr}_{g_0}[\nabla^{\partial, i}\nabla^{\partial, j} u] + \mathcal{O}(\epsilon^3)=\Delta_{g_0}^\partial u
\end{align}
Here, $\nabla^\partial$ denotes the covariant derivative w.r.t. to the basis vectors induces from $\phi_{0\to t}$ as a parmeterization. Expanding further, we can neglect and non-linear contributions in $u$ and identify $\nabla^\partial$ with the Levi-Civita-connection on $\partial\Omega_t$, ignoring contributions containing $u$ and derivatives thereof at least quadratically. $\Delta_{g_0}^\partial$ is the Laplace-Beltrami operator on $\partial\Omega_0$. Last, we need a matrix $\mathsf{J}$ which is anti-symmetric, $\mathsf{J}_{ij}=-\mathsf{J}_{ji}$ and that satisfies the normalization condition $\text{Tr}((g_0^{-1}\mathsf{J})^2)=1$. If $g_0$ is the Euclidean metric and $n=3$, we may take e.g. the $2\times 2$ matrix with $-1$ on the upper to the diagonal and $1$ on the lower to the diagonal. Then we define the anti-symmetric matrix $\text{Tr}_{g_0}[\mathsf{II}]\mathsf{J} = \Lambda$ and call the object \emph{mean curvature form} on $\partial\Omega_t$. We can put the stationary contribution to the geometrical action together. Introduce a dimensionful parameter $\mu_{H}$ and set
\begin{align}
\mathsf{V}\equiv T_0\dfrac{\sqrt{\det\left(g_0\vert_{\partial\Omega_t} + 2\mu_H\Lambda\right)}}{\sqrt{\det g_0\vert_{\partial\Omega_t}}}.
\end{align}
Taylor-expansion around $g_0$ up to quadratic order in $\mu_H$ gives us, using the definition of $\mathsf{J}$,
\begin{align}
\mathsf{V}\circ\phi_{0\to t}=T_0\left(1 + \dfrac{\mu_H^2}{2}(\text{Tr}_{g_0}(\mathsf{II}))^2\right) = T_0 + \dfrac{\mu}{2}(\Delta_{g_0}u)^2,
\end{align}
on $\partial\Omega_0$ instead of $\partial\Omega_t$. The stationary contribution to the action functional is then given by
\begin{align}
S_{geom,stat} &= T_0\int_{\partial\Omega_t}d^{n-1}y_t\sqrt{\vert g_0\vert}\dfrac{\sqrt{\det\left(g_0\vert_{\partial\Omega_t} + 2\mu_H\Lambda\right)}}{\sqrt{\det g_0\vert_{\partial\Omega_t}}}\\
&= T_0\int_{\partial\Omega_0}dt d^{n-1}y\,\sqrt{\vert g_0\vert}\left(1+\dfrac{g_0^{ij}}{2}\nabla^\partial_i u\nabla_j^\partial u + \dfrac{\mu}{2}(\Delta_{g_0}^\partial u)^2\right),
\end{align}
where we have pulled back $\partial\Omega_t$ to $\partial\Omega_0$ by means of $\phi_{0\to t}$ and expanded the pullback metric $g = g_0+\delta g$. Contributions of order $\epsilon^3$ have been ignored. The base coordinate $t$ has not been pulled back because we have to treat the kinetic energy separately because of a different pre-factor $\sigma_m=\rho_m d$. Let us integrate the stationary contribution w.r.t. the base space coordinate $t_0$ of the reference fiber bundle $\mathcal{M}_0$. The first contribution is a constant volume term which has no physical information and is discarded. Noting that $G_0$ is always in block-diagonal form, with $G_0^{tt}=-1$ on $\partial\mathcal{M}_0$, we obtain the following formulation for the stationary contribution to the geometrical action functional.
\begin{align}
S_{geom, stat.} = T_0\int_{\partial\mathcal{M}_0}dt d^{n-1}y\,\sqrt{-\vert G_0\vert}\left(\dfrac{g_0^{ij}}{2}\nabla^\partial_i u\nabla_j^\partial u + \dfrac{\mu}{2}(\Delta_{g_0}^\partial u)^2\right)
\end{align}
Observe that the integrand of the functional is co-variant. I.e., we can now drop the assumption that $g_0$ on $\partial\Omega_0$ is diagonal. We notice that since $u:\Gamma(\partial\mathcal{M}_0\to\mathbb{R})$, the Levi-Civita connection acting on $u$ is simply the usual partial derivative. Next, we look for the dynamical contribution. We treat the case of a conservative perturbation bundle first and afterwards focus on the more interesting and relevant case of a dissipative perturbation bundle.  At the end of this sub-paragraph, we want to state where the idea for choosing this 'potential' comes from. Fiber bundles and operators acting on fiber bundles have been investigated since the 1960s using index theorems, \cite{nakahara}. The most prominent example of such an index theorem is the Atiyah-Singer-theorem which uses a characteristic class called Chern character. The Chern character is defined in terms of a simpler characteristic class, the so-called (total) Chern class. Chern, \cite{gauss1, gauss2} together with Weyl considered so-called invariant polynomials and inserted, instead of a real number $x\in\mathbb{R}$ a Lie-algebra valued, i.e., matrix-valued for many practical purposes, $2$-form, the curvature form $\Omega$ of the principal fiber bundle associated to a given vector bundle which we take to be the tangent bundle of a closed and oriented Riemannian manifold $M$, i.e., $E=TM$. Then, the total Chern class has been defined by the polynomial in the (deRham) cohomology sequence of the vector bundle $E=TM$,
\begin{align}
p_{\text{Chern-Weyl}}(t)\det\left(\mathds{1}_n + t\dfrac{\Omega}{2\pi i}\right)\in H_{dR}^\ast(E).
\end{align}
This definition can be formally expanded in the parameter $t$ and yields elements, the $2k$-th Chern class $c_{2k}(TM)$, in $H^{2k}(TM)$ for all $k\in\mathbb{N}_0$ such that $2k\leq n$. In particular, if $n$ is even the highest Chern class is equal to the Euler class $e(M)=e(TM)$. The integral over the Euler-class then reproduces the Gauss-Bonnet-Chern theorem. If $n$ is odd, the Euler-characteristic of a closed oriented Riemannian manifold $M$ of dimension $n$ vanishes. But, the corresponding Euler-class is also equal to $0$, since there are only non-zero Chern class with even indices. These action functionals have also found there application in physics, e.g., for the Born-Infeld action and the  Dirac-Born-Infeld action functional in AdS/CFT or non-linear electrodynamics, \cite{gauss1, gauss2, metric2, gauss1} or to explain the Aharanov-Bohm effect, \cite{nakahara, gauss1, gauss2}. The curvature form $\Omega$ is anti-symmetric and this is also needed throughout the calculations. Namely, for $k=1$, we have $c_{2}=0$ because $\text{Tr}(\Omega)=0$. From a more utilitaristic view-point, we have introduced the matrix $\mathsf{J}$ the way we did precisely following this logic to reproduce the mean-curvature squared term obtained by \cite{membranleo}.
\item\textbf{Conservative dynamics: }Before we tackle the case of dissipation, i.e., $\vert\partial_t u\partial ^t u\vert^2 < c^2_m\vert\partial_i u\partial^i u\vert$, we handle the conservative case, i.e., $\vert\partial_t u\partial^t u\vert = c^2_m\vert\partial^i u\partial_i u\vert$. We use $\rho_m$, the volume mass density of the membrane matching with conceptual basis of \cite{christine, anupam1, anupam2}. $\rho_m$ has units $[\rho_m]=1\text{kgm}^{-n}$ in $n=\dim\Omega_0=\dim\Omega_t$ spatial dimensions. In order to achieve consistency in terms of the SI unit system, \cite{christine, anupam1, anupam2} introduced the experimentally well-known membrane thickness $d$ and defined implicitly the surface mass density
\begin{align}
\sigma_m\equiv \rho_m d.
\end{align}
We will use this equation with our quantities as well. Exploiting diagonality of $G_0$ once more, the kinetic term in lowest non-trivial order in $\delta G\vert_{\partial\mathcal{M}_0}$ thus is given by
\begin{align}
S_{geom}\supset\sigma_m\int_{\partial\mathcal{M}_0}dtd^{n-1}y\, G_0^{tt}\partial_t u\partial_t u
\end{align}
Higher order terms in $u$ can be neglected because the integrand attains already the maximum order in $u$,namely quadratic order, and the metric perturbations $\delta G$ would introduce additionally quadratic contributions in $u$. We notice the convenience that $\mathcal{M}_0$ is the trivial fiber bundle $\mathbb{R}^+_0\times\Omega_0$: We can integrate by parts without having to respect a time-dependence of the boundary $\partial\mathcal{M}=\partial\mathcal{M}(t)$.
\item\textbf{Conservative action: }We can put the pieces together and obtain the geometrical action functional $S_{geom}$ modeling the dynamics of the perturbed fibers $(\Omega_t)_{t>0}$ relative to the unperturbed fiber $\Omega_0$ as
\begin{align}
S_{geom}=\int_{\partial\mathcal{M}}dtd^{n-1}y\,\sqrt{-\vert G_0\vert_{\partial\mathcal{M}_0}\vert}\mathcal{L}_{geom},
\end{align}
where the geometrical Lagrangian density $\mathcal{L}$ is given by
\begin{align}
\mathcal{L}=\mathcal{T}(u)+\mathcal{V}(u),
\end{align}
where the plus sign results from the signature of the Minkowskian metric $G_0$ on $\partial\mathcal{M}_0$. The kinetic energy $\mathcal{T}$ is given by
\begin{align}
\mathcal{T}=\sigma_m G_0^{tt}\partial_t u\partial_t u,
\end{align}
the potential $\mathcal{V}$ can be decomposed into an intrinsic part $\mathcal{V}_{int}$ and an extrinsic part $\mathcal{V}_{ext}$. The intrinsic contribution models changes in the surface area of the boundary relative to the equilibrium boundary, i.e., $(\text{Vol}_{n-1}(\partial\Omega_t)-\text{Vol}_{n-1}(\partial\Omega_0))/(\text{Vol}_{n-1}(\partial\Omega_0))$. The extrinsic contribution stems from the mean curvature form $\Lambda$, more precisely $\text{Tr}_{g_0}(\Lambda^2)$. The overall pre-factor is set equal to the membrane tension, c.f., \cite{anupam1}.
\begin{align}
\mathcal{V}&= T_0\left(\dfrac{g_0^{ij}}{2}\nabla^\partial_i u\nabla_j^\partial u + \dfrac{\mu}{2}(\Delta_{g_0}^\partial u)^2\right)
\end{align}
Putting this together in the integral and setting $\dim\Omega_0=n$ i.e., $n-1=\dim\partial\Omega_0$, we obtain the complete geometrical action functional $S_{geom}$ in the conservative case,
\begin{align}
S_{geom} &= \dfrac{1}{2}\int_{\partial\mathcal{M}_0}dtd^{n-1}y\sqrt{-\vert G_0\vert}\,\left(\sigma_m G_0^{tt}\partial_tu\partial_t u + T_0G_0^{ij}\partial_i u\partial_j u +T_{curv}(\Delta_{g_0}^\partial u)(\Delta_{g_0}^\partial u)\right).
\end{align}
The conservative action on $\partial\mathcal{M}_0$ is the starting point for introducing a damping term by a null-set modification, i.e., a modification only on the $\text{Vol}_{n+1}$-null-set $\partial\mathcal{M}_0$, of the metric $G_0$ on $\mathcal{M}_0$.
\item\textbf{Dissipative case: }The dissipative case is more important than the conservative case for two reasons. Firstly, the boundary vibrations in physical reality naturally dissipate due to frictional effects in the membranes themselves and due to frictional effects of the membranes and the surrounding media, e.g., air. On the bio-level, one has one a much larger time-scale also the biochemical effect of degradation of the constituent bio-molecules of the membranes. Thus, dissipation of the boundary vibrations' energy is more realistic than their conservation. Secondly, recall that the perturbation bundle must be physical as well as $\infty$-periodic. While physicality requires the boundary vibrations $u$ to satisfy $\vert\partial_t u\vert^2\leq c^2_m\vert \partial_i u\partial^i u\vert$, the property of the perturbed bundle $\mathcal{M}$ to be $\infty$-periodicity requires two limits. The first limit $\lim_{t\to 0}\Omega_t=\Omega_0$ is clear:, On the boundary level by properness of the diffeomorphisms $\psi_{0\to t}$, $\lim_{t\to 0}\partial\Omega_t= \partial\Omega_0$. The second limit, i.e., $\lim_{t\to\infty^-}\Omega_t=\Omega_0$, is on the boundary level $\lim_{t\to\infty}\partial\Omega_t=\partial\Omega_0$. We exemplify this by setting $u/\epsilon=\mathcal{N}\sin\left(g^{0}_{ij}k_0^ix^j\right)\sin(\omega_0 t)$ where $\mathcal{N}$ denotes a normalization constant such that the right hand side is normalized to $1$ on $\partial\Omega_0$ and $\mathbf{k}=k^i_0\partial_i$ and $\omega = \omega_0\partial_0$ are given vector fields in $\mathcal{V}(\partial\mathcal{M}_0)$ such that $u$ satisfies the conservative case in the definition of physicality. Symbolically, we can write
\begin{align}
\partial\Omega_t = (\text{id}_{\partial\Omega_0}+u\mathbf{n}_{\partial\Omega_0})\cdot\partial\Omega_0 \equiv (1+u(t))\partial\Omega_0.
\end{align}
The short-hand notation indicates that $\partial\Omega_t$ is just the boundary $\partial\Omega_0$ perturbed by the local displacement given by the boundary vibrations $u$. If we let $t\to\infty$, the limit
\begin{align}
\lim_{t\to\infty}\partial\Omega_t=\lim_{t\to\infty}(\partial\Omega_0+u(t,\partial\Omega_0))=!?!?,
\end{align}
is ill-defined due to periodicity of the sine functions in $u$. If there was a function, say $D=D(t)$ such that $\lim_{t\to 0}D=1$ and $\lim_{t\to\infty}D=0$, and the boundary vibrations $u$ would be given by $u/\epsilon = \mathcal{N}D(t)\sin(\omega t)\sin(g^0_{ij}x^ik^j)$ instead, we would have
\begin{align}
\lim_{t\to 0}\partial\Omega_t = \lim_{t\to 0}(\partial\Omega_0 + u(t,\partial\Omega_0))=\partial\Omega_0,
\end{align}
because $\lim_{t\to 0}D(t)=1$ and $\lim_{t\to 0}\sin(\omega_0 t)=0$ and
\begin{align}
\lim_{t\to\infty}\partial\Omega_t = \lim_{t\to\infty}(\partial\Omega_0+u(t,\partial\Omega_0))=\partial\Omega_0,
\end{align}
because $\vert\lim_{t\to\infty}u/D\vert \leq \text{const}. < \infty$ stays finite and $\lim_{t\to 0}D(t)=0$ by requirement on $D$. Thus, $\infty$-periodicity is saved at the price of destroying the conservation of energy by introducing a dissipative contribution $D(t)$. Since $\infty$-periodicity is needed to assure existence of the solution, we have demonstrated the following theorem.\newline
\newline
\textbf{Theorem: }\emph{Let $\mathcal{M}$ be a proper and physical perturbation bundle. $\mathcal{M}$ is not $\infty$-periodic if it is conservative.}\newline
\newline
The contraposition is easier to understand.\newline
\newline
\textbf{Theorem: }\emph{Let $\mathcal{M}$ be a proper and physical perturbation bundle. $\mathcal{M}$ is not conservative if it is $\infty$-periodic.}\newline
\newline
Notice that this doesn't mean that the perturbation bundle is dissipative because the dissipation has to take place at every point $\mathbf{y}\in\partial\Omega_0$ whereas the bundle is already non-conservative if there is one point where the conservation equality in the definition of a physical perturbation bundle is not satisfied. E.g., there might be a time $T$ such that for $0<t< T$ the bundle is conservative and at $t=T$ it starts to be dissipative for all $t\geq T$. By the smoothness requirements on the setting, it follows that:\newline
\newline
\textbf{Theorem: }\emph{Let $\mathcal{M}$ be a proper and physical perturbation bundle and $\mathbb{R}^+_0\ni T\geq 0$ be given. If for all $t\geq T$ the perturbation bundle $\mathcal{M}$ satisfies the dissipative version of the physicality condition, it is $\infty$-periodic.}\newline
\newline
As we are interested in calculational method, we need to include damping in the geometrical action functional. For this reason, we notice that in the wave equations of interest \cite{christine, anupam1, anupam2}, dissipation can be achieved in the following way. For a given dissipation function $D(t)$, we introduce a second function $\Sigma(t)=1/(D(t))^2$. 
\begin{align}
\partial_t^2 u + \partial_t\log (\Sigma(t))\partial_t u -c^2_m\Delta_{\partial\Omega_0}u = f(t,\mathbf{y}),
\end{align}
where $f$ is a given source term for our purposes. We can rearrange this as
\begin{align}
\dfrac{1}{\Sigma(t)}\dfrac{1}{\partial t}\left(\Sigma(t)\dfrac{\partial u}{\partial t}\right) - c^2_m\Delta_{\partial\Omega_0} = f(t,\mathbf{y}).
\end{align}
We require for this structure to be valid $D(t)>0$ for all $t>0$, $D:\mathbb{R}^+_0\to\mathbb{R}^+$ and $\partial_t D < 0$ for all $t>0$. Together with the constraint $D(t=0)=1$, we have the following ordinary differential equation for $D$,
\begin{align}
\partial_t D(t) = -\alpha f(D,t) + g(t). 
\end{align}
where we constrain ourselves to $f:\mathbb{R}^+_0\times\mathbb{R}^+_0\to\mathbb{R}^+$ and $g:\mathbb{R}^+\to\mathbb{R}^-$ are Lipschitz-continuous and $\mathcal{C}^1$-functions. In the case $f(D,t)=f_1(t)f_2(D)$ where $f_1$ and $f_2$ are positive and further $f_2$ even bijective, we have for $g=0$,i.e., for the homogeneous differential equation,
\begin{align}
\int_{1}^D\dfrac{dD}{f_2(D)} = -\alpha\int_{0}^t\dfrac{dt}{f_1(t)}.
\end{align}
We define the auxiliary functions $h_1, h_2$ by means of
\begin{align}
h_1' = \dfrac{h_1}{f_1}\text{ and }h_2' = \dfrac{h_2}{f_2}.
\end{align}
with $h_1(0)=0=h_2(1)$. We can solve the ordinary differential equations to obtain
\begin{align}
h_1(t)=\exp\left(\int_{0}^{t}\dfrac{d\tau}{f_1(\tau)}\right)\quad h_2(D)=\exp\left(\int_1^D\dfrac{d\tilde{D}}{f_2(D)}\right).
\end{align}
We notice that $h_2$ is also bijective because of bijectivity of $f_2$. Rearranging the differential equations for $h_1,h_2$ yields
\begin{align}
\dfrac{1}{f_1(t)}=\partial_t\log(h_1(t))\text{ and }\dfrac{1}{f_2(D)}=\partial_D\log(h_2(D))
\end{align}
Substituting in the integral formulation of the ordinary differential equation we are interested in yields
\begin{align}
\int_{1}^DdD\partial_D\log h_2(D) = -\alpha\int_{0}^t d\tau\partial_\tau\log h_1(\tau).
\end{align}
Using the initial conditions $h_1(0)=1$ and $h_2(1)=1$ and composing the equation from the left with the bijective exponential function $\exp :\mathbb{R}\to\mathbb{R}^+$, we find an implicit equation for $D$ in terms of $t$,
\begin{align}
h_2(D) = (h_1(t))^{-\alpha}\Leftrightarrow D = h_2^{-1}(h_1(t)),
\end{align}
by requirement of bijectivity on $f_1,f_2$ and thus on $h_2$. If $g\neq 0$, we can use variation of constants principle to obtain with $c\in\mathbb{R}$ used for matching with the overall initial condition $D(0)=1$,
\begin{align}
D = D_{hom}+D_{inh}\text{ with }D_{inh}=cD_{hom}(t)\int_{0}^t\dfrac{d\tau\, g(\tau)}{D_{hom}(\tau)}.
\end{align}
$D_{hom}$ denotes the previously derived solution to the homogenized problem. The easiest model is given by $f_1(x)=1,\, x=f_2(x)$ and results in linear damping,
\begin{align}
D = \exp(-\alpha t)
\end{align}
This results in $\Sigma(t)=e^{2\alpha t}$. In the general case, i.e., when $D$ is given by the ordinary differential equation $\partial_t D = f(D,t)+g(t)$, with $f(D,t)=f_1(t)f(D)$ with suitable $f_1,f_2$ and $g$, we obtain by the previously outlined method one unique solution $D$ on a maximal interval of existence, $I\subseteq\mathbb{R}^+_0$. The technical issue to resolve in our theory is the self-adjointness of $\square^\Sigma_{\partial,G_0}$, defined by
\begin{align}
\square^\Sigma_{\partial,G_0}\equiv\dfrac{1}{\Sigma(t)}\dfrac{\partial}{\partial t}\left(\Sigma(t)\dfrac{\partial}{\partial t}\right)-c^2_m\Delta_{\partial\Omega_0}.
\end{align}
Self-adjointness can be restored most easily by modifying the metric $G_0$ on $\mathcal{M}_0$ and $\mathcal{M}_{ref}$. The acoustic wave equation on the other hand does not include damping contributions. Thus, the modification of the metric $G_0$ must take place only on the boundary $\partial\mathcal{M}_0^\circ = \bigcup_{t\geq 0}\lbrace t\rbrace\times\partial\Omega_0$. This is a Lebesgue Null-set which does not affect the acoustic wave equation if the modification takes only place in the base space component $G^0_{tt}$. One way to achieve this is to deform the metric $G_0$ in $\mathcal{M}\setminus\partial\mathcal{M}$ in a way to match the metric $G_0^\partial$ at $s=1$, i.e., at the boundary $\partial\mathcal{M}$
\begin{align}
G_0^\partial = \left(\begin{array}{ccc}-(\Sigma(t))^{-2}dt^2 & 0 & \mathbf{0}^T_{n-1}\\ 0 & 0 &\mathbf{0}^T_{n-1}\\
\mathbf{0}_{n-1} & \mathbf{0}_{n-1} & g^0_{ij}\vert_{\partial\Omega_0}\end{array}\right).
\end{align}
The function $(\Sigma(t))^{-1}$ with arguments $0<t<\infty$ is called \emph{time-lapse-function} in the ADM formalism in gravitational and black hole physics. The matching with the full unperturbed bundle metric $G_0$ can be achieved by setting
\begin{align*}
\Sigma^2(t,s)=\left[\lim_{l_c\to 0^+}\left[(\exp(-\vert s-1\vert/L_c)-\exp(-\vert s-1\vert/l_{c}))+\exp(-\vert s-1\vert/l_c)\Sigma^2(t)\right]\right]^{-1}
\end{align*}
where we let the \emph{time lapse correlation length} $l_c\to 0^+$ and $L_c$ is the \emph{causal correlation length} defined by $L_c\equiv\sqrt[n]{\text{Vol}^2_n(\Omega_0)}/l_c$. The pre-factor depending on $\text{Vol}_n(\Omega_0)$ just serves to restore the physical units of length correctly, i.e., to assure $[l_c]=1\,\text{m}=[L_c]$. Mathematically, the $l_c$-dependent exponential corresponds to insertion of a delta function centered at $s=1$ and normalized to unity at $s=1$ and the $L_c$-dependent exponential corresponds to an alternative way to express $1$, namely $1=\lim_{l_c\to 0^+}\exp(-\vert s-1\vert/L_c(l_c))$. Physically this means that the damping stored in the second contribution to $\Sigma(t,s)$ is only defined on the boundary $\partial\mathcal{M}_0$ but not present in the rest of the perturbation bundle $\mathcal{M}_0$. Since the boundary $\partial\mathcal{M}_0$, it does not affect the derivation of the acoustic wave equation. In other words, when we derive an equation on $\mathcal{M}_0$ we can still use the original metric $G_0$ instead of $G_0^\Sigma$ defined by
\begin{align}
G_0^\Sigma = \left(\begin{array}{cc}-\Sigma^2(t,s)dt^2 & \mathbf{0}^T_n\\
\mathbf{0}_n & g_{\mu\nu}^{0}(\mathbf{x}=(s,\mathbf{y}))\end{array}\right).
\end{align}
Notice that in the limit $l_c\to 0^+$, the exponential functions approach a Dirac delta function normalized to $1$ at $s=1$ centered on the boundary $\partial\Omega_0$ at $s=1$. In order to appeal to intuition, we note down the structure of the \emph{boundary-time-lapse-function} $\Sigma(t,s)$
\begin{align}
\Sigma(t,s)=\dfrac{1}{(1-\delta(s-1))+\delta(s-1)\Sigma^2(t)}.
\end{align}
With respect to a volume integration over $\mathcal{M}$, the contributions of $G_0$ at $s=1$ are just valid on a Lebesgue Null-set w.r.t. the Lebesgue-Borel integration measure $\text{Vol}_{n+1}:\mathcal{B}(\mathbb{M}_{n+1})\to\mathbb{R}^+_0$ on $\mathcal{M}_0$. Likewise for a given $t$, the boundary contribution live on $\lbrace t\rbrace\times\partial\Omega_0$ which are Lebesgue Null-sets w.r.t. the Lebesgue-Borel integration measure $\text{Vol}_n:\mathcal{B}(\mathbb{R}^n)\to\mathbb{R}^+_0$ on $\Omega_0$. Physically, the correlation lengths $l_c,L_c$ in the boundary-time-lapse-function $\Sigma(t,s)$ correspond to an ultra-short-distance modification of the metric element $G^{0,\Sigma}_{tt}$ which is de-correlated from the rest of the metric $G_{0,\Sigma}(s\neq 1)=G_0$ on $\mathcal{M}\setminus\partial\mathcal{M}$. The metric on $\mathcal{M}_0\setminus\partial\mathcal{M}_0$, i.e., effectively $G_0$, correlates all points of the perturbation bundle $\mathcal{M}_0$. Put in lax words, the astounding behavior of the metric $G_0^\Sigma$ at $s=1$ is not noticed by the points $(t,\mathbf{x}=(s,\mathbf{y}))$ with $s\neq 1$ on $\mathcal{M}_0$ and - after continuation of $G_0^\Sigma$ by the radial extrapolation introduced in the first section - the whole Minkowskian imbedding manifold $\mathbb{M}_{n+1}\hookleftarrow\mathcal{M}_0,\mathcal{M},\mathcal{M}_{ref}$. For the fluid action $S_{fluid}$ and acoustic action $S_{acous}$, the modification of the metric $G_0\to G_0^\Sigma$ is thus irrelevant, for the geometrical action defined on $\partial\mathcal{M}_0$, i.e. $s=1$, $S_{geom}$ the modification $G_0\to G_0^\Sigma$ yields to the inclusion of a damping term. More precisely, we have the modifications $\sqrt{-\vert G_0\vert_{\partial\mathcal{M}_0\vert}}\to \sqrt{-\vert G_0\vert_{\partial\mathcal{M}_0\vert}}(\Sigma(t))^{-1}$ and $G_0^{tt}\partial_t u\partial_t u=-\partial_t u\partial_t u$ becomes $G^{tt}_{0,\Sigma}\partial_t u\partial_t u=-\Sigma^2(t)\partial_t u\partial_t u$. The boundary-time-lapse-function $\Sigma(t,s)$ has been plotted in the example $\Sigma(t)=\exp(2\alpha t)$ in Fig. 11.
\begin{figure}
\begin{center}
\includegraphics[width = 0.67\textwidth]{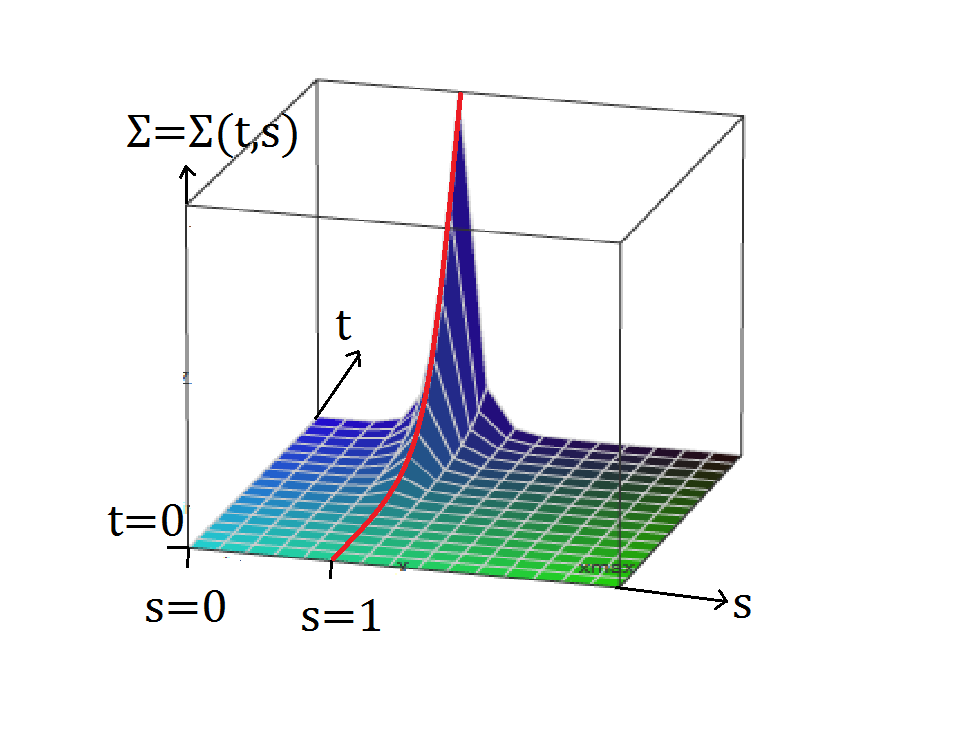}
\end{center}
\caption{The function $-G_{tt}=\Sigma(t,s)$ has been plotted for $s\in(0,2)$ and $\Sigma(t)=\exp(2\alpha t)$ with $\alpha = 1$ in non-dimensional units. The horizontal axis depicting $\Sigma$ has an offset $+1$. Further, $l_c=0.01,\,L_c = 10^6$ has been set in non-dimensional units. One sees that $\Sigma(t,s)=1$ except when $s=1$, where it is $\Sigma(t,s=1)=\Sigma(t)=\exp(2\alpha t)$ with $\alpha=1$. The fall-offs can be smoothed out by letting $l_c$ approach $0^+$. In the plots, $l_c=10^{-2}$ has produced the neatest figure.}
\end{figure}
\item\textbf{Dissipative action:} We summarize the results for the stationary contribution to the geometrical action functional and the dynamical contribution to the geometrical action functional. By definition of a perturbation bundle, we have homogeneous initial conditions for the boundary vibrations, $u(t=0)=0$ and $\partial_t u(t=0) = 0$. The first initial condition is a consequence of the perturbation bundle evaluated at the base space point $t=0$ is just equal to the reference bundle $\mathcal{M}_0$ evaluated at $t=0$ by definition of a perturbation bundle. The second initial condition is a consequence of properness and dissipativity of the perturbation bundle. Properness states that $\phi_{0\to t}$ maps up to an $\epsilon$ a point $\mathbf{x}\in\partial\Omega_0$ to $\mathbf{x}$ again. and that the same holds true for the linearization of $\phi_{0\to t}$. I.e., if $c=c(r)$ is a regular smooth curve passing through $\mathbf{x}$ in $\partial\Omega_0\subset\mathbb{R}^n$ at $r=0$, the tangential at that point, $\dot{c}(0)$, is mapped by the differential of $\phi_{0\to t}$ at $\mathbf{x}\in\partial\Omega_0$ almost to the same vector in $\mathbb{R}^n$, the absolute deviation being the perturbation strength $\epsilon$ again. Dissipativity in turn allows us to identify suitable combinations of derivatives of $u$, i.e., by definition of $u$, derivatives of $\phi_{0\to t}-\text{id}_{\partial\Omega_0}$ with $\partial_t u$. At $t=0$, we have $\phi_{0\to t}=\text{id}_{\partial\Omega_0}$ such that by the identification of $\phi_{0\to t}$ with its differential by properness of the perturbation bundle, we can conclude $\partial_t u\vert_{t=0}=0$ as well. We note down the result of our considerations
\begin{align}
u(t=0,\mathbf{y})=0=\partial_t(t=0,\mathbf{y})\text{ for all }\mathbf{y}\in\partial\Omega_0.
\end{align}
Physically, this states that the reference bundle $\mathcal{M}_0$ will not deform its fibers $\Omega_0$ in time $t$ by itself, but the formation of a perturbation bundle $\mathcal{M}$ needs an external stimulation. In view of the principle of least action, this is reasonable from the physical viewpoint. We turn to summarizing our considerations to build up the geometrical action functional for the boundary vibrations and explaining the physical intuition and meaning of the contributions stored in the symbolic expression. Denoting by $G_0^\Sigma$ the modified metric tensor on $\mathcal{M}_0$ to account for dissipativity of the perturbation bundle $\mathcal{M}_0$, we have found in the previous sub-paragraphs,
\begin{align}
\begin{split}
&S_{geom.,diss.}[u]\\
&=\dfrac{1}{2}\int_{\partial\mathcal{M}_0}\sqrt{-\vert G_0^\Sigma\vert}\left(\sigma_m (G_0^{\Sigma})^{tt}(\partial_t u)(\partial_t u)+T_0(G_0^{\Sigma})^{ij}(\partial_i u)(\partial_j u)+d^2T_{curv}(\Delta_{g_0}^\partial u)(\Delta_{g_0}^\partial u)\right).
\end{split}
\end{align}
The quantity $\sigma_m = \rho_m d$ is the boundary vibrations' mass density per $(n-1)$-dimensional area. It is expressed via the volume mass density of the boundary vibrations $\rho_m$ and a characteristic intrinsic length scale of the boundary vibrations $u$. In \cite{anupam2} this is the thickness of the tympanic membranes, denoted by $d$. Mathematically, we could have worked with the symbol $\sigma_m$ straight away, but experimentally $\rho_m$ and $d$ can be determined, whereas $\sigma_m$ is a derived quantity not directly accessible to our experimental collaborators. The next constant is the tension of the boundary vibrations, $T_0$. The same letter denotes in \cite{anupam2} the membrane tension of the tympanic membranes. It is a material constant that is needed to derive the phase velocity of the membrane's flexural waves. The next constant is new and called $T_{curv}$. In \cite{membranleo}, a $n=3$-model for bio-membranes had been derived from a geometrical viewpoint starting form the equations of elasticity theory. We interpret $T_{curv}$ here as a curvature tension that describes how much the membranes resist to changes in mean curvature. If $T_{curv}/(T_{0})\gg 1$, the changes in curvature dominate over the flexural membrane vibrations and we could idealize the boundary vibrations as a damped Kirchhoff-plate that gives bending waves predominantly. Effectively, the boundary vibrations equation would then reduce to a damped higher-dimensional analogue of the bio-membrane equation of motion derived in \cite{membranleo}. If we have $T_{curv}/T_0 \ll 1$ on the contrast, the bending waves propagate much slower than the flexural waves. This means that we can neglect the bending contribution already in the action functional $S_{geom}$ and consider only the flexural vibrations of the boundary vibrations. We arrive at a higher-dimensional analog of the damped membrane equation considered in \cite{christine, anupam1, anupam2} that has been investigated as a special case in \cite{david1}. In the following sub-paragraph we will be concerned with three issues. Firstly, we want to localize the boundary vibrations. In biophysics or classical acoustics as the main physical motivation for the theory, one usually has not a vibration of the entire boundary of $\partial\Omega_t$ but only some, pathwisely unconnected parts, say $\lbrace\Gamma_i\rbrace_{1\leq i\leq N}$ vibrate. In the biophysics of animal hearing, this intuitively clear. If e.g. lizards want to localize the sounds a prey makes, not the entire interaural cavity in the lizards head starts to vibrate but only the tympanic membranes. Geometrically, the membranes are disconnected from each other and thus two membrane equations for two tympanic membrane displacements must be solved. The second issue  is a consequence of the first issue: By definition, the boundary $\partial\Omega_t$ of each fiber $\Omega_t$ are topologically closed with $\partial^2\Omega_t = \emptyset$. For the pairwisely in $\partial\Omega_0$ unconnected parts $\lbrace\Gamma_i\rbrace_{1\leq i\leq N}$, this is not the case if $N > 1$: They will generically have a boundary $\partial\Gamma_i\neq\emptyset$. Thus, we need to specialize boundary conditions to the boundary vibrations equations. The third issue to address asks why the boundaries $\partial\Omega_t$ should actually start oscillating. Mathematically, this can be reformulated and we can ask what source term to the boundary vibrations we should specialize.
\item\textbf{Source term and boundary conditions: }We have to give the boundary vibrations boundary and initial conditions as well as an external source term,say $\Psi=\Psi(p,p_{ex})$. One way around boundary conditions is to impose a periodicity condition on $u$ by arguing that $\partial(\partial\Omega_0)=\emptyset$ such that an integration by parts w.r.t. the coordinate $\lbrace y^i\rbrace_{1\leq i\leq n,i\neq s}$ results in no boundary term. This has been done in \cite{membranleo} for closed bio-membranes such as cellular membranes. However, in \cite{anupam1, anupam2, christine} tympanic membranes, i.e., membranes which do not cover all of $\partial\Omega_0$, have been considered. In the set-up the tympanic membranes do not bound all of the interaural cavity, i.e., in our language $u\neq 0$ only on connected components of $\partial\Omega_0$. By compactness of $\Omega_0$ in $\mathbb{R}^n$, the boundary manifold $\partial\Omega_0$ is relatively compact in $\mathbb{R}^n$. I.e., a there are $N,M\in\mathbb{N}_0$ such that
\begin{align}
\partial\Omega_0\equiv\biguplus_{i=1}^N\Gamma_i\uplus\biguplus_{i=N+1}^{N+M}\Gamma_i,
\end{align}
is a decomposition of the $(n-2)$-connected $\partial\Omega_0$ in $(n-2)$-connected and retractible components $\Gamma_i, 1\leq i\leq N+M$ such that $\text{Vol}_{n-1}(\Gamma_i\cap\Gamma_j)=0$ for $i\neq j$. I.e., the $\Gamma_i$'s are allowed to share at most a Lebesgue Null-set w.r.t. the Lebesgue-Borel integration measure $\text{Vol}_{n-1}:\mathcal{B}(\partial\Omega_0)\to\mathbb{R}^+_0$ on the Borel-algebra of the boundary $\partial\Omega_0$ of the fiber $\Omega_0$ of the unperturbed bundle $\mathcal{M}_0=\mathbb{R}^+_0\times\Omega_0$. We define $\Gamma_i$ for $1\leq i\leq N$ by being a decomposition of $\partial\Omega_0$ in smooth, retractible, bounded $(n-1)$-dimensional sub-manifolds $\Gamma_i$,
\begin{align}
\biguplus_{i=1}^N\Gamma_i \equiv \bigcup_{t\geq 0}\text{supp}_{\partial\Omega_0}(u) \equiv \bigcup_{t\geq 0}\overline{\left\lbrace\mathbf{y}\in\partial\Omega_0 \vert: u(t,\mathbf{y})\neq 0\right\rbrace}=\overline{\left\lbrace\mathbf{y}\in\partial\Omega_0 \vert\exists t\geq 0: u(t,\mathbf{y})\neq 0\right\rbrace} .
\end{align}
The smoothness and manifold properties follow from smoothness of $u$ by smoothness of the $\psi_{0\to t}$'s for $t\geq 0$ stored in the definition of the unperturbed bundle $\mathcal{M}_0$ and the perturbed bundle $\mathcal{M}$. The over-line denotes topological closure in $\partial\Omega_0$ topologized by the relative topology $\tau_{\partial\Omega_0}(\Omega_0)$ where we recall that $\Omega_0$ has been topologized by the relative topology $\tau_{\Omega_0}(\mathbb{R}^n)$, using that $\Omega_0\hookrightarrow\mathbb{R}^n$ by the embedding $\iota_0$, and the topology is w.r.t. the norm $\Vert .\Vert_{g_0}$ induced by the metric $g_0$ extended from $\Omega_0$ to $\mathbb{R}^n$ by radial extrapolation. In turn $\Gamma_i$ for $N+1\leq i\leq N+M$ is defined as a decomposition in $\Gamma_i$'s of the same properties as the $\Gamma_i,\,1\lq i\leq N$,
\begin{align}
\biguplus_{i=N+1}^{N+M}\Gamma_i\equiv\overline{\left(\biguplus_{i=1}^N\Gamma_i\right)^\complement}\equiv\overline{\partial\Omega_0\setminus\left(\biguplus_{i=1}^N\Gamma_i\right)}=\overline{\partial\Omega_0\setminus\overline{\left\lbrace\mathbf{y}\in\partial\Omega_0 \vert\exists t\geq 0: u(t,\mathbf{y})\neq 0\right\rbrace}}.
\end{align}
The operation $\complement$ denotes complement in $\partial\Omega_0$and $\uplus$ stands as before for disjoint union modulo Lebesgue Null-sets w.r.t. the Lebesgue-Borel integration measure $\text{Vol}_{n-1}:\mathcal{B}(\partial\Omega_0)\to\mathbb{R}^+_0$. 
\end{itemize}
In the following, we will summarize our results on the equations we are interested in. As such, we decompose the action functional $S_{geom,diss}$ into $N$ action functionals for the $u_i\equiv u\vert_{\Gamma_i}$,
\begin{align}
S_{geom,diss}[u] = S_{geom,diss}\left[\sum_{i=1}^N u_i\right] = \sum_{i=1}^NS_{geom,diss}[u_i],
\end{align}
by pairwise disjointness modulo a Null-set of the $(\Gamma_i)_{1\leq i\leq N}$. \emph{For the derivation of the boundary conditions, let $i\in\lbrace 1,...,N\rbrace$ be arbitrary. We will also suppress the index $i$ of the $u_i$'s notationally in order to avoid confusion with the coordinate index $i$.} Using the previously obtained expressions for the sectional curvatures $K_{ij}$ and mean curvature $H$ in terms of the Levi-Civita connection $\nabla=\nabla^{g_0}$ and the Laplace-Beltrami operator $\Delta_{g_0}$,
\begin{align*}
H &= Tr_{g_0}[\nabla_i\nabla_j u] + \mathcal{O}(\epsilon^2)
\end{align*}
we can start integration by parts. We choose the normal vector $\mathbf{n}(t=0)=-\partial_t$ for the boundary term at $t=0$ and the normal vector $\mathbf{n}=\mathbf{n}_{\partial\Omega_0}\vert_{\partial\Gamma_i}$ for the boundary term at $\partial\Gamma_i$ for $1\leq i\leq N$. We choose the Dirichlet boundary conditions $u_i(t,\mathbf{y})=0$ if $\mathbf{y}\in\partial\Gamma_i$ for all $1\leq i\leq N$. We the first and second contribution to the geometrical action function by parts once, We integrate twice by parts the third contribution containing the Laplace-Beltrami acting on $u_k$ squared. The first boundary term resulting from the third contribution to the geometrical action functional, i.e., the first integral over $\partial\Gamma_k$ containing only covariant derivatives of $u$ is integrated by parts once more using $\partial^2\Gamma_k=0$. This allows us to obtain a boundary term that contains a factor $u_k$ to be evaluated on $\partial\Gamma_k$ without derivatives. Let us calculate,
\begin{align*}
&S_{geom, diss.}[\lbrace u_i\rbrace_{1\leq k\leq N}]\\
&= \dfrac{1}{2}\sum_{k=1}^N\int_{\mathbb{R}^+_0\times\Gamma_k}dtd^{n-1}\mathbf{y}\sqrt{-\vert G_0^\Sigma\vert}\left(\sigma_m G_0^{\Sigma,tt}(\partial_t u_k)^2+T_0(\partial_i u_k)^2 + (T_{curv}d^2)(\Delta_{g_0}^\partial u_k)^2\right)\\
&= \dfrac{\sigma_m}{2}\sum_{k=1}^N\int_{\lbrace 0\rbrace\times\Gamma_k}d^{n-1}\mathbf{y}\,\sqrt{-\vert G_0^{\Sigma}\vert}u G_0^{tt}\partial_t u + \dfrac{\sigma_m}{2}\sum_{k=1}^N\int_{\mathbb{R}^+_0\times\Gamma_k} d^{n-1}\mathbf{y}\,\sqrt{-\vert G_0^\Sigma\vert}u_k\Sigma^{-1}\partial_t(\Sigma\partial_t u_k)\\
&+ \dfrac{T_0}{2}\sum_{k=1}^N\int_{\mathbb{R}^+_0\times\partial\Gamma_k} d^{n-2}\mathbf{y}\,\sqrt{-\vert G_0^\Sigma\vert} u_k(G^{\Sigma}_0)^{ij}\partial_{\mathbf{n}} u_k - \dfrac{T_0}{2}\sum_{k=1}^{N}\int_{\mathbb{R}^+_0\times\Gamma_k} d^{n-1}\mathbf{y}\,\sqrt{-\vert G_0^\Sigma\vert} u_k\Delta_{g_0}^\partial u_k)\\
&+ \dfrac{T_{curv}d^2}{2}\sum_{k=1}^N\int_{\mathbb{R}^+_0\times\partial\Gamma_k}dtd^{n-2}\mathbf{y}\,\sqrt{-\vert G_0^\Sigma\vert} n_j\partial^j u_k(\Delta_{g_0}^\partial u_k) - \dfrac{T_{curv}d^2}{2}\sum_{k=1}^N\int_{\mathbb{R}^+_0\times\partial\Gamma_k}dtd^{n-2}\mathbf{y}\,\sqrt{-\vert G_0^\Sigma\vert} u_k n_j\partial^j(\Delta_{g_0}^\partial u_k)\\
&+\dfrac{T_{curv} d^2}{2}\sum_{k=1}^N\int_{\mathbb{R}^+_0\times\Gamma_k}dtd^{n-1}\mathbf{y}\,\sqrt{-\vert G_o^\Sigma\vert}\, u_k(\Delta_{g_0}^\partial)^2 u_k\\
&=\dfrac{\sigma_m}{2}\sum_{k=1}^N\int_{\lbrace 0\rbrace\times\Gamma_k}d^{n-1}\mathbf{y}\,\sqrt{-\vert G_0^{\Sigma}\vert}u G_0^{tt}\partial_t u + \dfrac{\sigma_m}{2}\sum_{k=1}^N\int_{\mathbb{R}^+_0\times\Gamma_k} d^{n-1}\mathbf{y}\,\sqrt{-\vert G_0^\Sigma\vert}u_k\Sigma^{-1}\partial_t(\Sigma\partial_t u_k)\\
&+ \dfrac{T_0}{2}\sum_{k=1}^N\int_{\mathbb{R}^+_0\times\partial\Gamma_k} d^{n-2}\mathbf{y}\,\sqrt{-\vert G_0^\Sigma\vert} u_k(G^{\Sigma}_0)^{ij}\partial_{\mathbf{n}} u_k - \dfrac{T_0}{2}\sum_{k=1}^{N}\int_{\mathbb{R}^+_0\times\Gamma_k} d^{n-1}\mathbf{y}\,\sqrt{-\vert G_0^\Sigma\vert} u_k\Delta_{g_0}^\partial u_k)\\
& - T_{curv}d^2\sum_{k=1}^N\int_{\mathbb{R}^+_0\times\partial\Gamma_k}dtd^{n-2}\mathbf{y}\,\sqrt{-\vert G_0^\Sigma\vert} u_k n_j\partial^j(\Delta_{g_0}^\partial u_k)+\dfrac{T_{curv} d^2}{2}\sum_{k=1}^N\int_{\mathbb{R}^+_0\times\Gamma_k}dtd^{n-1}\mathbf{y}\,\sqrt{-\vert G_0^\Sigma\vert}\, u_k(\Delta_{g_0}^\partial)^2 u_k\\
&= \dfrac{\sigma_m}{2}\sum_{k=1}^N\int_{\mathbb{R}^+_0\times\Gamma_k} d^{n-1}\mathbf{y}\,\sqrt{-\vert G_0^\Sigma\vert}u_k\Sigma^{-1}\partial_t(\Sigma\partial_t u_k)\\
& - \dfrac{T_0}{2}\sum_{k=1}^{N}\int_{\mathbb{R}^+_0\times\Gamma_k} d^{n-1}\mathbf{y}\,\sqrt{-\vert G_0^\Sigma\vert} u_k\Delta_{g_0}^\partial u_k)\\
&+\dfrac{T_{curv} d^2}{2}\sum_{k=1}^N\int_{\mathbb{R}^+_0\times\Gamma_k}dtd^{n-1}\mathbf{y}\,\sqrt{-\vert G_0^\Sigma\vert}\, u_k(\Delta_{g_0}^\partial)^2 u_k\\
&= \dfrac{1}{2}\sum_{k=1}^N\int_{\mathbb{R}^+_0}dt d^{n-1}\mathbf{y}\sqrt{-\vert G_0^\Sigma\vert}\, u_k\left(\sigma_m\Sigma^{-1}\partial_t (\Sigma u_k)-T_0\Delta_{g_0}^\partial u_k + T_0d^2(\Delta_{g_0}^\partial )^2 u_k\right).
\end{align*}
We have used the homogeneous initial conditions and the homogeneous Dirichlet boundary conditions for $u_k$ for all $k,\,1\leq k\leq N$ to set the boundary terms equal to $0$. The functional derivative of the geometrical action function w.r.t. the $\lbrace u_k\rbrace_{1\leq k\leq N}$ now is easy. Let $k$ be arbitrary but given. Hamilton's principle of least action gives us the necessary condition for a minimum,
\begin{align}
\dfrac{\delta S_{geom}}{\delta u_k}\stackrel{!}{=}0,
\end{align}
which is fulfilled if
\begin{align}
\sigma_m\Sigma^{-1}\partial_t (\Sigma u_k)-T_0\Delta_{g_0}^\partial u_k + T_0d^2(\Delta_{g_0}^\partial )^2 u_k = 0.
\end{align}
Division by $\sigma_m$ gives us with the definitions $c^2_m = T_0\sigma_m$ and $c^2_H = T_{curv}/\sigma_m$ of the flexural wave propagation velocity $c_m$ and the bending wave propagation velocity $C_h$ a non-standard damped wave equation,
\begin{align}
\Sigma^{-1}\partial_t (\Sigma u_k)-c^2_m\Delta_{g_0}^\partial u_k + c^2_Hd^2(\Delta_{g_0}^\partial )^2 u_k
\end{align}
We restrict ourselves to the cases, where the terminology wave equation is appropriate, i.e., we require $c^2_m\Vert\Delta_{g_0}^\partial\Vert_{L^2\to H^{2,2}} > c^2_H d^2\Vert\Delta_{g_0}^2\Vert_{L^2\to H^{2,2}}$. Using multiplicativity of the operator norm \cite{zeidler1} for a self-adjoint operator squared,
\begin{align*}
\Vert\Delta_{g_0}^\partial\Vert_{L^2\to H^{2,2}}<\frac{c^2_m}{c^2_H}\dfrac{1}{d^2}.
\end{align*}
Since $\Vert\Delta_{g_0}\Vert_{L^2\to H^{2,2}}\leq 1$, we obtain 
\begin{align}
c^2_m > c^2_Hd^2
\end{align}
This means that flexural waves of the boundary vibrations propagate faster than bending waves of the boundary vibrations. Experimentally, the condition can be verified by claculating the measuring the eigenfrequencies of the elastic structure in question. For the ICE model the condition is fulfilled, c.f., the model \cite{anupam1, anupam2, christine}. Introducing the polynomial $p=c^2_m x - c^2_H d^2 x^2$ and using the Bochner functional calculus \cite{zeidler2}, we can re-write the homogeneous boundary vibrations equation in the more compact form ,
\begin{align}
\Sigma^{-1}\partial_t(\Sigma\partial_t u_k)-p(\Delta_{g_0}^\partial) u_k = 0.
\end{align}
We will be concerned with reducing this non-standard wave equation to a simpler equation on $\mathcal{M}_0$ in short. The last issue to address is the source term. Let us re-consider the ICE model \cite{anupam1, anupam2, christine}. The boundary vibrations there, the membrane displacements, are pressure difference receivers. They respond to the difference of an external acoustic wave hitting the membranes from the outside, i.e., at $\partial\mathcal{M}^+$ in the embedding space $\mathbb{M}^{n+1}\supset\mathcal{M}$ and from the interior, i.e., at $\partial\mathcal{M}_-\subset\mathcal{M}$. Note that $\partial\mathcal{M}$ is the boundary of the perturbation bundle $\mathcal{M}$ and not the unperturbed bundle $\mathcal{M}_0$. In the ICE model, the acoustic pressure hitting the tympana from the outside of $\Omega_t$ for a given $t\geq 0$ is prescribed and denoted by $p_{ex}$. The idea is that there is a sound source localized outside the lizard's head such that the tympanic membranes undergo vibrations caused by the external sound stimulus. We will apply this idea as well. Since we prescribe $p_{ex}=p_{\partial\mathcal{M}_+}$, we introduce a discontinuity, i.e., $p\vert_{\partial\mathcal{M}^+}\neq p\vert_{\partial\mathcal{M}_-}$ in general. In the context of the ICE model, the directional hearing of lizards is then explained as a reaction of the membrane system coupled by the acoustic pressure in the interaural cavity to this external stimulus. We use the minimal definition for the external pressure $p_{ex}$ \cite{david1},
\begin{align}
p_{ex} = \left\lbrace\begin{array}{c}p_0e^{i\omega t}\text{ on }\partial\mathcal{M}_+\\0\text{ elsewhere }\end{array}\right.
\end{align}
In this definition, $p_0\in\mathbb{R}, \omega >0$ are real numbers which are assigned the physical dimensions of a pressure, $[p_0]=1\text{ Nm}^{-(n-1)}$ and $[\omega]=1\text{ Hz}$ for consistency with SI units. In analogy to the ICE model, we define the source term $\Psi$ to be
\begin{align}
\Psi = \dfrac{\Phi_{0\to t}^\ast(p\vert_{\partial\mathcal{M}_+-p\vert_{\partial\mathcal{M}_-}})}{\sigma_m} = \dfrac{\Phi_{0\to t}^\ast(p_{ex}\vert_{\partial\mathcal{M}}-p\vert_{\partial\mathcal{M}})}{\rho_m d}.
\end{align}
The appearance of $\Phi_{0\to t}$ acting as a pull-back on the pressure difference in brackets is due to the geometry. The differential equation for the boundary vibrations $u$ lives on the reference bundle $\mathcal{M}_0$, whereas the acoustic pressure $p$ lives on the perturbed bundle $\mathcal{M}$. The bundle diffeomorphism $\Phi_{0\to t}$ allows us to pull objects from $\mathcal{M}$ back to objects on $\mathcal{M}_0$, i.e., to deform mathematically the perturbation bundle to the reference bundle again. Later on, we will see that we just introduce errors of quadratic order in the perturbation strength, if we replace the acoustic wave equation on $\mathcal{M}$ by an acoustic wave equation on $\mathcal{M}_0$. The definition of the source term to the equations modeling the boundary vibrations is visualized in Fig. 12. The boundary vibration $u$ responds to the difference of external pressure $p\vert_{\partial\mathcal{M}_+}$ and (internal) acoustic pressure $p\vert_{\partial\mathcal{M}_-}$ evaluated at $\partial\mathcal{M}^+$, i.e.,letting $s\to 1^+$ to obtain $p\vert_{\partial\mathcal{M}_+}$ and evaluated at $\partial\mathcal{M}_-$, i.e., letting $s\to 1^-$ to obtain $p\vert_{\partial\mathcal{M}_-}.$
\begin{figure}
\includegraphics[width = 1\textwidth]{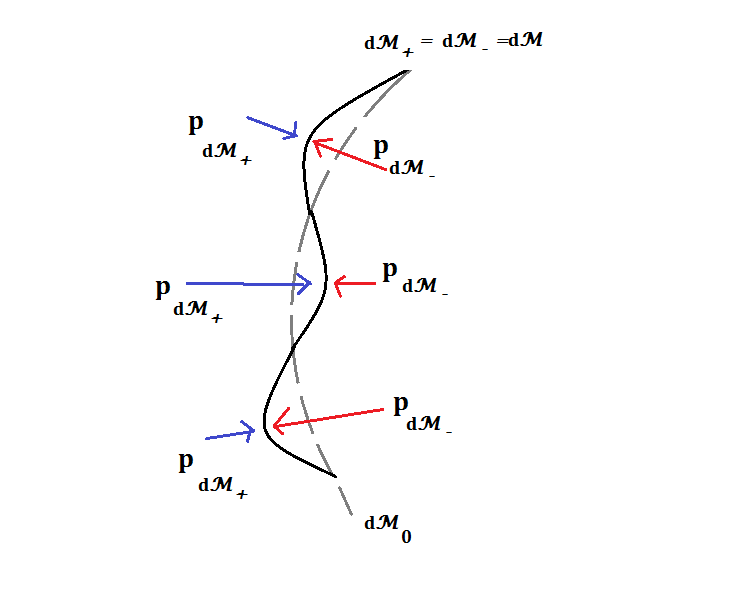}
\caption{Visualization of the source term driving the boundary vibrations: The boundary vibrations $u$ are driven by the difference of $p_{ex}\equiv p\vert_{\partial\mathcal{M}_+}$ and $p\vert_{\partial\mathcal{M}_+}$, i.e., the difference of the acoustic pressure in $\mathbb{R}^n\setminus\Omega_t$ and in $\Omega_t$ acting as a force density on the graph of the boundary vibrations.}
\end{figure}\newline
\newline
\textbf{Intermediate result: }We summarize the governing equations of the class of models we are interested in and afterwards comment to which class of equations they belong in the language of mathematicians. The two coupled equations read
\begin{align}
\partial_t^2p - c^2\Delta_{g_0}p &= \rho_0c^2\partial_t^2 u\circ\Phi_{0\to t}^{-1}\vert_{\partial\mathcal{M}}\delta((t,\mathbf{x})\in\partial\mathcal{M})\text{ on }\mathcal{M}\\
\Sigma^{-1}\partial_t(\Sigma\partial_t u)-p(\Delta_{g_0}^\partial) u &= \Psi\text{ on }\partial\mathcal{M}_0.
\end{align}
The two equations have homogeneous initial conditions,
\begin{align}
p(t=0,\mathbf{x})&=0=\partial_t p(t=0,\mathbf{x})\text{ on }\Omega_{t=0}=\Omega_0\\
u(t=0,\mathbf{y})&=0=\partial_t u(t=0,\mathbf{y})\text{ on }\partial\Omega_0.
\end{align}
The differential operators $\Delta_{g_0}$ and $\Delta_{g_0}^\partial$ on the fiber $\Omega_t$ of the perturbation bundle $\mathcal{M}$ and the boundary of the unperturbed fiber $\partial\Omega_0$of the reference bundle $\mathcal{M}_0$ have homogeneous Neumann resp. periodic boundary conditions because $\partial^2\Omega_0=\emptyset$,
\begin{align}
\partial_{\mathbf{n}}p = 0\text{ on }\partial\mathcal{M}.
\end{align}
The decomposition of $u$ in the localized boundary vibrations $\lbrace u_k\rbrace_{1\leq k\leq N}$ from the previous sub-paragraph gives us in total $N+1$ equations: An acoustic wave equation on the perturbed bundle $\mathcal{M}$ for $p$ and $N$ equations describing the dynamics of the localized boundary vibrations $\lbrace u_k\rbrace_{1\leq k\leq N}$
\begin{align}
\partial_t^2p - c^2\Delta_{g_0}p &= \rho_0c^2\sum_{k=1}^N\partial_t^2 u_k\circ\Phi_{0\to t}^{-1}\vert_{\mathbb{R}^+_0\times\Gamma_k}\delta((t,\mathbf{x})\in\mathbb{R}^+_0\times\Gamma_k)\text{ on }\mathcal{M}\\
\Sigma^{-1}\partial_t(\Sigma\partial_t u_k)-p(\Delta_{g_0}^\partial) u_k &= \Psi\vert_{\Gamma_k}\text{ on }\mathbb{R}^+_0\times\Gamma_k\subset\partial\mathcal{M}_0. 
\end{align}
The symbol $\Psi\vert_{\Gamma_k}$ denotes restriction of the source term for the boundary vibrations $u$ to $\mathbb{R}^+_0\times\Gamma_k$, where for each $k\in\lbrace 1,...,N\rbrace$ the boundary vibration $u_k$ is localized. The initial conditions for both equations stay unchanged, the Neumann boundary condition to the acoustic wave equation stays unchanged as well, but the Laplace-Beltrami operator $\Delta_{g_0}^\partial$ is now defined for all $\Gamma_k$'s individually and breaks the periodic  boundary conditions that we amended to the Laplace-Beltrami operator on $\partial\Omega_0$. The derivation from the previous paragraph demonstrated that we can take Dirichlet boundary conditions for the localized boundary vibrations equations, i.e., the equations for $\lbrace u_k\rbrace_{1\leq k\leq N}$,
\begin{align}
u_k\vert_{\partial\Gamma_k} = 0\text{ on }\mathbb{R}^+_0\times\partial\Gamma_k.
\end{align}
If we pull back the acoustic wave equation on $\mathcal{M}$ to the well-understood reference bundle $\mathcal{M}_0$ we introduce the perturbation operator $\mathsf{W}$ but we arrive at equations that live on the reference bundle $\mathcal{M}_0$ and its boundary $\partial\mathcal{M}_0$, denoting $\tilde{p}(\Phi_{0\to t}(t,\mathbf{x}))=p(t,\mathbf{x}_t)$ also by the symbol $p$,
\begin{align}
\partial_t^2 p - c^2\Delta_{g_0}p &= c^2\mathsf{W}[u]p+\rho_0c^2\sum_{k=1}^N\partial_t^2 u_k\delta((t,\mathbf{x})\in\mathbb{R}^+_0\times\Gamma_k)\\
\Sigma^{-1}\partial_t(\Sigma\partial_t u_k)-p(\Delta_{g_0}^\partial) u_k &= \Psi\vert_{\Gamma_k}\text{ on }\mathbb{R}^+_0\times\Gamma_k\subset\partial\mathcal{M}_0.
\end{align}
The Neumann boundary conditions for the acoustic pressure now become $\partial_{\mathbf{n}}p=0$ on $\partial\mathcal{M}_0$. For illustration purposes, the mutual acoustic feedback between the boundary vibrations and the acoustic pressure is shown in Fig. 13.
\begin{figure}
\includegraphics[width = 1\textwidth]{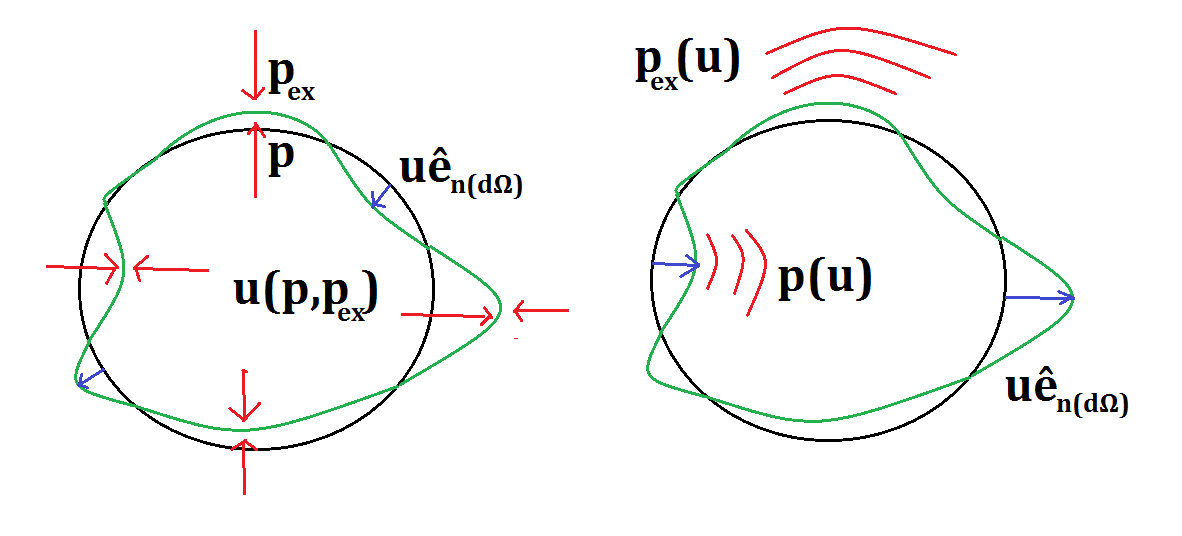}
\caption{The two physical processes described by the model equations. Left panel: The difference of pressure inside $\Omega_t$ and outside $\Omega_t$ drives the boundary vibrations $u$ for all $t\geq 0$. Right panel: The vibrations of the boundary lead effectively to a local change in the mass of air per unit volume in $\Omega_t$ and thus stimulate an acoustic pressure inside $\Omega_t$. Likewise, also the mass of air per unit volume changes close to $\partial\mathcal{M}^+$, i.e., for a fixed $t\geq 0$ the boundary of $\mathbb{R}^n\setminus\Omega_t$. This in turn also feedbacks the acoustic pressure outside $\Omega_t$. Since this feedback contributions propagates away from $\partial\Omega_t$, it does not contribute the pressure signal hitting the graph of the boundary vibrations from $\partial\mathcal{M}_+$ upon neglection of interference phenomena.}
\end{figure}
Let us comment on the differential equations adopting for a moment a mathematically more precise language. The \newline
\newline
\textbf{Eigenfunctions revisited: }We investigate the eigenfunctions again. By the assumptions on the geometry of $\Omega_t$ for all $t\geq 0$, namely smoothness, compactness, we can apply Lichernowicz' theorem to find for all individual $t$ eigenfunctions for the Neumann Laplace-Beltrami operator, i.e., we can find the sequence of pairs  $\lbrace(\lambda_n(t),\Psi_n(t))\rbrace_{n\in\mathbb{N}}$ consisting of pairwisely orthornormal eigenfunctions $\Psi_{n}(t)$ for $\Delta_{g_0}$ on $\Omega_t$ with Neumann eigenvalues $\lambda_n(t)$,
\begin{align}
\Delta_{g_0}\Psi_{n}(t) = -\lambda_n(t)\Psi_n(t)\text{ on }\Omega_t,
\end{align}
and $\partial_{\mathbf{n}}\Psi_{n}(t)=0$ on $\partial\Omega_t$. Likewise, we can solve the Neumann-eigenvalue problem for the Laplacian $\Delta_{g_0}$ on the reference fiber $\Omega_0$ using Lichernowicz theorem once again. The relevant equations are given by
\begin{align}
\Delta_{g_0}\Psi_n = -\lambda_n\Psi_n\text{ on }\Omega_0,
\end{align}
with the Neumann boundary conditions $\partial_\mathbf{n}\Psi_n = 0$ on $\partial\Omega_0$. We can use the global diffeomorphism $\psi_{0\to t}$ to pull back the eigenvalue problem on $\Omega_t$ to an equation of $\Omega_0$, namely
\begin{align}
\psi_{0\to t}^\ast (\Delta_{g_0}\Psi_n(t)) &= \psi_{0\to t}^\ast(-\lambda_n(t)\Psi_n(t))\\
\Leftrightarrow \psi_{0\to t}^\ast \Delta_{g_0}\Psi_n(\psi_{0\to t}(\mathbf{x})) &= -\lambda_n(t)\Psi_n(\psi_{0\to t}(\mathbf{x})).
\end{align}
Notice, that the functional dependencies of $\Psi_n(t)$ on the coordinate $t$ on the base space $\mathbb{R}^+_0$, i.e., the time coordinate $t$, can be traced back to $t$-dependencies of $\psi_{0\to t}$. Because the pull-back is a linear operation and the eigenvalues $\lambda_n(t)$ do not depend on spatial arguments, i.e., coordinates on the fibers $(\Omega_t)_{t> 0}$, they stay unaffected by the pull-back operation. On the other hand, we can use the definition of the perturbation operator $\mathsf{V}$,
\begin{align}
\mathsf{V}\equiv\psi_{0\to t}^\ast\Delta_{g_0}-\Delta_{g_0}.
\end{align}
By an estimate in the previous section, we found that the  norm of $\mathsf{V}$ relative to $\Delta_{g_0}$ on $\Omega_0$ satisfies
\begin{align}
\Vert\mathsf{V}\Vert = \mathcal{O}(\epsilon^2),
\end{align}
where $\epsilon\ll 1$ is the perturbation strength. With the definitions, the pull-back of the eigenvalue problem from $\Omega_t$ to $\Omega_0$ can be brought into the form of a perturbation problem. Denoting by $\tilde{\Psi}_n(t)=\Psi_n\circ\psi_{0\to t}$ the pull-back of the eigenfunctions $\lbrace\Psi_n(t)\rbrace_{n\in\mathbb{N}}$ on one $\Omega_t$ to $\Omega_0$, we have
\begin{align}
\Delta_{g_0}\tilde{\Psi}_n(t)+\mathsf{V}\tilde{\Psi}_n(t)&= -\lambda_n(t)\tilde{\Psi}_n(t)\text{ on }\Omega_0,
\end{align}
together with the Neumann boundary conditions $\partial_{\mathbf{n}}\tilde{\Psi}_n(t)=0$ on $\partial\Omega_0$. Now, we can make the Ansatz
\begin{align}
\tilde{\Psi}_{n}(t) &= \sum_{k=0}^{\infty}\epsilon^{2k}c_{nm}\Psi_m,
\end{align}
to derive a Schrödinger-like perturbation theory for the eigenfunctions $\tilde{\Psi}_n$. Notice that $\tilde{\Psi}_n(t)$ is normalized by assumption! Working only in lowest order, i.e., only considering contributions up to order $\epsilon^2$ inclusively, insertion of the Ansatz and comparing powers of $\epsilon^2$ gives us
\begin{align}
\tilde{\Psi}_n(t)&=\Psi_n + \sum_{m=0,\,m\neq n}^{\infty}\dfrac{\langle \Psi_m^0\vert\mathsf{V}(t)\vert\Psi_n\rangle_{L^2_{g_0}}}{-\lambda_n + \lambda_m}\Psi_m + \mathcal{O}(\epsilon^4)
\end{align}
Likewise, we obtain the spectral correction
\begin{align}
\delta\lambda_n\equiv \lambda_n(t)-\lambda_n = -\langle \Psi_n\vert\mathsf{V}\vert\Psi_n\rangle_{L^2_{g_0}}+\mathcal{O}(\epsilon^4).
\end{align}
Iterating one more time, we obtain the spectral corrections in fourth order in $\epsilon$,
\begin{align}
\delta\lambda_n = -\langle \Psi_n\vert\mathsf{V}\vert\Psi_n\rangle_{L^2_{g_0}} - \sum_{m=0,\,n\neq m}^{\infty}\dfrac{\left\vert\langle\Psi_m\vert\mathsf{V}\vert\Psi_n\rangle_{L^2_{g_0}}\right\vert^2}{-\lambda_n +\lambda_m}+\mathcal{O}(\epsilon^6).
\end{align}
For the eigenfunctions, we can simply invert the pull-back to recover $\Psi_n(t)$ from $\tilde{\Psi}_n(t)$. Denote by $\mathbf{x}_t$ local coordinates on the perturbed fibers $\Omega_t$
\begin{align}
\Psi_n(t) = \tilde{\Psi}_n\circ\psi_{0\to t}^{-1} = \Psi_n(\psi^{-1}_{0\to t}(\mathbf{x}_t))+\sum_{m=0,\,m\neq n}^{\infty}\dfrac{\langle \Psi_m^0\vert\mathsf{V}(t)\vert\Psi_n\rangle_{L^2_{g_0}}}{-\lambda_n + \lambda_m}\Psi_m(\psi_{0\to t}^{-1}(\mathbf{x}_t))+\mathcal{O}(\epsilon^4).
\end{align}
Expanding gives us on $\Omega_0\cap\Omega_t$,
\begin{align}
\Psi_n(t;\mathbf{x}_t)=\Psi_n(\mathbf{x})+\mathcal{O}(\epsilon).
\end{align}
In other words, a first for the eigenfunctions $\Psi_n(t)$ on $\Omega_t$ is given by $\Psi_n$ on $\Omega_0$. By regularity of the eigenfunctions on $\Omega_0$, we can continue the eigenfunctions $\Psi_n$ to be agree up to an error of $\epsilon$ with $\Psi_n(t)$, i.e., we have on $\Omega_t$ and not just on $\Omega_0\cap\Omega_t$,
\begin{align}
\Psi_n(t;\mathbf{x}_t)=\Psi_n(\mathbf{x}_t)+\mathcal{O}(\epsilon).
\end{align}
Let us compare the method with already existing mathematics. For stationary perturbations, i.e., $\psi_{0\to t_1}=\psi_{0\to t_2}$ for all $t_1,t_2>0,\,t_1\neq t_2$, this reproduces the perturbation theories obtained by \cite{brillouin, feshbach, frohlich, cabrera}. A special case of our setup has been investigated by \cite{DengLi} who concentrate on corrections to the eigenvalues of the Neumann Laplace-Beltrami operator on $\Omega_0$ vs. the Neumann Laplace-Beltrami operator on $\Omega_t$. They find that the corrections scale as $\epsilon^2$ in a cubic model which is reproduced by our approach and extended to higher order corrections. Last, we reproduce the model setup in \cite{beale1, beale2, beale3} who idealized the domain $\Omega_0$ to be the correct geometric location for the acoustic wave equation with acoustic boundary conditions (ABC) and derived results on properties of an operator matrix formulation of the acoustic boundary conditions problem in a non-dynamical setup.
\newline
\newline
\textbf{A comment on the interpretation: }\textbf{Physical interpretation: }After having derived the model equations, we need to re-think the meaning of the notion \"{}perturbation" in the context of or model. In order to achieve thus, we leave the mathematical realm and use physical reasoning instead. Suppose at first that our perturbation bundle $\mathcal{M}$ and the reference bundle $\mathcal{M}_0$ were such that every $\epsilon > 0$ could be called perturbation strength of the perturbation bundle $\mathcal{M}$. If this is true, we have $\langle\mathbf{n},\psi_{0\to t}-\text{id}_{\Omega_0}\rangle_{g_0,\mathbb{R}^n}\vert< \epsilon$ by definition of $\epsilon$ as perturbation strength. Consequently also the restriction of $\psi_{0\to t}$ to the boundary $\partial\Omega_0,$ of the unperturbed fiber, $\phi_{0\to t}=\psi_{0\to t}\vert_{\partial\Omega_0}$ and the restriction of the identity map $\text{id}_{\partial\Omega_0}=\text{id}_{\Omega_0}\vert_{\partial\Omega_0}$ satisfy the expression bounded from above by $\epsilon$. By definition of the boundary vibrations, $u(t,\mathbf{y})=\langle\mathbf{n}_{\partial\Omega_0},\phi_{0\to t}-\text{id}_{\partial\Omega_0}\rangle_{g_0,\mathbb{R}^n}$, the bound from above shows that $0\leq \vert u(t,\mathbf{y})\vert <\epsilon$. Since $\epsilon>0$ was arbitrary, we have $u(t,\mathbf{y})\equiv 0$. However, the acoustic wave equation pulled-back by means of the bundle diffeomorphism $\Phi_{0\to t} : \mathcal{M}_0\to\mathcal{M}$ from the perturbation bundle $\mathcal{M}$ to the reference bundle $\mathcal{M}_0$ features a contribution from the boundary displacements $u$ which is exclusively of linear order in $u$ and thus in $\epsilon$, namely the first contribution on the right hand side in
\begin{align*}
\partial_t^2p - c^2\Delta_{g_0}p = \rho_0c^2\partial_t^2 u\delta((t,\mathbf{x})\in\partial\mathcal{M}_0)+c^2\mathsf{W}[p].
\end{align*}
In the case investigated above, i.e., if $u\equiv 0$, both contributions vanish and the evolution of the perturbed fibers can be described by the constant map $t\to\Omega_t = \Omega_0$. As soon as the minimal $\epsilon>0$ that we can take by Zorn's Lemma as perturbation strength, does not become arbitrarily small, i.e., $\epsilon\neq 0$, the map $t\to\Omega_t$ is no longer a constant map from $\mathbb{R}^+_0$ to the category of sub-manifolds, $\textsf{SubMan}(\mathbb{R}^n)$, of $\mathbb{R}^n$. In particular $u\neq 0$ such that we have the two contributions to the acoustic wave equation pulled back from the perturbation bundle $\mathcal{M}$ to the reference bundle $\mathcal{M}_0$. Namely, one $\sim\rho_0\partial_t^2 u$, which is of linear order in $u$ and thus of linear order in $\epsilon$, and a second contribution that is quadratic in $u$ and first derivatives thereof, namely $\mathsf{W}\sim\mathcal{O}(u^2,u\partial u,(\partial u)^2)=\mathcal{O}(\epsilon)$. In linear perturbation theory in $\epsilon$, it makes sense to regard the first contribution as a perturbation operator as well, i.e., set
\begin{align}
\mathsf{W}_{lin}[p]+w(t,\mathbf{x})\delta((t,\mathbf{x})\in\partial\mathcal{M}_0)=\rho_0c^2\partial_t^2 u\delta((t,\mathbf{x})\in\partial\mathcal{M}_0).
\end{align}
The $w$ is a known function i.e., a source term. The operator can be shown to be \emph{affine} in $p$, by formally solving the equations for the boundary vibrations $u$ for $u$ and substituting the expression which contains $p$ linearly back in the acoustic wave equation as replacement for $u$. The constant contribution in the affine operator is given by $p_{ex}$ which we can treat as a source term. Using the (informal)\footnote{It is informal, because we have not yet inverted the hyperbolic partial differential operator on $\partial\mathcal{M}_0$. We postpone this to the next section.} abbreviation $u=\mathsf{L}[p\vert_{\partial\mathcal{M}_0}-p_{ex}\vert_{\partial\mathcal{M}_0}]$ with a linear operator $\mathsf{L}\in\mathsf{LinOp}(H^{1,2;2,2}_{0,G_0}(\partial\mathcal{M}_0)\to H^{1,2;4,2}_{0,G_0}(\partial\mathcal{M}_0))$, we can express $\mathsf{W}_{lin}$ as
\begin{align*}
\mathsf{W}_{lin}[p]=\rho_0c^2\partial_t^2\mathsf{L}[p\vert_{\partial\mathcal{M}_0}]\delta((t,\mathbf{x})\in\partial\mathcal{M}_0).
\end{align*}
The contribution from $p_{ex}$ has been identified as the source term $w$. The linear contribution has the form of a $\delta$-perturbation that is familiar from graduate quantum mechanics courses \cite{dirac}: We formulate the time-dependent Schr\"{o}dinger equation on the unperturbed bundle $\mathcal{M}_0$ in order to ease the accessibility of the analogy. We investigate the Hamilton operator $\mathsf{H}$ given as the sum of an tractable Hamilton operator $\mathsf{H}_0$ and a perturbation operator $\mathsf{V}=\kappa f(t,\mathbf{x})\delta((t,\mathbf{x})\in\partial\mathcal{M}_0)$ with $\kappa\ll 1$ and $f\in L^2_{g_0}(\mathcal{M}_0)$ bounded,
\begin{align*}
i\hbar\partial_t\psi(t,\mathbf{x})=-\dfrac{\hbar^2}{2m}\Delta_{g_0}\psi(t,\mathbf{x})+\mathsf{V}[\psi] =-\dfrac{\hbar^2}{2m}\Delta_{g_0}\psi(t,\mathbf{x})+ \kappa f(t,\mathbf{x})\delta((t,\mathbf{x})\in\partial\mathcal{M}_0)\psi(t,\mathbf{x}).
\end{align*}
This is nothing but the scattering/perturbation problem for a free Schr\"{o}dinger particle on a $\delta$-potential that has variable strength, $\kappa f(t,\mathbf{x})$. However, the potential $\mathsf{V}$ is physically treated as a perturbation to the unperturbed Hamilton operator $\mathsf{H}_0=-\hbar^2/(2m)\Delta_{g_0}$ for free particles\footnote{Of course we could have also taken a Hamilton operator for a hydrogen atoms or any other Hamiltonian that we can solve and that is in the operator norm big compared to the perturbation operator $\mathsf{V}$ used in this paragraph.}. Physically, \cite{dirac} this perturbation problem is solved by a Dirac perturbation theory with a perturbation operator that is only defined geometrically on $\partial\mathcal{M}_0$. This corresponds to keeping the boundary in our acoustics problem fixed \cite{beale1, beale2, beale3} and using $\mathcal{M}_0\approx\mathcal{M}$. In our perturbation theory, this means that we ignore the contributions form $\mathsf{W}\sim\mathcal{O}(\epsilon^2).$ The contributions from $\mathsf{W}_{lin}$to the acoustic wave equation given in the beginning of this paragraph stem from the translation of the boundary conditions to the acoustic wave equation to a source term. By Cauchy-Kowalewskaja's theorem, it is ensured that the solution to the a priori different problems, i.e., one problem with trivial source term but non-trivial boundary conditions and one problem with trivial boundary conditions but non-trivial source term, agree if they exist (for acoustic applications see e.g. \cite{pan1, pan2}. We can term the contributions from $\mathsf{W}_{lin}$ as \emph{acoustic boundary conditions dynamics} (ABCD) building upon the approach by Beale and Rosencrans \cite{beale1, beale2, beale3}. The non-linear contributions stored in $\mathsf{W}$ are called \emph{geometrical perturbation dynamics}. They are due to geometrical effects that arise because we pull back the perturbation bundle $\mathcal{M}$ to the reference bundle $\mathcal{M}_0$. The contribution $w$ includes the external pressure $p_{ex}=p_{\partial\mathcal{M}_0^+}$ and is for our purposes just a source term to the acoustic wave equation which ensures that the solution $p$ is non-trivial. If it were not present, we would just have trivial solutions because we chose homogeneous initial conditions as well as homogeneous boundary conditions to both the acoustic wave equation for $p$ and the equation for the dynamics of the boundary vibrations $u$. We summarize our considerations on the physical interpretation schematically as follows
\begin{align*}
\mathsf{W}_{lin}&\Leftrightarrow\text{Linear perturbation operator}\\
&\Leftrightarrow\text{Acoustic boundary conditions dynamics}\\
\mathsf{W}&\Leftrightarrow\text{Non-linear perturbation operator}\\
&\Leftrightarrow\text{Geomtrical perturbation dynamics}\\
w &\Leftrightarrow\text{Source term including the external source}p_{ex}\\
&\Leftrightarrow\text{Physical cause of boundary vibrations and cavity pressure, }u\text{ and }p.
\end{align*}
The goal of the next section is to find perturbative solutions to the coupled partial differential equations in linear order in $\epsilon$. Qualitatively, we will use a two-fold series expansion based on the Banach's fixed point theorem and Duhamel's principle. The first series expansion is by a Magnus-Dirac like perturbation theory: We combine the ideas of the Dirac perturbation theory with the Magnus series expansion \cite{magnus1}. It will allow neglect the contributions from $\mathsf{W}$ against the contributions from the acoustic boundary condition dynamics and see how an inclusion of the geometrical perturbation dynamics would have affected the solution $p$ in terms of the perturbation strength $\epsilon$. The second series expansion uses a decoupling argument based on Banach's fixed point theorem again, or more precisely, the technique of Picard iterations for operator differential equations. In order to truncate the iterations, we will introduce and interpret an additional small parameter, the \emph{coupling strength} $\mathfrak{g}=\rho_0/\rho_m$, i.e., the ratio of mass density of air and the mass density of the boundary vibrations. In realistic applications, \cite{david1, kriegsmann1, kriegsmann2} we have $\mathfrak{g}\ll 1$. In our method, it will play the role of Lipschitz constant.

\section{Derivation of the Perturbation Theory}
\textbf{Convention: }\emph{In the following let $I=[0,a),\,a\in\bar{\mathbb{R}}^+$ be the intersection of the maximal interval of existence of a solution to a system of first order ordinary differential equations, $I_{max}$, intersected with all non-negative reals, $I=I_{max}\cap\mathbb{R}^+_0$.}\newline
\newline
\textbf{Duhamel's principle - I: }Duhamel's principle is a method to calculate the solution to a liner first order differential equation. We start from the ordinary differential equations case for the reader's convenience. Let $\mathbf{g}=\mathbf{g}(t)\in C^1(I\to\mathbb{R}^n)$ be the vector-valued function which is sought for and let $\mathsf{A}(t)\in\mathfrak{gl}(n,C^{0}(I\to\mathbb{R}))$. Last,let $\mathbf{f}\in C^0(I\to\mathbb{R}^n)$ be a given vector-valued function. Consider the following ordinary differential equation with initial conditions $\mathbf{g}(0)=\mathbf{0}$,
\begin{align}
\dfrac{d\mathbf{g}}{dt}(t) = \mathsf{A}(t)\mathbf{g}(t)+\mathbf{f}(t)
\end{align}
As a first step, we assume that $\mathsf{A}$ is time-independent and set $\mathsf{A}(t)=\mathsf{A}$. Duhamel's principle states that there is a unique solution to the initial value problem, given by
\begin{align}
\mathbf{g}(t) = \int_{0}^td\tau\,\exp\left(\mathfrak{G}(t-\tau)\right)\mathbf{f}(\tau),
\end{align}
where $\mathfrak{G}(t-\tau)=\mathfrak{G}\cdot (t-\tau)$ and $\mathfrak{G}=\mathsf{A}$ is referred to as the generator of the smoothly parameterized group $GL(n\mathbb{R})\ni\exp(t\mathfrak{G}), t\in\mathbb{R}$. The definition uses the correspondence of the exponential map for a Lie-group $G$ and its derivation, i.e., its Lie-algebra $\mathfrak{g}$,
\begin{align}
\exp:\mathbb{R}^+_0\times\mathfrak{g}\to G,\,(t,\mathsf{A})\to\exp(t\mathsf{A}),
\end{align}
where the exponential is for finite-dimensional Lie-groups $G$ in one of their associated matrix representation\footnote{One could distinguish the matrix representation of a Lie group and the Lie group itself. However, we will identify the two notions mainly since for most practical purposes physicists resort directly to (the adjoint or fundamental) matrix representation.} just the ordinary matrix exponentiation rewritten in terms of the Laurent series expansion,
\begin{align}
\exp(t\mathsf{A})=\sum_{k=0}^\infty\dfrac{1}{k!}t^k\mathsf{A}^k,
\end{align}
where the $t$ could be pulled in front of the powers of $\mathsf{A}$ because $\mathsf{A}$ defines a linear mapping $\mathbb{R}^n\to\mathbb{R}^n$. Indeed, we have respecting that we also have to differentiate the integral and not only the integral in the sense of parameter integrals,
\begin{align*}
\dfrac{d\mathbf{g}}{dt} &= \dfrac{d}{dt}\int_{0}^td\tau\,\exp\left(\mathfrak{G}(t-\tau)\right)\mathbf{f}(\tau)\\
&= \mathfrak{G}\int_{0}^td\tau\,\exp\left(\mathfrak{G}(t-\tau)\right)\mathbf{f}(\tau)+\exp(\mathfrak{G}(t-t))\mathbf{f}(t)\\
&= \mathsf{A}\mathbf{g}+\mathsf{f}.
\end{align*}
Thus Duhamel's principle yields indeed a solution to the ordinary differential equation system under consideration.
\newline
\newline
\textbf{Duhamel's principle - II: }Let us now modify the Lie-algebra a bit to obtain an algebra over a commutative ring, namely the ring $C^\infty(I\to\mathbb{R})$\footnote{Such an algebra $\mathfrak{a}$ is basically the ring analog of a vector space over a field, i.e., a module over the ring with a multiplication mapping $\cdot:\mathfrak{a}\times\mathfrak{a}\times\mathfrak{a}$. That is all we need conceptually for our purposes.}. Let us now assume $\mathsf{A}\in\mathfrak{g}$ where 
\begin{align*}\mathfrak{gl}_c=\mathfrak{gl}(n,C^\infty(I\to\mathbb{R}))/[\mathfrak{gl}(n,C^\infty(I\to\mathbb{R})), \mathfrak{gl}(n,C^\infty(I\to\mathbb{R}))], 
\end{align*}
i.e., $\mathsf{A}$ is a representative of the quotient algebra of elements in $\mathfrak{gl}(n)$ with matrix coefficients in $C^\infty(I\to\mathbb{R})$ modulo elements in the commutator algebra
\begin{align*}
[\mathfrak{gl}(n,C^\infty(I\to\mathbb{R})), \mathfrak{gl}(n,C^\infty(I\to\mathbb{R}))] \equiv \lbrace[\mathsf{A}(t_1),\mathsf{A}(t_2)];t_1,t_1\in\mathbb{R},\mathsf{A}(t)\in\mathfrak{gl}(n,C^\infty(I\to\mathbb{R}))\rbrace .
\end{align*}
We have ensured that given $t_1,t_2\in\mathbb{R}$, the commutator $\text{ad}_{\mathsf{A}(t_1)}\mathsf{A}(t_2)=[\mathsf{A}(t_1),\mathsf{A}(t_2)]=-\text{ad}_{\mathsf{A}(t_2)}\mathsf{A}(t_1)$ vanishes, i.e., $[\mathsf{A}(t_1),\mathsf{A}(t_2)]=0$ for all $t_1,t_2\in\mathbb{R}$. If we didn't use the quotient algebra $\mathfrak{gl}_c$ but the full $\mathfrak{gl}(n,C^\infty(I\to\mathbb{R}))$ instead, we would have $[\mathsf{A}(t_1),\mathsf{A}(t_2)]=0$ in general for $t_1\neq t_2$. So the usage of the quotient algebra $\mathfrak{gl}_c$ ensures that we have an algebra, in the present case, an algebra, which is fully commutative w.r.t. its inner multiplication mapping, $\cdot:\mathfrak{gl}_c\times\mathfrak{gl}_c\to\mathfrak{gl}_c$. To make the distinctions between groups, rings and fields a bit clearer: The group is the $\text{GL}(n,C^\infty(I\to\mathbb{R}))$ which takes coefficients in the matrix representation not in a field, say $\mathbb{R}$, as usual, but this time a ring, $C^\infty(I\to\mathbb{R})$. The ring is a more general structure than a field, but somewhat weaker in the sense that for a given ring element, there is not necessarily an inverse ring element w.r.t. the multiplication mapping in the ring. E.g. let $I=[-1,1]$ and the function $\text{id}_{I\to\mathbb{R}}:I\to\mathbb{R},x\mapsto x$ is $\mathcal{C}^\infty$ and also in $C^\infty(I\to\mathbb{R})$. However, it's inverse with respect to the multiplication mapping on $C^\infty(I\to\mathbb{R})$, i.e., the pointwisely defined scalar multiplication in $\mathbb{R}$ is not in $C^\infty(I\to\mathbb{R})$. We would have $1/\text{id}_{I\to\mathbb{R}}=1/x$ which is discontinuous at $x=0$ and thus not even differentiable at $x=0$, let alone smooth. Thus the inverse of $\text{id}_{I\to\mathbb{R}}$ w.r.t. the scalar multiplication mapping is not in $C^\infty(I\to\mathbb{R})$. The Lie-algebra $\mathfrak{gl}(n,\mathbb{R})=\text{End}_{\mathbb{R}}(\mathbb{R}^n\to\mathbb{R}^n)$, i.e., all quadratic $n\times n$ matrices with coefficients in $\mathbb{R}$ form a $n^2$-dimensional vector space over $\mathbb{R}$, taking as basis vectors the elementary matrices $e_{i_0j_0}$ with components $(e_{i_0j_0})_{ij}=e_{ij;i_0j_0}=\delta_{ii_0}\delta_{jj_0}$ for all $i_0,j_0\in\lbrace 1,...,n\rbrace,$. The module\footnote{It is unital in this paper although some authors might use the terminology unital module instead of module.} is just a vector space with the field used in the scalar-vector multiplication mapping replaced by the ring. For us, this means $\cdot_{scalar}:C^\infty(I\to\mathbb{R})\times\mathfrak{gl}_c\to\mathfrak{gl}_c$ is the scalar multiplication in the module, whereas for the vector space $\cdot_{scalar}:\mathbb{R}\times\mathfrak{gl}(n;\mathbb{R})\to\mathfrak{gl}(n;\mathbb{R})$ is the appropriate scalar multiplication. The algebra over a ring is now just like the algebra over a field, substituting vectors space over a field by module over a commutative ring. We consider the first order ordinary differential equation system with $\mathbf{g}(0)$
\begin{align}
\dfrac{d\mathbf{g}}{dt} = \mathsf{A}(t)\mathbf{g}(t)+\mathbf{f}(t),
\end{align}
with $\mathbf{g}:I\to\mathbb{R}^n$ being sought after and $\mathbf{f}:I\to\mathbb{R}$ being given and $\mathsf{A}(t)\in\mathfrak{gl}_c$. Duhamel's principle now states that the system has a unique solution, given by
\begin{align}
\mathbf{g}(t)=\int_0^t d\tau\,\exp(\mathfrak{G}(t-\tau)\mathbf{f}(\tau),
\end{align}
where the generator $\mathfrak{G}=\mathfrak{G}(t)$now depends on time and is given by
\begin{align}
\mathfrak{G}(\tau)=\int_0^\tau d\tau'\,\mathsf{A}(\tau').
\end{align}
Indeed,following the procedure outlined in the Picard-Lindel\"{o}ff-theorem and symmetrizing the resulting iteration scheme, we can compare with the definition of the matrix exponential defining here $\tau=\tau_{-1}=\tau_0$, $0/0=0$ and $(\mathsf{A}(t))^0=\mathds{1}_n$,
\begin{align}
\exp\left(\int_0^\tau d\tau\,\mathsf{A}(\tau')\right) &= \sum_{k=0}^\infty\dfrac{1}{k!}\left(\int_{0}^\tau\,d\tau'\mathsf{A}(\tau')\right)^k\\
&= \sum_{k=0}^\infty\prod_{i=0}^k\int_{0}^{\tau_{i-1}:=\tau(1-i/k)}d\tau_i\,\mathsf{A}(\tau_i)
\end{align}
The last equation is the $\mathsf{A}$-dependent part in the Picard-iteration, i.e., the application of the Banach fixed point theorem in the proof of the Picard-Lindel\"{o}ff theorem. The integrals could be symmetrized with the result of equal integration domain $[0,\tau]$ and a symmetrization factor $(k!)^{-1}$. Notice that only in the symmetrization procedure we need the commutativity of the $t$-dependent matrices: In our setting, we have for all $t_1,t_2\in\mathbb{R}$ that $[\mathsf{A}(t_1,),\mathsf{A}(t_2)]=\mathsf{0}$. In other words, we could commute the $t$-dependent matrices just as we do for ordinary functions depending on $t$. We check that Duhamel's principle yields a solution to the ordinary differential equation system under consideration
\begin{align*}
\dfrac{d\mathbf{g}}{d t}&=\dfrac{d}{dt}\int_0^t d\tau\,\exp(\mathfrak{G}(t-\tau)\mathbf{f}(\tau)\\
&=\int_0^t d\tau\,\dfrac{d\mathfrak{G}(t-\tau)}{dt}\exp(\mathfrak{G}(t-\tau)\mathbf{f}(\tau)+\mathbf{f}(t)\\
&=\int_0^td\tau\,\left.\dfrac{d}{d(t-\tau)}\right\vert_{\tau=0}\left(\int_{0}^{t-\tau}d\tau'\,\mathsf{A}(\tau')\right)\exp(\mathfrak{G}(t-\tau)\mathbf{f}(\tau)+\mathbf{f}\\
&=\int_0^td\tau\,\dfrac{d}{dt}\left(\int_{0}^{t}d\tau'\,\mathsf{A}(\tau')\right)\exp(\mathfrak{G}(t-\tau)\mathbf{f}(\tau)+\mathbf{f}\\
&= \int_0^td\tau\,\mathsf{A}(t)\exp(\mathfrak{G}(t-\tau)\mathbf{f}(\tau)+\mathbf{f}\\
&=\mathsf{A}(t)\int_0^td\tau\,\exp(\mathfrak{G}(t-\tau)\mathbf{f}(\tau)+\mathbf{f}\\
&=\mathsf{A}(t)\mathbf{g}+\mathbf{f},
\end{align*}
using that $t=(t-\tau)\vert_{\tau=0}$ and $d(t-\tau)=dt$. Also in this case, Duhamel's principle works. We have summarized the geometrical idea of Duhamel's principle in Fig. 13. A Lie algebra element $\mathsf{A}\in \mathfrak{gl}(n,\mathbb{R})$ or in $\mathfrak{gl}_c/[\mathfrak{gl}_c,\mathfrak{gl}_c]$ can be used to define a straight line $\mathfrak{G}(t)=\int_0^t d\tau\,\mathsf{A}$ which is a curve in the Lie algebra. By means of the exponential map, we can in a neighborhood of $\mathsf{0}\in\mathfrak{gl}(n\mathbb{R})$ or $\mathfrak{gl}_c/[\mathfrak{gl}_c,\mathfrak{gl}_c]$, and by suitable translations of the origin also elsewhere, map $\mathfrak{G}(t)$ to an element $\exp\left(\int_0^t d\tau\,\mathsf{A}\right)$ in the Lie group $GL(n,\mathbb{R})$ or $GL_c/[GL_c,GL_c]$. The inverse exponential is the logarithm, $\log$, $\exp^{-1}=\log$.
\begin{figure}
\includegraphics[width = 1\textwidth]{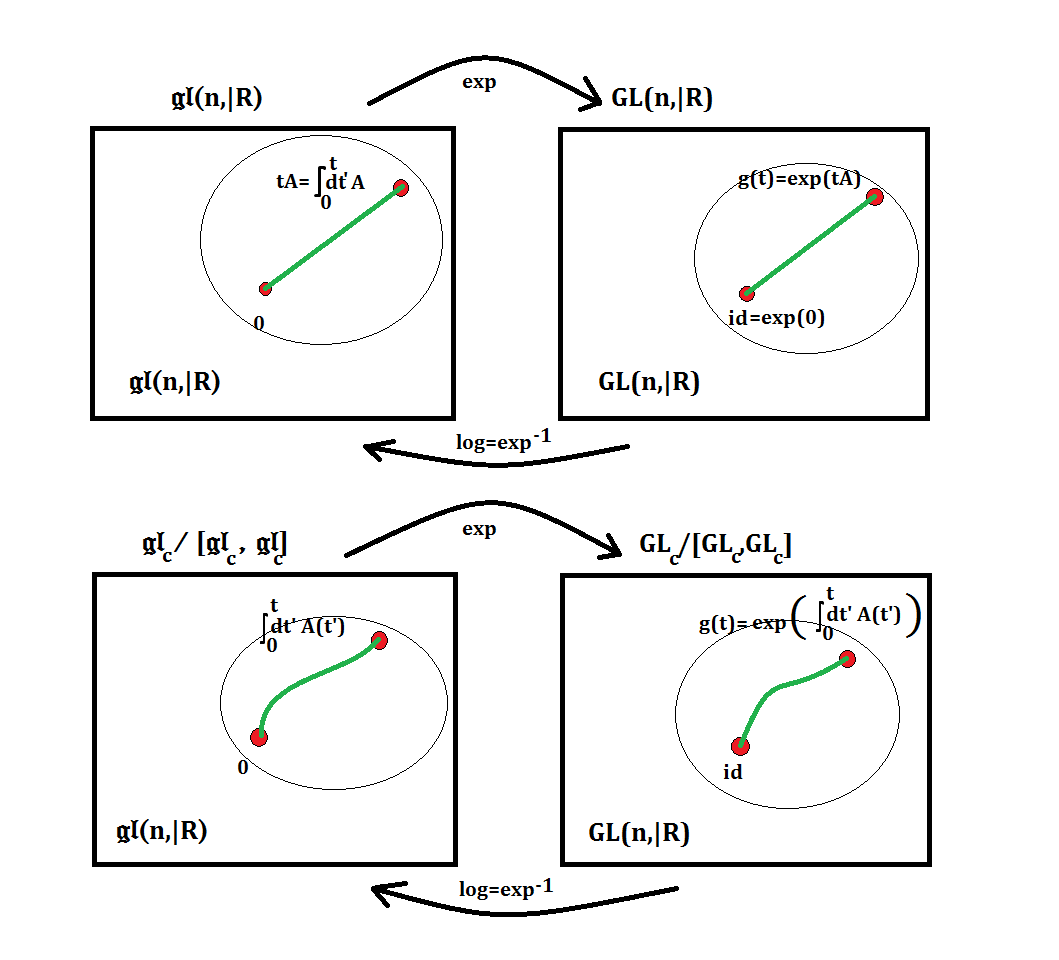}
\caption{On the geometrical interpretation of Duhamel's principle.}
\end{figure}\newline
\newline
\textbf{Duhamel's principle - III: }The important question is what to do when we are using the full algebra $\mathfrak{gl}(n,C^\infty(I\to\mathbb{R}))$ instead of its abelianization. Duhamel's principle can in this case be combined with the Magnus series expansion, a technique used e.g. in quantum electrodynamics. We explain this for a matrices. At first let $\mathfrak{gl}$ denote our non-commutative algebra. We now analogize to manifolds. Recall that $\text{GL}(n,\mathbb{R})$ is a Lie group and that $\mathfrak{gl}(n,\mathbb{R})=T_{\text{id}}\text{GL}(n,\mathbb{R})$ is the tangential space. We can define a curve $t\to g(t)$ with $g(t)\in \text{GL}(n,\mathbb{R})$ for all $t\geq 0$ in $\text{GL}(n,\mathbb{R})$ and define the associated velocity field by covariant derivative w.r.t. $t$, $D/Dt g(t)=\mathfrak{G}(t)$. Conversely, given an element $\mathsf{A}\in\mathfrak{gl}(n,\mathbb{R})$, we obtain a curve $g(t)$ by $t\to \int_{0}^td\tau\mathsf{A}= t\mathsf{A}$ in $\mathfrak{g}$ such that the exponential map $\exp:\mathfrak{G}(t)\to g(t)$ is a local diffeomorphism between the curves $\mathfrak{G}(t)$ in $\mathfrak{gl}(n,\mathbb{R})$ and $g(t)$ in $\text{GL}(n,\mathbb{R})$. As a slight generalization, we let $g(t)\in\text{GL}(n,C^\infty(I\to\mathbb{R}))$ be given and denote by $t$ the argument of the smooth function which form the coefficients of $\mathsf{A}(t)$. We can interpret one $g(t)$ as a curve in $\text{GL}(n,\mathbb{R})$ and define the associated velocity field, $\mathfrak{G}(t)=D/Dt g(t)$ w.r.t. the covariant derivative on $\text{GL}(n,\mathbb{R})$. From the differential geometry on fiber bundles course \cite{baum}, it is known that $\exp:\mathfrak{g}\to G$ is a local diffeomorphism, i.e., for each $t$ sufficiently small such that we can approximately linearly, $\exp(\mathfrak{G}(t))=g(t)$ and, by smoothness of the coefficients of $\mathsf{A}$, we have a curve $t\to\int_{0}^td\tau\,\mathsf{A}(t)=\mathfrak{G}(t)$ in $\mathfrak{gl}.$ The only thing that is different from the previous case is that $\mathfrak{G}(t)$ now can be curved in $\mathfrak{gl}$ instead of being a straight line as in the $\mathfrak{gl}(n,\mathbb{R})$ case. Now let us calculate the derivative of $g(t)$ exploiting the regularity of $g(t)$ and $\mathfrak{G}(t)$ to exchange differentiation, integration and limits whenever needed. $D_{\ast}/Dt$ denotes the covariant derivative in $\mathfrak{gl}$, which maps in $T_{(g(t),\mathfrak{G}(t))}\mathfrak{gl}\simeq\mathfrak{gl}$ for given $t$. Practically, we obtain just the velocity field of $\mathfrak{G}(t)$ which is again in $\mathfrak{gl}$.
\begin{align}
\dfrac{Dg}{Dt}&=\dfrac{D\exp(\mathfrak{G}(t))}{Dt}\\
&= \lim_{N\to\infty}\dfrac{D}{Dt}\left(1+\dfrac{\mathfrak{G}(t)}{N}\right)^N\\
&= \lim_{N\to\infty}\sum_{k=1}^N\left(1+\dfrac{\mathfrak{G}(t)}{N}\right)^{N-k}\dfrac{D_{\ast}\mathfrak{G}}{Dt}\left(1+\dfrac{\mathfrak{G}(t)}{N}\right)^{k-1}.
\end{align}
This expression can in general not be simplified further by non-commutativity of $\mathfrak{G}(t)$ and $D_{\ast}\mathfrak{G}(t)/Dt$ in $\mathfrak{gl}$. Now, we divide the interval $[0,1]$ into $N$ pieces of length $d\tau = 1/N$ and set $\tau = k /N$. Then we can turn the sum into a Riemann integral always keeping $t$ fixed. Recall that the group $GL(n,\mathbb{R})$ in which the $g(t)$'s live can act on its algebra $\mathfrak{gl}(n,\mathbb{R})$ by the adjoint action,
\begin{align}
\text{Ad}_{g(t)}:\mathfrak{gl}\to\mathfrak{gl},\,\mathfrak{H}(t)\mapsto (g(t))^{-1}\mathfrak{G}(t)g(t).
\end{align}
The differential of $\text{Ad}_{g(t)}$ w.r.t. $g(t)$ is the adjoint action of the Lie algebra on itself, i.e.,
\begin{align}
\text{ad}_{\mathfrak{G}(t)}(\mathfrak{H}(t))=[\mathfrak{G}(t),\mathfrak{H}(t)]
\end{align}
in terms of the matrix commutator to be evaluated pointwisely for each $t$. A useful relationship, \cite{baum}, states that the exponential map and adjoint actions are commutative in the following sense
\begin{align}
\text{Ad}_{\exp(\mathfrak{g}(t))} = \exp(\text{ad}_{\mathfrak{g}(t)}).
\end{align}
The $\mathbb{R}$ bi-linearity of $\text{ad}$ is clear from its representation as the Lie bracket or commutator between two elements of the Lie algebra. We turn to our main calculation in this paragraph,
\begin{align}
\dfrac{Dg(t)}{Dt} &= \int_{0}^1d\tau\,\exp((1-\tau)\mathfrak{G}(t))\dfrac{D_{\ast}\mathfrak{G}(t)}{Dt}\exp(\tau\mathfrak{G}(t))\\
&=\exp(\mathfrak{G}(t))\int_{0}^1d\tau\,\exp(-\tau\mathfrak{G}(t))\dfrac{D_{\ast}\mathfrak{G}(t)}{Dt}\exp(\tau\mathfrak{G}(t))\\
&=\exp(\mathfrak{G}(t))\int_0^1d\tau\,\text{Ad}_{\exp(-s\mathfrak{G}(t))}\left[\dfrac{D_{\ast}\mathfrak{G}(t)}{Dt}\right]\\
&=\exp(\mathfrak{G}(t))\int_0^1d\tau\,\exp(-s\text{ad}_{\mathfrak{G}(t)})\left[\dfrac{D_\ast\mathfrak{G}(t)}{Dt}\right]\\
&=\exp(\mathfrak{G}(t))\left(\dfrac{1-\exp(-\text{ad}_{\mathfrak{G}(t)})}{\text{ad}_{\mathfrak{G}(t)}}\right)\left[\dfrac{D_{\ast}\mathfrak{G}}{Dt}\right].
\end{align}
In practical notation, we have with the function $\chi$ defined by $\chi(x)=x^{-1}(1-\exp(-x))$ that
\begin{align}
\partial_t g(t)=e^{\mathfrak{G}(t)}\chi(\text{ad}_{\mathfrak{G}(t)})\partial_t\mathfrak{G}(t).
\end{align}
We seek to make contact with the Magnus series expansion. We have to relate $e^{\mathfrak{G}(t)}$ to $\mathsf{A}(t)$. For this, we calculate
\begin{align}
-\mathsf{A}(t)&\stackrel{!}{=}(g(t))\dfrac{\partial (g(t))^{-1}}{\partial t}\\
&=e^{\mathfrak{G}(t)}\dfrac{\partial e^{-\mathfrak{G}(t)}}{\partial t}\\
&=-\exp(\mathfrak{G}(t))\exp(-\mathfrak{G}(t))\left(\dfrac{1-\exp(\text{ad}_{\mathfrak{G}(t)})}{-\text{ad}_{\mathfrak{G}(t)}}\right)\dfrac{\partial\mathfrak{G}}{\partial t}\\
&=-\left(\dfrac{\exp(\text{ad}_{\mathfrak{G}(t)})-1}{\text{ad}_{\mathfrak{G}(t)}}\right)\dfrac{\partial\mathfrak{G}}{\partial t}
\end{align}
Solving the equation for $\partial_t\mathfrak{G}$ yields
\begin{align}
\dfrac{\partial\mathfrak{G}(t)}{\partial t} = \left(\dfrac{\text{ad}_{\mathfrak{G}}}{\exp(\text{ad}_{\mathfrak{G}})-1}\right)\mathsf{A}(t) = \dfrac{1}{\chi(-\text{ad}_{\mathfrak{G}(t)})}\mathsf{A}(t):=B(\text{ad}_{\mathfrak{G}(t)})\mathsf{A}(t).
\end{align}
Now we use some special numbers, the so-called Bernoulli numbers $\lbrace B_k\rbrace_{k\in\mathbb{N}_0}$, defined by the Taylor-expansion coefficients of $B(x)=1/\chi(-x)$,
\begin{align}
B(x) = \dfrac{x}{e^x-1} = \sum_{k=0}^\infty\dfrac{B_k}{k!}x^k,
\end{align}
Then, we can write
\begin{align}
\dfrac{\partial\mathfrak{G}(t)}{\partial t} = \sum_{k=0}^\infty\dfrac{B_k}{k!}\text{ad}_{\mathfrak{G}(t)}^k[\mathsf{A}(t)].
\end{align}
Denoting by the product of $\text{ad}_{\mathfrak{G}_i}$'s in the following equation composition, the full $\mathfrak{G}(t)$ can be obtained by means of recursion formulas,
\begin{align}
\mathfrak{G} &= \sum_{k=1}^{\infty}\mathfrak{G}_{(k)}(t)\\
\mathfrak{G}_{(1)} &= \int_{0}^td\tau,\mathsf{A}(\tau)\\
\mathfrak{G}_{(n)} &= \sum_{j=1}^n\dfrac{B_j}{j!}\sum_{\sum_{i=1}^{n-1}k_i = n-1,\, k_i\geq 1\forall 1\leq i\leq n-1}\int_0^t d\tau\,\left(\prod_{i=1}^{n-1}\text{ad}_{\mathfrak{G}_{(k_i)}(\tau)}\right)\mathsf{A}(\tau).
\end{align}
Practically, we have for the first three $\mathfrak{G}_{(i)}$'s the following expressions involving the commutator of $\mathsf{A}(t)$ evaluated at distinct $t$'s,
\begin{align}
\mathfrak{G}_{(1)}(t)&=\int_{0}^td\tau\,\mathsf{A}(\tau)\\
\mathfrak{G}_{(2)}(t)&=\dfrac{1}{2}\int_{0}^td\tau_1\int_{0}^{\tau_1}d\tau_2\,[\mathsf{A}(\tau_1),\mathsf{A}(\tau_2)]\\
\mathfrak{G}_{(3)}(t)&=\dfrac{1}{6}\int_{0}^td\tau_1\int_{0}^{\tau_1}d\tau_2\int_0^{\tau_2}d\tau_3\,\left([\mathsf{A}(\tau_1),[\mathsf{A}(\tau_2),\mathsf{A}(\tau_3)]]+[\mathsf{A}(\tau_3),[\mathsf{A}(\tau_2),\mathsf{A}(\tau_1)]]\right)
\end{align}
We notice that this reproduces the conventional Duhamel expression if $\mathsf{A}(t)\in\mathfrak{gl}_c=\mathfrak{gl}/[\mathfrak{gl},\mathfrak{gl}]$, i.e., it is a representative of the abelianization algebra $\mathfrak{gl}_c$ of $\mathfrak{gl}$. We can now formulate the answer to the initial question in this subparagraph, namely, whether the following first order system of ordinary differential equations, has a solution
\begin{align}
\dfrac{d\mathbf{f}}{dt} = \mathsf{A}(t)\mathbf{f}(t)+\mathbf{g}(t),
\end{align}
with $\mathbf{f}(0)=\mathbf{0}$ and $\mathsf{A}(t)\in\mathfrak{gl}(n;C^\infty(\mathbb{R}^+_0\to\mathbb{R}))$ and $f:\mathbb{R}^+_0\to\mathbb{R}^n$ being sought and given $\mathbf{g}:\mathbb{R}^+_0\to\mathbb{R}^n$ suitably regular. The solution exists and is uniquely given by the \emph{Magnus expansion}
\begin{align}
\begin{split}
\mathbf{f}(t) &= \int_{0}^t d\tau\,\exp(\mathfrak{G}(t-\tau))\mathbf{g}(\tau)\\
\mathfrak{G}(t) &= \sum_{k=1}^{\infty}\mathfrak{G}_{(k)}(t)\\
\mathfrak{G}_{(1)}(t) &= \int_{0}^td\tau\,\mathsf{A}(\tau)\\
\mathfrak{G}_{(n)}(t) &= \sum_{j=1}^n\dfrac{B_j}{j!}\sum_{\sum_{i=1}^{n-1}k_i = n-1,\, k_i\geq 1\forall 1\leq i\leq n-1}\int_0^t d\tau\,\left(\prod_{i=1}^{n-1}\text{ad}_{\mathfrak{G}_{(k_i)}(\tau)}\right)\mathsf{A}(\tau).
\end{split}
\end{align}
The $B_j$'s are the Bernoulli numbers as introduced above. Unfortunately, the convergence of the expansion is widely unclear. Magnus himself \cite{magnus1} obtained a sufficient criterion for the convergence of the series which has been improved by Moan \cite{magnus7}: It is \emph{sufficient} for the series to converge for $t'\in[0,t)$ if
\begin{align}
\int_0^t d\tau\,\Vert\mathsf{A}(\tau)\Vert_{\text{Frob}} < \pi,
\end{align}
in the Frobenius matrix norm and $\pi \approx 3.14159 \cdots$ the usual $\pi$. It must be emphasized that the criterion is only sufficient and not necessary. In particular, sharper bounds, see e.g. \cite{magnus2, magnus3, magnus4, magnus5} have been obtained for various cases. The observation of Iserles et al. \cite{magnus6, magnus7, magnus8} was that a Magnus series like expansion preserves the Lie-algebra structure. In terms of the lower central series, the decomposition of the Magnus generator $\mathfrak{G}$ as a sum corresponds to choosing elements as,
\begin{align}
\mathfrak{gl}_c &= \dfrac{\mathfrak{gl}_c(t_1)}{[\mathfrak{gl}_c(t_1),\mathfrak{gl}_c(t_2)]}\oplus\dfrac{[\mathfrak{gl}_c(t_1),\mathfrak{gl}_c(t_2)]}{[\mathfrak{gl}_c(t_1)[\mathfrak{gl}_c(t_2),\mathfrak{gl}_c(t_3)]]}\oplus\cdots\\
\mathfrak{G} &= \mathfrak{G}_{(1)}+\mathfrak{G}_{2}+\cdots ,
\end{align}
where $[\mathfrak{gl}_c(t_1),\mathfrak{gl}_c(t_2)]$ is the Lie-sub-algebra of $\mathfrak{gl}_c$ generated by expressions of the form $\mathsf{A}(t_1)\mathsf{A}(t_2)-\mathsf{A}(t_2)\mathsf{A}(t_1)$ with $\mathsf{A}(t_i)\in\mathfrak{gl}_{c}(t_i)$ for $i\in\lbrace 1,2\rbrace$, i.e., the Lie sub-algebra of $\mathfrak{gl}_c(t)$ containing the elements of the Lie-algebra which modulo commutators with elements of $\mathfrak{gl}_c$. We denote by by $\text{ad}_{\mathfrak{gl}_c}^k$ the Lie sub-algebra of $\mathfrak{gl}_c$ generated by $k$-fold commutators, e.g., $\text{ad}_{\mathfrak{gl}_c}^3[\mathfrak{gl}_c]$ consists of expressions like $[\mathsf{A}(t_1),[\mathsf{A}(t_2),[\mathsf{A}(t_3),\mathsf{A}(t_4)]]]$. Observe that the $\mathsf{A}=\mathsf{A}(t)$ stays always fixed! Let us drop the arguments $t_1,t_2,t_3,...$ in the following again. For more specialized Lie algebras, the decomposition
\begin{align}
\mathfrak{gl}_c = \bigoplus_{k=0}^{\infty}\dfrac{\text{ad}_{\mathfrak{gl}_c}^k[\mathfrak{gl}_c]}{\text{ad}_{\mathfrak{gl}_c}^{k+1}[\mathfrak{gl}_c]},
\end{align}
may truncate after the $N$-th factor in the direct sum. In that case, the Lie-algebra would be solvable and convergence of the Magnus series would follow from finiteness of the sum $\mathfrak{G}=\sum_{k\in\mathbb{N}}\mathfrak{G}$. In general however, the series stays infinite. One can ask how to re-obtain from the element $\exp(\mathfrak{G})\in\text{GL}(n,C^\infty(I\to\mathbb{R})):=\text{GL}_c$ the element $\mathsf{A}(t)\in\mathfrak{gl}_c$. This is achieved by a convenient property of the covariant\footnote{in the functorial sense} Lie-functor $\mathfrak{L}$, namely that it respects the lower central series decomposition for the Lie-group and the Lie-algebra. More precisely,
\begin{align}
\mathfrak{L}:\text{GL}_c=\bigoplus_{k=0}^{\infty}\dfrac{\text{Ad}_{\text{GL}_c}^k[\text{GL}_c]}{\text{Ad}_{\text{GL}_c}^{k+1}[\text{GL}_c]}\mapsto\mathfrak{gl}_c=\bigoplus_{k=0}^{\infty}\dfrac{\text{ad}_{\mathfrak{gl}_c}^k[\mathfrak{gl}_c]}{\text{ad}_{\mathfrak{gl}_c}^{k+1}[\mathfrak{gl}_c]},
\end{align}
preserves the individual factors where $\text{Ad}_{\text{GL}_c}$ denotes the action of the $\text{GL}_c$ on itself by conjugation. Moreover, the operators that we are interested in, namely Laplacians are unbounded operators in the $L^2$ and the $H^{1,2}$ norm. However, when restricting to $H^{2,2}$, i.e., another Sobolev-space, which we can also equip with the $L^2$-norm, the Laplacians can be bounded by the $H^{2,2}$-norms of the functions they are acting on. At the end of this paragraph let us briefly recapitulate what the Magnus series does: In the non-commutative case of the full $\mathfrak{gl}_c$, we cannot use the exponential straight away to map from the Lie algebra to the Lie group, c.f., the upper row in Fig. 16. However, we can define a curve $t\to\mathfrak{G}(t)$ in the Lie-algebra such that the exponential maps $\mathfrak{G}(t)$ to an element $S(t)$ of the Lie-group such that its differential logarithm, $d\exp^{-1} S(t)= S'(t)/S(t)=\mathsf{A}(t)$. It is more useful in the context of differential equation solving to interpret $G(t)$ for $t\geq 0$ as a matrix semigroup, c.f. \cite{magnus1, magnus2, magnus3, magnus4, magnus5, magnus6}. The process is depicted in the lower half of Fig. 16.
\begin{figure}
\includegraphics[width = 1\textwidth]{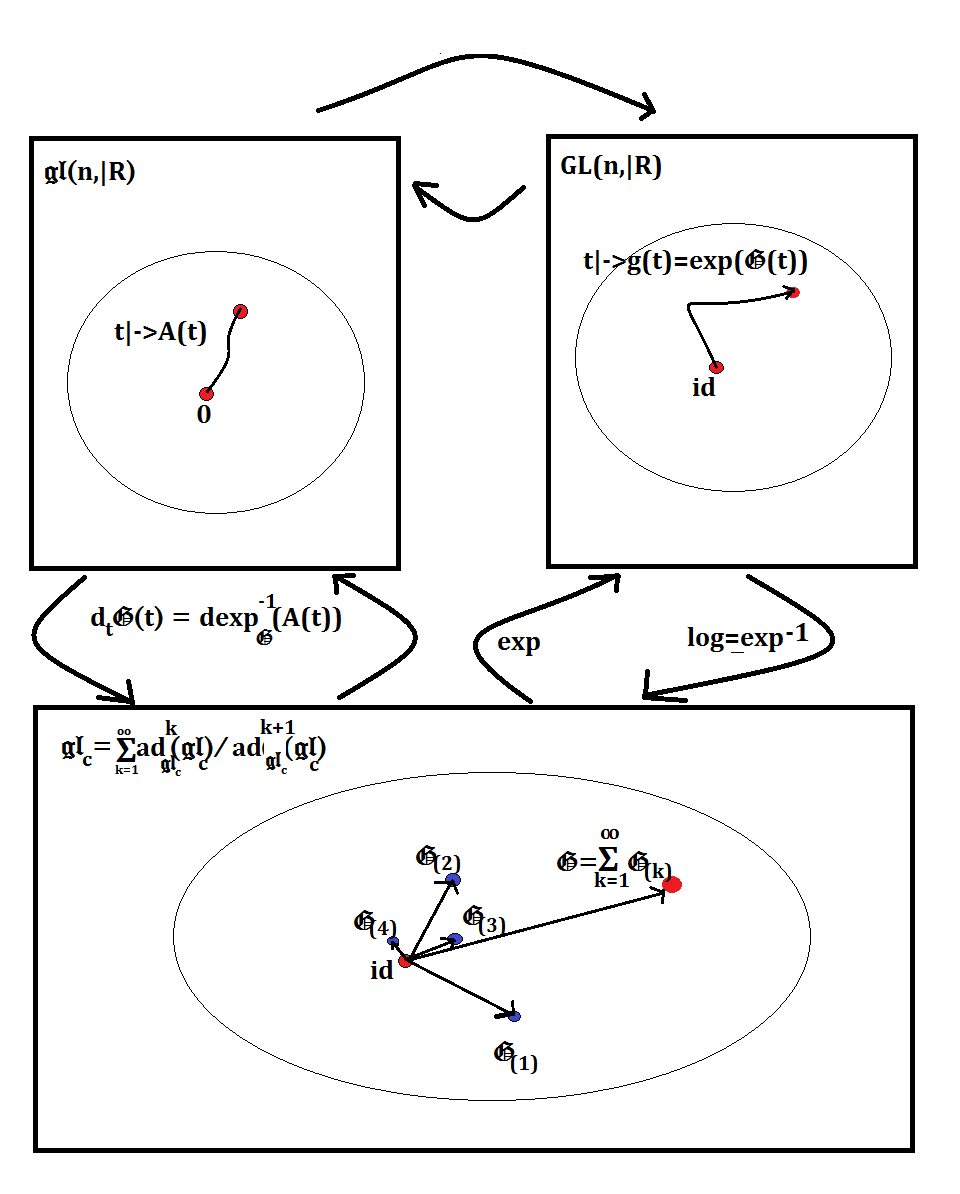}
\caption{The Magnus series and the exponential map $\exp:\mathfrak{gl}_c\to GL_c$.}
\end{figure}
\newline
\newline
\textbf{Duhamel IV} Duhamel's principle and its one of the iteration schemes derived thereof - the Magnus series - can be generalized to operator, \cite{magnus1, zeidler1, zeidler2}. The analogy in physics used to give sense to this process is the quantum mechanical imagination of (linear) operators as infinite-dimensional, but else well-behaved  matrices mapping from one infinite-dimensional vector space with an inner product, the so-called Hilbert space, to another. The discussion can be extended to Banach spaces as a more general class of function spaces, \cite{jost1, zeidler1, zeidler2}, but we only need Hilbert spaces which have more convenient structures on them. From the more mathematical angle, the analogy is useful but not totally correct. We will be interested in two specific issues, namely boundedness and compactness of the operators. We will use the notion of an operator as a synonym for an operator between Hilbert spaces and the notion matrix as a synonym for a linear map between finite-dimensional vector spaces. Recall, that an operator or a matrix, $\mathsf{A}:(V_1,\Vert .\Vert_1)\to (V_2,\Vert .\Vert_2)$ is called \emph{bounded} if for a $c\in\mathbb{R}^+_0$, the relative operator norm of $\mathsf{A}$ is bounded by $c$, i.e.,
\begin{align}
\Vert\mathsf{A}\Vert_{1\to 2}:=\sup_{v\in V_1}\left[\dfrac{\Vert\mathsf{A}v\Vert_2}{\Vert v\Vert_1}\right] = c<\infty.
\end{align} 
In the finite dimensional case, i.e., when $\mathsf{A}$ is a matrix, we need not distinguish between two different relative norms $\Vert .\Vert_{1\to 2}$ and $\vert .\vert_{1\to 2}$ on $\text{Hom}_{\mathbb{C}}(V_1,V_2)$. The norms are said to be equivalent. In the infinite-dimensional case, i.e., when $V_1,\,V_2$ are Hilbert spaces and the operator spaces is the space of $\mathbb{C}-$linear operators between $V_1$ and $V_2$, $\text{LinOp}_\mathbb{C}(V_1,V_2)$, the distinction becomes crucial. We can have one relative operator norm $\Vert .\Vert_{1\to 2}$ on $\text{LinOp}_\mathbb{C}(V_1,V_2)$ where an operator $\mathsf{A}$ is bounded, but another norm $\vert.\vert_{1\to 2}$ on $\text{LinOp}_\mathbb{C}(V_1,V_2)$ where it is unbounded. With the definition of the operator norm in the context of boundedness, an appropriate norm choices can be broken down to an appropriate selection of norms on the domain Hilbert space $V_1$ of the operator $\mathsf{A}$ and an appropriate norm on the target Hilbert space $V_2$. The second issue is compactness. Recall that an operator (or a matrix) between two vector spaces with inner product, $(V_1,\Vert .\Vert_1)$ and $(V_1,\Vert .\Vert_2)$, is called \emph{compact} if it can be approximated by finite-rank operators, i.e., a sequence $(\mathsf{A}_n)_{n\in\mathbb{N}}$ in $\text{LinOp}_{\mathbb{C}}(V_1,V_2)$ such that $\dim\text{Im}(\mathsf{A}_n)<\infty$ for all $n\in\mathbb{N}$ and such that for all $v\in\text{Dom}(\mathsf{A})$ with $\Vert v\Vert_1 < \infty$, the operators converge $\lim_{n\to\infty}\Vert\mathsf{A}v-\mathsf{A}_n v\Vert_2 =0$. Geometrically, one can reformulate the definition equivalently as, operators are compact if and only if the image of the unit ball $B_1(V_1)$ in $V_1$ is mapped to a compact subset of $V_2$. In finite-dimensional, the unit ball $B_1^n(\mathbf{0})$ is always compact. So matrices are compact. But in infinite dimensions, this breaks down. We show as an exercise in foundations of functional analysis that the unit ball in $C([0,1])$ endowed with the max norm $\Vert .\Vert_\infty$ is not compact. In $C([0,1])$ endowed with the max norm, $\Vert .\Vert_\infty$, the sequence $(x^n)_{n\in\mathbb{N}}$ diverges to a function that is $0$ on $[0,1)$ and $1$ at $\partial[0,1]\setminus\lbrace 0\rbrace$. Since $(C([0,1]=\Vert .\Vert_1)$ is normed, compactness is equivalent to each sequence in $B_1(C([0,1]))$ sequentially convergent in $B_1(C([0,1]))$. Since we have found an example of a divergent sequence, we cannot have compactness of the unit ball $B_1(C([0,1]))$ in $C([0,1])$. The example is quite generic since the method-of-proof can be carried over to other function spaces as well. What does compactness do for us? Compactness ensures that we can think of our (linear) operator as a matrix. If $\mathsf{A}$ is compact then $\mathsf{A}$ is bounded. Recall that $\mathsf{A}$ is linear. If $\mathsf{A}$ is bounded then it is also continuous. The converse also holds true. In total, compactness of an operator gives us what we want to have for the quantum-mechanical analogy of operators and matrices. However, generic operators are nor compact. We explain in our setting that and how compactness can be achieved. The linear operators we have in our model  are the Laplace-Beltrami operators $\Delta_{g_0}^\partial$ and $\Delta_{g_0}$ on $\partial\Omega_0$ and $\partial\Omega_t$ or $\Omega_0$ and $\Omega_t$ depending on whether we work on the unperturbed bundle $\mathcal{M}_0$ and its boundary $\partial\mathcal{M}_0$ or the perturbation bundle $\mathcal{M}$ and its boundary $\partial\mathcal{M}$. By definition of the perturbation and reference bundle, $\Omega_0,\Omega_t$ is smooth and in particular compact in $\mathbb{R}^n$. The same holds true for the bounding manifolds $\partial\Omega_0$ and $\partial\Omega_t$ in $\mathbb{R}^n$. In a previous section, we have constructed the Sobolev spaces $H^{1,2}_{g_0}(X)$ with $X\in\lbrace\Omega_0,\Omega_t,\partial\Omega_0,\partial\Omega_t\rbrace$. As Hilbert-spaces, the Sobolev spaces are reflexive by Riesz' representation theorem. Using the Gelfand triplet with compact and dense continuous (by Rellich's embedding theorem) imbeddings,
\begin{align*}
H^{1,2}_{g_0}(X)\hookrightarrow L^2_{g_0}(X)\hookrightarrow (H^{1,2}_{g_0}(X))^\ast = H^{-1,2}_{g_0}(X)\simeq H^{1,2}_{g_0}(X),
\end{align*}
we have shown self-adjointness of the Laplace-Operators $\Delta_{g_0}$ and $\Delta_{g_0}^\partial$ as linear maps $H^{1,2}_{g_0}(X)\to H^{-1,2}_{g_0}(\Omega_0)$. On the right of the Gelfand-triplet, the equality follows from definition of $H^{-1,2}_{g_0}(X)$ and the $\simeq$ means isometric equivalence by the induced pairing between the Sobolev spaces $H^{1,2}_{g_0}(X)$ and $H^{-1,2}_{g_0}(X)$ the  $\langle .,.\rangle_{L^2_{g_0}(X)}$. Self-adjointness of a linear operator $\mathsf{A}$ consists of two ingredients, namely firstly that $\text{Dom}(\mathsf{A})=\text{Im}(\mathsf{A})$, i.e., the domain Hilbert space and the target Hilbert space are identical plus secondly that we have the Hermiticity property satisfied, that is
\begin{align*}
\mathsf{A}=\mathsf{A}^\dagger ,
\end{align*}
where $^\dagger$ is the physics notation for the adjoint operator. For the Laplacians, we have both. If we interpret the Laplace-Beltrami operators as linear self-adjoint operators,
\begin{align*}
&\Delta_{g_0}:(H^{1,2}_{g_0}(\Omega_0),\Vert .\Vert_{H^{1,2}_{g_0}(\Omega_0)})\to (H^{-1,2}_{g_0}(\Omega_0),\Vert .\Vert_{L^2_{g_0}(\Omega_0)}\\
&\Delta_{g_0}:(H^{1,2}_{g_0}(\Omega_t),\Vert .\Vert_{H^{1,2}_{g_0}(\Omega_t)})\to (H^{-1,2}_{g_0}(\Omega_t),\Vert .\Vert_{L^2_{g_0}(\Omega_t)}\\
&\Delta^\partial_{g_0}:(H^{1,2}_{0,g_0}(\partial\Omega_0),\Vert .\Vert_{H^{1,2}_{0,g_0}(\partial\Omega_0)})\to (H^{-1,2}_{0,g_0}(\partial\Omega_0),\Vert .\Vert_{L^2_{g_0}(\partial\Omega_0)}\\
&\Delta_{g_0}:(H^{1,2}_{0,g_0}(\partial\Omega_t),\Vert .\Vert_{H^{1,2}_{g_0}(\partial\Omega_t)})\to (H^{-1,2}_{0,g_0}(\partial\Omega_t),\Vert .\Vert_{L^2_{g_0}(\partial\Omega_t)},
\end{align*}
they are unbounded as can be quickly verified by a trivial calculation. The central idea of the proof is an exploitation of the fact that the $\Vert .\Vert_{1,2}$ norm on $H^{1,2}_{(0),g_0}((\partial) X)$ forgets about second order regularity, i.e., we don't know on the mathematical level whether the second order derivatives of our functions are well-behaved. By construction of the Sobolev spaces as norm closure of $\mathcal{C}^\infty_c$-functions w.r.t. the Sobolev norms, it is clear that Sobolev spaces that contain only functions with desirable regularity properties can be embedded densely and compactly by a continuous embedding in Sobolev spaces with nice but not so useful regularity properties. More concretely, we need second order regularity in spatial arguments, i.e.,arguments on the fiber $\Omega_0$ if we tackle the acoustic wave equation or fourth order in spatial arguments, i.e., arguments on the boundary of the unperturbed fiber, $\partial\Omega_0$. if we tackle the boundary vibrations equations which contains the square of the Laplace-Beltrami operator. The Weierstrass approximation theorem tells us that by denseness, we can approximate functions of less regularity arbitrarily well by functions of higher regularity, e.g., by a three-dimensional Taylor series, we can approximate on $L^2(\mathbb{R}^3)$ the function $\exp(-\vert x+y+z\vert)=\exp(-\sqrt{(x+y+z)^2})$ which is just in $C^0(\mathbb{R}^3)$, but the Taylor series is a limit of multinomials, i.e., elements of $\mathbb{R}[x,y,z]$, which are $\mathcal{C}^\infty$. If our Laplace-Beltrami operators were bounded, they were by linearity automatically continuous, i.e., we could also approximate the eigenfunctions of the Laplace-Beltrami operators, given by the spectral theorem, in general on $H^{1,2}$-spaces by functions from $H^{2,2}$- or $H^{2,4}$-spaces using the Weierstrass argument. The continuity enters implicitly because the eigenfunctions to the Laplace-Beltrami operators come together with associated eigenvalues. The continuity of the Laplace-Beltrami operators guarantees that we can reproduce these eigenvalues. If we specialize even further and use compactness of the Riemannian manifolds, Lichernowicz' theorem, a curvy analog of the spectral theorem that is even valid intrinsically, i.e., without reference to an embedding space of the Riemannian manifolds, guarantees that the set of eigenfunctions and eigenvalues are discrete, i.e., at most countable. Thus, we can express in Dirac notation the operator $\mathsf{A}$ which stands for one the Laplace-Beltrami operators in terms of eigenfunctions and eigenvalues,
\begin{align*}
\mathsf{A}=-\sum_{n\in\mathbb{N}}\vert\Psi_{n}^{\mathsf{A}}\rangle\lambda_n^{\mathsf{A}}\langle\Psi_n^{\mathsf{A}}\vert .
\end{align*}
So, as a preliminary result of our considerations, we only use densely defined Laplace-Beltrami operators, i.e.,
\begin{align*}
&\Delta_{g_0}:(H^{2,2}_{g_0}(\Omega_0)\subset H^{1,2}_{g_0}(\Omega_0),\Vert .\Vert_{H^{2,2}_{g_0}(\Omega_0)})\to (H^{-1,2}_{g_0}(\Omega_0),\Vert .\Vert_{L^2_{g_0}(\Omega_0)}\\
&\Delta_{g_0}:(H^{2,2}_{g_0}(\Omega_t)\subset H^{1,2}_{g_0}(\Omega_t),\Vert .\Vert_{H^{1,2}_{g_0}(\Omega_t)})\to (H^{-1,2}_{g_0}(\Omega_t),\Vert .\Vert_{L^2_{g_0}(\Omega_t)}\\
&\Delta^\partial_{g_0}:(H^{4,2}_{0,g_0}(\partial\Omega_0)\subset H^{1,2}_{0,g_0}(\partial\Omega_0),\Vert .\Vert_{H^{4,2}_{0,g_0}(\partial\Omega_0)})\to (H^{-1,2}_{0,g_0}(\partial\Omega_0),\Vert .\Vert_{L^2_{g_0}(\partial\Omega_0)}\\
&\Delta_{g_0}:(H^{4,2}_{0,g_0}(\partial\Omega_t)\subset H^{1,2}_{0,g_0}(\partial\Omega_t),\Vert .\Vert_{H^{4,2}_{g_0}(\partial\Omega_t)})\to (H^{-1,2}_{0,g_0}(\partial\Omega_t),\Vert .\Vert_{L^2_{g_0}(\partial\Omega_t)},
\end{align*}
As a side remark, we note that the Gelfand triplet construction works of course also for the $H^{2,2}$ and $H^{2,4}$ spaces. For the Laplacian on unbounded domains, the limit in the definition of compactness cannot be given sense since the spectrum is continuous and not discrete. In the above equation, the sum needs to be replaced by an integral over the spectrum. However, in our setting, we have the sum as in the above eigenfunction equations and we can define the sequence $(\mathsf{A}_N)_{N\in\mathbb{N}}$ as consistent truncations of the above eigenfunction expansion, that is,
\begin{align*}
\mathsf{A}_N = -\sum_{n=1}^N\vert\Psi_{n}^{\mathsf{A}}\rangle\lambda_n^{\mathsf{A}}\langle\Psi_n^{\mathsf{A}}\vert .
\end{align*}
Now, we let $f\in L^2$. This is by our high-choice of regularity, namely second order resp. fourth regularity possible. First, we can approximate firstly the eigenfunctions by higher-regularity functions from $H^{2,2}$- and $H^{2,4}$-spaces. We assume that we have done so. Lichernowicz theorem gives us that we have a complete and orthonormal set of eigenfunctions. Completeness of the eigenfunctions on $H^{1,2}$-spaces and denseness of $H^{1,2}$-spaces in $L^2$-spaces implies that we can expand $f$ by an eigenfunction expansion,
\begin{align*}
f = \sum_{n\in\mathbb{N}}f_n\vert\Psi^{\mathsf{A}}_n\rangle
\end{align*}
where more precisely the equality is up to a Lebesgue null set w.r.t. the Lebesgue-Borel integration measure on $X,\, X\in\lbrace\partial\Omega_0,\partial\Omega_t,\Omega_0,\Omega_t\rbrace$. The expansion coefficient $f_n$ is given by the equation $f_n=\langle\Psi^{\mathsf{A}}_n\vert f\rangle_{L^2_{g_0}(X)}$ where completeness and orthonormality entered. Since $f\in L^2(X)$, we have together with Parseval's equation
\begin{align*}
\infty>\Vert f\Vert^2_{L^2_{g_0}(X)} = \sum_{n\in\mathbb{N}}\vert f_n\vert^2\equiv \sum_{n\in\mathbb{N}}a_n.
\end{align*}
Now, clearly $a_n\geq 0$ for all $n\in\mathbb{N}$. The sequence $(a_n)_{n\in\mathbb{N}}$ clearly is non-negative by the appearance of the modulus in the definition of each part of the sequence and is by the above argument square-sum-able, i.e.,  $(a_n)_{n\in\mathbb{N}}\in l^2(\mathbb{R})$. Since the series over the $a_n$''s converges, real analysis tells us that $a_{n}\to 0$ as $n\to \infty$, i.e., the $(a_n)_{n\in\mathbb{N}}$ is a null sequence. Now, we set $f = \mathsf{A}g$ for $g\in H^{2,2}_{g_0}(X)$ for $X\in\lbrace\Omega_0,\Omega_t\rbrace$ and $H^{2,4}$ for $X\in\lbrace\partial\Omega_0,\partial\Omega_t\rbrace$. Actually, $H^{2,2}$ suffices in both cases for the proof of compactness but we will need the higher regularity for $X\in\lbrace\partial\Omega_0,\partial\Omega_t\rbrace$ later on. Further, we set $f_N =\mathsf{A}_N g$. Let $\epsilon>0$ and $N$ such that $\sum_{n=N}^\infty a_n \leq\epsilon$ which is possible by convergence of the series. Then, we have
\begin{align*}
\Vert f-f_N\Vert_{L^2_{g_0}(X)}^2 = \sum_{n=N+1}^{\infty}\vert f_n\vert^2\leq  \sum_{n=N}^{\infty}a_n <\epsilon.
\end{align*}
Rewritten in terms of operators $\mathsf{A},\mathsf{A}_N$ and the function $g\in H^{2,2}_{g_0}(X)$, we have compactness. Notice that the limit operations depended heavily on the compactness of the $X\in\lbrace\Omega_0,\Omega_t,\partial\Omega_0,\partial\Omega_t\rbrace$. Boundedness is now easy, we have for $g\in H^{2,2}_{g_0}(X)$ resp. $g\in H^{2,4}_{g_0}(X)$ the following short calculation $\Vert \mathsf{A} g\Vert_{L^2} \leq \Vert g\Vert_{H^{2,2}} (\leq \Vert g\Vert_{H^{4,2}})$, i.e., $\Vert\mathsf{A}\Vert_{L^2_{g_0}(X)\to H^{2,2}_{g_0}(X)}\leq 1$ for $X\in\lbrace\Omega_0,\Omega_t\rbrace$ and $\Vert\mathsf{A}\Vert_{L^2_{g_0}(X)\to H^{2,4}_{0,g_0}(X)}\leq 1$ for $X\in\lbrace\partial\Omega_0,\partial\Omega_t\rbrace$. Let us summarize. For the Laplace-Beltrami operators in question,
\begin{align}
&\Delta_{g_0}:(H^{2,2}_{g_0}(\Omega_0)\subset H^{1,2}_{g_0}(\Omega_0),\Vert .\Vert_{H^{2,2}_{g_0}(\Omega_0)})\to (H^{-1,2}_{g_0}(\Omega_0),\Vert .\Vert_{L^2_{g_0}(\Omega_0)}\\
&\Delta_{g_0}:(H^{2,2}_{g_0}(\Omega_t)\subset H^{1,2}_{g_0}(\Omega_t),\Vert .\Vert_{H^{1,2}_{g_0}(\Omega_t)})\to (H^{-1,2}_{g_0}(\Omega_t),\Vert .\Vert_{L^2_{g_0}(\Omega_t)}\\
&\Delta^\partial_{g_0}:(H^{4,2}_{0,g_0}(\partial\Omega_0)\subset H^{1,2}_{0,g_0}(\partial\Omega_0),\Vert .\Vert_{H^{4,2}_{0,g_0}(\partial\Omega_0)})\to (H^{-1,2}_{0,g_0}(\partial\Omega_0),\Vert .\Vert_{L^2_{g_0}(\partial\Omega_0)}\\
&\Delta_{g_0}:(H^{4,2}_{0,g_0}(\partial\Omega_t)\subset H^{1,2}_{0,g_0}(\partial\Omega_t),\Vert .\Vert_{H^{4,2}_{g_0}(\partial\Omega_t)})\to (H^{-1,2}_{0,g_0}(\partial\Omega_t),\Vert .\Vert_{L^2_{g_0}(\partial\Omega_t)},
\end{align}
we have compactness, boundedness, continuity, Hermiticity and by a Gelfand-construction essential self-adjointness and linearity. We will use the short-hand notation $\Delta_{g_0}$ and $\Delta^\partial_{g_0}$ again. In particular, we have,
\begin{align}
&\Delta_{g_0}\in\text{LinOp}_{\mathbb{C}}^{c,ess.-sa}(H^{2,2}_{g_0}(\Omega_0),L^2_{g_0}(\Omega_0))\\
&\Delta_{g_0^\partial}\in\text{LinOp}_{\mathbb{C}}^{c,ess.-sa}(H^{4,2}_{g_0}(\partial\Omega_0),L^2_{g_0}(\partial\Omega_0))
\end{align}
Notice that since the perturbation operator $\mathsf{W}$ only contains first order partial derivatives, it is for fixed $u$, also bounded and by eigenfunction expansion also compact, i.e., it is a compact perturbation in the sense of the Kato-Rellich theorem. That it includes time-derivatives is unproblematic since they are only of first order and our solutions $p$ for the acoustic wave equation will live in $H^{1,2}_0(\mathbb{R}^+;H^{2,2}_{g_0}(\Omega_0))\equiv H^{1,2;2;2}(\mathcal{M}_0)$ where the semicolon indicates distinction between regularity in base space $\mathbb{R}^+_0$ and fiber space $\Omega_0$ of the unperturbed bundle $\mathcal{M}_0=\mathbb{R}^+_0\times\Omega_0$. It is however, not self-adjoint. The perturbation operator $\mathsf{V}$ is also compact in our setting.\newline
\newline
\textbf{Duhamel V} In this paragraph we want to compare three methods that can be used to tackle the time-dependent perturbations we encounter in the vibro-acoustic setting. The first method dates back to Paul Dirac, \cite{dirac}, and uses a variation of constants argument. The second method dates back to Dyson, \cite{dyson1, dyson2}, and also uses a variation if constants arguments but uses Duhamel's principle to calculate the evolution family of operators. The third method is the Magnus expansion, \cite{magnus1, magnus2, magnus11}, which uses Duhamel's principle again and - as does the Dyson method - calculate the evolution family of the perturbed problem, but does not use the time ordering operator $\mathcal{T}$. We notice that the methods yield the same result. For the sake of matching the notation with the problem we are interested in, we will consider the problem
\begin{align*}
\dfrac{\partial\mathbf{f}}{\partial t} = \mathsf{A}\mathbf{f}+\mathbf{g},
\end{align*}
where $\mathbf{f}=(f_1(t,\mathbf{x}),f_2(t,\mathbf{x}))^T, \mathbf{g}=(g_1(t,\mathbf{x}),g_2(t,\mathbf{x}))^T$ and the matrix $\mathsf{A}$ is in $\mathfrak{gl}(2,\mathsf{W})$, i.e., the Lie-algebra $\mathfrak{gl}_2$ with coefficients in a von-Neumann algebra $\mathsf{W}$ of bounded operators. von-Neumann algebras are studied in \cite{neumann1, neumann2, neumann4, neumann3}. We specialize to the case $\mathbf{f}(t=0)=\mathbf{0}=(0,0)^T$ and the matrix consisting of a time-independent part and a small perturbation part $\mathsf{V}_A(t)$, i.e., $\Vert\mathsf{V}\Vert(t) \ll \Vert\mathsf{A}_0\Vert$ for all $t\geq 0$. We have then $\mathsf{A}=\mathsf{A}_0 + \mathsf{V}_{A}(t)$. The norm $\Vert .\Vert$ is the Frobenius-operator norm,
\begin{align}
\Vert\mathsf{A}\Vert = \sqrt{\sum_{i=1}^2\sum_{j=1}^2\Vert\mathsf{A}_{ij}\Vert^2},
\end{align}
and for the class of models we are interested in, we use our previous considerations on properties of the Laplace-Beltrami-operator and specialize on operators bounded in the $H^{2,2}\to L^2$-norm, i.e., $\Vert\mathsf{A}_{ij}\Vert = \Vert\mathsf{A}_{ij}\Vert_{H^{2,2}_{g_0}\to L^2_{g_0}}$. Further, by boundedness, we observe that the unperturbed operator $\mathsf{A}_0$ generates not only an evolution family, but also a $\mathcal{C}^0$ semi-group, $\mathsf{S}_0(t,\tau),\,t\geq\tau\geq 0$. We assume that $[\mathsf{A}_{0;i,j},\mathsf{A}_{0,i',j'}]=0$ for all $i,i',j,j'\in\lbrace 1,2\rbrace$,i.e., that we can obtain a joint set of eigenfunctions for the entries of the operator matrix $\mathsf{A}_0$. For the cases of interest, this will hold true.
\begin{itemize}
\item\textbf{Dirac's method: }Dirac's method consisted in splitting the system of partial differential equations from above into two parts. A perturbation part and a conventionally solvable inhomogeneous equation,
\begin{align}
\dfrac{\partial\mathbf{f}}{\partial t} = \mathsf{A}_0\mathbf{f}+\mathbf{g} + \mathsf{V}_a\mathbf{f}.
\end{align}
Let us use the Banach fixed point theorem to obtain a recursive equation. Since $\Vert\mathsf{V}_A\Vert\ll\Vert\mathsf{A}_0\Vert <\infty$, we have
\begin{align}
\dfrac{\partial\mathbf{f}^{(k+1})}{\partial t} = \mathsf{A}_0\mathbf{f}^{(k+1)}+\mathbf{g}+\mathsf{V}_A\mathbf{f}^{(k)},
\end{align}
where $\mathbf{g}$ stays unaffected by the iteration because it is assumed to be a known suitably regular $\mathbb{R}^2$-valued function. Next, we use Duhamel's principle. The equation,
\begin{align*}
\dfrac{\partial\mathbf{f}^{(k)}_{h}}{\partial t} = \mathsf{A}_0\mathbf{f}^{(k)}_h+\mathbf{g}
\end{align*}
is uniquely solvable by the Picard-iteration technique. By our pre-considerations, $\mathsf{A}_0$ generates a $\mathcal{C}^0$ semigroup of operators, $\mathsf{S}_0(t,\tau)$, or, more explicitly,
\begin{align*}
\mathsf{S}_0(t,\tau)= \exp((t-\tau)\mathsf{A}_0).
\end{align*}
By boundedness, we can apply the spectral theorem together with the spectral mapping theorem and give sense to the operator exponential. It reduces for each vector of eigenfunctions for either one of the operators, i.e.,
\begin{align*}
\mathsf{S}_0(t,\tau)&=\exp\left(\int_{\sigma(\mathsf{A}_{11})}d\mu(\lambda\,)\left\langle\left(\begin{array}{c}\vert\Psi_{\lambda}\rangle\\ \vert\Psi_{\lambda}\rangle\end{array}\right),\left(\begin{array}{cc}\lambda_{11}=\lambda & \lambda_{12}(\lambda)\\ \lambda_{21}(\lambda) & \lambda_{22}(\lambda)\end{array}\right)\left(\begin{array}{c}\langle\Psi_{\lambda}\vert\\ \langle\Psi_{\lambda}\vert\end{array}\right)\right\rangle_{\mathbb{R}^2}\right)\\
&=\int_{\sigma(\mathsf{A}_{11})}d\mu(\lambda\,)\left\langle\left(\begin{array}{c}\vert\Psi_{\lambda}\rangle\\ \vert\Psi_{\lambda}\rangle\end{array}\right),\exp\left(\left(\begin{array}{cc}\lambda & \lambda_{12}(\lambda)\\ \lambda_{21}(\lambda) & \lambda_{22}(\lambda)\end{array}\right)\right)\left(\begin{array}{c}\langle\Psi_{\lambda}\vert\\ \langle\Psi_{\lambda}\vert\end{array}\right)\right\rangle_{\mathbb{R}^2}.
\end{align*}
Since the initial conditions have been settled to zero, Duhamel's principle allows to obtain the integral representation for $\mathbf{f}_h$,
\begin{align*}
\mathbf{f}^{(k)}_h = \int_0^t d\tau\,\mathsf{S}_0(t,\tau)\mathbf{g}(\tau).
\end{align*}
The starting point of the Dirac perturbation theory is to partially invert the problem using Picard iteration, in terms of a resolvent approach in physical notation,
\begin{align*}
\mathbf{f}&=(\partial_t-\mathsf{A}_0-\mathsf{V}_A)^{-1}\mathbf{g}\\
&=(\partial_t-\mathsf{A}_0)^{-1}\mathbf{g} + (\partial_t-\mathsf{A}_0)^{-1}\mathsf{V}_A(\partial_t -\mathsf{A}_0)^{-1}\mathbf{g}\\
&+(\partial_t-\mathsf{A}_0)^{-1}\mathsf{V}_A(\partial_t-\mathsf{A}_0)^{-1}\mathsf{V}_A(\partial_t-\mathsf{A}_0)^{-1}\mathbf{g}+...\\
&= (\partial_t-\mathsf{A}_0)^{-1}\sum_{k=0}^{\infty}\prod_{j=1}^{k}\left(\mathsf{V}_A(t_j)(\partial_t-\mathsf{A}_0)^{-1}\right)\mathbf{g}.
\end{align*}
In terms of the semi-group $\mathsf{S}_0(t,\tau)$, we have the more practical expression where $\prod_{j=1}^{0}=1$,
\begin{align*}
\mathbf{f} &= \int_{0}^{\infty}d\tau_0\,S_0(t,\tau_0)\sum_{k=0}^\infty\prod_{j=1}^{k}\left(\int_{0}^{\infty}d\tau_j \mathsf{V}_A(\tau_{j-1})S_{0}(\tau_{j-1},\tau_j)\right)\mathbf{g}(\tau_k).
\end{align*}
Here, we use that $\mathsf{S}_{0}(t,\tau)=\mathsf{0}$ of $t<\tau$ and $\mathsf{S}_0(\tau,\tau)=\mathsf{1}$. I.e., the nested integrals span in the contribution from the zeroth order of $\mathsf{V}_A$ from $[0,t]$, in the first order contribution from the inner integral to the outer integral from $[0,\tau_0]$ and $[0,t]$, in the contribution of second order in $\mathsf{V}_A$, the nested integrals span over $[0,\tau_1],\,[0,\tau_0]$ and $[0,t]$ from the inner nested integral going to the outer of the nested integrals. By introduction of the time-ordering symbol, $\mathsf{T}$, it is possible to symmetrize the integrands such that the product sign cancels. The $t$ integration is stretched to $[0,t]$ in all of the nested integrals. Further, we have to limit the outer integration to $[0,t]$ instead of $\mathbb{R}^+_0$. Combination with the Neumann summation, yields the Dirac series in time-ordered form
\begin{align}
\mathbf{f} = \mathcal{T}\left(\exp\left(\int_{0}^t d\tau\,\mathsf{A}(\tau)\right)\right)\mathbf{g}.
\end{align}
The expression looks compact, however only the first order term is quite useful. The $\mathcal{T}$-product is used for symmetrization but for practical calculations it is not really useful. In linear order in $\mathsf{V}_A$, i.e., neglecting terms of order $\delta^2:=\Vert\mathsf{V}_A\Vert^2 /\Vert\mathsf{A}_0\Vert^2$, we have the approximate solution,
\begin{align}
\mathbf{f}(t)=\int_0^t d\tau\,\mathsf{S}_0(t,\tau)\mathbf{g}(\tau)+\int_0^t d\tau\,\mathsf{S}_0(t,\tau)\int_{0}^{\tau}d\tau'\mathsf{V}_A(\tau)\mathsf{S}_0(\tau,\tau')\mathbf{g}(\tau')+\mathcal{O}(\delta^2).
\end{align}
The equation is useful but has the conceptual drawback that it hides today's operator-theory entering the solution theory of the above non-autonomous Cauchy problem. The Dyson series was a step more in the direction that modern operator theory follows.
\item\textbf{Dyson's method: }The method invented by Dyson \cite{dyson1} is somewhat more modern. In operator theory one says the differential equation
\begin{align}
\dfrac{\partial\mathbf{f}}{\partial t} = \mathsf{A}_0\mathbf{f}+\mathbf{g} + \mathsf{V}_a\mathbf{f}
\end{align}
is solvable if it is well-posed and there is a $\mathcal{C}^0$ semi-group $(S(t,\tau))_{t\geq\tau\geq 0}$ of operators which solves the homogeneous equation,
\begin{align}
\dfrac{d\mathsf{S}(t,\tau)}{dt} = \mathsf{A}(t)\mathsf{S}(t,\tau).
\end{align}
By Duhamel's principle the solution to the differential equation for $\mathbf{f}$is then given by
\begin{align}
\mathbf{f}(t,\mathbf{x}) = \mathsf{S}(t,0)\mathbf{f}(0,\mathbf{x})+\int_0^t d\tau\,\mathsf{S}(t,\tau)\mathbf{g}(\tau) = \int_0^t d\tau\,\mathsf{S}(t,\tau)\mathbf{g}(\tau).
\end{align}
The subtlety is that we don't know what $\mathsf{S}$ is. The Dyson series expansion uses the perturbation lemma \cite{magnus3, goldbart, philips, kato} that if $\mathsf{S}_0(t,\tau)$ is an evolution family and $\mathsf{V}_A$ is a perturbation, then $\mathsf{S}(t,\tau)$ is an evolution as well, given by the integral equation
\begin{align}
\mathsf{S}(t,\tau) = \mathsf{S}_0(t,\tau) + \int_\tau^t d\tau' \mathsf{S}_0(t,\tau')\mathsf{V}_A(\tau')\mathsf{S}(\tau',\tau).
\end{align}
This is Duhamel's principle applied to the operator equation for the evolution family $\mathsf{S}(t,\tau)$ treating $\mathsf{V}_A\mathsf{S}$ as the inhomogeneity by Banach fixed point theorem. The proof of existence uses Gronwall's lemma, \cite{gronwall}. Since the operators $\mathsf{A},\mathsf{A}_0$ are bounded, Gronwall gives
\begin{align}
\Vert\mathsf{S}(t,\tau)\Vert\leq\Vert\mathsf{S}_0(t,\tau)\Vert\left\Vert\exp\left(\int_\tau^t d\tau'\,\mathsf{V}_A(\tau')\right)\right\Vert \leq \Vert\mathsf{S}_0(t,\tau)\Vert\exp\left(\int_\tau^t d\tau\,\left\Vert\mathsf{V}_A(\tau)\right\Vert\right).
\end{align}
If $\Vert\mathsf{V}_A(t)\Vert\ll 1$, the series converges. Bounds can be found e.g. in \cite{magnus2, engel}. Mathematicians work on improvement of bounds. In the case we are interested in, we have $\Vert\mathsf{V}_A\Vert\ll 1$ such that locally these series converge. The summation problem that Dyson worked on stays the same as for the Dirac series, except that now the operators $\mathsf{S}$ are to be build up from $\mathsf{S}_0$ and $\mathsf{V}$. Dyson found the expression
\begin{align}
\mathsf{S}(t,\tau)&=\int_{0}^{\infty}d\tau_0\,S_0(t,\tau_0)\sum_{k=0}^\infty\prod_{j=1}^{k}\left(\int_{0}^{\infty}d\tau_j \mathsf{V}_A(\tau_{j-1})S_{0}(\tau_{j-1},\tau_j)\right)\\
&\equiv \mathcal{T}\left(\exp\left(\int_\tau^t d\tau'\,\mathsf{A}(\tau')\right)\right).
\end{align}
using the time-ordering symbol $\mathcal{T}$ for symmetrization of the Neumann series again.
\item\textbf{Magnus' method: }Magnus \cite{magnus1} proceeded in a different way and circumvented the formal time-ordering symbol. The main issue lies in the fundamental theorem of calculus, no longer giving the familiar result for the scalar quantities,
\begin{align}
\dfrac{d\exp\left(\int_\tau^t d\tau'\,a(\tau')\right)}{dt} = a'(t)\exp\left(\int_\tau^t d\tau'\,a(\tau')\right)= \exp\left(\int_\tau^t d\tau'\,a(\tau')\right)a'(t),
\end{align}
in the case of $\mathsf{A}(t)$ being an element of a Lie-algebra $\mathfrak{g}$ or more generally of an associative Banach algebra turned into a Lie-algebra by endowment with a Lie bracket,
\begin{align}
\dfrac{\partial \exp(\int_\tau^t d\tau'\,\mathsf{A}(\tau))}{\partial t} \stackrel{!?!?}{=} \mathsf{A}'(t)\exp\left(\int_0^t d\tau\,\mathsf{A}(\tau)\right) \stackrel{!?!?}{=} \exp\left(\int_0^t d\tau\,\mathsf{A}(\tau)\right) \mathsf{A}'(t).
\end{align}
In general, we will have equality if for all $t_1,t_2\in\mathbb{R}^+_0$
\begin{align}
[\mathsf{A}(t_1),\mathsf{A}(t_2)]=\mathsf{0}
\end{align}
or, slightly weaker, if for all $t\in\mathbb{R}^+_0$
\begin{align}
\left[\mathsf{A}(t),\int_0^t d\tau\,\mathsf{A}(\tau)\right]=\mathsf{0}
\end{align}
As usual, the integral over $\mathsf{A}(t)$ is to be understood as a Bochner integral for Banach space valued functions, see \cite{emmerich} or \cite{engel} for an introduction to the theory of Bochner integration. We have to properties that help us further. Firstly, the semi-group property still is valid if $\mathsf{V}_A(t)$ has suitably well-behaved coefficient functions in the spectral expansions in terms of eigenfunctions of $\mathsf{A}_0$. In particular, they should be $\alpha$-Lipschitz continuous and differentiable in $t$. The semi-group property states that for all $0\leq \tau\leq \tau'\leq t$,
\begin{align}
\mathsf{S}(t,\tau)=\mathsf{S}(t,\tau')\mathsf{S}(\tau',\tau)
\end{align}
and $\mathsf{S}(\tau,\tau)=0$ for all $\tau\in\mathbb{R}^+$. For the inverse, we can use that the resolvent of $\mathsf{S}-\lambda\mathsf{1}$ is an entire function in $\lambda,$ and define a formal inverse by a contour integral over a Jordan curve $\Gamma\subset\mathbb{C}$ that encloses the origin $z=0$, 
\begin{align*}
\mathsf{S}(t,\tau)=(\mathsf{S}(\tau,t))^{-1}=\dfrac{1}{2\pi i}\oint_{S^1_{\epsilon}(z=0)}\dfrac{dz\,\mathsf{R}(\mathsf{S}(\tau,t),z)}{z-0}
\end{align*}
if $t\leq \tau$ and $\mathsf{S}_0$ is invertible, we obtain a group-like structure. The second property we can exploit is the Baker-Campbell-Hausdorff formula which is valid not only for finite-dimensional Lie algebras but also for associative Banach algebras. The associativity of the multiplication map, i.e., composition of matrices of operators, is needed to endow the Banach algebra with a Lie bracket and thus \cite{bch1,bch2,bch3,bch4,bch4a,bch5,bch6} turn it into a Lie algebra. Denoting $[\mathsf{A}_1,\mathsf{A}_2]=\text{ad}_{\mathsf{A}_1}[\mathsf{A}_2]$ and the $k$-fold nested Lie bracket $[\mathsf{A}_1,...,[\mathsf{A}_1,\mathsf{A}_2]...] = \text{ad}_{\mathsf{A}_1}^k[\mathsf{A}_2]$, the Baker-Campbell-Hausdorff formula is given by
\begin{align}
\exp(\mathsf{A}_1)\exp(\mathsf{A}_2) = \exp(\mathsf{A}_1+B(-\text{ad}_{\mathsf{A}_1}[\mathsf{A}_2]),
\end{align}
where $B(x)$ is the generating function of the Bernoulli numbers $\lbrace B_k\rbrace_{k\in\mathbb{N}_0}$ with $B_0=1$,
\begin{align}
B(x) = \dfrac{x}{e^x-1}= \sum_{k\in\mathbb{N}_0}\dfrac{B_k}{k!}x^k.
\end{align}
The series expansion of $B$ has a convergence radius of $\rho(B)=2\pi$, i.e., for $x\in (-2\pi,2\pi)$, the function $B(x)$ is analytic and can be used for composition with matrices and by the spectral mapping theorem for functional calculus also for composition with operators such as $\mathsf{A}_1,\mathsf{A}_2\in\mathfrak{gl}(2,\mathfrak{W})$ with a von-Neumann algebra $\mathfrak{W}$. The Baker-Campbell-Hausdorff formula now reduces to
\begin{align}
\exp(\mathsf{A}_1)\exp(\mathsf{A}_2) = \exp(\mathsf{A}_1+\sum_{k\in\mathbb{N}_0}\dfrac{B_k}{k!}(-\text{ad}_{\mathsf{A}_1})^k[\mathsf{A}_2]).
\end{align}
We can investigate convergence. Since we have $\Vert\text{ad}_{\mathsf{A}_1}\Vert\leq 2\Vert\mathsf{A}_1\Vert$ by the triangle inequality, we have convergence if 
\begin{align}
\Vert\mathsf{A}_1\Vert \leq \pi ,
\end{align}
because of the generating functions for the Bernoulli-numbers $\lbrace B_k\rbrace_{k\in\mathbb{N}_0}$, $B=B(x)$, having convergence radius $\rho(B)=2\pi$. Based on the definition of the exponential functional, we have derived the equation for the differential of the exponential map, $d\exp_{\mathsf{A}_1}:\mathfrak{gl}(2,\mathfrak{W})\to\mathfrak{gl}(2,\mathfrak{W})$, where domain and range become clear from the algebra property of the formula,
\begin{align}
d\exp_{\mathsf{A}_1} = \dfrac{\exp(\text{ad}_{\mathsf{A}_1})-\mathsf{1}}{\text{ad}_{\mathsf{A}_1}} = \dfrac{1}{B(\text{ad}_{\mathsf{A}_1})}.
\end{align}
The series converges everywhere. If the eigenvalues of $\text{ad}_{\mathsf{A}_1}$ are not integer multiples of $2\pi i$, one can invert the differential of the exponential. For the case of compact operators we find by inspection that if $\mathsf{A}_1$ has the eigenvalues $\lbrace \lambda_n\rbrace_{n\in\mathbb{N}}$, then $\text{ad}_{\mathsf{A}_1}$ has the eigenvalues $\lbrace \lambda_n -\lambda_m\rbrace_{(n,m)\in\mathbb{N}^2}$. In particular if the spectrum of $\mathsf{A}_1$ is purely real, we won't have trouble performing the inversion. The inverse of the differential of the exponential map is given by,
\begin{align}
d\exp^{-1}_{\mathsf{A}_1}=(d\exp_{\mathsf{A}_1})^{-1}=B(-\text{ad}_{\mathsf{A}_1}) = \dfrac{-\text{ad}_{\mathsf{A}_1}}{\exp(-\text{ad}_{\mathsf{A}_1})-\mathsf{1}} = \sum_{k\in\mathbb{N}_0}\dfrac{B_k}{k!}\text{ad}_{\mathsf{A}_1}^k .
\end{align}
We can now formulate the Magnus expansion for inhomogeneous linear operator evolution equations. Namely,
\begin{align}
\dfrac{\partial\mathbf{f}}{\partial t} = \mathsf{A}(t)\mathbf{f}+\mathbf{g},
\end{align}
with homogeneous initial conditions is solved by
\begin{align}
\mathbf{f}(t) = \int_{0}^t d\tau\,\mathsf{S}(t,\tau)\mathbf{g}(\tau),
\end{align}
where $\mathsf{S}(t,\tau)$ is the $\mathcal{C}^0$ semi-group generated by the Magnus generator $\mathfrak{G}$ that is the solution of the operator Riccati-like equation
\begin{align}
\dfrac{\partial\mathfrak{G}}{\partial t} = d\exp^{-1}_{\mathfrak{G}}[\mathsf{A}(t)] = B(\text{ad}_{\mathfrak{G}})[\mathsf{A}(t)],
\end{align}
the calculation following the steps as in the matrix case considered in the preceding paragraphs on Duhamel's principle. The differential equation for $\mathfrak{G}$ is unfortunately non-linear. Since at $t=0$ we have $\mathsf{S}(0,0)=\mathds{1}$, we set $\mathfrak{G}(t=0)=\mathsf{0}$ and apply the Banach fixed-point theorem in order to obtain an iteration scheme for $\mathfrak{G}$. More precisely, we use the following reasoning \cite{magnus11}. Rewrite the differential equation in the form
\begin{align}
\dfrac{\partial\mathbf{f}}{\partial t} = \lambda\mathsf{A}_1\mathbf{f}+\mathbf{g},
\end{align}
where $\lambda\in\mathbb{R}$ is a (not necessarily small) dimensionful parameter. In the end, $\lambda=1$ to recover the original equation. Further, standard operator semi-group theory tells us that we need to solve the homogeneous problem in order to obtain the semi-group $\mathsf{S}(t,\tau)=\exp(\mathfrak{G}(t,\tau))$. Picard iteration yields in the interval $[t,t+dt]$ where $dt\ll 1$ to ensure local existence and uniqueness by standard theory of ordinary differential equations,
\begin{align}
\mathfrak{G}(t+dt,t) = \mathfrak{G}(t,t) + \int_{t}^{t+dt}dt\,\lambda\mathsf{A}_1(\tau)+\mathcal{O}((dt)^2) = dt\lambda\mathsf{A}_1(t)+\mathcal{O}((dt)^2)
\end{align}
as usual in terms of a Bochner integral over $\mathsf{A}_1\in\mathfrak{gl}(2,\mathfrak{W})$. The semi-group $\mathsf{S}$ generated by the Magnus generator $\mathfrak{G}$ should fulfill the group-property, i.e., $\mathsf{S}(t+dt,\tau)=\mathsf{S}(t+dt,t)\mathsf{S}(t,\tau)$. Using that the Magnus generator is defined in terms of the semi-group via the equation $\mathsf{S}(t,\tau)=\exp(\mathfrak{G}(t,\tau))$, its exponential satisfies the group property in the limit $\delta t\to 0$ as well,
\begin{align}
\exp(\mathfrak{G}(t+dt,\tau)) &= \exp(\mathfrak{G}(t+dt,t))\exp(\mathfrak{G}(t,\tau))\\
&=\exp(dt \lambda\mathsf{A}_1)\exp(\mathfrak{G}(t,\tau)).
\end{align}
We can now apply the Baker-Campbell-Hausdorff formula and truncate at lowest order in $dt$ and use the inverse of the exponential, $\exp^{-1}=\log$. By $\mathbb{C}$-linearity of $\text{ad}_{\mathfrak{G}}$, we can pull the constant $\lambda\cdot dt$ in front of the sum,
\begin{align}
\mathfrak{G}(t+dt,\tau)=\mathfrak{G}(t,\tau)+\lambda\cdot dt\sum_{k=1}^\infty\dfrac{(-1)^k B_k}{k!}\text{ad}_{\mathfrak{G}}^k[\mathsf{A}_1]+\mathcal{O}((dt)^2).
\end{align}
By definition of the partial derivative w.r.t. $t$, we find that the Magnus generator $\mathfrak{G}$ satisfies the a non-linear differential equation of first order. Since we need to ensure the semi-group property of $\mathsf{S}(t,\tau)=\exp(\mathfrak{G}(t,\tau))$, we obtain the initial condition $\mathfrak{G}(\tau,\tau)=\mathsf{0}$. By Picard-Lindel\"{o}ff's theorem for operator differential equations, the initial condition and the following differential equation determine the Magnus generator uniquely on a maximal local interval of existence
\begin{align}
\dfrac{\partial\mathfrak{G}(t,\tau)}{\partial t} = \lambda\sum_{k=1}^\infty\dfrac{(-1)^k B_k}{k!}\text{ad}_{\mathfrak{G}}^k[\mathsf{A}_1].
\end{align}
Since this is a differential equation of Riccati-like type, we need to obtain the solution recursively. We use the technique of Picard iterations for small enough $t$, namely such that the sufficient convergence condition of the Magnus expansion, $\int_{\tau}^t d\tau'\,\Vert\mathsf{A}_1(\tau')\Vert < \pi$ holds true for the pair $(\tau,t)\in(\mathbb{R}_0^+)^2$. Afterwards, we can use continuation to patch local solutions, i.e., solutions on a suitable bounded interval $I\subsetneq\mathbb{R}^+_0$ together. Using the series expansion of the generator in terms of the parameter $\lambda$, we set
\begin{align}
\mathfrak{G}(t,\tau)=\sum_{k=1}^\infty \lambda^k\mathfrak{G}_{(k)}.
\end{align}
Insertion into the differential equation for $\mathfrak{G}(t,\tau)$ allows decoupling the equation into an infinite system of trivially integrable differential equations for the Magnus coefficients $\lbrace\mathfrak{G}_{(k)}\rbrace_{k\in\mathbb{N}}$. We obtain the equations for the individual $\mathfrak{G}_{(k)},\, k\in\mathbb{N}$ that we have already found in the matrix case. For all $k\in\mathbb{N}$, we have the recursive formula for the determination of $\mathfrak{G}_{(k)}$,
\begin{align}
\mathfrak{G}_{(k)}(t,\tau) = \sum_{j=1}^{k-1}\dfrac{B_{j}}{j!}\sum_{\sum_{i=1}^{j}k_i=n-1; k_i\geq 1\forall i}\int_{0}^t d\tau\,\prod_{i=1}^{j}\left(\text{ad}_{\mathfrak{G}_{k_j}(\tau)}\right)[\mathsf{A}_1(\tau)].
\end{align}
This could be expressed in terms of $\mathsf{A}_1$ exclusively. The $k$-th Magnus coefficient $\mathfrak{G}_{(k)}$ then is a $k$-fold nested integral over $(k-1)$ commutators of $\mathsf{A}_1$. Since for our purposes, the lowest order coefficient, i.e., $\mathfrak{G}_{(1)}$ suffices, we will just give this coefficient. It reduces to the first coefficient that we also have in the Dyson series expansion and if we use the expansion to act on the source term $\mathbf{g}$ also in the Dirac series expansion of our differential equation system
\begin{align}
\mathfrak{G}_{(1)}(t,\tau) = \int_\tau^t d\tau'\,\mathsf{A}_1(\tau').
\end{align}
The higher order expressions which we derives in the matrix case transfer after replacing matrices of reals with matrices of elements of the von-Neumann algebra to the operator case. If we set $\lambda=1$ in the series expansion $\mathfrak{G}=\sum_{k\in\mathbb{N}}\lambda^k\mathfrak{G}_{(k)}$, we recover the original problem, i.e., $\dot{\mathbf{f}}(t)=\mathsf{A}_1(t)\mathbf{f}(t)+\mathbf{g}(t)$ since by construction $\mathfrak{G}$ is analytic in $\lambda\in\mathbb{R}$. We set $\lambda=$ and turn to Duhamel's principle again. Then we can solve the originally inhomogeneous problem by
\begin{align}
\mathbf{f}(t)=\int_0^t d\tau\,\mathsf{S}(t,\tau)\mathbf{g}(\tau)=\int_0^t\,d\tau\,\exp(\mathfrak{G}(t,\tau))\mathbf{g}(\tau),
\end{align}
where $\mathfrak{G}(t,\tau)$ is the Magnus generator given as a sum over the Magnus coefficients. We notice that the the $n$-th coefficient consists of $\sim\mathcal{O}(2^n/n)$ contributions. Numerically, the convergence of the Magnus series might not be so fast, \cite{magnus2}. The relationship between the Dyson series and the Magnus series has been detailed in \cite{magnus2}. Like for Feynman diagrams, a graphical method is presented as well.
\end{itemize}
The Magnus series provides a convenient approximation tool, however, it need not exist globally, \cite{magnus2}. Batkai \cite{magnus3} has used a perturbative approach to show convergence of the Magnus expansion in $H^{2,2}$-norms for a more general class of operators than in \cite{magnus2, magnus12}.\newline
\newline
\textbf{A comment on the literature: }Unlike in \cite{magnus12} the operators which we want to investigate live on $H^{1,2;2,2}_{;0}(\mathcal{M}_0)$ or $H^{1,2;4,2}_{0;0}(\partial\mathcal{M}_0)$ spaces as densely defined operators. That we need to include regularity properties for derivatives w.r.t. the base coordinate $t$ as well is due to the fact that $\mathsf{W}$ contains $t$-derivatives of the boundary vibrations $u$ as well and furthermore the differential operator $\partial_t$. The problematic operator is $\Delta_{g_0}+\mathsf{W}$. Since $\mathsf{W}$ is a partial differential operator of first order, the operator will not be normal since we have $[\Delta_{g_0},\mathsf{W}]\neq 0$ unless $u$ depends only on $t$ (see the section on the piston bundles below), the result obtained in \cite{magnus12} is not applicable, which states convergence of the Magnus series for normal bounded operators. Inspection of the proofs leading to the result shows that the operator $\Delta_{g_0}+\mathsf{W}$ being normal is convenient but not required compulsorily. The proof uses a theorem that states that if there is a certain $0<\lambda < \pi$ such that $\Re(\langle(\Delta_{g_0} +\mathsf{W})f,f\rangle)/\Vert f\Vert^2\leq \lambda$ and $\Re(\langle(\Delta_{g_0} +\mathsf{W})^\ast f,f\rangle)/\Vert f\Vert^2\leq \lambda$ for $f\in H^{1,2;2,2}(\mathcal{M})_0$ in the $L^2$-norms, we have $\sigma(\Delta + \mathsf{W}) \subset B_{\gamma}:=\lbrace \mathbf{z}\in\mathbb{C}: z = \vert z\vert\exp(i\mu), 0\leq\vert\mu\vert\leq \lambda\rbrace$, i.e., that $\Delta +\mathsf{W}$ is sectorial with angle $\lambda$, \cite{magnus12} Lemma 3.3 and the comment before. For normal operators, the requirement on the adjoint operator is satisfied trivially. Since the (densely defined) $\Delta_{g_0}$ dominates in the relative norm over $\mathsf{W}$ as we have seen before, it is natural to ask whether, since the self-adjoint $\Delta_{g_0}$ part necessarily is normal. Since we know that the spectrum $\sigma(\Delta_{g_0})$ is purely real, if $\mathsf{W}=0$, we would have $\lambda =0$. Since the $L^2$-product is sesqui-linear in the first argument, i.e., $\langle zf,f\rangle = \bar{z}\langle f,f\rangle$, the perturbation $\mathsf{W}$ only results in $0<\lambda\ll\pi$ because the perturbation $\mathsf{W}\ll \Delta_{g_0}$on the domain of $\Delta_{g_0}+\mathsf{W}$. Thus, for self-adjoint operators with small perturbations, the requirements of Lemma 3.3. from \cite{magnus12} are fulfilled. Since this is the only point where the requirement on normal operators enters the proof of the main theorem 3.4  \cite{magnus12}, we use the estimate of this theorem, i.e., we have convergence of the Magnus series on $[\tau, t]$ if
\begin{align}
\int_\tau^t\,\Vert\mathsf{A}\Vert < \pi .
\end{align}
Loosely, small perturbations to a self-adjoint operator call the need for the Magnus expansion but do not deteriorate the convergence results.\newline
\newline
\textbf{Decoupling the model -- Duhamel's principle and the Banach fixed point theorem: }The strategy to solve the coupled system of partial differential equations consists of using Duhamel's principle in its two formulations, namely in the perturbative form for the calculation of the Magnus generator with its original formulation to calculate the solution to an inhomogeneous differential equation. The smallness of the perturbations, i.e., the smallness of the boundary vibrations $u$ by definition of the perturbation bundles allows an application of the Banach-fixed point theorem. Let us recall the model equations first.
\begin{align}
\partial_t^2 p -c^2\Delta_{g_0}p &= \rho_0c^2\delta((t,\mathbf{x})\in\mathcal{M}_0) + c^2\mathsf{W}[u,p],\\
\Sigma^{-1}\partial_t(\Sigma\partial_t u)-p(\Delta_{g_0}^\partial) u &= \sigma_m^{-1}(p-p_{ex}).
\end{align}
We can introduce the vector $\mathbf{X}_1\in H^{1,2;2,2}(\mathcal{M}_0)\times H^{0,2;2,2}(\mathcal{M}_0)\times H^{2,2;4,2}(\partial\mathcal{M}_0)\times H^{1,2;4,2}(\partial\mathcal{M}_0)$ and rewrite the above coupled system of partial differential equations as a non-linear system of first order coupled operator differential equations. $\mathbf{X}_1$ is given by the choice,
\begin{align*}
\mathbf{X}_1=\left(\begin{array}{c}p\\ \partial_t p\\ u \\ \partial_t u\end{array}\right).
\end{align*}
Likewise, we can specialize to localized boundary vibrations $\lbrace u_i\rbrace_{1\leq i\leq N}$ and introduce the vector $\mathbf{Y}_{N}$ given by
\begin{align*}
\mathbf{Y}_N=\left(\begin{array}{c}p\\ \partial_t p\\ u_1 \\ \partial_t u_1 \\ u_2 \\ \partial_t u_2 \\ \vdots \\ u_N \\ \partial_t u_N\end{array}\right).
\end{align*}
By our previous considerations on regularity of (weak) solutions and convergence properties of the Magnus expansion, we need $\mathbf{Y}_N$ to be element of the following product Hilbert space,
\begin{align*}
\mathbf{Y}_N\in Y:=H^{1,2;2,2}(\mathcal{M}_0)\times H^{0,2;2,2}(\mathcal{M}_0)\times\left(\prod_{i=1}^N\left(H^{2,2;4,2}(\mathbb{R}^+_0\times\Gamma_i)\times H^{1,2;4,2}(\mathbb{R}^+_0\times\Gamma_i)\right)\right)
\end{align*}
Expanding the time-derivatives in the equation for the boundary vibrations, we  can bring our problem in the form
\begin{align}
\dfrac{\partial\mathbf{Y}_N}{\partial t} = \mathsf{A}_N\mathbf{Y}_N + \mathbf{f}_N,
\end{align}
where the quadratic $2(N+1)\times 2(N+1)$ matrix $\mathsf{A}_N$ has operator values entries, i.e., $\mathsf{A}_N\in\mathfrak{gl}(2,\mathsf{W}_{big})$ where $\mathsf{W}_{big}$ is a suitable Neumann-algebra of operators to be determined now. The object $\mathsf{f}_N\in Y$ is a source term. The form of the objects $\mathsf{A}_N$ and $\mathbf{f}_N$ allows some simplifications. The matrix $\mathsf{A}_N$ is chosen to be in block-diagonal form, more precisely,
\begin{align}
\mathsf{A}_N = \left(\begin{array}{ccccc}\mathsf{M}_0 & \mathsf{0}_{2\times 2} & \mathsf{0}_{2\times 2} & \cdots & \mathsf{0}_{2\times 2}\\
\mathsf{0}_{2\times 2} & \mathsf{M}_1 & \mathsf{0}_{2\times 2} & \cdots & \mathsf{0}_{2\times 2}\\
\vdots & \vdots & \vdots &\ddots & \vdots \\
\mathsf{0}_{2\times 2} & \mathsf{0}_{2\times 2} & \mathsf{0}_{2\times 2} & \cdots & \mathsf{M}_N\end{array}\right),
\end{align}
where the matrices $\lbrace \mathsf{M}_{j}\rbrace_{0\leq j\leq N}$ are $2\times 2$-matrices such that $\mathsf{M}_j\in\mathfrak{gl}(2,\mathsf{W}_j),\,0\leq j\leq N$ where $\mathsf{W}_j$ is for all $j,\,0\leq j\leq N$ a $\mathsf{W}^\ast$-algebra of operators, i.e., a $\mathsf{C}^\ast$-algebra of bounded operators such that the $\mathsf{W}^\ast$-algebra $\mathsf{W}_{big}$ can be taken as,
\begin{align*}
\mathsf{W}_{big}=\bigoplus_{j=0}^N\mathsf{W}_j.
\end{align*}
A posteroi the direct sum decomposition explains why we used $\mathsf{A}_N\in\mathfrak{gl}(2,\mathsf{W}_{big})$ instead of a $\mathfrak{gl}(2N+2)$-algebra. In total, $\mathsf{A}_N$ lives in the associative $\mathsf{B}^\ast$-, i.e., Banach, algebra,
\begin{align}
\mathsf{A}_N\in\bigoplus_{j=0}^{N}\mathfrak{gl}(2,\mathsf{W}_j).
\end{align}
Endowing the associative Banach algebra $\mathfrak{gl}(2,\mathsf{W}_{big})$ with a Lie-Bracket, $[\odot,\heartsuit] := \odot\circ\heartsuit - \heartsuit\circ\odot \in\mathfrak{gl}(2,\mathsf{W}_{big})$ by block-diagonality of $\heartsuit,\odot\in\mathfrak{gl}(2,\mathsf{W}_{big})$, we have a Lie bracket on the associative Banach algebra. This allows us to apply the Magnus series expansion, the Baker-Campbell-Hausdorff formula and its reverse, the Zassenhaus formula to be given below. By the choice of our model, the $\mathsf{W}^\ast$-algebras can be specialized further. Namely, we choose for $0< 1\leq j\leq N$ and a smooth function $q_j\in C^\infty(\mathbb{R}^+_0)$ the $\mathsf{W}^\ast$-algebra $\mathsf{W}_j$ to be the minimal $\mathsf{W}^\ast$-algebra that contains operators of the form $\Delta_{g_0,\Gamma_j}^\partial+q_j(t)$, i.e.,
\begin{align}
\mathsf{W}_j = \mathsf{W}^\ast(\Delta_{g_0,\Gamma_j}^\partial + q_j(t))
\end{align}
for all $1\leq j\leq N$ and $t$ denotes the time coordinate, resp. base space coordinate of the reference bundle $\mathcal{M}_0=\mathbb{R}^+_0\times\Omega_0$. We will see below that with this choice $\mathsf{W}_j$ is a commutative, associative Banach-algebra for $1\leq j\leq N$ such that the Magnus series truncates for the special case of the $\mathsf{M}_j$, $1\leq j\leq N$ we are interested in. The $\mathsf{W}^\ast$-algebra $\mathsf{M}_0$ is given by the minimal $\mathsf{W}^\ast$ algebra that contains $\Delta_{g_0}+\mathsf{W}$ for, i.e.,
\begin{align}
\mathsf{W}_0=\mathsf{W}^\ast(\Delta_{g_0}+\mathsf{W}).
\end{align}
Then, we have for the block matrices $\lbrace\mathsf{M}_j\rbrace_{0\leq j\leq N}$ in the definition of $\mathsf{A}_N$,
\begin{align}
\mathsf{M}_0=\left(\begin{array}{cc}0 & 1\\ c^2\Delta_{g_0}+c^2\mathsf{W} & 0\end{array}\right)\text{ and }\mathsf{M}_j = \left(\begin{array}{cc}0 & 1\\ p(\Delta_{g_0,\Gamma_j}^\partial) & -\partial_t\log\Sigma\end{array}\right).
\end{align}
In order to match the operator evolution equation form of our problem with the partial differential equation formulation, we need to take for the source term $\mathbf{f}_N$ the expression
\begin{align*}
\mathbf{f}_1 = \left(\begin{array}{c}0\\ \rho_0c^2\partial_t^2u \delta((t,\mathbf{x})\in\partial\mathcal{M}_0)\\ 0 \\ \sigma_m^{-1}(p-p_{ex})\end{array}\right),
\end{align*}
where $\sigma_m=\rho_m d$ for abbreviation. The above equation gives the source term in the case of the boundary vibrations $u\in H^{2,2;4,2}(\partial\mathcal{M}_0)$ with spatial arguments in $\partial\Omega_0$, i.e., the boundary of the fiber $\Omega_0$ of the reference bundle $\mathcal{M}_0$. In the case of localized boundary vibrations $\lbrace u_i\rbrace_{1\leq i\leq N}$, we have the source term $\mathbf{f}_n$ with $2(N+1)$ rows,
\begin{align*}
\mathbf{f}_N = \left(\begin{array}{c}0 \\ \rho_0c^2\sum_{j=1}^N\partial_t^2 u_j \delta((t,\mathbf{x})\in\mathbb{R}^+_0\times\Gamma_j)\\ 0 \\ \sigma_m^{-1}(p-p_{ex})\vert_{\Gamma_1}\\ 0 \\ \sigma_m^{-1}(p-p_{ex})\vert_{\Gamma_2}\\ \vdots \\ 0 \\ \sigma_m^{-1}(p-p_{ex})\vert_{\Gamma_N}\end{array}\right),
\end{align*}
identifying $\Gamma_i \simeq\Gamma_i\times\lbrace s=0\rbrace = \text{pr}_{2}\vert_{\Gamma_i}(\partial\mathcal{M}_0)\subset\Omega_0 = \text{pr}_2(\mathcal{M}_0)$ for notational brevity. That we require $H_0^{2,2;4,2}$-regularity for the localized boundary vibrations is now clear: The operator $p(\Delta_{g_0}^\partial)$ is a fourth order partial differential operator, i.e., we would like to have that also fourth derivatives of $u_i$ are bounded. That also the second derivatives of the localized boundary vibrations should be bounded is not necessary when we treat the boundary vibrations equation separately, i.e., for a fixed model choice of $p-p_{ex}=\Psi$, but treating the localized boundary vibrations $\lbrace u_i\rbrace_{1\leq i\leq N}$ in conjunction with the acoustic pressure $p$, the second-time derivatives $\partial_t^2 u_i$ enter in the source term $\mathbf{f}_N\cdot\hat{e}_2\supset \partial_t^2 u_i\delta((t,\mathbf{x})\in\mathbb{R}^+_0\times\Gamma_i)$. I.e., we ensure that our solutions do not blow up. The form of the matrix $\mathsf{A}_N$ allows us, in contrast to \cite{beale1, beale2, beale3, beale4, beale5, beale6, beale8, beale9} to obtain $(N+1)$ $2$-dimensional operator evolution equations instead of a $2(N+1)\times 2(N+1)$-dimensional generator $\mathsf{A}_N$. We have for the projection on the components relevant for the acoustic pressure $p$,
\begin{align}
\dfrac{\partial}{\partial_t}\left(\begin{array}{c}p \\ \partial_t p\end{array}\right)&= \left(\begin{array}{cc}0 & 1\\ c^2\Delta_{g_0}+c^2\mathsf{W}& 0\end{array}\right)\left(\begin{array}{c}p \\ \partial_t p\end{array}\right) + \left(\begin{array}{c}0 \\ \rho_0 c^2\partial_t^2 u\end{array}\right)\delta((t,\mathbf{x})\in\partial\mathcal{M}_0).
\end{align}
Likewise, we have for the boundary vibrations $u$,
\begin{align}
\dfrac{\partial}{\partial t}\left(\begin{array}{c}u\\ \partial_t u\end{array}\right) &= \left(\begin{array}{cc}0 & 1\\ p(\Delta_{g_0}^\partial) & -\partial_t\log\Sigma\end{array}\right)\left(\begin{array}{c}u \\ \partial_t u\end{array}\right) + \left(\begin{array}{c}0 \\ \sigma_m^{-1}(p-p_{ex})\vert_{\partial\mathcal{M}_0}\end{array}\right).
\end{align}
Anologously, we have in the case of localized boundary vibrations $\lbrace u_i\rbrace_{1\leq i\leq N}$ for the acoustic pressure $p$ and $u_i$ for $1\leq i\leq N$,
\begin{align}
\dfrac{\partial}{\partial_t}\left(\begin{array}{c}p \\ \partial_t p\end{array}\right)&= \left(\begin{array}{cc}0 & 1\\ c^2\Delta_{g_0}+c^2\mathsf{W}& 0\end{array}\right)\left(\begin{array}{c}p \\ \partial_t p\end{array}\right) + \sum_{i=1}^N\left(\begin{array}{c}0 \\ \rho_0 c^2\partial_t^2 u_i\end{array}\right)\delta((t,\mathbf{x})\in\mathbb{R}^+_0\times\Gamma_i)\\
\dfrac{\partial}{\partial t}\left(\begin{array}{c}u_i\\ \partial_t u_i\end{array}\right) &= \left(\begin{array}{cc}0 & 1\\ p(\Delta_{g_0,\Gamma_i}^\partial) & -\partial_t\log\Sigma\end{array}\right)\left(\begin{array}{c}u_i \\ \partial_t u_i\end{array}\right) + \left(\begin{array}{c}0 \\ \sigma_m^{-1}(p-p_{ex})\vert_{\Gamma_i}\end{array}\right).
\end{align}
The approach we choose is based on block-diagonality of the matrix $\mathsf{A}_N$ and the observation that perturbations to the homogeneous equations are small compared to the Laplace-Beltrami-operators involved. In the case of neglection of $\mathsf{W}$, there is another method available: It is also possible to use another method, e.g., in \cite{beale1, DengLi} and investigate an extended matrix of operators which accounts for all the source terms directly. In this case, one doesn't have an operator differential equation system with a source term like ours but rather an inhomogeneous operator differential equation system which is sourced exclusively by $p_{ex}$. The advantage of the other method above is that one does not need to use perturbative arguments, i.e., it is more elegant than ours. The disadvantage is that it is impractical for explicit calculations compared to our method. We define the \emph{coupling strength} $\mathfrak{g}$,
\begin{align}
\mathfrak{g}=\dfrac{\rho_0}{\rho_m}.
\end{align}
$\rho_0$ denotes the mass density per $n$-dimensional volume of air and $\rho_m$ denotes the mass density of the boundary vibrations $u$. In the ICE model, we have typical values $\mathfrak{g}=\rho_0/\rho_m=\mathcal{O}(10^{-3})$ whereas $\epsilon\simeq U/L = \mathcal{O}(10^{-6})-\mathcal{O}(10^{-7})$, so we conclude
\begin{align}
\mathfrak{g}^2\approx \epsilon .
\end{align}
The values are characteristic in view of applications in the sense that the boundary vibrations $u$ typically describe the displacement of physical objects such as membranes, plates, pistons, etc. from their equilibrium position. These objects are massive and consist of a solid material which has mass density $\rho_m\gg\rho_0$. \cite{kriegsmann1, kriegsmann2} modeled an infinitely-thin baffled piston with surface mass density $\sigma_m$ vibrating in half-space: In our notation, an artificial length scale $d$ for the thickness of the piston has been introduced there and the parameter $\mathfrak{g}=\rho_0/(\sigma_m d)=\rho_0/\rho_m$ has been introduced to decouple the equations of motion for the half-space acoustic pressure and the piston. We use a similar approach that differs in the observation that one can determine the thickness of the objects involved already a priori by measurements, as has been done in \cite{anupam1, anupam2, christine} and used in \cite{david1}. The observation that $\mathfrak{g}\ll 1$ is the so-called \emph{light-fluid-assumption} in vibrational acoustics, \cite{howe1, howe2, howe3, howe4}. From the formal viewpoint, $\mathfrak{g}$ takes the role of a Lipschitz constant in the subsequent argument. Let us assume the existence of $\mathcal{C}^0$-semi-groups $\hat{\mathsf{S}}_0(t,\tau)$ $\hat{\mathsf{T}}(t,\tau)$ associated to the unperturbed differential operators, i.e., such that
\begin{align}
\partial_t\hat{\mathsf{S}} = \mathsf{A}_0\hat{\mathsf{S}}_0\text{ and }\partial_t\hat{\mathsf{T}} = \mathsf{B}\hat{\mathsf{T}},
\end{align}
where we define the operator-valued matrices $\mathsf{A}$ and $\mathsf{B}$ by
\begin{align}
\mathsf{A}_0=\left(\begin{array}{cc}0 & 1\\ c^2\Delta_{g_0}\end{array}\right)\text{ and }\mathsf{B}=\left(\begin{array}{cc}0 & 1\\ c^2 p(\Delta_{g_0}^\partial) & -\partial_t\log\Sigma\end{array}\right).
\end{align}
Duhamel's principle in the form of a perturbation lemma tells us that there is an evolution family $\hat{\mathsf{S}}(t,\tau)$, which is a $\mathcal{C}^0$-semi group by the regularity of $t$-dependence of $\mathsf{W}$ such that for $\delta\mathsf{A}=\delta\mathsf{A}(t)\equiv\mathsf{M}_0-\mathsf{A}_0$ it solves the following inhomogeneous Volterra-like operator integral equation
\begin{align}
\hat{\mathsf{S}}(t,\tau) = \hat{\mathsf{S}}_0(t,\tau)+\int_{\tau}^t d\tau'\,\hat{\mathsf{S}}_0(t,\tau')\delta\mathsf{A}(\tau')\hat{\mathsf{S}}(\tau',\tau).
\end{align}
The integral is to be understood as a Bochner-integral. On the other hand, the $\mathcal{C}^0$-semi-group $\hat{\mathsf{S}}(t,\tau)$ is precisely the semi-group used in the Magnus expansion for the system to be solved for $\mathbf{P}=(p,\partial_t p)^T,$ with source term $\mathbf{f}=(0,\rho_0c^2\partial_t^2 u\delta((t,\mathbf{x})\in\partial\mathcal{M}_0)$ and homogeneous initial conditions $\mathbf{P}(t=0)=(0,0)^T$, i.e., $\mathbf{P}$ shall solve
\begin{align}
\dfrac{\partial}{\partial t}\mathbf{P}=\mathsf{M}_0\mathbf{P}+\mathbf{f}.
\end{align}
Duhamel's principle for inhomogeneous dynamical system allows us to find the solution for this differential equation in terms of a convolution integral with kernel $\hat{\mathsf{S}}(t,\tau)$, i.e., the Magnus exponential, corresponding to the operator $\mathsf{M}_0\in\mathfrak{gl}(2,\mathsf{W}_0)$,
\begin{align}
\mathbf{P}(t)=\int_0^t d\tau\,\hat{\mathsf{S}}(t,\tau)\mathbf{f}(\tau).
\end{align}
Using the perturbative formulation of Duhamel's principle, we substitute instead of $\mathsf{S}(t,\tau)$ the right hand side involving the semi-group $\hat{\mathsf{S}}_0(t,\tau)$,
\begin{align}
\mathbf{P}(t)=\int_0^t d\tau\,\hat{\mathsf{S}}_0(t,\tau)\mathbf{f}(\tau)+\int_0^t d\tau\,\int_{\tau}^{t} d\tau'\,\hat{\mathsf{S}}_0(t,\tau')\delta\mathsf{A}(\tau')\hat{\mathsf{S}}(\tau',\tau)\mathbf{f}(\tau).
\end{align}
Likewise, we want $\mathbf{U}=(u,\partial_t u)^T$ with $\mathbf{U}(t=0)=(0,0)^T$ and the source term $\mathbf{g}=(0,\sigma_m^{-1}(p-p_{ex})\vert_{\partial\mathcal{M}_0})$ to solve the dynamical system formulation of the boundary vibrations equation,
\begin{align}
\partial_t\mathbf{U}=\mathsf{B}\mathbf{U}+\mathbf{g}.
\end{align}
By Duhamel's principle this can be phrased in terms of the $\mathcal{C}^0$ semi-group $\hat{\mathsf{T}}(t,\tau)$ corresponding to $\mathsf{B}\in\mathfrak{gl}(2,\mathsf{W}_1)$,
\begin{align}
\mathbf{U}(t) = \int_0^t d\tau\,\hat{\mathsf{T}}(t,\tau)\mathbf{g}(\tau).
\end{align}
Since we are only interested in $p$ and $u$, we multiply the integral representations for $\mathbf{P}$ and $\mathbf{U}$ from the left with $\langle\hat{e}_1,\heartsuit\rangle_{\mathbb{R}^2}$ and observe that $\langle\hat{e}_1,\mathbf{g}\rangle_{\mathbb{R}^2}=0=\langle\hat{e}_1,\mathbf{f}\rangle_{\mathbb{R}^2}$ we have
\begin{align}
p &= \rho_0c^2\left(\int_0^t d\tau\,\mathsf{S}_0(t,\tau)+\int_0^t d\tau\,\int_{\tau}^{t} d\tau'\,\mathsf{S}_0(t,\tau')\left[\delta\mathsf{A}(\tau')\hat{\mathsf{S}}(\tau',\tau)\right]_{22}\right)[\partial_t^2 u\delta((t,\mathbf{x})\in\partial\mathcal{M}_0)]\\
u &= \rho_0^{-1}d^{-1}\mathfrak{g}\int_0^t d\tau\,\mathsf{T}(t,\tau)(p-p_{ex})
\end{align}
where $\mathsf{S}_0(t,\tau)=\langle\hat{e}_1,\hat{\mathsf{S}}(t,\tau),\hat{e}_2\rangle_{\mathbb{R}^2}$ and $\mathsf{T}(t,\tau)=\langle\hat{e}_1,\hat{\mathsf{T}}(t,\tau),\hat{e}_2\rangle_{\mathbb{R}^2}$ denote the $(1,2)$-components of the $\mathcal{C}^0$ semi-groups $\hat{\mathsf{S}}(t,\tau),\hat{\mathsf{T}}(t,\tau)$. The operation $[]_{22}$ denotes the $(2,2)$-component of the operator matrix that is inside the brackets. Observe that $\delta\mathsf{A}$ has by definition only a $(2,1)$-component. Since the source term $\mathbf{f}$ only features a $2$-component, only the $(2,2)$-component of the object in brackets contributes non-trivially. We recall that by dissipativity of the perturbation bundle $\mathcal{M}$, $u$ shall solve a damped wave equation. The $\mathcal{C}^0$ semi-group, $\hat{\mathsf{T}}(t,\tau)$ then is a dissipative semi-group, i.e., it is a contraction over a suitable product Hilbert space. This implies that also $\mathsf{T}(t,\tau) = \langle\hat{e}_1,\hat{\mathsf{T}}(t,\tau),\hat{e}_2\rangle_{\mathbb{R}^2}$, i.e., the $(1,2)$-component of the matrix $\mathsf{T}(t,\tau)$ is a contraction. Namely, we can use convergence of the Magnus exponential by having restricted to densely defined operators, and partially decouple the two equations: Using that the Sobolev spaces in question are dense in each other words $\mathsf{T}:H^{1,2;2;2}_{g_0}(\partial\mathcal{M}_0)\to H^{2,2;2;4}_{g_0}(\partial\mathcal{M}_0)\subset H^{1,2;2,2}_{g_0}(\partial\mathcal{M}_0)$, satisfies $\Vert\mathsf{T}\Vert < 1.$ in the Sobolev norm. This allows us to apply the Banach fixed point theorem to the two coupled equations. Indeed, substituting the integral expression for $u$ in the above expression for $p$, we find that $p$ is the solution of a non-linear integral equation which involves $\mathsf{T}(t,\tau)$. By convergence of the Magnus exponential \cite{magnus3} we can apply the Banach fixed-point theorem, the mapping $p=\mathsf{N}[p]$ contracts as well and we can apply the Banach-fixed point theorem to decouple the two integral equations. Starting at $t=0$, the starting values are $u^{(0)}=0$ and $p^{(0)}=0$ and $\hat{S}^{(0)}=\mathsf{0}_{2\times 2}$ we have the iteration scheme
\begin{align}
p^{(k+1)}&=  \rho_0c^2\left(\int_0^t d\tau\,\mathsf{S}_0(t,\tau)+\int_0^t d\tau\,\int_{\tau}^{t} d\tau'\,\mathsf{S}_0(t,\tau')\left[\delta\mathsf{A}^{(k)}(\tau')\hat{\mathsf{S}^{(k)}}(\tau',\tau)\right]_{22}\right)[\partial_t^2 u^{(k)}\delta((t,\mathbf{x})\in\partial\mathcal{M}_0)]\\
u^{(k+1)}&= \rho_0^{-1}d^{-1}\mathfrak{g}\int_0^t d\tau\,\mathsf{T}(t,\tau)(p^{(k)}-p_{ex})\\
\hat{\mathsf{S}}^{(k+1)}(t,\tau) &= \hat{\mathsf{S}}_0(t,\tau)+\int_{\tau}^t d\tau'\,\hat{\mathsf{S}}_0(t,\tau')\delta\mathsf{A}^{(k)}(\tau')\hat{\mathsf{S}}^{(k)}(\tau',\tau)
\end{align}
The $\delta\mathsf{A}^{(k)}$ means that instead of the full boundary vibrations, we have to insert the $k$-th approximation obtained from iteration scheme $u^{(k)}$ instead of $u$. Since we are interested in perturbation theory up to order $\epsilon$, we only need the first iterates. We introduce the surface mass density of air, $\rho_0 d=\sigma_0$. Then the coupling strength $\mathfrak{g}$ is a measure for the strength of the effect how strong the coupling between the motion of air molecules and the boundary vibrations is. Since in application, the vibrating boundaries are made out of a heavy (compared to air) solid material, we typically have $\mathfrak{g}\ll 1$. $\mathfrak{g}=1$ if the boundary vibrations were composed of air. For the experimental values given in \cite{anupam1, anupam2}, we have $\mathfrak{g}\approx\sqrt{\epsilon}\approx 10^{-3}$,
\begin{align}
u^{(1)}&=-\sigma_0^{-1}\mathfrak{g}\int_0^t d\tau\,\mathsf{S}(t,\tau)p_{ex}(\tau)\\
p^{(1)}&=0\\
u^{(2)}&=-\sigma_0^{-1}\mathfrak{g}\int_0^t d\tau\,\mathsf{S}(t,\tau)p_{ex}(\tau)\\
p^{(2)}&=\rho_0 c^2\mathfrak{g}\left(\int_0^t d\tau\,\mathsf{S}_0(t,\tau)+\int_0^t d\tau\int_\tau^t d\tau'\,\mathsf{S}_0(t,\tau)\left[\delta\mathsf{A}^{(1)}\right]_{2,2}\right)\partial_t^2 u\delta((t,\mathbf{x})\in\partial\mathcal{M}_0).
\end{align}
We observe that the second contribution in $p^{(2)}$ scales as $\sim u^3\sim\mathfrak{g}^3$ if $p_{ex}\sim 1$. Since $p_{ex}$ is an acoustic quantity as well, it must be of the order of acoustic quantities as well, i.e., of order $p_{ex}=\mathcal{O}(\varepsilon)$. We can choose the linearization parameter $\varepsilon$ from the derivation of the acoustic wave equation from Euler's equation in curved space-time to satisfy,
\begin{align*}
\varepsilon = \mathcal{O}(\mathfrak{g})\approx 10^{-3}.
\end{align*}
We see that we have the scaling
\begin{align*}
u^{(2)}=\mathcal{O}(\mathfrak{g}^2=\epsilon)\\
p^{(2)}=\mathcal{O}(\mathfrak{g}^2=\epsilon),
\end{align*}
as we need for consistency because $\max\vert u\vert = \epsilon$ by definition of the perturbation bundle $\mathcal{M}$. We can stop in linear order perturbation theory and just notice that we have by mathematical induction for all $k\in\mathbb{N}$ additional contributions,
\begin{align}
\epsilon\mathfrak{u}_k\mathfrak{g}^k:=u^{(k+2)}-u^{(k+1)}&=\mathcal{O}(\epsilon\mathfrak{g}^k)\\
\epsilon\mathfrak{p}_k\mathfrak{g}^k:=p^{(k+2)}-p^{(k+1)}&=\mathcal{O}(\epsilon\mathfrak{g}^k),
\end{align}
i.e., $\mathfrak{u}_k=\mathcal{O}(1)=\mathfrak{p}_k$ as we would expect from a perturbation theory. In particular, we have for the fixed points of the iteration scheme $u$ and $p$ expressed as a telescope sum over $\mathfrak{u}_k$ and $\mathfrak{p}_k$,
\begin{align}
u &= \sum_{k=0}^{\infty}\left(u^{(k+1)}-u^{(k)}\right)=\epsilon\sum_{k=1}^\infty\mathfrak{u}_k\mathfrak{g}^k\\
p &= \sum_{k=0}^\infty \left(p^{(k+1)}-p^{(k)}\right)=\epsilon\sum_{k=1}^\infty\mathfrak{p}_k\mathfrak{g}^k,
\end{align}
noting that $u^{(1)}=u^{(2)}$ and $u^{(0)}=0=p^{(0)}=p^{(1)}$. Using that $\mathfrak{g}^2\approx\epsilon$, we can set in linear order perturbation theory in $\epsilon$,
\begin{align}
p^{(2)}= p\text{ and }u^{(2)}=u.
\end{align}
For the acoustic pressure $p$ engineers are typically doing so, see e.g. the textbooks for many examples \cite{howe1, howe2, howe3, howe4}. In the next paragraph, we will be concerned with the investigation of the relation to the Magnus series and, most importantly, what we can say about the entries $\mathsf{S}_0(t,\tau)$ and $\mathsf{T}(t,\tau)$ of the $\mathcal{C}^0$ semi-groups $\hat{\mathsf{S}}_0(t,\tau)$ and $\hat{\mathsf{T}}(t,\tau)$ in terms of explicit, practical equations.\newline
\newline
\textbf{Boundary vibrations: }Recall that the differential equation describing the boundary vibrations $\lbrace u_i\rbrace_{1\leq i\leq N}$ is given by
\begin{align}
\Sigma^{-1}\partial_t(\Sigma\partial_t u_i)-p(\Delta_{g_0}^\partial)u_i = \left.\dfrac{p\vert_{\partial\mathcal{M}_+}-p\vert_{\partial\mathcal{M}_-}}{\rho_m d}\right\vert_{\Gamma_i} \equiv \Psi(p_+,p_-;\Gamma_i)
\end{align}
where we take $\Psi:=\Psi(p_+(t,\mathbf{y}),p_-(t,\mathbf{y});\Gamma_i)$ as a known source term to the differential equation for the present. The differential equation for $u_i$ is, by construction, valid on $\Gamma_i\subset\partial\Omega_i$ and the Laplace-Beltrami-operator $\Delta_{g_0}$ is assigned the Dirichlet boundary conditions $u_i\vert_{\Gamma_i}=0$ for all $i,1\leq i\leq N$ which follow from the construction of $\lbrace\Gamma_i\rbrace_{1\leq i\leq N}$. Furthermore, we specialize to the initial conditions $u_i(t=0,\mathbf{y})=0$ for all $\mathbf{y}\in\Gamma_i$ and $i\in\lbrace 1,...,n\rbrace$. The goal of this paragraph is to convert the partial differential equation for $u_i$ in an integral representation for $u_i$, involving an integration kernel acting on the source term $\Psi$. The first step consists of investigating the $t$-dependent first term to the differential equation, i.e.,
\begin{align}
\dfrac{1}{\Sigma}\dfrac{\partial}{\partial t}\left(\Sigma \dfrac{\partial u}{\partial t}\right) = \partial_t^2 u + \Sigma^{-1}\partial_t\Sigma\partial_t u.
\end{align}
Since this expression involves a damping term $\sim\partial_t u$, we have to find a transformation of the differential equation, such that we can bring the damped wave equation for $u_i$ in a differential equation for an auxiliary function $w_i$ satisfying the same regularity properties as $u_i$ for all $i\in\lbrace 1,...,n\rbrace$ such that the differential equation for $w_i$ is a \emph{generalized Klein-Gordon equation}, i.e.,
\begin{align}
\partial_t^2 w_i -(p(\Delta_{g_0}^\partial)-q(t))w_i = \Psi_{f},
\end{align}
The function $q(t)$ is a yet to be determined mass squared which will be $\alpha^2$ if $\Sigma(t)=\exp(2\alpha t)$ for constant damping with damping function $D(t)=\exp(-\alpha t)$. For general damping, $q(t)$ depends on time. $\Psi_f$ is the result of the composition of transformation that related $w_i$ and $u_i$ (in this order) and the original source term $\Psi$. Let us drop the index $i$ in the following calculation for notational simplification. In order to relate $w_i$ and $u_i$, we choose a conformal transformation, i.e., we set $u=\exp(f)w$ with another non-negative function $f$. The function $f=f(t)$ will satisfy a differential equation of order $2-1=1$ such that the linear damping term $\partial_t\log\sqrt{\Sigma}\partial_t$ is canceled by suitable derivatives of $f$ w.r.t. the base space coordinate or, physically, time $t$, Insertion of the Ansatz in the differential equation for $u$ and only investigate the part of the differential equations that involves the time-derivatives of $u$. Thus should the be equal to the $\partial^2_tw+q(t)w$ contribution to the massive generalized Klein-Gordon equation given before, i.e.,
\begin{align}
&\partial_t^2 w \exp(f)+2f'\partial_t w\exp(f)+\partial_t\log\Sigma\partial_t w\exp(f) + ((f')\partial_t\log\Sigma+f''+(f')^2)w\exp(f)\\
&\stackrel{!}{=}\partial_t^2 w \exp(f)+q(t)w\exp(f).
\end{align}
Comparison of the coefficients of $\partial_t^k w$ for $k=0,1,2$ yields the identifications for the still unknown functions $f=f(t)$ and $q=q(t)$. For the latter function, we find an equation which still involves the conformal factor $f$,
\begin{align}
q=(f')\partial_t\log\Sigma+(f')^2+f'' = (f')\partial_t\log\Sigma+(f')^2+f''
\end{align}
The determining ordinary differential equation for $f$ is obtained by comparing coefficients of $\partial^k_t w$ for $k=1$ and is given by
\begin{align}
2f' = -\partial_t\log\Sigma\Leftrightarrow f' =- \partial_t\log\sqrt{\Sigma}\leftrightarrow f(t)=-\log\sqrt{\Sigma(t)}.
\end{align}
As initial condition to first order elementarily integrable ordinary differential equation for $f$, we must take $f(0)=0$ to mirror the fact that $\Sigma(0)=1/D^2(0)=1/1^2=1$ by definition of a damping function. We observe that the right-hand-side is of the differential equation is well-defined since $0<D\leq 1$ and $\Sigma = 1/D^2\geq 1$ for all $t\geq 0$. Thus, the argument of the logarithm does not diverge to $\pm\infty$. Solving the ordinary differential equation for $f$ results in the equation right to the equivalence arrow, i.e., $f=\log\sqrt{\Sigma}$. We are now in the position to find an explicit expression of the function $q$ in terms of the time-lapse function $\Sigma$,
\begin{align}
q(t)=\partial_t^2\log\sqrt{\Sigma}-(\partial_t\log\sqrt{\Sigma})^2.
\end{align}
The whole derivation works exclusively because we have $[\Delta_{g_0}^\partial \mathsf{C}]=0,$ where $\mathsf{C}[w]=\exp(f(t))[w]$ is a linear multiplication operator acting on $w$. The commutativity of the Laplacian $\Delta_{g_0}^\partial$ and $\mathsf{C}$ is ensured by the time-dependence of $f$ and the time-independence of $\Delta_{g_0}^\partial$ plus the dependence of $\Delta_{g_0}^\partial$ on the fiber coordinates $\lbrace y^i\rbrace_{1\leq i\leq n}$ restricted from $\partial\Omega_0$ to $\Gamma_i$ and the independence of $\mathsf{C}$ on $\lbrace y^i\rbrace_{1\leq i\leq n}$. In other words, $\mathsf{C}$ is a constant w.r.t. differentiation w.r.t. the coordinates $\lbrace y^i\rbrace_{1\leq i\leq n}$. It remains to relate the source term $\Psi$ in the differential equation for $u$ to the source term $\Psi_f$ in the differential equation for $w$. This is attained by acting with $\mathsf{C}^{-1}=\exp(-f)$ on the differential equation for $u$ from the left. Since the Ansatz for $u$ with the previous identifications recasts the right hand side of the differential equation for $u$ in $\exp(f)$ times the generalized Klein-Gordon operator acting on $w$, the $\mathsf{C}^{-1}$ cancels the redundant contribution from the conformal factor $f$ and leaves us with,
\begin{align}
\partial_t^2w -(p(\Delta_{g_0})w-q(t))=\Psi\exp(-f)\equiv\Psi_{f}.
\end{align}
Naturally, the approach has its limitations, namely we have to require a well-behaved damping. The well-behavedness is to be understood s.t. $p(\Delta_{g_0})-q(t)\geq 0$, i.e., $p(-\gamma_1)\geq q(t)$ for all $t\geq 0$ where $\gamma_1$ denotes the smallest eigenvalue of the Dirichlet Laplacian $-\Delta_{g_0}^\partial$ in the sense of the Lichnernowicz theorem. $\gamma_1$ is strictly positive. Let us define the operators $\mathsf{D}^2(t)=p(\Delta_{g_0}^\partial)-q(t)$. We now recast the second order partial differential equation in a first order system of ordinary operator equations by defining $W = \partial_t w$ and rewriting the above equation in the form of a first order system of ordinary differential equations which are in fact ordinary operator equations. The substitution yields
\begin{align}
\dfrac{\partial\mathbf{w}}{\partial t}\equiv \dfrac{\partial}{\partial t}\left(\begin{array}{c}w\\ W\end{array}\right) &= \left(\begin{array}{cc}0 & 1\\ \mathsf{D}^2(t) & 0\end{array}\right)\left(\begin{array}{c}w \\ W\end{array}\right) + \left(\begin{array}{c}0 \\ \Phi_f\end{array}\right)\equiv\mathsf{A}(t)\mathbf{w} + \vec{\Psi}_f .
\end{align}
We have finished the necessary preliminary work to apply Duhamel's principle to the differential equation for $\mathbf{w}$. We let $t_1,t_2\geq 0$ and $g\in H^{1,2}(\Gamma_i)$. Let us calculate the commutator $[\mathsf{D}^2(t_1),\mathsf{D}^2(t_2)]$, where we will suppress the composition sign $\cdot$ for operators.
\begin{align*}
[\mathsf{D}^2(t_1),\mathsf{D}^2(t_2)]g &= \mathsf{D}^2(t_1)\mathsf{D}^2(t_2)-\mathsf{D}^2(t_2)\mathsf{D}^2(t_1)g\\
&=\left(p(\Delta_{g_0}^\partial)p(\Delta_{g_0}^\partial) - p(\Delta_{g_0}^\partial)q(t_2)-q(t_1)p(\Delta_{g_0}^\partial)+q(t_1)q(t_2)\right)g\\
&-\left(p(\Delta_{g_0}^\partial)p(\Delta_{g_0}^\partial) - q(t_2)p(\Delta_{g_0}^\partial)-p(\Delta_{g_0}^\partial)q(t_1)+q(t_2)q(t_1)\right)g\\
&=\left(p(\Delta_{g_0}^\partial)p(\Delta_{g_0}^\partial)-p(\Delta_{g_0}^\partial)p(\Delta_{g_0}^\partial)\right)g -(q(t_1)p(\Delta_{g_0}^\partial)-p(\Delta_{g_0}^\partial)q(t_1))g\\
&+(q(t_2)p(\Delta_{g_0}^\partial)-p(\Delta_{g_0}^\partial)q(t_2))g + (q(t_1)q(t_2)-q(t_1)q(t_2))g\\
&=-(q(t_1)p(\Delta_{g_0}^\partial)-p(\Delta_{g_0}^\partial)q(t_1))g +(q(t_2)p(\Delta_{g_0}^\partial)-p(\Delta_{g_0}^\partial)q(t_2))g\\
&= \mathsf{0}[g],
\end{align*}
where $\mathsf{O}$ is the zero operator, sending each function it acts on, identically to zero. In the fifth step, we have used our previous observation that $p(\Delta_{g_0}^\partial)q(t)=0$. By arbitrariness of $g\in H^{1,2}(\Gamma_i)$, we have found the result
\begin{align}
[\mathsf{D}^2(t_1),\mathsf{D}^2(t_2)] = \mathsf{0}.
\end{align}
Let us concentrate on the $\mathsf{D}^2(t)$'s a family of operators. The function $q(\tau)$ can is bounded for all $t\geq 0$ because $\Sigma(t=0)=1$ excludes the nasty cases where $\Sigma(t)$ is hyper-exponential, i.e., a function like $\Sigma(t)=\exp(\exp(\alpha(t)))$ where $\alpha(t):\mathbb{R}^+_0\to\mathbb{R}$ is w.l.o.g. $\mathcal{C}^\infty$. However, the Laplacian $\Delta_{g_0}^\partial$ is unbounded in the norm $\Vert .\Vert_{H^{1,2}_{g_0}}$ on $H^{1,2}(\Gamma_i)$. However, if we require our functions to be a bit more regular, namely to be in $H^{2,2}(\Gamma_i)$ as well, we assure that $\Delta_{g_0}^\partial: (H^{2,2}(\Gamma_i),\Vert .\Vert_{L^2_{g_0}}\to (H^{2,2}(\Gamma_i,\Vert .\Vert_{L^{2}_{g_0}})$ is bounded. The restriction of the domain $\text{dom}(\Delta_{g_0}^\partial)$ are necessary and sufficient to ensure self-adjointness of $\Delta_{g_0}^\partial$ w.r.t. the $L^2_{g_0}$ norm restricted to $\Gamma_i$. Indeed, we have for $g\in H^{2,2}(\Gamma_i)$,
\begin{align}
\Vert\Delta_{g_0}^\partial g \Vert_{L^2_{g_0}}\leq \Vert g\Vert_{H^{2,2}_{g_0}} < \infty,
\end{align}
by restriction of $\Delta_{g_0}^\partial$ to $H^{2,2}(\Gamma_i)$. Then we have in the graph norm applied to $\Delta_{g_0}^\partial$,
\begin{align*}
\Vert.\Vert_{gr}\equiv\Vert\Delta_{g_0}^\partial g\Vert_{gr} = \sqrt{\Vert g\Vert_{L^2_{g_0}}+\Vert\Delta_{g_0}\Vert_{L^{2}_{g_0}}} \leq \Vert g \Vert_{H^{2,2}_{g_0}} < \infty .
\end{align*}
With suitable restrictions, we can turn the on all of $L^2(\Gamma_i)$ unbounded Dirichlet Laplace-Beltrami operator $\Delta_{g_0}^\partial$ into a bounded operator. Notice that the imbedding $H^{2,2}(\Gamma_i)\hookrightarrow H^{1,2}(\Gamma_i)$ is a compact linear operator by the Rellich imbedding theorem and in $H^{2,2}(\Gamma_i)$ is by construction dense in both $H^{1,2}(\Gamma_i)$ and $L^2(\Gamma_i)$ such that under the above restrictions $\Delta_{g_0}^\partial$ is still densely defined with $\text{Dom}(\Delta_{g_0}^\partial)$. Choosing our Laplacian with the suitable restrictions, we have using linearity of $\Delta_{g_0}^\partial$ and boundedness that $\Delta_{g_0}^\partial$ is a continuous linear operator and moreover completely continuous, i.e., for a weakly convergent sequence $(g_n)_{n\in\mathbb{N}}$ in $(H^{2,2}(\Gamma_i),\Vert .\Vert_{L^2_{g_0}})$, the sequence $(\Delta_{g_0}^\partial g_n)_{n\in\mathbb{N}}$ converges in $(H^{2,2}(\Gamma_i),\Vert .\Vert_{L^2_{g_0}})$ in the norm topology. The compactness follows from the complete continuity of the linear operator $\Delta_{g_0}^\partial$ the reflexivity of $H^{2,2}$, equipped with the $L^2_{g_0}$-norm, as Banach spaces and. We seek to relate the compactness of $\Delta_{g_0}^\partial$ under our restrictions to the compactness of $\mathsf{D}^2(t)$. By boundedness of $q$ for all $t$ and as a function of $t$, we have that the operators $\mathsf{D}^2(t)= p(\Delta_{g_0}^\partial)-q(t): (H^{2,2}(\Gamma_i),\Vert .\Vert_{L^2_{g_0}})\to (H^{2,2}(\Gamma_i),\Vert .\Vert_{L^2_{g_0}})$ are linear, bounded, continuous and compact operators for all $t\geq 0$. This allows us to define suitable algebras of operators containing the family $(\mathsf{D}^2(t))_{t\geq 0}$. Indeed, the family of operators $(\mathsf{D}^2(t))_{t\geq 0}$ generates a Banach-algebra, and, by self-adjointness of each $\mathsf{D}^2(t)$, a $C^\ast$ algebra $\mathsf{C}^\ast(\lbrace\mathsf{D}^2(t)\rbrace_{t\geq 0})$, which is minimal in the sense that if $\mathtt{C}^\ast$ denotes all $C^\ast$-algebras $C^\ast$ containing the family $(\mathsf{D}^2(t))_{t\geq 0}$, we have
\begin{align}
\mathfrak{W}\equiv\mathsf{C}^\ast((\mathsf{D}^2(t))_{t\geq 0}) = \bigcap_{\mathsf{C}^\ast\in\mathtt{C}^\ast}\mathsf{C}^\ast .
\end{align}
By boundedness of the $\Delta_{g_0}$'s subject to the suitable restrictions, the $\mathsf{C}^\ast$-algebra $\mathfrak{W}$ is even a $\mathsf{W}^\ast$-algebra, i.e., a von-Neumann algebra. Since it is a von-Neumann algebra over the real (and thus also complex) Hilbert space $H^{2,2}(\Gamma_i)$, endowed with with the $L^2_{g_0}$ norm restricted to $\Gamma_i$, $\Vert .\Vert_{L^2_{g_0}}$, we can apply von-Neumann's theorem on the decomposition of von-Neumann algebras. Observing that $[\mathsf{D}^2(t_1),\mathsf{D}^2(t_2)]=\mathsf{0}$ for all $t_1,t_2\geq 0$ and by minimality of $\mathfrak{W}$, the $\mathsf{W}^\ast$-algebra $\mathfrak{W}$ is even commutative,i.e., for all $\mathsf{A},\mathsf{B}\in\mathfrak{W}$, we have $[\mathsf{A},\mathsf{B}]=\mathsf{0}$. As a commutative $\mathsf{W}^\ast$-algebra, $\mathfrak{W}$ has a trivial decomposition in terms of commutant sub-algebras. In the decomposition, we have
\begin{align}
\mathfrak{W} = \bigoplus_{k=0}^\infty \dfrac{ad^k_{\mathfrak{W}}[\mathfrak{W}]}{ad^k_{\mathfrak{W}}[\mathfrak{W}]} = \dfrac{\mathfrak{W}}{\text{ad}_{\mathfrak{W}}[\mathfrak{W}]},
\end{align}
by commutativity of $\mathfrak{W}$. We have define iteratively $\text{ad}_{\mathfrak{W}}^{k+1}[\mathfrak{W}]=\text{ad}_{\mathfrak{W}}[\text{ad}_{\mathfrak{W}}^{k}[\mathfrak{W}]]$ with $\text{ad}_{\mathfrak{W}}^{0}[\mathfrak{W}]=\mathfrak{W}$ and $\text{ad}_{\mathfrak{W}}[\mathfrak{W}] = \lbrace \mathsf{A}\in\mathfrak{W}:[\mathsf{A},\mathfrak{B}]=\mathsf{0} \text{ for all }\mathsf{B}\in\mathsf{W}\rbrace$ denotes the centralisator of $\mathfrak{W}$ w.r.t. $\mathfrak{W}$ and denote by $\text{ad}_{\mathfrak{W}}^{k+1}[\mathfrak{W}]$ the $(k+1)$-th centralisator of $\mathfrak{W}$ w.r.t. $\mathfrak{W}$ or, equivalently, the centralisator of $\text{ad}_{\mathfrak{W}}^{k}[\mathfrak{W}]$ w.r.t. $\mathfrak{W}$. Thus in the Magnus expansion, the generator $\mathfrak{G}(\tau)$ only consists the first contribution $\mathfrak{G}_1(\tau)$ resulting from the first factor of the decomposition of the $\mathsf{W}^\ast$-algebras $\mathfrak{W}$, i.e., in practical calculus-like notation again, we have the exact expression,
\begin{align}
\mathfrak{G}(\tau)=\mathfrak{G}_{(1)}(\tau) = \int_{0}^\tau d\tau\,\mathsf{A}(\tau) = \int_{0}^\tau d\tau'\,\left(\begin{array}{cc}0 & 1\\ \mathsf{D}^2(\tau') & 0\end{array}\right) = \left(\begin{array}{cc}0 & \tau\\ \int_0^\tau d\tau'\,\mathsf{D}^2(\tau') & 0\end{array}\right)
\end{align}
It remains to evaluate $\exp(\mathfrak{G}(\tau))$ and replace in the solution formula given above $\tau\to t-\tau$ because we use Duhamel's principle to solve an inhomogeneous partial differential equation. We have using commutativity of the $\mathsf{W}^\ast$ algebra, the following two equations for $k\in\mathbb{N}_0$
\begin{align}
\mathfrak{G}^{2k}(\tau) = \left(\begin{array}{cc}\tau^k\left(\int_0^\tau d\tau' \mathsf{D}^2(\tau')\right)^k & 0\\ 0 & \tau^k\left(\int_0^\tau d\tau' \mathsf{D}^2(\tau')\right)^k\end{array}\right)
\end{align}
and
\begin{align}
\mathfrak{G}^{2k+1}(\tau) = \left(\begin{array}{cc}0 &\tau^k\left(\int_0^\tau d\tau' \mathsf{D}^2(\tau')\right)^{k+1}  \\ \tau^{k+1}\left(\int_0^\tau d\tau' \mathsf{D}^2(\tau')\right)^k & 0 \end{array}\right)
\end{align}
We can now apply the Taylor series expansion of the exponential functions of compact self-adjoint operators,
\begin{align}
\exp(\mathfrak{G}(\tau)) &= \sum_{k=0}^{\infty}\dfrac{1}{k!}\mathfrak{G}^k(\tau)\\
&= \sum_{k=0}^{\infty}\dfrac{1}{(2k)!}\mathfrak{G}^{2k}(\tau) + \sum_{k=0}^\infty\dfrac{1}{(2k+1)!}\mathfrak{G}^{2k+1}(\tau)\\
&\equiv\exp_{even} + \exp_{odd}.
\end{align}
Noting that $\mathsf{D}^2$ is a negative operator for all $t\geq 0$, the sums over even and odd summation indices $k$ can be expressed in terms of sine and cosine functions with operator values arguments. For the sum over even indices, we have
\begin{align}
\exp_{even} &= \left(\begin{array}{cc}\cos\left(\sqrt{\tau\int_{0}^\tau d\tau'(-\mathsf{D}^2)(\tau')}\right) & 0\\ 0 & \cos\left(\sqrt{\tau\int_{0}^\tau d\tau'(-\mathsf{D}^2)(\tau')}\right)\end{array}\right)\\
&= \left(\begin{array}{cc}\cos\left(\tau\sqrt{\int_{0}^1 d\zeta(-\mathsf{D}^2)(\tau\cdot\zeta)}\right) & 0\\ 0 & \cos\left(\tau\sqrt{\int_{0}^1 d\zeta(-\mathsf{D}^2)(\zeta)}\right)\end{array}\right),
\end{align}
where we have rescaled in the last step in the integration variable $\tau' = \tau\zeta$ and used $\tau\geq 0$ to pull $\tau$ out of the square root. In case $\mathsf{D}^2$ is time-independent this gives the well-known formulas for the constantly damped wave-equation.
For the sum over odd indices, we use again $\tau' = \zeta\tau$ as a transformation and find
\begin{align}
\exp_{odd} = \left(\begin{array}{cc} 0& \sqrt{\int_{0}^{1}d\zeta\,(-\mathsf{D}^2)(\tau\zeta)}\sin\left(\tau\sqrt{\int_{0}^{1}d\zeta\,(-\mathsf{D}^2)(\tau\zeta)}\right)\\ \sqrt{\int_{0}^{1}d\zeta\,(-\mathsf{D}^2)(\tau\zeta)}^{-1}\sin\left(\tau\sqrt{\int_{0}^{1}d\zeta\,(-\mathsf{D}^2)(\tau\zeta)}\right) & 0 \end{array}\right).
\end{align}
In total, we have evaluated the \emph{Magnus exponential} $\mathcal{G}(\tau)\equiv\exp(\mathfrak{G}(\tau))$ with the result
\begin{align}
\mathcal{G}(\tau) = \left(\begin{array}{cc} \cos\left(\tau\sqrt{\int_{0}^1 d\zeta(-\mathsf{D}^2)(\tau\cdot\zeta)}\right)& \sqrt{\int_{0}^{1}d\zeta\,(-\mathsf{D}^2)(\tau\zeta)}\sin\left(\tau\sqrt{\int_{0}^{1}d\zeta\,(-\mathsf{D}^2)(\tau\zeta)}\right)\\ \sqrt{\int_{0}^{1}d\zeta\,(-\mathsf{D}^2)(\tau\zeta)}^{-1}\sin\left(\tau\sqrt{\int_{0}^{1}d\zeta\,(-\mathsf{D}^2)(\tau\zeta)}\right) & \cos\left(\tau\sqrt{\int_{0}^1 d\zeta(-\mathsf{D}^2)(\tau\cdot\zeta)}\right) \end{array}\right).
\end{align}
For the partial differential equation describing the dynamics of $w$, we only need the $(1,2)$ entry of the matrix in Duhamel's principle because $\mathbf{w}(0)=0$ as $w(0)=0=\partial_t w(0)$. The integral formula for $w$ is then given by
\begin{align}
w = \int_{0}^t d\tau\,\dfrac{\sin\left((t-\tau)\sqrt{\int_{0}^{1}d\zeta\,(-\mathsf{D}^2)((t-\tau)\zeta)}\right)}{\sqrt{\int_{0}^{1}d\zeta\,(-\mathsf{D}^2)((t-\tau)\zeta)}}\Psi_{f}(\tau).
\end{align}
We now have to do two things. Namely, we have to express the solution involving functions of operators in terms of the eigenfunctions of $\mathsf{D}^2$. Since $q(t)$ is just a conventional function of a real variable and $p(\Delta_{g_0}^\partial)$ is just a polynomial in $\Delta_{g_0}$ and because $\mathsf{D}^2(t)$ has for all $t\geq 0$ the same set of complete and orthornomal eigenfunctions $\lbrace\Phi_k\rbrace_{k\in\mathbb{N}_0}$ corresponding to the eigenvalues $\lbrace -\gamma_k\rbrace_{k\in\mathbb{N}}$, namely those of $\Delta_{g_0}^\partial$, we obtain a relativity simple expression for $u(t,\mathbf{y}) = \exp(f(t))w(t,\mathbf{y})$ with source term $\Psi_f(\tau) = \exp(-f(\tau))$. Insertion of the corresponding resolution of the identity and the spectral theorem applied to $-\mathsf{D}^2(\tau)$ and functions thereof
\begin{align}
\begin{split}
&u(t,\mathbf{y})\\
&=\sum_{k=1}^\infty\int_{0}^t \dfrac{d\tau\,\sqrt{\Sigma(\tau)}}{\sqrt{\Sigma(t)}}\dfrac{\sin\left((t-\tau)\sqrt{\int_{0}^{1}d\zeta\,(-(p(-\gamma_k)-q((t-\tau)\zeta)))}\right)}{\sqrt{\int_{0}^{1}d\zeta\,(-(p(-\gamma_k)-q((t-\tau)\zeta)))}}\langle\Phi_k(\mathbf{y})\vert\Psi(t,\mathbf{y})\rangle_{L^2_{g_0}(\Gamma_i)}\Phi_k(\mathbf{y}),
\end{split}
\end{align}
where the function $q$ is given in terms of the time-lapse function $\Sigma$ as $q=-(\partial_t\log\sqrt{\Sigma})^2+\partial_t^2\log\sqrt{\Sigma}$. We want to check the this equation reproduces for $p(-\Delta_{g_0}^\partial)=c^2_m\Delta_{g_0}^\partial$ the analog expression to our previous paper \cite{david1}. This goal requires us to set furthermore $D=\exp(-\alpha t)$, i.e., $\Sigma=\exp(2\alpha t)$. We find
\begin{align*}
u_{exp}=\sum_{k=1}^\infty\int_{0}^td\tau e^{-\alpha(t-\tau)}\dfrac{\sin\left((t-\tau)\sqrt{-c^2_m\gamma_k -\alpha^2}\right)}{\sqrt{-c^2_m\gamma_k -\alpha^2}}\langle\Phi_{k}\vert\Psi\rangle_{L^2_{g_0}(\Gamma_i)}\Phi_k,
\end{align*}
which is precisely the result we hoped to obtain.\newline
\newline
\textbf{Acoustics: }Recall that the acoustic wave equation of interest on the unperturbed bundle $\mathcal{M}_0$ is given by
\begin{align}
\partial_t^2p - c^2\Delta_{g_0}p = c^2\mathsf{W}[u,p] + c^2\rho_0\partial_t^2 u\delta(\mathbf{x}\in\partial\mathcal{M}_0).
\end{align}
$\delta(\mathbf{x}\in\partial\mathcal{M}_0)$ is again the surface Dirac delta distribution introduced above. For our purposes, it suffices to ensure that $\Delta_{g_0}$ and $\mathsf{W}$ are defined densely on $H^{1,2}_0(\mathcal{M}_0)$, i.e., we can also take $H^{2,2}_0(\mathcal{M}_0)$ for $\text{Dom}(\Delta_{g_0})$ and $\text{Dom}(\mathsf{W})$.  We further use the decomposition of the boundary vibrations $u$ in non-trivial components $\lbrace u_i\rbrace_{1\leq i\leq N}$,
\begin{align}
u = \sum_{k=1}^N u_i\delta(\mathbf{x}\in\mathbb{R}^+_0\times\lbrace s=1\rbrace\times\Gamma_i).
\end{align}
Next, we define the operator $\mathsf{O}^2(t)=\Delta_{g_0}+\mathsf{W}$. We reformulate the acoustic wave equation by means of setting $P=\partial_t p$ as
\begin{align}
\dfrac{\partial\mathbf{p}}{\partial t} \equiv \dfrac{\partial}{\partial t}\left(\begin{array}{c}p\\P\end{array}\right) = \left(\begin{array}{cc}0 & 1\\ c^2\mathsf{O}^2(t)& 0\end{array}\right)\left(\begin{array}{c}p\\P\end{array}\right)+\left(\begin{array}{c}0\\ \rho_0c^2\partial_t^2 u\end{array}\right)\equiv \mathsf{B}(t)\mathbf{p}+\mathbf{h}.
\end{align}
By non-commutativity of $\Delta_{g_0}$ and $\mathsf{W}$, $[\Delta_{g_0},\mathsf{W}]$ in general and $[\mathsf{O}(t_1),\mathsf{O}(t_2)]\neq\mathsf{0}$ for general $t_1,t_2\in\mathbb{R}^+_0$, we also have $[\mathsf{B}(t_1),\mathsf{B}(t_2)]\neq\mathsf{0}$ for $t_1,t_2\in\mathbb{R}^+_0$ typically. This means, that the $\mathsf{W}^\ast$-algebra generated by $\mathsf{O}(t), t\geq 0$, say $\mathfrak{W}$, cannot be decomposed using the Lie lower central series such that the $\mathfrak{W}$ is nil-potent, i.e., such that the series truncates after $M\in\mathbb{N}$ factors. By the previous considerations, the endowment of $\mathfrak{gl}(2,\mathfrak{W})$ with the usual Lie-bracket $[.,.]$ actually turns $\mathfrak{gl}(2,\mathfrak{W})$ into a Lie algebra. The acoustic wave equation has the form of an operator evolution equation, and $\mathsf{O}$ is linear and continuous, as well as bounded in the $H^{2,2}$-norm: For $f\in H^{2,2}_0(\Omega_0)$, we have for $\Delta_{g_0}:(H^{2,2}_0(\mathcal{M}_0),\Vert .\Vert_{H^{2,2}_{g_0}})\to(H^{2,2}_0(\mathcal{M}_0),\Vert .\Vert_{L^{2}_{g_0}})$
\begin{align}
\Vert\Delta_{g_0}f\Vert_{L^{2}_{g_0}} \leq \Vert f\vert_{H^{2,2}_{g_0}}\Rightarrow \Vert\Delta_{g_0}\Vert_{H^{2,2}_{g_0}\to L^2_{g_0}} \leq 1 .
\end{align}
Thus, we can apply the Magnus series and our previous considerations for the abstract problem Magnus problem ensure convergence of the Magnus series. We make the usual Ansatz and notice $\mathbf{p}(t=0)=\mathbf{0}$ by our specification of initial conditions $p(t=0)=0=\partial_t p(t=0)$,
\begin{align}
\mathbf{p} = \int_{0}^{t}d\tau\,\exp(\mathfrak{G}(t-\tau))\mathbf{h}(\tau),
\end{align}
by Duhamel's principle. By Magnus' theorem, the Magnus generator $\mathfrak{G}\in\mathfrak{gl}(2,\mathfrak{W})$ satisfies the following evolution equation,
\begin{align}
\dfrac{\partial\mathfrak{G}}{\partial t} = d\exp_{\mathfrak{G}}^{-1}(\mathsf{B}) = \sum_{k=0}^{\infty}\dfrac{B_k}{k!}\text{ad}_{\mathfrak{G}}^k[\mathsf{B}],
\end{align}
where $B_k$ denotes the $k$-th Bernoulli number and $\text{ad}_{\mathfrak{G}}$ denotes the left adjoint action of the Lie-algebra $\mathfrak{gl}(2,\mathfrak{W})$ on itself. Using the above convergence properties in the $(H^{2,2}_0(\mathcal{M}_0))$-Frobenius norm, a Banach fixed-point argument \cite{magnus1} resulted in the following Picard iteration scheme ($0\leq\tau_n\leq\tau_{n-1}\leq\cdots\leq\tau_1\leq\tau_0=\tau$)
\begin{align}
\mathfrak{G} &= \sum_{k=0}^{\infty}\mathfrak{G}_{(k)}\\
\mathfrak{G}_{(1)}(\tau) &= \int_{0}^{\tau}d\tau'\,\mathsf{B}(\tau')\\
\mathfrak{G}_{(n)}(\tau=\tau_0) &= \sum_{j=1}^{n-1}\dfrac{B_j}{j!}\sum_{\sum_{i=1}^j k_i=n-1;k_i\geq 1}\prod_{i=1}^{j}\left(\int_{0}^{\tau_{i-1}}d\tau_i\,\text{ad}_{\mathfrak{G}_{k_i}(\tau_i)}\right)\int_{0}^{\tau_{n-1}}d\tau_n\,[\mathsf{B}(\tau_{n})].
\end{align}
We define $\mathfrak{G}_0 :=\mathfrak{G}_{(1)}$ and $\delta\mathfrak{G}_0:=\sum_{k>1}\mathfrak{G}_{(k)}$. Then, the Magnus exponential can be factorized using the Zassenhaus \cite{bch1, bch2, bch3} product formula, i.e., the dual of the Baker-Campbell Hausdorff formula \cite{bch4, bch5, bch6},
\begin{align}
\exp(\mathfrak{G}_0+\delta\mathfrak{G}_0) = \exp(\mathfrak{G}_0)\exp(\delta\mathfrak{G}_0)\prod\exp(\mathcal{O}(\epsilon^2))
\end{align}
where the product indicates contributions stemming from commutants of $\mathfrak{G}_0$ and $\delta\mathfrak{G}_0$. We can neglect them since the perturbation operator scales as $\mathsf{W}=\mathcal{O}(\epsilon^2)$ and is the source of non-commutativity, i.e., also $[\mathsf{B}(t_1),\mathsf{B}(t_2)]=\mathcal{O}(\epsilon^2)$ and inductively also for higher commutants, such that $\mathfrak{G}_{k>2}$ scales as $\sim\epsilon^4$ because it involves squares of the perturbation operator $\mathsf{W}$. In particular, we have for $\mathbf{f}\in H^{2,2}(\mathcal{M}_0)\times H^{1,2;2,2}(\mathcal{M}_0)$
\begin{align}
&\,\,[\mathsf{B}(\tau_1),\mathsf{B}(\tau_2)]\mathbf{f}\\
&= \left[\left(\begin{array}{cc}0 & 1\\ c^2\mathsf{O}^2(\tau_1)&0\end{array}\right)\left(\begin{array}{cc}0 & 1\\ c^2\mathsf{O}^2(\tau_2)&0\end{array}\right)-\left(\begin{array}{cc}0 & 1\\ c^2\mathsf{O}^2(\tau_2)&0\end{array}\right)\left(\begin{array}{cc}0 & 1\\ c^2\mathsf{O}^2(\tau_1)&0\end{array}\right)\right]\mathbf{f}\\
&= c^2\left(\mathsf{O}(\tau_1)-\mathsf{O}(\tau_2)\right)\left(\begin{array}{cc}-1 & 0\\0 & 1\end{array}\right)\mathbf{f}\\
&= c^2\left(\mathsf{W}(\tau_1)-\mathsf{W}(\tau_2)\right)\left(\begin{array}{cc}-1 & 0\\0 & 1\end{array}\right)\mathbf{f}.
\end{align}
For the higher commutator contributions, i.e., $\mathfrak{G}_{(k>2)}$, we observe that only contributions quadratic and higher in $\mathsf{W}$ can survive
Thus, also $\delta\mathfrak{G}_0=\mathcal{O}(\epsilon^2)$ because it just a sum of the $\mathfrak{G}_{(k)}'s$, converging by convergence of the Magnus series. Thus, each commutant which contains at least one contribution from the commutator $[\delta\mathfrak{G}_0,\mathfrak{G}_0]$, scales as the perturbation strength squared, i.e., as $\epsilon^2$. By the Zassenhaus formula, we would evaluate products of $\exp(\mathcal{O}(\epsilon)^2)$, each of which we can approximate as $1$. The truncated generator $\mathfrak{G}_0$ on the other hand, scales as $\epsilon^0$. Thus,
\begin{align}
\exp(\mathfrak{G}_0+\delta\mathfrak{G}_0)=\exp(\mathfrak{G}_0)+\mathcal{O}(\epsilon^2),
\end{align}
which we can safely neglect in linear perturbation theory in $\epsilon$. Thus, we have
\begin{align}
\mathbf{p}(t) = \int_{0}^t dt\,\exp(\mathfrak{G}_0(t-\tau))\mathbf{h}(\tau)+\mathcal{O}(\epsilon^2),
\end{align}
and the truncated generator $\mathfrak{G}_0$ is of a form such that we can evaluate the exponential symbolically,
\begin{align}
\mathfrak{G}_0(\tau)=\int_{0}^{\tau}d\tau'\,\mathsf{B}(\tau')= \left(\begin{array}{cc}0 & \tau\\ \int_{0}^\tau d\tau'\,c^2\mathsf{O}^2(t) & 0\end{array}\right).
\end{align}
We use the Taylor series representation for the exponential by functional calculus,
\begin{align}
\exp(\mathfrak{G}(\tau)) = \sum_{k=0}^\infty\dfrac{1}{k!}\mathfrak{G}_0^k.
\end{align}
As we did for the boundary vibrations, we obtain the following two matrix identities valid for $k\in\mathbb{N}$. For even powers of $\mathfrak{G}_0$ in the exponential series, we have after substituting the integration variable $\tau'$ by $\tau'=\tau\zeta$ with $\zeta\in[0,1]$,
\begin{align}
\mathfrak{G}_0^{2k} = \left(\begin{array}{cc}\tau^k\left(c^2\int_0^\tau d\tau'\,\mathsf{O}^2(\tau')\right)^k & 0\\ 0 & \tau^k\left(c^2 \int_0^\tau d\tau'\,\mathsf{O}^2(\tau')\right)^k\end{array}\right) = \left(\begin{array}{cc}\tau^{2k}\left(c^2\int_0^1 d\zeta\,\mathsf{O}^2(\tau\zeta)\right)^k & 0\\ 0 & \tau^{2k}\left(c^2\int_0^1 d\zeta\,\mathsf{O}^2(\tau\zeta)\right)^k\end{array}\right).
\end{align}
For odd powers of $\mathfrak{G}_0$ in the exponential series, we find using the previous identity and evaluating one further matrix product,
\begin{align}
\mathfrak{G}_0^{2k-1} = \left(\begin{array}{cc}0 & \sqrt{c^2\int_0^1 d\zeta\,\mathsf{O}^2(\tau\zeta)}\tau^{2k-1}\sqrt{c^2\int_0^1 d\zeta\,\mathsf{O}^2(\tau\zeta)}^{2k-1}\\ \sqrt{c^2\int_0^1 d\zeta\,\mathsf{O}^2(\tau\zeta)}^{-1}\tau^{2k-1}\sqrt{c^2\int_0^1 d\zeta\,\mathsf{O}^2(\tau\zeta)}^{2k-1} & 0\end{array}\right).
\end{align}
Decomposing the exponential in even and odd powers, we are finally left with,
\begin{align}
&\exp(\mathfrak{G}_0(\tau)) = \sum_{k=0}^\infty\dfrac{\mathfrak{G}_0^k}{k!}= \sum_{k=0}^{\infty}\dfrac{\mathfrak{G}_0^{2k}}{(2k)!} + \sum_{k=1}^{\infty}\dfrac{\mathfrak{G}_0^{2k-1}}{(2k-1)!}\\
&= \left(\begin{array}{cc}\cos\left(\tau\sqrt{-c^2\int_0^1 d\zeta\,\mathsf{O}^2(\tau\zeta)}\right) & \sqrt{-c^2\int_0^1 d\zeta\,\mathsf{O}^2(\tau\zeta)}\sin\left(\tau\sqrt{-c^2\int_0^1 d\zeta\,\mathsf{O}^2(\tau\zeta)}\right)\\ \sqrt{-c^2\int_0^1 d\zeta\,\mathsf{O}^2(\tau\zeta)}^{-1}\sin\left(\tau\sqrt{-c^2\int_0^1 d\zeta\,\mathsf{O}^2(\tau\zeta)}\right) & \cos\left(\tau\sqrt{-c^2\int_0^1 d\zeta\,\mathsf{O}^2(\tau\zeta)}\right)\end{array}\right).
\end{align}
Since only the second column of $\mathbf{h}$ is non-zero, and we are only interested in the acoustic pressure $p$, the relevant matrix entry is given by the $(1,2)$-entry of the exponential. Insertion of $\hat{e}_2\mathbf{h}=\rho_0c^2\partial_t^2 u\delta(\mathbf{x}\in\mathcal{M}_0)$ and projecting on the first components, gives us an integral representation of the solution $p$ for the acoustic wave equation from the beginning of the paragraph valid up to corrections of order $\epsilon^4$ from truncation of the Zassenhaus formula. We have
\begin{align}
p = \rho_0c^2\int_0^t d\tau\,\dfrac{\sin\left((t-\tau)\sqrt{-c^2\int_0^1 d\zeta\,\mathsf{O}^2((t-\tau)\zeta)}\right)}{\sqrt{-c^2\int_0^1 d\zeta\,\mathsf{O}^2((t-\tau)\zeta)}}\partial_t^2 u\delta(\mathbf{x}\in\partial\mathcal{M}_0).
\end{align}
Let us now investigate the integral in the integral equation more closely. By definition of $\mathsf{O}^2(t)$ and using the previous result $\Vert W\Vert/\Vert\Delta_{g_0}\Vert =\mathcal{O}(\epsilon^2)$, we have
\begin{align}
\sqrt{-c^2\int_0^1 d\zeta\,\mathsf{O}^2((t-\tau)\zeta)} &= \sqrt{-c^2\int_0^1 d\zeta\,\Delta_{g_0}\left(1+\dfrac{\mathsf{W}}{\Delta_{g_0}}\right)}\\
&= \sqrt{-c^2\Delta_{g_0}}\sqrt{1+\int_0^{1}d\zeta\,\dfrac{\mathsf{W}}{\Delta_{g_0}}}\\
&= \sqrt{-c^2\Delta_{g_0}}\left(1+\dfrac{1}{2}\dfrac{-c^2\int_0^1 d\zeta\,\mathsf{W}(\zeta\tau)}{-c^2\Delta_{g_0}}\right)+\mathcal{O}(\epsilon^2),
\end{align}
where we only work up to order $\epsilon^2$ in the familiar Taylor expansion of $\sqrt{1+x}\simeq 1+ 1/2 x + \mathcal{O}(x^2)$. Functional calculus assures that in the $\Vert .\Vert_{2,2}$-norm, we can actually perform this expansion. Next, we need two further identities. The addition theorem of the sine function and the geometric series. For the operator sine function in the integral equation for $p$, we use the following addition theorem and the power series representation of the sine and cosine function excluding orders from $\epsilon^4$ on. Let $x,y\in\mathbb{R}^+,\,y/x = \mathcal{O}(\epsilon^2)$ and use the addition theorem,
\begin{align}
\sin(x+y) =\sin x \cos y + \sin y \cos x,
\end{align}
to obtain
\begin{align}
\sin(x+y) = \sin x + y\cos x + \mathcal{O}(\epsilon^4).
\end{align}
Next, we use the geometric series for $x,y$ as in the above identity,
\begin{align}
\dfrac{1}{\sqrt{x+y}} = \dfrac{1}{\sqrt{x}}\dfrac{1}{\sqrt{1+(y/x)}} = \dfrac{1}{\sqrt{x}}\dfrac{1}{1-(-y/(2x))}=\dfrac{1}{\sqrt{x}}\sum_{k=0}^{\infty}\left(-\dfrac{y}{2x}\right)^k = \dfrac{1}{\sqrt{x}}-\dfrac{1}{2}\dfrac{y}{\sqrt{x}^3}+\mathcal{O}(\epsilon^4),
\end{align}
since $y/x = \mathcal{O}(\epsilon^2)$ by assumption. By functional calculus, we can make the substitutions $(x_1,y_1)$ for the the operator sine function and the substitutions $(x_2,y_2)$ for the denominator involving the square root of the integral under consideration,
\begin{align}
x_1 = (t-\tau)\sqrt{-c^2\Delta_{g_0}}&\text{ and }y_1 = \dfrac{t-\tau}{2}\cdot\dfrac{-c^2\int_0^1 d\zeta\,\mathsf{W}}{\sqrt{-c^2\Delta_{g_0}}},\\
x_2 = \sqrt{-c^2\Delta_{g_0}}&\text{ and }y_2 = \dfrac{y_2}{\sqrt{x_2}^3}=\dfrac{-c^2\int_{0}^1 d\zeta\,\mathsf{W}}{\sqrt{-c^2\Delta_{g_0}}^3}.
\end{align}
This results in the following integral representation for $p$ which is again valid up to order $\epsilon^4$,
\begin{align}
p &= \rho_0c^2\int_{0}^t d\tau\,\dfrac{\sin((t-\tau)\sqrt{-c^2\Delta_{g_0}})}{\sqrt{-c^2\Delta_{g_0}}}\partial_t^2 u\delta(\mathbf{x}\in\partial\mathcal{M}_0)\\
&+\dfrac{\rho_0c^4}{2}\int_{0}^t d\tau\,\dfrac{\sin((t-\tau)\sqrt{-c^2\Delta_{g_0}})}{\sqrt{-c^2\Delta_{g_0}}^3}\left(\int_0^1 d\zeta\mathsf{W}((t-\tau)\zeta)\right)\partial_t^2 u\delta(\mathbf{x}\in\partial\mathcal{M}_0)\\
&- \dfrac{\rho_0c^4}{2}\int_{0}^t d\tau\,\dfrac{\cos((t-\tau)\sqrt{-c^2\Delta_{g_0}})}{\sqrt{-c^2\Delta_{g_0}}^2}\left((t-\tau)\int_0^1 d\zeta\mathsf{W}((t-\tau)\zeta)\right)\partial_t^2 u\delta(\mathbf{x}\in\partial\mathcal{M}_0)\\
&+\mathcal{O}(\epsilon^4)
\end{align}
The advantage is that we now have the perturbation operator acting only on functions to its right. Notice that the integral over $\mathsf{W}$ can be given a mathematical sense when inserting resolutions of the identify operator in terms of the (time-independent) complete and orthornomal set of eigenfunctions of the Laplace-Beltrami operator $\Delta_{g_0}$ on the fiber $\Omega_0$ of the unperturbed bundle $\mathcal{M}_0=\mathbb{R}^+_0\times\Omega_0$,
\begin{align}
\mathsf{1} = \sum_{n\in\mathbb{N}_0}\vert\Psi_n^{(0)}\rangle\langle\Psi_n^{(0)}\vert ,
\end{align}
to the left and right of the operator sine function and the perturbation term. From the previous section, we know that we can approximate up to order $\epsilon^2$ the eigenfunctions of the perturbed operator $\mathsf{O}^2$, assuming they existed, by the eigenfunctions of the unperturbed Laplace-Beltrami operator $\Delta_{g_0}$, which we know to exist by Lichernowicz theorem. Last, we have to be consistent with the orders of the perturbation theory. Since we seek to do perturbation theory in linear order in $\epsilon$, the contributions involving the perturbation operator $\mathsf{W}$ can be neglected. This results in
\begin{align}
p = \rho_0c^2\int_{0}^t d\tau\,\dfrac{\sin((t-\tau)\sqrt{-c^2\Delta_{g_0}})}{\sqrt{-c^2\Delta_{g_0}}}\partial_t^2 u\delta(\mathbf{x}\in\partial\mathcal{M}_0).
\end{align}
This demonstrates that the local volume change caused by the boundary vibrations contributes in quadratic order in the acoustic wave equation and its solution. I.e., the only error that we make by passing from the perturbed bundle $\mathcal{M}_0$ to the unperturbed bundle $\mathcal{M}_0$ is of order $\epsilon^2$. This agrees with the result found by Li et al, \cite{DengLi}, in the context of a cube-like structure endowed with one locally reacting surface in $\mathbb{R}^3$ by a purely operator-theoretic and non-geometric argument.\newline
\newline
\textbf{Convergence and analytic vectors: }So far, we have spoken of convergence of a series with operator-valued summands and integrals over operators. More precisely, we mean by convergence of a series for a closed symmetric operator between two Hilbert spaces $\mathsf{O}\in\text{SymOp}(\text{Dom}(\mathsf{O})\subseteq X\to X)$ that there is an analytic vector $f\in V^\omega(\mathsf{O})$, i.e., $\Vert\sum_{k\geq 0}(k!)^{-1}\mathsf{O}^k f\Vert < \infty$. Further, if $V^\omega(\mathsf{O})\stackrel{\text{dense}}{\hookrightarrow}X$, Nelson's theorem \cite{nelson} states that $\mathsf{O}$ has a unique self-adjoint extension, i.e., $\mathsf{O}$ is essentially self-adjoint. The perturbation theory has yielded that the boundary vibrations $u$ are sourced by the external pressure $p_{ex}$ and the boundary vibrations $u$ source the acoustic pressure $p$. Analytic vectors are not directly applicable in the method because the Magnus generator $\mathfrak{G}$ is not symmetric. However, if we work on the perturbation bundle $\mathcal{M}$ instead of the reference bundle $\mathcal{M}_0$, the Neumann Laplace-Beltrami operator $\Delta_{g_0,t}:H^{2,2}_{g_0}(\text{pr}_2(\mathcal{M})(t))\to L^2_{g_0}(\text{pr}_2(\mathcal{M})(t))$ is symmetric and bounded on its domain for all fixed $t,\,t\in\mathbb{R}^+_0$. We arrived at the perturbation by defining the perturbation operator $\mathsf{W}$ as the difference of the pull-back by $\Phi_{0\to t}:\mathcal{M}_0\to\mathcal{M}$ of the Helmholtz differential operator $\square_{G_0,t}=\partial_t^2 - c^2\Delta_{g_0,t}:H^{1,2;2,2}_{G_0}(\mathcal{M})\to L^{2}_{G_0}(\mathcal{M})$ and the Helmholtz differential operator $\square_{G_0}=\partial_t^2-c^2\Delta_{g_0}: H^{1,2;2,2}_{G_0}(\mathcal{M}_0)\to L^{2}_{G_0}(\mathcal{M}_0)$. By pull-back, we obtained $\Phi_{0\to t}^\ast\square_{G_0}=\square_{G(t)}:H^{1,2;2,2}_{G(t)}(\mathcal{M}_0)\to L^2_{G(t)}(\mathcal{M})$ Last, we defined the perturbation operator $\mathsf{W}$ as an operator $\text{LinOp}(H^{1,2;2,2}_{G_0}(\mathcal{M})_0\to L^2_{G_0}(\mathcal{M}_0))$ in the metric $G_0$ on the reference bundle $\mathcal{M}_0$, i.e., in the system that our experimentator performs measurements in,
\begin{align*}
\mathsf{W}(t)=\Phi_{0\to t}^\ast(\square_{G_0,t})-\square_{G_0,0} = \square_{G(t)}-\square_{G_0}
\end{align*}
The perturbation theory is bases on the (essential) self-adjointness of the Laplace-Beltrami operators $\Delta_{g_0}^\partial$ and $\Delta_{g_0}$ because only then, we have that the mild solutions obtained by Duhamel's principle actually converge to classical solutions. However, by the regularity restrictions, we achieved that $\Delta_{g_0}^\partial$ and $\Delta_{g_0}$ are bounded but no longer self-adjoint. As a symmetric operators, they are closable and we denote the closure of $\Delta_{g_0}$ and $\Delta_{g_0}^\partial$ by $\Delta_{g_0}^\partial$ and $\Delta_{g_0}$ again. Since the matrices of operators $\mathsf{A}_0$ and $\mathsf{B}$ have only the Laplace-Beltrami operators and the identity operators $\mathsf{1}$ times $t$-dependent functions as entries, they are closable themselves. Letting them act on two-vectors of functions in their domain, i.e., the domain of the Laplace-Beltrami operators, $(H^{2,2}_{g_0}(\Omega_0))^2$ and $(H^{2,4}_{g_0}(\partial\Omega_0))^2$ and multiplying from the right w.r.t. the inner product on $(H^{2,2}_{g_0}(\Omega_0))^2$ and $(H^{2,4}_{g_0}(\partial\Omega_0))^2$ induced by the $L^2$-inner product on the Sobolev-spaces and the standard Euclidean inner product on $\mathbb{R}^2$, we see that $\mathsf{A}_0$ and $\mathsf{B}$ are symmetric. This generalizes to case of the localized boundary vibrations, i.e., the matrices of operators $\lbrace\mathsf{A}_0\rbrace\cup\lbrace\mathsf{M}_j\rbrace_{1\leq j\leq N}$ are symmetric operators such that the block-diagonal $\mathsf{A}_{N,0}$, i.e., $\mathsf{A}_N$ with $\mathsf{A}_0$  instead of $\mathsf{M}_0$ is on its domain, the $2(N+1)$-dimensional product space of Sobolev-spaces given above, also symmetric. Thus, it is closable and we denote its closure by $\mathsf{A}_{N,0}$ again. The perturbation lemma \cite{engel} aided at the construction of a Magnus semi-group by the fact that $\mathsf{W}$ is closed for fixed $u$. The perturbation lemma assures that if we are having a semi-group generated by the block-diagonal and symmetric $\mathsf{A}_{N,0}$, then also $\mathsf{A}_{N}$ generates a $\mathcal{C}^0$-semi-group given the perturbation is bounded. The explicit expressions for the semi-groups derived in the previous paragraph contained only entire functions consisting of square-roots of non-negative operators in the denominator and operator sine resp. cosine functions. Even more, the only eigenvalue that causes problems is the eigenvalue $\lambda_0=0$ for the Neumann Laplace-Beltrami operator $\Delta_{g_0}^\partial$. But L'Hopital's theorem applied to the functions $f,g,,h$ for $t> 0$ and $x\in\mathbb{R}^+_0$
\begin{align*}
x\to \dfrac{\sin(t\cdot x)}{x}:=f(t,x)\, x\to x\sin(t\cdot x)=:g(t,x)\, x\to\cos(t\cdot x)=:h(t,x)
\end{align*}
assures that the limit $\lim_{x\to 0}f(t,x)=t$ exists for all $t\geq 0$ because for $t=0$, $f\equiv 0$. Likewise $\lim_{x\to 0}g(t,x)=0$ and $\lim_{x\to 0}h(t,x)=1$ by the standard calculational rules for operator sine and cosine functions. This means that because the Laplace-Beltrami operators are bounded on their domains, also the functions of the functions of the Laplacians should be bounded linear operators on $H^{2,2}_{g_0}(\Omega_0)$ and $H^{2,4}_{g_0}(\partial\Omega_0)$. This is the case if $\mathsf{A}_{N,0}$ is essentially self-adjoint. We use Nelson's theorem \cite{nelson} insetad of the Cayley transform, \cite{beale1}. Further, the symmetry of the Laplacians carries over to functions of the Laplacians. So, for all $t\geq\tau\geq 0$ and $1\leq j\leq N$ the index of the localized boundary vibrations $\lbrace u_j\rbrace_{1\leq j\leq N}$, the blocks of the Magnus exponential of $\mathsf{A}_{N,0}$ exist, as does the exponential because the Magnus expansion truncates after the first contribution, i.e., the familiar $\hat{\mathsf{S}}_{0,j}(t,\tau)$ and $\mathsf{T}(t,\tau)$ are bounded symmetric linear operators on the relevant Sobolev-spaces. Since $H^{2,2}_{g_0}(\Omega_0)$ is densely and thus continuously embedded in $L^{2}_{g_0}(\Omega_0)$ by Rellich's theorem and $H^{2,4}_{0,g_0}(\partial\Omega_0)$ is also densely and thus continuously embedded in $L^{2}_{0,g_0}(\partial\Omega_0)$. We conclude that $\mathsf{A}_{N,0}\in\text{End}((L^{2}_{g_0}(\Omega_0)^2)\oplus\bigoplus_{j=1}^N(L^2_{0,g_0}(\Gamma_j))^2)$ is essentially self-adjoint since the set of analytic vectors for $\mathsf{A}_{N,0}$ contains a subset which is dense,
\begin{align*}
V^\omega(\mathsf{A}_{N,0})\supset (H^{2,2}_{g_0}(\Omega_0))^2\oplus\bigoplus_{i=1}^{N}(H^{2,4}_{0,g_0}(\Gamma_k))^2\stackrel{\text{dense}}{\hookrightarrow}(L^{2}_{g_0}(\Omega_0)^2)\oplus\bigoplus_{j=1}^N(L^2_{0,g_0}(\Gamma_j))^2.
\end{align*}
Thus, $\mathsf{A}_{N,0}$ generates a $\mathcal{C}^0$ semi-group. The perturbation lemma relating $\hat{\mathsf{S}}$ to $\hat{\mathsf{S}}_0$ now gives that, by boundedness of $\mathsf{W}$, also $\hat{\mathsf{S}}$ is an evolution family. The application of the Banach fixed-point theorem to the three equations, the acoustic wave equation, the boundary vibrations equation and the equation for the full $\mathsf{S}$ in terms of $u^{(k)}$ and $\mathsf{S}_0$ yields also the $\mathcal{C}^0$-property if $p_{ex}$ is $\mathcal{C}^\infty$. Since in our model $p_{ex}\propto\exp(i\omega t)$, the smoothness of $p_{ex}$ in spatial and temporal arguments is clear. We emphasize that it is crucial that the source term is suitably well-behaved, ideally smooth but $H^{2,2;2,4}_{g_0}(\partial\Omega_0)$ regularity suffices. Otherwise, the Magnus series might not converge (in the norm sense) due to the radius of convergence of Bernoulli numbers being only $<\pi$, c.f. \cite{magnus3}!

\section{Associated Piston Bundles}
\textbf{Poincaré's inequality and piston bundles: }In \cite{anupam1, anupam2}, the piston approximation has been introduced arguing that for the ICE model, high vibration frequencies of the acoustic are physically negligible because they are outside the audible frequency range of the geckos. A physical back-of-the-envelope argument suggested that indeed only plane-wave modes for $p$ inside $\Omega_t$ are dominant. In \cite{david1}, the spinning mode expansion has been introduced as generalization of the modal cut-off criterion for evanescent modes \cite{howe1, howe2, howe3, howe4}. In the generalized setup, the spinning mode series expansion can be performed as well, but it is of little use when no empirical data for an investigation of the individual contributions are available. From a more formal viewpoint, the theory of partial differential equations \cite{jost1, jost2, zeidler2} features a result due to H. Poincaré which can be interpreted as a piston approximation in a more general context - the Poincaré inequality. The versions that we will be interested in give an upper bound on the $L^2$-deviation of $u\in H^{2,2}(\partial\mathcal{M}_0)$, interpreted $u=u_t\in H^{2,2}_0(\partial\Omega_0)$ as an (almost) smooth $1$-parameter family of functions defined on $\partial\Omega_0$, from its geometrical mean $\langle u\rangle_{\partial\Omega_0}$ defined by
\begin{align}
\langle u\rangle_{\partial\Omega_0}=\langle u\rangle_{\partial\Omega_0}(t)\equiv\dfrac{1}{\text{Vol}_{n-1}(\partial\Omega_0)}\int_{\partial\Omega_0}d\text{Vol}_{n-1}(\partial\Omega_0)\,u(t,\mathbf{y}).
\end{align}
As shorthand notation, we will make use of the abbreviations $d\text{Vol}_{n-1}(\partial\Omega_0)=d(\partial\Omega_0)$ and $\text{Vol}_{n-1}(\partial\Omega_0)=\vert\partial\Omega_0\vert$ in the following derivation.  By the partial differential equation that $u$ satisfies, we need to have $u\in H^{2,2}_0(\partial\mathcal{M})\subsetneq H^{1,2}_0(\Gamma_i)$. By our decomposition of the $u$ in the finite family $\lbrace u_i\rbrace_{1\leq i\leq N}$ which is non-constant regarded as a $t$-parameterized family of functions on $\Gamma_i$, we have,
\begin{align}
\infty &>\dfrac{1}{\vert\Gamma_i\vert}\int_{\Gamma_i} d\Gamma_i\, \vert\Delta_{g_0}^\partial u_i\vert\\
& = \dfrac{1}{\vert\Gamma_i\vert}\int_{\Gamma_i}d\Gamma_i\,\left\vert\Delta_{g_0}^\partial\sum_{k\in\mathbb{N}}\mathcal{U}_k(t)\Phi_k(\mathbf{y})\right\vert\\
&\geq \dfrac{1}{\vert\Gamma_i\vert}\int_{\Gamma_i}d\Gamma_i\,\left\vert\gamma_1\sum_{k\in\mathbb{N}}\mathcal{U}_k(t)\Phi_k(\mathbf{y})\right\vert\\
& = \dfrac{\gamma_1}{\vert\Gamma_i\vert}\int_{\Gamma_i}d\Gamma_i\,\vert u_i\vert\\
&= \gamma_1\langle\vert u_i\vert\rangle_{\Gamma_i}.
\end{align}
Next, define $\delta u_i:=u_i - \langle u_i \rangle_{\Gamma_i}$. We are interested in bounding $\Vert \delta u_i\Vert_{L^2_{g_0}(\Gamma_i)}^2$. This is achieved by considering the Rayleigh quotient of the (positive!) $-\Delta_{g_0}^\partial$ which is familiar either from the theory of partial differential equations or from variational approximation method in quantum mechanics (Rayleigh-Ritz-variation procedure for multi-atomic molecules) and inserting $\delta u_i$ as a test function. Observing that $\langle u_i\rangle_{\Gamma_i}$ has no dependencies on the fiber coordinates $\lbrace y_i\rbrace_{1\leq i\leq N}$ any longer and that $u_i$ satisfies homogeneous Dirichlet boundary conditions on $\partial\Gamma_i$, $u_i\vert_{\partial\Gamma_i}=0$, we have
\begin{align}
\dfrac{\langle \delta u_i\vert -\Delta_{g_0}^\partial\vert\delta u_i\rangle_{L^2_{g_0}(\Gamma_i)}}{\langle \delta u_i\vert\delta u_i\rangle_{L^2_{g_0}(\Gamma_i)}} \stackrel{\text{i.b.p.}}{=}\dfrac{\langle \nabla^\partial_{g_0}\delta u_i\vert\nabla_{g_0}^\partial \delta u_i\rangle_{L^2_{g_0}(\Gamma_i)}}{\langle \delta u_i\vert\delta u_i\rangle_{L^2_{g_0}(\Gamma_i)}} = \dfrac{\langle \nabla^\partial_{g_0} u_i\vert\nabla_{g_0}^\partial u_i\rangle_{L^2_{g_0}(\Gamma_i)}}{\langle \delta u_i\vert\delta u_i\rangle_{L^2_{g_0}(\Gamma_i)}}.
\end{align}
On the other hand, we can also expand $\langle u_i\rangle_{\Gamma_i}$ in eigenfunctions $\lbrace\Phi_k\rbrace_{k\in\mathbb{N}}$ of the Laplacian. By self-adjointness of $\Delta_{g_0}$ on $H^{2,2}_0(\Gamma_i)$ and non-degeneracy of the $L^2_{g_0}(\Gamma_i)$ inner product, we have
\begin{align}
\dfrac{\langle \delta u_i\vert -\Delta_{g_0}^\partial\vert\delta u_i\rangle_{L^2_{g_0}(\Gamma_i)}}{\langle \delta u_i\vert\delta u_i\rangle_{L^2_{g_0}(\Gamma_i)}} = \dfrac{\langle \delta u_i\vert -\Delta_{g_0}^\partial\vert u_i\rangle_{L^2_{g_0}(\Gamma_i)}}{\langle \delta u_i\vert\delta u_i\rangle_{L^2_{g_0}(\Gamma_i)}} &\geq \gamma_1\dfrac{\langle \delta u_i\vert \delta u_i\rangle_{L^2_{g_0}(\Gamma_i)}}{\langle \delta u_i\vert\delta u_i\rangle_{L^2_{g_0}(\Gamma_i)}} = \gamma_1.
\end{align}
Putting the two calculations together, we have
\begin{align}
\gamma_1 \leq \dfrac{\langle \nabla^\partial_{g_0} u_i\vert\nabla_{g_0}^\partial u_i\rangle_{L^2_{g_0}(\Gamma_i)}}{\langle \delta u_i\vert\delta u_i\rangle_{L^2_{g_0}(\Gamma_i)}} &= \dfrac{\Vert \nabla^\partial_{g_0}u_i\Vert^2_{L^2_{g_0}(\Gamma_i)}}{\Vert \delta u_i\Vert^2_{L^2_{g_0}(\Gamma_i)}} .
\end{align}
By non-constancy of $u$ and $\delta u_i$, the numerator and denominator stay both strictly positive and we can re-arrange and take the square-root. By non-negativity of the $L^2_{g_0}$-norms, we have the inequalities
\begin{align}
0<\dfrac{\Vert \delta u_i\Vert_{L^2_{g_0}(\Gamma_i)}}{\Vert u_i\Vert_{L^2_{g_0}(\Gamma_i)}} \leq \sqrt{\dfrac{1}{\gamma_1}}\dfrac{\Vert \nabla^\partial_{g_0} u_i\Vert_{L^2_{g_0}(\Gamma_i)}}{\Vert u_i\Vert_{L^2_{g_0}(\Gamma_i)}}.
\end{align}
We substitute the definition $\delta u_i = u_i-\langle u_i\rangle_{\Gamma_i}$ back to obtain a bound on the relative error that we make if we replace the boundary vibrations $\lbrace u_i\rbrace_{1\leq i\leq N}$ by their geometrical means $\lbrace \langle u_i\rangle\rbrace_{1\leq i\leq N}$,
\begin{align}
\dfrac{\Vert u -\langle u\rangle_{\Gamma_i}\Vert_{L^2_{g_0}(\Gamma_i)}}{\Vert u\Vert_{L^2_{g_0}(\Gamma_i)}} \leq \sqrt{\dfrac{1}{\gamma_1}}\dfrac{\Vert\nabla_{g_0}^\partial u\Vert_{L^2_{g_0}(\Gamma_i)}}{\Vert u\Vert_{L^2_{g_0}(\Gamma_i)}}.
\end{align}
In the geometrical setting of the ICE model, \cite{anupam1, anupam2, membranleo}, the expression on the right hand side can be shown to be $\ll 1$ by insertion of the model parameters and using the observation that \cite{christine} was able to fit with the $5$ lowest membrane eigenmodes the real behavior of the membranes up to $95\%$ accuracy. If taking all modes and working in non-dimensional units, the quotient on the right-hand-side on the left inequality stays $<\sqrt{\gamma_1}^{-1}$ because of properness of the perturbation bundle $\mathcal{M}$.\newline
\newline
\textbf{Definition: }We call the perturbation bundle with the boundary vibration $u$ replaced with $\langle u\rangle_{\partial\Omega_0}$, defined by
\begin{align}
\langle u\rangle_{\partial\Omega_0}:=\dfrac{1}{\vert\partial\Omega_0\vert}\int_{\partial\Omega_0}d(\partial\Omega_0)\,u ,
\end{align}
the \emph{piston bundle}, $\langle\mathcal{M}\rangle_{\partial\Omega_0}$ to the perturbation bundle $\mathcal{M}_0$. The substitution $u\to\langle u\rangle_{\partial\Omega_0}$ is correspondingly called \emph{piston approximation}. We say that the piston approximation is \emph{leading order} if for the lowest eigenvalue $\gamma_1(\partial\Omega_0)$ of $-\Delta^\partial_{g_0}: H^{2,2}_0(\partial\mathcal{M}_0)\to H^{2,2}_0(\partial\mathcal{M}_0)$, we have the inequality
\begin{align}
\dfrac{\Vert u - \langle u\rangle_{\partial\Omega_0}\Vert_{L^2_{g_0}(\partial\Omega_0)}}{\Vert u\Vert_{L^2_{g_0}(\partial\Omega_0)}}\leq\sqrt{\dfrac{1}{\gamma_1(\partial\Omega_0)}}\dfrac{\Vert\nabla^\partial_{g_0}u\Vert_{L^2_{g_0}(\partial\Omega_0)}}{\Vert u \Vert_{L^2_{g_0}(\partial\Omega_0)}} < C_{piston}\epsilon.
\end{align}
where $C_{piston}/\epsilon = \mathcal{O}(\epsilon^\delta)$ with $\mathbb{R}\ni\delta > -1$.\newline
\newline
\textbf{Explanation: }In physical language, the geometrical mean is equivalent to replacing a local membrane $u_i$ by flat pistons. Effectively, any local information such as the curvature of $\text{graph}(u_i)$ relative to $\partial\Omega_0$ is neglected. The geometrical intuition is that the boundary vibrations $u$ can be replaced by pistons $\langle u\rangle_{\partial\Omega_0}$ at the cost of a relative error according to the above estimate. The calculational advantage is that concrete models become analytically more tractable. The drawback is that local information about the boundary vibrations are lost. As in the ICE model, it may in some application be desirable to risk this additional error. The notion "piston" and its geometrical relation to the Poincaré inequality is depicted in Fig. 16.
\begin{figure}
\includegraphics[width = 1\textwidth]{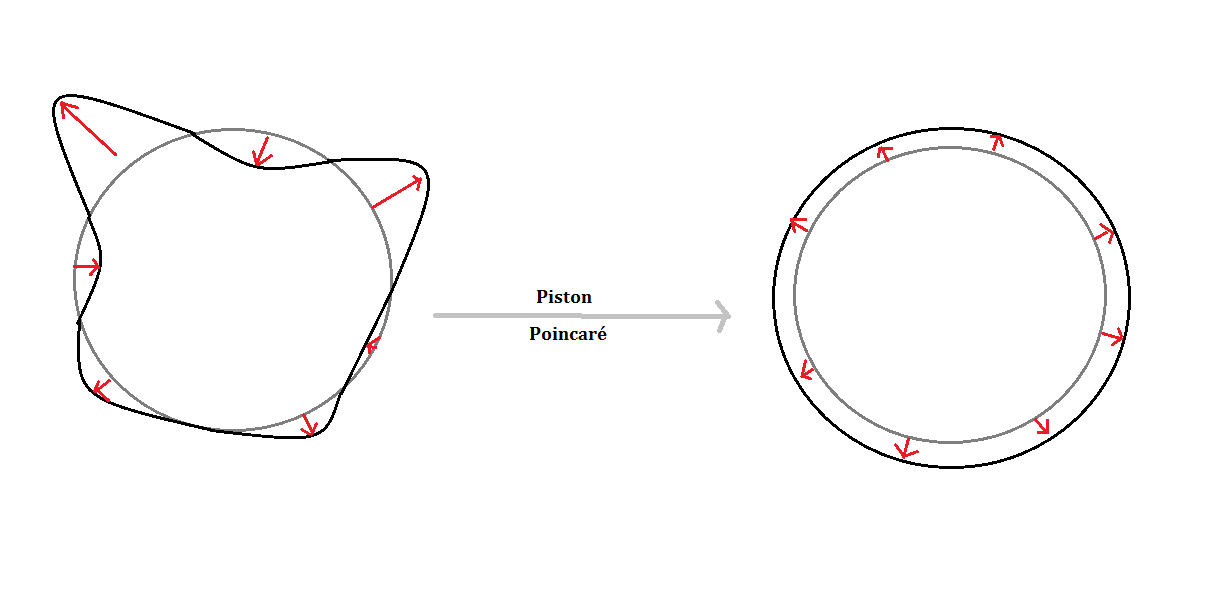}
\caption{The effect of the piston approximation and the acoustic interpretation of the Poincaré inequality.}
\end{figure}\newline
\newline
\textbf{Implication: }The crucial part in the definition is the bound on the error estimate given by $C\epsilon$ with $C$ being of order $\epsilon^{\delta}$ with $\delta > -1$. In a perturbation theory that only works up to including contributions of order $\epsilon$, the boundary vibration $u$ can be replaced on all of $\partial\Omega_0$ by $\langle u\rangle_{\partial\Omega_0}$ if one stays in order $\epsilon$ throughout the calculations. For bio-acoustic models the piston concept has been used widely, c.f. \cite{fletcher} for an introduction to the methodology, and \cite{anupam1, anupam2} for an application of the approximation.\newline
\newline
\textbf{Mean curvature vibrations and piston bundles: }The definition of a piston bundle has the advantage of being general but the disadvantage of being practically difficult to apply. One would have to know the functional form of $u$ before one applies it. In biophysical applications, one can however a priori obtain information on the mean curvature $H(u)$ of the boundary vibration $u$, or more precisely on the sub-manifold $\text{graph}(u)(t,.)\subsetneq\mathbb{R}^n$. Notice that in the theory of perturbation bundles, we have because $u\sim\partial u=\mathcal{O}(\epsilon)$,
\begin{align}
H(u) = \dfrac{1}{2}\dfrac{\text{Tr}_{g_0}[\nabla_{i,g_0}^\partial\nabla^\partial_{g_0,j}u]}{\sqrt{1+\Vert\nabla_{i,g_0}^\partial u\Vert_{\partial,g_0}^2}}= \dfrac{\text{Tr}_{g_0}[\nabla_{i,g_0}^\partial\nabla^\partial_{g_0,j}u]}{2}+ \mathcal{O}(\epsilon^3)=\dfrac{\Delta_{g_0}^\partial u}{2}+\mathcal{O}(\epsilon^3).
\end{align}
This allows us to re-write in linear order in $\epsilon$,
\begin{align}
\Delta_{g_0}^\partial = 2\mathsf{H}[u] = 2H(u),
\end{align}
and call $\mathsf{H}$ the \emph{mean curvature operator} acting on $u$. The partial differential equation for the boundary vibrations now takes the form
\begin{align}
\dfrac{1}{\Sigma(t)}\left(\Sigma(t)\dfrac{\partial u}{\partial t}\right)-p(2\mathsf{H})u = \Psi
\end{align}
\textbf{Digression into mean curvature flow - the dynamics of geometry: }In order to give a geometrical meaning to this equation, we digress into the theory of mean curvature flow. Assume that the sub-manifolds $\mathcal{S}_t\equiv\text{graph}(S(t,.))$ for a sufficiently regular, i.e., $\mathcal{C}^2$, function $S:\mathbb{R}^+_0\times \bar{U}\to\mathbb{R}^+_0$ with $\bar{U}$ a $\mathcal{C}^2$-bounded domain in $\mathbb{R}^{n-1}$ satisfy the following integral condition
\begin{align}
S_{mcf}[S] = \int_{\bar{U}}d\text{Vol}_{n-1}(U)(\partial_t S + 2D_{mcf}\mathsf{H} S)=\text{min.}!
\end{align}
By non-negativity of $S$ and positivity as well as monotony of the integral, we receive the following differential equation,
\begin{align}
\partial_t S = -2D_{mcf}\mathsf{H}S.
\end{align}
Defining $\mathbf{j}=2\mathbf{n}_{\text{graph}(S)}$, this can be recast in the form of a conservation equation,
\begin{align}
\partial_t S = -D\nabla\mathbf{j}.
\end{align}
The integral formulation states that the velocity that $S$ changes in time is proportional to the net flux of the curvature stream density $\mathbf{j}$ through the boundary $\partial\text{graph}(S)$. The proportionality constant $D_{mcf}$ is called the \emph{mean curvature diffusion constant} in the following. For weakly curved hypersurfaces $\text{graph}(S)$, i.e., hypersurfaces with $\Vert\nabla S\Vert_2\ll 1$, the mean curvature flow equation reduced to a diffusion equation for $S$,
\begin{align}
\partial_t S = -D\Delta S.
\end{align}
The equation has been extensively studied by mathematicians and physicists alike such that we restrict ourselves to discussing two special cases. The first special case is when $\partial_t S = 0$ such that the surface does not evolve in time. Then the mean curvature flow model tells us that $2D\mathsf{H}S=0$, i.e., $\mathsf{H}[S]=0$. This means that $S$ is a so-called \emph{minimal surface} and $S=S(x,y)$. On the other hand, if $2\mathsf{H}S=2\bar{\gamma} S$ for a (positive because $\partial\text{graph}S\neq\emptyset$ by assumption) $\bar{\gamma}>0$, we have the ordinary differential equation
\begin{align}
\partial_t S = -2D\bar{\gamma}S,
\end{align}
Then, the mean curvature flow states that $S=S(t,x,y)=S_0(x,y)\exp(-2D\bar{\gamma}t)$ where we have chosen the dummy initial condition $S(0,x,y)=S_0(x,y).$ In other words, a surface with constant mean curvature $\bar{\gamma}$ reduces in the limit $t\to\infty$ to a flat surface with mean curvature $0$.\newline
\newline
\textbf{Mean curvature vibrations: }The boundary vibrations satisfy a damped wave equation such that for constant mean curvature $\mathsf{H}[u]=\bar{\gamma}$ limit, we would have $u\to 0$ as $t\to\infty$ as well. This agrees with the $\infty$-periodicity from above. Furthermore, it is known experimentally \cite{membranleo, anupam1, anupam2, chris1, chris2, chris3, chris4} that curvature effects in bio-membranes are typically negligibly small. Physically this is due to large curvature effects costing the membranes a large amount of energy, contradicting the principle of least equation. As a matter of fact, \cite{anupam1, anupam2, christine, david1} only considered a special case of the model for our boundary vibrations, namely a model where $\mathsf{H}[u]=\mathcal{O}(\epsilon)$. As a consequence, we can ask what happened if we defined $\delta\mathsf{H}=\mathsf{H}u-\bar{\gamma}u$ and required a \emph{locally constant mean curvature} for all $t\geq 0$. The locally constant mean curvature property is defined that the relative error made by replacing the mean curvature operator $\mathsf{H}$ by the operator $\bar{\gamma}\mathsf{1}$ is of order $\epsilon$, i.e.,
\begin{align}
\dfrac{\Vert \delta\mathsf{H}\Vert_{L^2_{g_0}(\partial\Omega_0)}}{\Vert\bar{\gamma}u\Vert_{L^{2}_{g_0}(\partial\Omega_0)}} = \mathcal{O}(\epsilon).
\end{align}
$\bar{\gamma}>0$ is called \emph{global mean curvature constant} and a consequence of the locally constant mean curvature property is the scaling $\bar{\gamma}u=\mathcal{O}(\mathsf{H}[u])=\mathcal{O}(\epsilon)$. Effectively this means that the boundary vibrations do not vary to wildly at all times $t\geq 0$ although we all for an overall non-vanishing but constant dominating contribution $\bar{\gamma}$ to the mean curvature $H(u)=\mathsf{H}[u]\simeq \Delta_{g_0}^\partial u+\mathcal{O}(\epsilon^2)$. Experimentally, this point of view has been validated for biological membranes, c.f. \cite{membranleo, anupam1, anupam2, chris1, chris2, chris3, chris4} and references therein. Thus a posteroi, the choice of choosing a damped wave equation to describe the boundary vibrations in \cite{anupam1, anupam2, christine, david1} as an effective model is justified on experimental grounds. Let us return to the differential equation that describes the boundary vibrations. We introduce the \emph{local curvature perturbation operator} (LCPO) $\mathsf{V}_{curv}$ in the following way. We use the assumption of $u$ having a locally constant mean curvature $2H(u)=2\mathsf{H}[u]=\Delta_{g_0}^\partial u+\mathcal{O}(\epsilon^2)$ with global mean curvature constant $\bar{\gamma}$ to simplify the difference $\mathsf{V}_{curv}:=p(2\mathsf{H})-p(2\bar{\gamma}\mathsf{1})$ and express it in terms of $\delta\mathsf{H}$,
\begin{align}
p(2\mathsf{H}[u])-p(2\bar{\gamma}u) &= c^2_m(2\mathsf{H})[u]-d^2c^2_K(2\mathsf{H})^2[u]-c^2_m(2\bar{\gamma})u+d^2c^2_K(2\bar{\gamma})^2u \\
&= 2c^2_m(\mathsf{H}-\bar{\gamma}\mathsf{1})u -4 c^2_Kd^2(\mathsf{H}^2-\bar{\gamma}^2\mathsf{1}^2)u\\
&= 2c^2_m\delta\mathsf{H}-8\bar{\gamma}c^2_k d^2\delta\mathsf{H}u + \mathcal{O}(\epsilon^2)\\
&\equiv \mathsf{V}_{curv}[u].
\end{align}
We ignore the error of order $\epsilon^2$ obtained by identifying $(\mathsf{H}+\bar{\gamma}\mathsf{1})u=2\bar{\gamma}u + \mathcal{O}(\epsilon^2)$ in the definition of $\mathsf{V}_{curv}$. The differential equation to describe the boundary vibrations $u$ can now be turned into another equation which we re-write in the form such that Banach's fixed point theorem can be applied again,
\begin{align}
\dfrac{1}{\Sigma(t)}\dfrac{\partial}{\partial t}\left(\Sigma(t)\dfrac{\partial u}{\partial t}\right)-p(2\bar{\gamma})u = \mathsf{V}_{curv}[u]+\Psi .
\end{align}
For the perturbative solution of our model equations, we have applied the Banach fixed point theorem to full boundary vibrations equation. This yielded the following an iterative integral equation which we rewrite in partial differential equation form using Fredholm's theorem (we work in $H^{2,2}_0$-spaces where the Laplacian is Fredholm!),
\begin{align}
\dfrac{1}{\Sigma(t)}\dfrac{\partial}{\partial t}\left(\Sigma(t)\dfrac{\partial u^{(k)}}{\partial t}\right)-p(2\bar{\gamma})u^{(k)} = \mathsf{V}_{curv}[u^{(k)}]+\Psi^{(k-1)}.
\end{align}
Since we have by the uniform local mean curvature assumption $\Vert\delta\mathsf{H}\Vert \leq \epsilon\Vert\bar{\gamma}\mathds{1}\Vert$ with $\epsilon\ll 1$ in the operator norm, the Banach fixed point theorem allows us do one further iteration to handle the mean curvature perturbation operator $\mathsf{V}_{curv}$ because by our re-definition up to an error of order $\epsilon^2$, we have for $\mathsf{V}_{curv}$ the following relative bound in the operator norm, $\Vert\mathsf{V}_{curv}\Vert / \Vert p(2\bar{\gamma}\mathds{1})\Vert = \mathcal{O}(\epsilon)$. The Banach fixed point theorem gives us the following iteration scheme,
\begin{align}
\dfrac{1}{\Sigma(t)}\dfrac{\partial}{\partial t}\left(\Sigma(t)\dfrac{\partial u^{(k,l)}}{\partial t}\right)-p(2\bar{\gamma})u^{(k,l)} = \mathsf{V}_{curv}[u^{(k,l-1)}]+\Psi^{(k-1,l)},
\end{align}
where $u^{(k,0)}=\lim_{l\to\infty}[u^{(k-1,l)}]$. We are now having an iteration scheme with vectorial iteration index $(k,l)\in\mathbb{N}_0^2$ instead of a scalar iteration index $k\in\mathbb{N}_0$. Notice that $u^{(k,l)}=u^{(k,l)}(t,\mathbf{y})\in H^{2,2}_0(\partial\mathcal{M}_0)$ still because the mean curvature operator $\mathsf{V}_{curv}$ introduces in higher iterations in the index $l$, i.e., for $k,l> 1$, dependencies of $u^{(k,l)}$ on the coordinates $\mathbf{y}=(\lbrace y^i\rbrace)_{1\leq i\leq n,i\neq s}$ of the boundary $\partial\Omega_0$ of the unperturbed fiber space $\Omega_0$. The choice of $\bar{\gamma}$ is fixed if we set 
\begin{align}
(\bar{\gamma}\mathds{1})[u] = (\bar{\gamma}\mathds{1})[u] = \dfrac{1}{\vert\partial\Omega_0\vert}\int_{\partial\Omega_0}d\text{Vol}_{n-1}(\partial\Omega_0)\, \mathsf{H}[u].
\end{align}
By the uniform local mean curvature assumption, we obtain
\begin{align}
u^{(k)} = \sum_{l=1}^{\infty}u^{(k,l)} = u^{(k,1)}+\mathcal{O}(\epsilon^2).
\end{align}
The $\epsilon^2$ is due to the observation that $u^{(k,1)}$ is already of order $\epsilon$ by definition of a proper dissipative perturbation bundle. In this perturbation theory, we know that $u^{(k)}$ exists and is in $H^{2,2}_0(\partial\Omega_0)$ because of the previous perturbative arguments for the full model equations.\newline
\newline
\textbf{Relation to piston bundles: }This means up to an error of order $\epsilon^2$, we have in the acoustic wave equation
\begin{align}
\partial_t^2p^{(k,l)} - c^2\Delta_{g_0}p^{(k,l)} &= -\rho_0c^2\partial_t^2 u^{(k-1,l)}\delta((t,\mathbf{x})\in\partial\mathcal{M}_0) + \mathcal{O}(\epsilon^2)\\
&= -\rho_0c^2\partial_t^2 u^{(k-1,1)}\delta((t,\mathbf{x})\in\partial\mathcal{M}_0) + \mathcal{O}(\epsilon^2).
\end{align}
Notice that by definition of the global mean curvature constant $\bar{\gamma}$ we have the identification
\begin{align}
u^{(k,1)}=u^{(k,1)}(t)=\int_{\partial\Omega_0}d\text{Vol}_{n-1}(\partial\Omega_0)u^{(k)}(t,\mathbf{y}) = \langle u^{(k)}\rangle_{\partial\Omega_0}(t).
\end{align}
In the acoustic wave equation, this means we can set $l=1$ for all iterations $(k,l)\in\mathbb{N}_0\times\mathbb{N}_0$ and solve up to errors of order $\epsilon^2$ (suppressed in the notation)
\begin{align}
\partial_t^2 p^{(k,1)} - c^2\Delta_{g_0}p^{(k,1)} = -\rho_0c^2\partial_t^2\langle u^{(k)}\rangle_{\partial\Omega_0}(t)\delta((t,\mathbf{x})\in\partial\mathcal{M}_0).
\end{align}
We will introduce the notation $p^{(k,1)}=\langle p^{(k)}\rangle_{\partial\Omega_0}(t,s)$ and give justification for it. The acoustic wave equation becomes an {"}averaged one",
\begin{align}
\partial_t^2 \langle p^{(k)}\rangle_{\partial\Omega_0}(t,s) - c^2\Delta_{g_0}\langle p^{(k)}\rangle_{\partial\Omega_0}(t,s) = -\rho_0c^2\partial_t^2\langle u^{(k)}\rangle_{\partial\Omega_0}(t)\delta((t,\mathbf{x})\in\partial\mathcal{M}_0).
\end{align}
The issue that we need to clarify is what the Neumann-Laplace-Beltrami operator $\Delta_{g_0}$ does with $p^{(k,1)}$ and why we can say that $p^{(k,1)}$ depends only on $t$ and $s$, but has no (non-trivial) dependencies on the boundary coordinates $\mathbf{y}$. For this purpose we need to investigate the separability of $\Delta_{g_0}$. By definition a positive operator $\mathsf{O}$, i.e., $-\Delta_{g_0}$ in our special case, is separable, if there are positive definite operators $\mathsf{P}_{k},\mathsf{Q}_{k}$ with $k\in\lbrace 1,2\rbrace$ such that we have
\begin{align}
\mathsf{O} = \mathsf{P}_1\otimes\mathsf{Q}_1+\mathsf{P}_2\otimes\mathsf{Q}_2.
\end{align}
Recall that by means of $\sigma_0: [0,1]\times\partial\Omega_0\to\Omega_0$ we obtain a global parameterization on $\Omega_0$ in terms of a radial parameter $s$ and the parameterization on $\partial\Omega_0$. By means of $\psi_0:\Omega_0\to B^n_1(\mathbf{0})$, we can relate this to conventional $n$-dimensional spherical coordinates on the unit ball $B^n_1(\mathbf{0})=\lbrace\mathbf{x}\in\mathbb{R}^n:\Vert\mathbf{x}\Vert_2 \leq 1 \rbrace$ w.r.t. the Euclidean norm $\Vert .\Vert_2:\mathbb{R}^n\to\mathbb{R}^+_0$ on $n$-dimensional Euclidean space $\mathbb{R}^n$. We denote the metric on $B^n_1(\mathbf{0})$ in $n$-dimensional spherical coordinates by $g_{B^n_1(\mathbf{0})}$. By means of the composition $\psi_0\circ\sigma_0: [0,1]\times\partial\Omega_0\to B^n_1(\mathbf{0})$ of diffeomorphisms $\psi_0,\sigma_0$, we can introduce a \"{}pseudo-spherical" parameterization on $\Omega_0$ with radial coordinate $s\in [0,1]$ and angular coordinates $\mathbf{y}=\lbrace y^i\rbrace_{1\leq i\leq n,i\neq s}$, i.e., the coordinates on the boundary $\partial\Omega_0$ of the unperturbed fiber $\Omega_0$. For the operator $-\Delta_{g_0}$, we have in the metric $g_0 = (\psi_{0}\circ\sigma_0)^\ast g_{B^n_1(\mathbf{0})}$ the analogous identifications as for the Neumann Laplace-Beltrami operator $-\Delta_{B^n_1(\mathbf{0})}$ on the unit ball $B^n_1(\mathbf{0})$. Namely, we can separate $-\Delta_{g_0}$ by the operators $\mathsf{P}_k,\mathsf{Q}_k$ with $k\in\lbrace 1,2\rbrace$ given by,
\begin{align}
&\mathsf{P}_1 = -\dfrac{1}{s^{n-1}}\dfrac{\partial}{\partial s}\left(s^{n-1}\dfrac{\partial}{\partial s}\right)\text{ and }\mathsf{Q}_1 = \text{1}\vert_{\partial\Omega_0}\\
&\mathsf{P}_2=s^{n-1}\mathsf{1}\vert_{s\in[-1,0]}\text{ and }\mathsf{Q}_2 = -\Delta_{g_0}^{N,\partial} .
\end{align}
The superscript $N$ in the Laplace-Beltrami operator $\Delta_{g_0}^\partial$ with derivatives w.r.t. the coordinates $\lbrace y^i\rbrace_{1\leq i\leq n,i\neq s}$ indicates that $\Delta_{g_0}^{\partial,N}$ contributes to an operator with Neumann boundary conditions. In the boundary vibrations equation we had Dirichlet boundary conditions for $\Delta_{g_0}^\partial$ without the superscript $N$. In practical notation, the separability reduces to a Laplace-Beltrami operator in {"}pseudo-spherical" coordinates,
\begin{align}
\Delta_{g_0} = \dfrac{1}{s^{n-1}}\dfrac{\partial}{\partial s}\left(s^{n-1}\dfrac{\partial}{\partial s}\right) + s^{n-1}\Delta_{g_0}^{N,\partial} .
\end{align}
A consequence of the separability of $\Delta_{g_0}$ is that the eigenvalue problem for $\Delta_{g_0}$ on the unperturbed fiber $\Omega_0$ becomes separable as well, i.e., we have using that $\mathbb{N}\simeq\mathbb{N}\times\mathbb{N}$ by means of a bijection $\mathbb{N}\to\mathbb{N}\times\mathbb{N}$, e.g. obtained via Cantor's diagonal argument,
\begin{align}
\Psi_n \equiv \Psi_{\mathbf{n}}\equiv\Psi_{n_s n_\partial} \equiv \chi_{n_s}\otimes\Phi^N_{n_\partial}.
\end{align}
The equation means that an eigenfunction $\Psi_n$ of the Neumann Laplace-Beltrami operator $\Delta_{g_0}^\partial$ can be expressed as a product of radial eigenfunctions $\chi_{n_s}$ and the eigenfunctions $\Phi^N_{n_\partial}$ for the (Neumann-)Laplace-Beltrami operator $\Delta_{g_0}^{N,\partial}$ on $\partial\Omega_0$. We remark that in general, we need \emph{not} have $\Phi^N_{n_\partial}=\Phi_{n_\partial}$ i.e., equality of eigenfunctions for $\Delta_{g_0}^{N,\partial}$ and $\Delta_{g_0}^\partial$. Using Lichernowicz theorem, equality of the complete and without loss of generality orthonormal eigenfunction sets $\lbrace \Phi^N_{n_\partial}\rbrace_{n_\partial\in\mathbb{N}}$ of $\Delta_{g_0}^{\partial,N}$ and $\lbrace \Phi_{n_\partial}\rbrace_{n_\partial\in\mathbb{N}}$ of $\Delta_{g_0}^{\partial}$ holds if and only if $\Delta_{g_0}^\partial$ and $\Delta_{g_0}^{N,\partial}$ are both assigned periodic boundary conditions on $\partial\Omega_0$. The separability of $\Delta_{g_0}$ in an boundary Laplace-Beltrami operator $\Delta_{g_0}^\partial$ and a radial part allows us to use two resolutions of the identity operator $\mathsf{1}_{\partial\Omega_0}$ acting on functions $\partial\Omega_0\to\mathbb{R}$, namely we have the two dyadic expressions
\begin{align}
\sum_{n_\partial\in\mathbb{N}}\vert\Psi_{n_\partial}^N\rangle\langle\Psi_{n_\partial}^N\vert = \mathsf{1}_{\partial\Omega_0} = \sum_{n_\partial\in\mathbb{N}}\vert\Psi_{n_\partial}\rangle\langle\Psi_{n_\partial}\vert,
\end{align}
because Lichernowicz' theorem guarantees completeness of the sets of eigenfunctions $\lbrace\Psi_{n_\partial}^N\rbrace_{n_\partial\in\mathbb{N}}$ of $\Delta_{g_0}^{\partial,N}$ and $\lbrace\Psi_{n_\partial}\rbrace_{n_\partial\in\mathbb{N}}$ of $\Delta_{g_0}^{\partial}$ of the two Laplace-Beltrami operators on $\partial\Omega_0$. The differential equation for $u^{(1,1)}$,
\begin{align}
\Sigma^{-1}\partial_t(\Sigma\partial_t u^{(1,1)})-p(2\bar{\gamma})u^{(1,1)} = \Psi[p_{ex},p^{(0,1)}=0]=\rho_0d^{-1}\mathfrak{g}p_{ex},
\end{align}
includes no dependencies on coordinates on $\partial\Omega_0$, i.e., we have $u^{(1,1)}=u^{(1,1)}(t)$. This allows us to express $u^{(1,1)}$ in the acoustic wave equation for $p^{(2,1)}$ by the partial eigenfunction method used also by Vossen et al. \cite{christine} in a different and much more specialized case,
\begin{align}
u^{(1,1)} = \sum_{n_\partial\in\mathbb{N}}\langle u\rangle_{\partial\Omega_0}\langle \Phi_{n_\partial}\vert 1\rangle_{L^2_{g_0}(\partial\Omega_0)}\vert\Phi_{n_\partial}\rangle = \sum_{n_\partial\in\mathbb{N}}\langle u\rangle_{\partial\Omega_0}\langle \Phi^N_{n_\partial}\vert 1\rangle_{L^2_{g_0}(\partial\Omega_0)}\vert\Phi^N_{n_\partial}\rangle,
\end{align}
The form of the coefficients now follows from Cauchy-Schwarz' and H\"{o}lder's inequality for $L^p$-spaces which becomes an equality for the orthonormal eigenfunction sets under consideration,
\begin{align*}
\langle\Phi^{N}_{n_\partial}\vert 1 \rangle_{L^2_{g_0}(\partial\Omega_0)}^2&=\Vert \Phi^{N}_{n_\partial}\Vert_{L^1_{g_0}}^2=\Vert 1 \Vert_{L^2_{g_0}(\partial\Omega_0)}^2\Vert \Phi^{N}_{n_\partial}\Vert_{L^2_{g_0}(\partial\Omega_0)}^2 = \text{Vol}_{n-1}(\partial\Omega_0)\\
\Rightarrow -\sqrt{\text{Vol}_{n-1}(\partial\Omega_0)}&\leq\langle\Phi^{N}_{n_\partial}\vert 1 \rangle_{L^2_{g_0}(\partial\Omega_0)}\leq\sqrt{\text{Vol}_{n-1}(\partial\Omega_0)}\text{ and }\\
\langle\Phi_{n_\partial}\vert 1 \rangle_{L^2_{g_0}(\partial\Omega_0)}^2&=\Vert \Phi_{n_\partial}\Vert_{L^1_{g_0}}^2 = \Vert 1 \Vert_{L^2_{g_0}(\partial\Omega_0)}^2\Vert \Phi_{n_\partial}\Vert_{L^2_{g_0}(\partial\Omega_0)}^2 = \text{Vol}_{n-1}(\partial\Omega_0)\\
\Rightarrow -\sqrt{\text{Vol}_{n-1}(\partial\Omega_0)}&\leq\langle\Phi_{n_\partial}\vert 1 \rangle_{L^2_{g_0}(\partial\Omega_0)}\leq\sqrt{\text{Vol}_{n-1}(\partial\Omega_0)}
\end{align*}
with the last step utilizing the normalization condition we imposed on $\Phi^N_{n_\partial}$ and $\Phi_{n_\partial}$ for all $n_\partial\in\mathbb{N}$ in the beginning. In concrete applications it turns out useful to artificially create a situation comparable to the one in \cite{anupam1, anupam2}, namely that there is a constant eigenfunction to $\Delta_{g_0}$. By having imposed Neumann boundary conditions, there is an $n_\partial\in\mathbb{N}$, say $n_\partial = 1$ such that
\begin{align}
\Phi^N_{n_\partial=1}(\mathbf{y})=\sqrt{\dfrac{1}{\text{Vol}_{n-1}(\partial\Omega_0)}}.
\end{align}
Let us now equip $L^2_{g_0}(\partial\Omega_0)$ with the following inner product for $f,g\in L^2_{g_0}(\partial\Omega_0)$,
\begin{align}
\langle f\vert g\rangle^{norm}_{L^2_{g_0}(\partial\Omega_0)} = \dfrac{1}{\text{Vol}_{n-1}(\partial\Omega_0)}\int_{\partial\Omega_0}d\text{Vol}_{n-1}(\partial\Omega_0) \bar{f}g,
\end{align}
such that the norms $\Vert .\Vert_{L^2_{g_0}(\partial\Omega_0)}$ and $\Vert .\Vert_{L^2_{g_0}(\partial\Omega_0)}$ are equivalent with equivalence constant $\sqrt{\text{Vol}_{n-1}(\partial\Omega_0)}$ and the norm equivalence inequality turning into an equality. In the new norm, we have $\Phi_{n_\partial=1}=1$ The general Parseval equality for $f\in L^2_{g_0}(\partial\Omega_0)$
\begin{align}
\Vert f\Vert_{L^2_{g_0}(\partial\Omega_0)}^2=\sum_{n_\partial\in\mathbb{N}}\left\vert\langle f\vert\Phi^N_{n_\partial}\rangle_{L^2_{g_0}(\partial\Omega_0)}\right\vert^2,
\end{align}
tells us for $f=1$ that all other expansion coefficients have to vanish, i.e., $\langle\Phi_{n_\partial\neq 1}\vert 1\rangle^{norm}_{L^2_{g_0}(\partial\Omega_0)}=0$ because the right side evaluates to $1$. Switching back from $\Vert .\Vert^{norm}_{L^2_{g_0}(\partial\Omega_0)}$, we see that $\langle 1,\Phi^N_{n_\partial=1}\rangle_{L^2_{g_0}(\partial\Omega_0)}=\sqrt{\text{Vol}_{n-1}(\partial\Omega_0)}$ In the case that $\Phi^N_{n_\partial}=\text{const.}$, we can rewrite the expansions of $u^{(1,1)}$ in terms of the complete orthonormal eigenfunction set $\lbrace\Psi_{n_\partial}^N\rbrace_{n_\partial\in\mathbb{N}}$ of $\Delta_{g_0}^{\partial,N}$ for the separation Laplace-Beltrami operators on $\partial\Omega_0$ in the more compact form
\begin{align}
\langle u\rangle_{\partial\Omega_0}=u^{(1,1)} = \sum_{n_\partial\in\mathbb{N}}\langle u\rangle_{\partial\Omega_0}\vert\Phi^N_{n_\partial}\rangle = \langle u_{\partial\Omega_0}\rangle + \mathcal{O}(\mathfrak{g}^2,\epsilon^2),
\end{align}
This can be used to solve the acoustic wave equation for $p^{(1,1)}=\langle p\rangle_{\partial\Omega_0}+\mathcal{O}(\mathfrak{g}^2,\epsilon^2)$. We have the integral representation
\begin{align}
\langle p\rangle_{\partial\Omega_0} = \dfrac{\mathfrak{g}c^2}{d}\sum_{n_s\in\mathbb{N}}\vert\chi_{n_s}(s)\rangle\langle\chi_{n_s}(0)\vert\int_{0}^t d\tau\,\dfrac{\sin\left((t-\tau)\sqrt{c^2\lambda_{n_s,n_\partial=1}}\right)}{\sqrt{c^2\lambda_{n_s,n_\partial=1}}}\partial_t^2\langle u\rangle_{\partial\Omega_0}(\tau)+ \mathcal{O}(\mathfrak{g}^2,\epsilon^2)
\end{align}
If $\langle u\rangle_{\partial\Omega_0}(t)=\langle u\rangle_{\partial\Omega_0}e^{i\omega t}$, we have
\begin{align}
\langle p\rangle_{\partial\Omega_0}=\dfrac{-\omega^2\mathfrak{g}c^2}{d}\sum_{n_s\in\mathbb{N}}\vert\chi_{n_s}(s)\rangle\langle\chi_{n_s}(0)\vert\int_{0}^t d\tau\,\dfrac{\sin\left((t-\tau)\sqrt{c^2\lambda_{n_s,n_\partial=1}}\right)}{\sqrt{c^2\lambda_{n_s,n_\partial=1}}}\langle u\rangle_{\partial\Omega_0}(\tau) + \mathcal{O}(\mathfrak{g}^2,\epsilon^2)
\end{align}
We evaluate the following definite integral
\begin{align*}
&\int_0^t d\tau\,\dfrac{e^{i\omega \tau}\sin((t-\tau)\sqrt{c^2\lambda_{n_s,n_\partial=1}})}{\sqrt{c^2\lambda_{n_s,n_\partial}}}\\
&= \dfrac{1}{2i}\int_0^t d\tau\,\dfrac{e^{i\omega \tau}e^{i(t-\tau)\sqrt{c^2\lambda_{n_s,n_\partial}}}}{\sqrt{c^2\lambda_{n_s,n_\partial}}}-\dfrac{1}{2i}\int_0^t d\tau\,\dfrac{e^{i\omega t}e^{-i(t-\tau)\sqrt{c^2\lambda_{n_s,n_\partial}}}}{\sqrt{c^2\lambda_{n_s,n_\partial}}}\\
&=\dfrac{e^{it\sqrt{c^2\lambda_{n_s,n_\partial}}}}{2i}\int_0^t d\tau\,\dfrac{e^{i(\omega - \sqrt{c^2\lambda_{n_s,n_\partial}})\tau}}{\sqrt{c^2\lambda_{n_s,n_\partial}}} - \dfrac{e^{-it\sqrt{c^2\lambda_{n_s,n_\partial}}}}{2i}\int_0^t d\tau\,\dfrac{e^{i(\omega + \sqrt{c^2\lambda_{n_s,n_\partial}})\tau}}{\sqrt{c^2\lambda_{n_s,n_\partial}}}\\
&=\dfrac{e^{i\omega t}}{2i^2\sqrt{c^2\lambda_{n_s,n_\partial=1}}(\omega-\sqrt{c^2\lambda_{n_s,n_\partial=1}})} - \dfrac{e^{i\omega t}}{2i^2\sqrt{c^2\lambda_{n_s,n_\partial=1}}(\omega+\sqrt{c^2\lambda_{n_s,n_\partial=1}})}\\
&-\dfrac{e^{i\sqrt{c^2\lambda_{n_s,n_\partial=1}} t}}{2i^2\sqrt{c^2\lambda_{n_s,n_\partial=1}}(\omega-\sqrt{c^2\lambda_{n_s,n_\partial=1}})} + \dfrac{e^{-i\sqrt{c^2\lambda_{n_s,n_\partial=1}} t}}{2i^2\sqrt{c^2\lambda_{n_s,n_\partial=1}}(\omega+\sqrt{c^2\lambda_{n_s,n_\partial=1}})}\\
&=-\dfrac{e^{i\omega t}-R_{n_s,n_\partial=1}(t)}{-\omega^2+c^2\lambda_{n_s,n_\partial}},
\end{align*}
introducing the \emph{resonance function} $R_{n_s,n_\partial=1}$ analogous to \cite{david1}. Insertion in the equation for $\langle p\rangle_{\partial\Omega_0}(t,s)$, we find recalling $\langle u\rangle_{\partial\Omega_0}(t)=\langle u\rangle_{\partial\Omega_0}e^{i\omega t}$,
\begin{align}
\langle p\rangle_{\partial\Omega_0}(t,s)=\dfrac{\omega^2\mathfrak{g}c^2\langle u\rangle_{\partial\Omega_0}}{d}\sum_{n_s\in\mathbb{N}}\vert\chi_{n_s}(s)\rangle\dfrac{e^{i\omega t}-R_{n_s,n_\partial=1}(t)}{-\omega^2+c^2\lambda_{n_s,n_\partial = 1}}\langle\chi_{n_s}(0)\vert + \mathcal{O}(\mathfrak{g}^2,\epsilon^2).
\end{align}
In the setup considered by \cite{anupam1, anupam2} the partial eigenfunction technique was not applicable whence Vedurmudi et al. resorted to a truncation method termed piston approximation. Albeit the correctness of the statement that only low eigenfrequencies need to be considered in the model setup, this not the most general reason for the validity of the piston approximation. Either one limits oneself to certain frequency spaces and uses the numerically easy to implement method presented in our previous paper \cite{david1} or one assesses by experimental means the model-dependent validity of the local uniform mean curvature assumption.\newline
\newline
\textbf{Remark: }(i) We have two things to discuss. The first is that in application, localized boundary vibrations are more present than global ones, i.e., $u=\sum_{i=1}^N u_i\delta(\mathbf{t,y}\in\mathbb{R}^+_0\times\Gamma_i)$ with the sets $\lbrace\Gamma_i\rbrace_{1\leq i\leq N}$ defined in the previous sections. Naturally, this breaks periodic boundary conditions on the closed boundary $\partial\Omega_0$ of the unperturbed fiber $\Omega_0$. We can however, solve individually the eigenvalue equations,
\begin{align}
\Delta_{g_0}^{\Gamma_i}\Phi_{n}^{\Gamma_i} = -\gamma^{\Gamma_i}_k\Phi_n^{\Gamma_i}
\end{align}
with Dirichlet boundary conditions on $\partial\Gamma_i$ for $1\leq i\leq N$ and the eigenvalue equation
\begin{align}
\Delta^{N,\partial}_{g_0}\Phi^N_n = -\gamma^{N}_n\Phi_n^{N},
\end{align}
with the separation Laplacian $\Delta_{g_0}^{\partial,N}$ on $\partial\Omega_0$, including the corresponding boundary conditions inherited from $\Delta_{g_0}$ on $\Omega_0$. Since there are only finitely many $\Gamma_i$'s, we can use a sequence of mollification functions defined on $\partial\Omega_0$ and denoted by $(m_{k,i}(t,\mathbf{y}))_{n\in\mathbb{N},1\leq i\leq N}$ with $\mathcal{C}^\infty$ regularity properties such that
\begin{align}
\lim_{n\to\infty}m_{n,i}(t,\mathbf{y}) = \delta((t,\mathbf{y})\in\mathbb{R}^+_0\times\Gamma_i)
\end{align}
and mollify $u$ on $\partial\mathcal{M}_0$ by convolution over $\Gamma_i$ for $1\leq i\leq N$,
\begin{align}
u = \lim_{n\to\infty}\left[\sum_{i = 1}^N (m_{n,i}\star_{\Gamma_i} u_i)(t,\mathbf{y})\right] = \sum_{i=1}^N u_i\delta(\mathbf{t,y}\in\mathbb{R}^+_0\times\Gamma_i)
\end{align}
The boundary vibrations $u$ are then defined globally on $\partial\Omega_0$ and sufficiently regular. We can then carry out the procedure in the derivation above with the new $u$ because Lichernowicz' theorem ensures the existence of a complete and orthonormal set of eigenfunctions for $\Delta_{g_0}^{\partial, N}$ on $\partial\Omega_0$.\newline
(ii) The second issue is that $\langle\mathcal{M}\rangle_{\partial\Omega_0}$ still needs to be obtained in terms of $\langle u\rangle_{\partial\Omega_0}$. By definition, we have to replace the boundary vibrations $u$ with their geometrical mean, $\langle u\rangle_{\partial\Omega_0}$. More haptically, the procedure works as follows: Take the diffeomorphisms $\phi_{0\to t}:\partial\Omega_0\to\partial\Omega_t\subset\mathbb{R}^n$ and use the componentwisely in terms of the canonical basis $\lbrace\hat{e}_\mu\rbrace_{1\leq\mu\leq n}$ on $\mathbb{R}^n$ defined geometrical mean,
\begin{align}
\langle\phi_{0\to t}\rangle_{\partial\Omega_0} = \dfrac{1}{\text{Vol}_{n-1}(\partial\Omega_0)}\int_{\partial\Omega_0}d\text{Vol}_{n-1}(\partial\Omega_0)\,\phi_{0\to t}.
\end{align}
This maps $\partial\Omega_0$ to $\partial\langle\Omega_t\rangle_{\partial\Omega_0} = \langle\partial\Omega_t\rangle_{\partial\Omega_0}$. Then, we use the Gaussian map to extrapolate from $\langle\partial\Omega_t\rangle$ by a radial coordinate $s\in[0,1]$ to the whole $\langle\Omega_t\rangle$,
\begin{align}
\langle\sigma_t\rangle_{\partial\Omega_0}:[0,1]\times\langle\partial\Omega_t\rangle_{\partial\Omega_0}\to\langle\Omega_t\rangle_{\partial\Omega_0}, (s,\mathbf{y})\to \mathbf{y}-s\langle\mathbf{y},\mathbf{n}_{\langle\partial\Omega_t\rangle_{\partial\Omega_0}}\rangle_{g_0,\mathbb{R}^n},
\end{align}
and define $\langle\Omega_t\rangle_{\partial\Omega_0}=\text{Im}(\langle\sigma_t\rangle_{\partial\Omega_0})$. Last, we define the piston bundle
\begin{align}
\langle\mathcal{M}\rangle_{\partial\Omega_0} = \bigcup_{t\geq 0}\lbrace t\rbrace\times\langle\Omega_t\rangle_{\partial\Omega_0}.
\end{align}
Obviously, the piston bundle $\langle\mathcal{M}\rangle_{\partial\Omega_0}$ associated to a perturbation bundle $\mathcal{M}$ is again a perturbation bundle with the same perturbation strength $\epsilon$ as the original perturbation bundle $\mathcal{M}$ to the reference bundle $\mathcal{M}_0$. The perturbation theory for general perturbation bundle then recovers for the geometrical mean $\langle u\rangle_{\partial\Omega_0}$ the perturbation equations with $u$ replaced by $\langle u\rangle_{\partial\Omega_0}$, i.e., the equations that we already obtained in the preceding derivation.\newline
\newline
\textbf{Piston bundle theorems: }In total our considerations have shown the validity of the piston bundle theorems which give criteria for when one may approximate the perturbation bundle $\mathcal{M}$ via the associated piston bundle. We give the geometrical and analytic formulation of the theorems.
\begin{itemize}
\item\textbf{Theorem: \emph{(Piston approximation - geometrical formulation)} } \emph{Let $\mathcal{M}$ denote the dissipative proper perturbation bundle for the reference bundle $\mathcal{M}_0$ and $\langle\mathcal{M}_0\rangle_{\partial\Omega_0}$ the piston bundle associated to $\mathcal{M}$. The boundary vibrations $u$ of $\mathcal{M}$ satisfy the local uniform mean curvature assumption with global mean curvature constant $\bar{\gamma}\geq 0$ if and only if the piston constant $C_{piston}(\geq 0)$ of the associated piston bundle $\langle\mathcal{M}_0\rangle_{\partial\Omega_0}$ satisfies $C_{piston}=\mathcal{O}(1)$ in powers of the perturbation strength $\epsilon$.}
\item\textbf{Theorem: \emph{(Piston approximation - analytic formulation)} }\emph{ Let $\mathcal{M}$ denote the dissipative proper perturbation bundle for the reference bundle $\mathcal{M}_0$ and $\langle\mathcal{M}\rangle_{\partial\Omega_0}$ the piston bundle associated to $\mathcal{M}$. Further, let $p(\Delta_{g_0}^\partial)=c^2_m\Delta_{g_0}^\partial - c^2_kd^2\Delta_{g_0}^\partial$. The differential equations for the acoustic pressure $p$ and the boundary vibrations $u$ on the unperturbed bundle $\mathcal{M}_0$, i.e.,
\begin{align}
\partial_t^2 p - c^2\Delta_{g_0}p &= \rho_0c^2\partial_t^2 u\delta((t,\mathbf{x})\in\mathcal{M}_0)\\
\Sigma^{-1}\partial_t(\Sigma\partial_t u)-p(\Delta_{g_0}^\partial)u &= \Psi ,
\end{align}
and the averaged differential equations for $\langle u\rangle_{\partial\Omega_0}$ and $\langle p\rangle_{\partial\Omega_0}$ on the piston bundle $\langle\mathcal{M}_0\rangle_{\partial\Omega_0}$, i.e.,
\begin{align}
\partial_t^2 \langle p\rangle_{\partial\Omega_0} - c^2\Delta_{g_0}\langle p\rangle_{\partial\Omega_0} &= \rho_0c^2\partial_t^2 \langle u\rangle_{\partial\Omega_0}\delta((t,\mathbf{x})\in\langle\mathcal{M}_0\rangle_{\partial\Omega_0})\\
\Sigma^{-1}\partial_t(\Sigma\partial_t \langle u\rangle_{\partial\Omega_0})-p(2\bar{\gamma})\langle u\rangle_{\partial\Omega_0} &= \langle\Psi\rangle_{\partial\Omega_0},
\end{align}
are equivalent up to an error of order $\epsilon^2$ if and only if the boundary vibrations $u$ of $\mathcal{M}$ satisfy the uniform local mean curvature assumption with global mean curvature constant $\bar{\gamma}\geq 0$.}
\end{itemize}
Physically speaking, the piston bundle theorems formalize that one can approximate the boundary vibrations $u$ by its geometrical mean $\langle u\rangle_{\partial\Omega_0}$ if and only if the graphs of the boundary vibrations have (up to an $\epsilon)$ constant mean curvature $H(u)$ globally on $\partial\mathcal{M}_0$. Even more concretely this means that one can approximate elastic structures as piston structures if and only if the vibrations change the local mean curvature of the structures by a small amount.

\section{Discussion}

It is now time to gather our results and view them in perspective,  summarizing the main steps carried out in this paper. First, we have defined the notion of perturbation bundles to a reference bundle. Intuitively speaking, the perturbation bundles has topologically the same features as the reference bundle has, but differs locally by non-stationary fibers. The fibers of the perturbation bundle originate from the (unperturbed) fiber of the reference bundle by small local perturbations of the boundary, the so-called boundary vibrations. The smallness of the perturbations was formalized in the perturbation strength. The overall goal was to obtain solutions to certain classes of models in linear order in the perturbation strength. In order to obtain a quantitative expression, we have used the Gauss map from extrinsic Riemannian differential geometry and a radial coordinate system on the whole fiber to relate the perturbations of the boundary to perturbations of the fibers. Since the boundary vibrations originate from diffeomorphisms between the fibers at different points in time, we could reconstruct the perturbation bundle from the reference bundle once the boundary vibrations and thus the bundle diffeomorphism is known. 

Next, we have derived the acoustic wave equation and a boundary vibrations equation including curvature contributions from a variational principle. For the acoustic wave equation, we have started from Euler's equation for ideal, irrotational and isentropic fluid motion and afterwards performed the acoustic linearization procedure to recover a scalar wave equation in curved space-time, i.e., on the perturbation bundle, for the acoustic pressure. For the boundary vibrations, we have started from a 'tinkered' combination of the action functional that has already been used to described the undulations of bio-membranes and an action functional that reproduces the conservative membrane equation. In order to ensure existence of solutions, we imposed a dissipativity requirement on the perturbation bundle. 

Subsequently, we found that the dissipativity requirement also enters during the derivation of a Banach iteration scheme in the form of giving a contractive mapping in the form of a dissipative strongly continuous semi-group of essentially self-adjoint operators. The dissipativity requirement has been accounted for in the derivation of the boundary vibrations equation by the introduction of time-lapsing in the boundary metric of the unperturbed bundle. Since then the second order partial derivative with respect to time takes the form of a mire general Sturm-Liouville operator, we could by a suitable choice of the time-lapse function, recover a generalized version of a damped membrane-plate equation. 

The spatial dependencies of the boundary vibrations have entered by variation of the geometrical contribution to the action functional. It gave a contribution in the form of a second order polynomial in the Laplace-Beltrami operator on the boundary of the unperturbed fiber. The contribution of polynomial contribution of degree one produced the membrane equation part of the boundary vibrations equation and the polynomial contribution of degree two reproduced a (clamped) plate equation contribution. For the purpose of applications, we have introduced the notion of localized boundary vibrations, i.e., we weakened the requirement that the entire boundary way supposed to vibrate in time to the case of only mutually disconnected sub-manifolds of the boundary of the unperturbed fiber undergoing small perturbations. 

In order to exclude the occurrence of a bi-harmonic operator for the sake of the occurrence of a $2$-Laplace-Beltrami operator, we derived clamping boundary conditions, that is, homogeneous Dirichlet boundary conditions for the localized boundary vibrations equation. An argument based on the Cauchy-Kowaleskaja theorem demonstrated that we can translate the inhomogeneous boundary condition to the acoustic wave equation into a source term. The solutions of the two acoustic wave equations agreed because we imposed homogeneous Neumann boundary conditions on the inhomogeneous acoustic wave equation. The perturbation operator to the acoustic wave equation included the boundary vibrations and the acoustic pressure. Since the perturbation operator is a first order differential operator, small relative to the unperturbed Laplacian, i.e., it does not affect the principal part of the partial differential operators governing the dynamics of acoustic pressure, we could use suitable Sobolev spaces as the domains of the wave equation. 

As an aside, we have checked our theory on a sound perturbation result in the acoustics research literature namely that the perturbations should only influence the eigenvalues of the Laplace-Beltrami operator quadratically in the perturbation strength. We observed that the acoustic wave equation and the boundary vibrations equation together have the shape of a non-autonomous and non-linear Cauchy problem in the two dynamical quantities, the acoustic pressure and the boundary vibrations. Since the non-linearities constituted the reason why the acoustic wave equation became non-autonomous when written in the form of an operator evolution equation and were all stored in the perturbation operator, we have resorted to perturbative methods. 

In so doing, we have recapitulated Duhamel's principle, also known as the variation of constants formula for autonomous dynamical systems, and afterwards extended the discussion to the Magnus expansion method for non-autonomous vector-valued dynamical system. In so doing, we have derived the Magnus series expansion for matrices using an analogy to the lower central series of matrix Lie algebras and Lie-theoretic methods pertaining to calculations. Furthermore, we have extended our discussion to matrix systems where the dynamical matrix takes values in certain operator algebras, namely the von Neumann algebra of bounded operators on suitable function spaces. These matrix algebras form well-understood associative Banach algebras for which there exist theorems that allow the derivation of a Magnus series expansion method as well. 

After a short introduction to operator evolution equations, we have compared three physical perturbation theories to handle time-dependent perturbations. The earliest method is due to Dirac and widely in use in quantum mechanics. The second one is due to Dyson and pervades modern quantum field theory. In contrast to the Dirac perturbation theory, it makes a statement about the semi-group solving the operator evolution equation rather than the solution to the differential equation itself as does the Dirac theory. It includes, however,  the time-ordering symbol which is difficult to handle in explicit calculations and which is rather inappropriate for the discussion of convergence issues. 

We have then discussed the Magnus series expansion method again from the perspective of semi-groups which overcomes the difficulties associated with the time-ordering symbol. In particular we have addressed convergence issues of the Magnus expansion for the Magnus generator. We have sketched a modification to a proof of a convergence result from the literature from normal operators to for small perturbations to otherwise symmetric operators. Upon replacing an everywhere defined Laplacian, we used a densely defined Laplacian instead such that the Laplacian is no longer self-adjoint but symmetric and compact due to discreteness of the eigenvalues of the Laplacian on compact suitably regular domains by Lichernowicz' theorem. 

Imposing regularity conditions on the external pressure, i.e., the overall source term to the coupled partial differential equations system, this ensured convergence of the series expansion. After the methodological discussion, we started to obtain a perturbation theory for our differential equation system based on Banach fixed point theorem. In total, we had to iterate three equations instead of two as one might have intuited at first. Besides the acoustic wave equation and the boundary vibrations equation, we also had to iterate the inhomogeneous Volterra integral equation -- resulting from Duhamel's principle once again -- for the semi-group of the full perturbed acoustic wave equation. 

Defining a Lipschitz-like constant, the coupling strength, we could truncate due to smallness of the coupling strength the iteration scheme such that we only consider linear perturbations in our perturbation parameter, the perturbation strength. It turned out that there is a phenomenological relation between the smallness of the perturbations, i.e., the perturbation strength and the smallness of the coupling of the acoustic pressure to the boundary vibrations, the coupling strength. We have found the scaling behavior of higher order contributions in the two perturbation parameters, namely the perturbation strength and the coupling strength. We then turned to obtaining explicit formulas for our model. 

Using the Magnus expansion, we derived an explicit expression for the operator sine function that governs the dynamics of the acoustic wave equation and calculated the lowest order effects of the perturbation operator in the integral formulation of the acoustic wave equation. For the boundary vibrations equations, we used the observation that the relevant operator algebra is commutative for all points in time and derived a closed form expression for the operator sine function that governs the dynamics of the solution to the boundary vibrations equation. We have checked that this expression reduces to the expression we have found in our previous papers concentrating exclusively on the treatment of the internally coupled ears (ICE) model. These equations can be used in the iteration scheme presented in the preceding paragraph to the explicit calculations. Examples of how the formalism interweaves in a more concrete setting can be found in our previous papers on acoustic boundary conditions dynamics (ABCD). 

The final point we have addressed for the derivation of the perturbation theory is the question of convergence of semigroups. Since partial differential operators are in general unbounded operators and the Laplace-Beltrami operators in particular are unbounded linear operators, we needed to restrict their domain such that they turn into bounded and by compactness of the geometrical domains of the operators even compact operators. Otherwise, we could not have used the already existent results on convergence of the Magnus expansion. By linearity of the Laplace-Beltrami operators it followed from boundedness that the Laplace-Beltrami operators are also continuous and thus closed linear operators. Because the Laplace-Beltrami operators are then no longer self-adjoint but just symmetric operators, we needed for the check on the relevant items of well-posedness of the problem, i.e., existence, uniqueness, and continuous dependence on the source term of the solution, at least essentially self-adjoint operators, i.e., symmetric differential operators that admit a unique (maximal) self-adjoint extension, the Friedrichs extension. 

We have used the notion of analytic vectors combined with the Sobolev spaces embedding theorems to perform the conceptually instructive exercise that the Laplace-Beltrami operators admit self-adjoint extensions. Investigating the regularity properties of the overall source term to the partial differential equation system, we could check convergence of the iteration scheme with a positive result. This agrees with a literature result on convergence properties of the Magnus expansion. 

After having finished the analysis of out geometric perturbation theory, we turned to the investigation of an applied question. Namely, one might want to simplify the solutions of the equation by neglecting the local dependencies of the boundary vibrations in the solution to the acoustic wave equation. This corresponds geometrically with replacing the boundary vibrations with pistons and is formalized in the notation of piston bundles associated to a perturbation bundle. 

The pistons are displaced from their equilibrium position by the geometrical mean of the boundary vibrations. While it is clear that such an approximation can be performed, we asked what the approximation means geometrically. We have found that the approximation corresponds to the mean curvature of the boundaries of the perturbed fibers differing from a constant mean curvature only by a small additional perturbation operator. 

From the experimental viewpoint, the local mean curvature can be assessed easily. One simply looks whether the mean curvature is approximately constant, i.e., the boundary vibrations (e.g. plates or membranes in the physical setting) only vibrate very mildly. More precisely, the local mean curvature should only differ from the local mean curvature constant by a quantity in the order of the perturbation strength. Under this circumstances, it is consistent with the perturbation theory developed before to replace the boundary vibrations by piston boundary vibrations or, in geometrical language, one can associate to the original perturbation bundle a piston perturbation bundle of the same perturbation strength such that one loses spatial dependencies of the perturbations of the fibers around the unperturbed fiber without deterioration of the analytical perturbation theory. 

Finally, we noted the similarity between the physical piston approximation and the Poincar\'e inequality from the theory of partial differential equations. Put simply, the piston approximation works well, if the upper bound given by the Poincar\'e approximation is almost - up to a quantity of the order of the perturbation strength - zero.

\section{Outlook}

First, it is natural to ask whether the definition of a perturbation bundle needs to be restrictive in the sense that we require the imbedding space to be $\mathbb{M}_{n+1}$ and that $t\in\mathbb{R}^+$ and whether the topological requirements do have to be that restrictive that the fibers are all properly diffeomorphic to $B^n_1(\mathbf{0})$. As an abstraction from the physical model, the answer is yes. In terms of generalizations of the acoustic boundary condition dynamics (ABCD) method, the answer is no. The Sard-Whitney imbedding theorem guarantees that $\mathbb{R}_{2n+1}$ is always a suitable imbedding space for $n$-dimensional manifolds $(\Omega_t)_{t\geq 0}$. Furthermore, the parameter $t$ could be equally well element of $\Omega_b$, a $m$-dimensional base manifold. Then using Sard-Whitney once more, the imbedding space of $\mathcal{M}$ and $\mathcal{M}_0$ defined analogously to the definition presented in the main text is at most of real dimension $2n+2m+2$. Then, a global Gauss map for $\partial\Omega_t$ goes to $G_{n-1,2n+1}$ where $G_{k,n}$ is the Grassmann manifold containing all plains of dimension $k$ in $n$ dimensional Euclidean space. Since the perturbation theory is purely local, the orientation requirement can be weakened by required that only locally, the $\Omega_t$'s are uniformly oriented. This holds true e.g. for perturbations of the finitely thick M\"{o}bius strip taken as reference fiber $\Omega_0$. The topological requirement of retractibility for $\Omega_t$ and $(n-2)$-connectedness of $\partial\Omega_t$ might also be relaxed: We can just require that in terms of homotopy, the homotopy sequences of the fibers agree $\pi^{\ast}(\Omega_{t})=\pi^{\ast}(\Omega_{t'})$ for $t,t'\in\Omega_b$. This means that the manifolds $\Omega_t$ are topologically equivalent for $t\in\Omega_b$. By the topological classification of differentiable manifolds, there is a reference manifold $\Omega_{ref}$ with the same topological properties as $\Omega_t,\,t\in\Omega_b$ which takes the role of $B^n_1(\mathbf{0})$. Compactness on the other hand is still required - otherwise the Lichnerowicz theorem and the existence and uniqueness theorems aren't applicable. Equally powerful alternatives do not exist to the authors' knowledge. We end the outlook by noting that the Minkowskian property of the imbeddings space $\mathbb{M}_{n+1}$ must be modified to match the overall signature of the symbol of the $\Psi$-differential operators. 

Second, from the operator-theoretic viewpoint, the interesting question is how to stretch the theory of perturbation bundles from the Laplace-Beltrami-operators as prototypical elliptical $\Psi$-differential operators to a more general class of operators. E.g., one might be interested in $\Psi$-differential operators on $\Omega_0, (\Omega_t)_{t\geq 0}$ which are of the shape $\mathsf{A}=\partial_\mu(a^{\mu\nu}(\mathbf{x})\partial_\nu)+b^\mu(\mathbf{x})\partial_\mu + c(\mathbf{x})$ where $(a_{\mu\nu})_{1\leq\mu ,\nu\leq n}$ is a matrix with - say $\mathcal{C}^\infty$-coefficients - and $(b_\mu)_{1\leq\mu\leq n}$ is a vector with $\mathcal{C}^\infty$-coefficients again and $c$ is a real-valued function $\Omega_0\to\mathbb{R}$ resp. $\Omega_t\to \mathbb{R}$ for all $t\geq 0$. The usage of the Sobolev-spaces is still possible but one obtains a larger class of differential operators, e.g., accounting for convective phenomena in the unperturbed case as well. 

Third, from the viewpoint of partial differential equations, it would be interesting to use conformal mappings, more precisely inversion on $S^n_1(\mathbf{0})$ and $\partial\Omega_0$ as well as $\partial\Omega_t$, to solve the exterior problem.

Fourth, from the viewpoint of applications, it is desirable to look for applications of the formalism outside continuum mechanics. Since the gauge gravity duality has inspired some of the work, it might be possible to use with suitable modifications parts of our formalism to study by an analytical perturbation theory models form the gauge gravity duality. The acoustic pressure $p$ could be regarded as a mediator field which propagates as in the associated piston bundles on the AdS scale and hits the boundary where a conformal field theory, say $u$ lives. The boundary vibrations equation then should be replaced on $\mathcal{M}_0$ with a differential equation for $U$ which also includes the AdS scale, denoted by $s$ in this article. 

Fifth, and as plausible suggestion,  the decoupling method by means of Banach's fixed point theorem might well provide a convenient way of handling electro-physiological models of neuronal networks: Modeling the propagation of spikes by a lossy cable equation, at each knot of the Kirchhoff network, we have Kirchhoff's law as conservation equations. Although the boundaries are not moving, the partial differential equations are still coupled via Kirchhoff's law which lives on the end-caps of the cables, i.e., gives natural boundary conditions for linear interactions between the cables.

Finally, our treatment of vibrating boundaries may well turn out to be useful in medical physics. It is a goal of modern radiation therapy to keep the radioactivity exposition of a patient seeking cancer treatment as low as possible. For lung cancer, the patient still breathes and clinicians basically have two options: Either they suppress the patient's breathing partially by medication \emph{or} the machine learns how to react to the patient's lung deforming during the respiratory cycle. 

Modeling the intensity field of the incident radiation by an electromagnetic wave equation, one encounters a geometrical setup comparable to ours. The patient is a stationary volume $\Omega_0$ and the patient's lungs are two time-dependent families of manifolds satisfies (upon the modeler's choice) the requirements of our setting, $t\to\Omega_t^l$ and $t\to\Omega_t^r$ for the left and right lung wing. In medical physics, one looks for solutions to a minimization problem in $\Omega_0\setminus(\Omega_0^{l}\cup\Omega_0^r)$ so that the intensity delivered to this volume should be as small as possible where the intensity delivered to stationary lungs $\Omega_0^l\cup\Omega_0^r$ is maximized. The solution of the stationary problem is then comparable to the solution for the acoustic pressure that we have found. The reasoning leading to the perturbation operator $\mathsf{W}$ in our setting transfers however to the radiation therapy problem. One deforms $\Omega_t^l,\Omega_t^r$ and $\Omega_0\setminus(\Omega_t^l\cup\Omega_t^r)$ to their stationary counterparts again and solves the perturbed minimization resp. maximization problem.
\nocite{*}
\bibliographystyle{plain}
\bibliography{abcd_part_3_lit_rev6}
\end{document}